\documentclass[12pt,letterpaper, bibliography=totoc,
    listof=totoc]{JHEP3}  
 
\usepackage{amscd,amsmath,amssymb,amsfonts,xspace,mathrsfs,amsthm}
\usepackage{xcolor} 

\usepackage{latexsym, slashed}     
\usepackage{graphicx}
\usepackage{dsfont}
\usepackage{color}   
\usepackage{longtable} 
\usepackage{tikz-cd}
\tikzset{>=latex}
\tikzcdset{arrow style=tikz}
\usetikzlibrary{shapes.geometric,positioning}
\usetikzlibrary{bending}
\usetikzlibrary{shapes}
\usetikzlibrary{calc}
\tikzset{stretch/.initial=1}
\newcommand\drawloop[4][]%
   {\draw[shorten <=0pt, shorten >=0pt,#1]
      ($(#2)!\pgfkeysvalueof{/tikz/stretch}!(#2.#3)$)
      let \p1=($(#2.center)!\pgfkeysvalueof{/tikz/stretch}!(#2.north)-(#2)$),
          \n1= {veclen(\x1,\y1)*sin(0.5*(#4-#3))/sin(0.5*(180-#4+#3))}
      in arc [start angle={#3-90}, end angle={#4+90}, radius=\n1]%
   }

 \usepackage{enumerate} 
\newcommand{\be}{\begin{equation}} 
\newcommand{\ee}{\end{equation}} 
\newcommand{\bes}{\begin{equation*}}
\newcommand{\ees}{\end{equation*}}
\newcommand{\sgn}{\mathrm{sgn}}

\newcommand{\CA}{\mathcal{A}} 
\newcommand{\CB}{\mathcal{B}} 
\newcommand{\CC}{\mathcal{C}}  
\newcommand{\cD}{\mathcal{D}}  
\newcommand{\CE}{\mathcal{E}}  
\newcommand{\CF}{\mathcal{F}} 
 
\newcommand{\CH}{\mathcal{H}}

\newcommand{\CJ}{\mathcal{J}}
\newcommand{\CK}{\mathcal{K}}
\newcommand{\CL}{\mathcal{L}} 
\newcommand{\CM}{\mathcal{M}}  
\newcommand{\CN}{\mathcal{N}}
\newcommand{\CO}{\mathcal{O}} 
\newcommand{\CP}{\mathcal{P}}

\newcommand{\CR}{\mathcal{R}}
\newcommand{\CS}{\mathcal{S}} 
\newcommand{\CT}{\mathcal{T}} 
\newcommand{\CU}{\mathcal{U}}
\newcommand{\CV}{\mathcal{V}}

\newcommand{\CZ}{\mathcal{Z}}

\newcommand{\BC}{\mathbb{C}}
\newcommand{\IC}{\mathbb{C}}

\newcommand{\BR}{\mathbb{R}}
\newcommand{\IR}{\mathbb{R}}

\newcommand{\BH}{\mathbb{H}}
\newcommand{\IH}{\mathbb{H}}

\newcommand{\BP}{\mathbb{P}}
\newcommand{\BZ}{\mathbb{Z}}
\newcommand{\IZ}{\mathbb{Z}}

\newcommand{\IQ}{\mathbb{Q}}

\newcommand{\fg}{\mathfrak{g}}

\newcommand{\fs}{\mathfrak{s}}

\newcommand{\bfb}{{\boldsymbol b}}
\newcommand{\bfc}{{\boldsymbol c}}

\newcommand{\bfl}{{\boldsymbol \ell}}

\newcommand{\bfx}{{\boldsymbol x}}
\newcommand{\bfy}{{\boldsymbol y}}
\newcommand{\bfrho}{{\boldsymbol \rho}}
\newcommand{\bfnu}{{\boldsymbol \nu}}

\newcommand{\bfk}{{\boldsymbol k}}
\newcommand{\bfz}{{\boldsymbol z}}
\newcommand{\bfmu}{{\boldsymbol \mu}}

\newcommand{\p}{\partial}

\newcommand{\uv}{{\rm uv}}

\newcommand{\SL}{\text{SL}(2,\BZ)}

\def\I{{\rm i}}
\title{Topological correlators of $SU(2)$, $\CN=2^*$ SYM on four-manifolds}
\author{Jan Manschot$^{1,2}$, Gregory W. Moore$^3$\\
{\it  $^1$ School of Mathematics, Trinity College, Dublin 2, Ireland}\\
{\it $^2$ Hamilton Mathematical Institute, Trinity College, Dublin 2,
  Ireland}\\
{\it $^3$ NHETC and Department of Physics and Astronomy, Rutgers University, 126 Frelinghuysen Rd., Piscataway NJ 08855, USA
}

}

\abstract{ 

We consider topologically twisted $\CN=2$, $SU(2)$ gauge theory with a
massive adjoint hypermultiplet on a smooth, compact four-manifold
$X$. A consistent formulation requires coupling the theory to a ${\rm Spin}^c$ structure, which is necessarily non-trivial if $X$ is non-spin. 
We derive explicit formulae for the topological correlation functions when $b_2^+\geq 1$. 
We demonstrate that, when the ${\rm Spin}^c$ structure is canonically determined by an almost complex structure and the mass is taken to zero, the path integral reproduces known results for the path integral of the $\CN=4$ gauge 
theory with Vafa-Witten twist. On the other hand, we reproduce results from Donaldson-Witten theory after taking a suitable infinite mass limit. 

The topological correlators are functions of the 
UV coupling constant $\tau_{\rm uv}$ and we confirm that they 
obey the expected $S$-duality transformation laws. 
The holomorphic part of the partition 
function is a generating function for the Euler numbers of the matter
(or obstruction) bundle over the instanton moduli space. For
$b_2^+=1$, we derive a non-holomorphic contribution to the path
integral, such that the partition function and correlation functions are mock modular forms rather than modular forms.

We comment on the generalization of this work to the large class of
$\CN=2$ theories of class $S$.\\
\\
\today
} 

\begin{document}

\section{Introduction}
This paper investigates applications of supersymmetric quantum field theory (SQFT)
to the differential topology of four-dimensional manifolds. The story of the
relation of SQFT and four-manifold topology is one of the paradigmatic examples of
the interplay between mathematics and questions raised by  fundamental 
physics, a subject sometimes called physical mathematics.
Indeed the invention of topological quantum field theory (TQFT) by  Witten was motivated by
the desire to give a QFT interpretation of the Donaldson invariants \cite{Witten:1988ze}.
The desire to use the SQFT interpretation to give an effective evaluation of the
Donaldson invariants was in turn one of the motivations for the ground-breaking work of
Seiberg and Witten on the low-energy dynamics of SQFT with $\CN=2$ supersymmetry \cite{Seiberg:1994rs, Seiberg:1994aj}.
(We henceforth refer to such theories as $d=4,\CN=2$ theories.) 
The discoveries of Seiberg and Witten led to the formulation of the Seiberg-Witten
invariants of four-manifolds, culminating in the formulation of the Witten conjecture
expressing the Donaldson invariants in terms of the Seiberg-Witten invariants \cite{Witten:1994cg}. (For a complete derivation of the Witten
conjecture using ideas from quantum field theory see \cite{Moore:1997pc}.)

Witten's original application of SQFT to Donaldson invariants made use of ``pure''
$\CN=2$ supersymmetric Yang-Mills (SYM) with a nonabelian gauge group of rank one, where the adjective ``pure''
means that the field content consists entirely of a vectormultiplet. However, the general twisting
procedure of \cite{Witten:1988ze} can in principle be applied to any $d=4, \CN=2$ field theory
and it can be interesting to investigate other such theories. (For much more about
this general point, see Section \ref{SecClassS}  below.) The present paper focuses on the
extension to $d=4, \CN=2$ rank one SYM with an adjoint-valued hypermultiplet with $\CN=2$
preserving mass term, a theory sometimes referred to as the $\CN=2^*$ theory.
We carry out a detailed investigation of the topological correlators of this theory
on a certain class of four-manifolds $X$.

Throughout this paper $X$ will be a smooth, compact, oriented four-dimensional manifold
without boundary.  For simplicity we
will assume $X$ is simply connected, although this condition could certainly
be relaxed. Let $b_2^+$ denote the dimension of the maximal positive definite
subspace of $H^2(X,\IR)$ under the intersection pairing.  We will assume that $b_2^+$ is positive and odd. When
$b_2^+$ is even and positive the topological correlators all vanish. When
$b_2^+=0$ important new considerations enter the story which are beyond the scope
of this paper. It is worth noting that the manifolds in the above class always
admit an almost complex structure (ACS). (For a detailed explanation see Section 10.1 of \cite{Scorpan}.)

In order to formulate the topological correlators of the twisted   $\CN=2^*$ theory  we will need the following
four pieces of data:

\begin{enumerate}

\item An ``ultraviolet coupling.'' Mathematically this is
encoded in a point $\tau_{\rm uv} \in \IH$, the
upper half of the complex plane. Following standard definitions we will denote
$q_{\rm uv} := e^{2\pi \I \tau_{\rm uv}}$.

\item A ``mass parameter'' $m \in \IC$ and a ``UV scale''  $\Lambda \in \IC$.
These quantities enter the topological correlators only through the dimensionless
ratio $t:=m/\Lambda$.

\item An ``ultraviolet'' ${\rm Spin}^c$-structure $\fs_{\rm uv}$ on $X$ endowed with
${\rm Spin}^c$ connection. The invariants only depend on this data through the
characteristic class of $\fs_{\rm uv}$, denoted $c_{\rm uv} \in H^2(X,\IZ)$.

\item A ``homology orientation,'' that is, an
orientation on $H^2(X,\IR)$. (Physically, this is used to orient the
Berezinian of fermion zero modes.)

\end{enumerate}

We use the adjective ``ultraviolet'' to characterize the ${\rm Spin}^c$-structure $\fs_{\rm uv}$
because (as was noted long ago \cite{Labastida:1997rg}) the presence of adjoint-valued hypermultiplet fields
in the $\CN=2^*$ theory requires a choice of ${\rm Spin}^c$-structure in order to formulate a well-defined
topologically twisted theory.
This  ${\rm Spin}^c$-structure will play a very important role in this paper. It should be
distinguished from the ``infrared''  ${\rm Spin}^c$-structures that appear in the LEET evaluation
of the path integral. For example  in the Witten conjecture, and its generalizations discussed below
the result is expressed as a sum over \underline{infrared}  ${\rm Spin}^c$-structures.

A generating function for the topological correlators can be written
as a formal power series of polynomial functions on the homology $H_*(X,\IC)$. See equation \eqref{eq:formalMathdef}
below. Formal path integral arguments suggest that, while the formulation of the QFT path integral makes use
of a choice of Riemannian metric on $X$, nevertheless, the result of the QFT path integral should
be independent of this choice of metric. Thus, formally, the topological correlators should be
smooth manifold invariants, much like the Donaldson invariants. However, careful analysis
reveals that the alleged metric-independence is not quite true: The path integral argument
only guarantees that the variation of the path integral with respect to the metric is an integral 
over field space of a total derivative.  
There can be ``BRST anomalies'' when the boundary contribution is nonzero. For $b_2^+>1$ the invariants
really are metric independent, but for $b_2^+=1$ there is metric dependence. The dependence
enters only through the \emph{period point} $J \in H^2(X,\IR)$. The period point $J$ is
defined to be the unique class in the forward light cone of $H^2(X,\IR)$ which
satisfies   $J=*J$ and $J^2=1$.   (The use of the forward light cone here is one place where
the homology orientation enters in the expressions.) The dependence on $J$ turns out to be not
a bug, but a feature. It is crucial in the derivation of the formulae for the invariants 
for manifolds with $b_2^+>1$,
as explained in Section \ref{SWcontributions}.

The topological correlators we study have a mathematical definition in terms of intersection theory on a
moduli space of equations that generalize both the instanton and the Seiberg-Witten equations. We will now
explain this in some detail. The relevant equations
are the ``nonabelian (adjoint) monopole equations'' or the ``nonabelian Seiberg-Witten equations.'' Their relevance in
our context was already noted long ago \cite{Labastida:1997rg, LoNeSha}. To formulate the invariants
we choose a principal $SO(3)$ bundle $P\to X$ with connection. Since $X$ has a ${\rm Spin}^c$ structure $\fs_{\rm uv}$
there is an associated rank 2 bundle $W^+ \to X$ of chiral spinors and we choose a section $M\in \Gamma(W^+ \otimes {\rm ad}P)$.
There is a natural ${\rm Spin}^c$-Dirac operator acting on $M$ and the nonabelian monopole equations state that $M$ satisfies the
Dirac equation while the self-dual curvature $F^+$ is equal to the hyperk\"ahler moment map of the action of the
group of gauge transformations on $W^+ \otimes {\rm ad} P$.  See \eqref{MonopoleEqs} below for the precise statement.

We note in passing that if we choose the ${\rm Spin}^c$ structure $\fs_{\rm uv}$ to be the one canonically determined by an
ACS then the adjoint-Seiberg-Witten equations are closely related to the renowned
Vafa-Witten (VW) equations. The VW equations are equations for a gauge connection on $P$ together with a
section $C \oplus B^+$ of $(L\oplus \Lambda^{2,+}T^*X)\otimes {\rm ad} P$ where $L$ is the trivial real line bundle.
 The essential observation is that, given an ACS, there is a canonical isomorphism
$\Lambda^{2,+}T^*X \cong  L \oplus K_{\IR}$ where  $K_{\IR}$ is the
underlying real 2-plane bundle of the canonical bundle $K$. On the other hand, one also has the natural
isomorphism $W^+ \cong L\otimes \IC \oplus K$. In this way we can identify  the fields $C, B^+$ of the
VW theory with the monopole fields $M$. For a K\"ahler manifolds, the standard VW  equation $D_\mu C + D^\nu B^+_{\mu\nu}=0$ is then equivalent to the Spin$^c$ Dirac equation, and the equations for the curvature become isomorphic as well. For an almost complex manifold, we find that the nonabelian Seiberg-Witten equations are equivalent to a deformation of the standard VW equations. The deformation involves a compact perturbation of the Vafa-Witten operator $d\oplus d^\dagger$ by 
a term involving the Nijenhuis tensor. Based on the explicit results of this paper, we have reasons to believe that the solution spaces of the original and deformed VW equations give rise to the same topological invariants.

Returning to the mathematical formulation of the invariants we recall that the isomorphism class of the bundle $P$ is
determined by the instanton number $k$ (see equation \eqref{kPs} for our normalization of the instanton number)
and the ``'t Hooft flux'' $\nu:=w_2(P)$.
\footnote{In this paper we usually represent the 't Hooft flux by a cohomology class $\bfmu \in H^2(X,\IQ)$ so that $2\bfmu$ has integral periods.  We
assume that $\nu$ has an integral lift $\bar \nu$ and set $\bfmu = \frac{1}{2} \bar \nu$ in $H^2(X,\IQ)$. Indeed, 
this is how the 't Hooft flux naturally enters in the sum over fluxes in the LEET, and in subsequent sections we will in fact write $Z_{\bfmu}$ rather than $Z_{\nu}$.  With our definition, the topological correlators are independent of the choice of lift. However, in the infinite mass limit to the $N_f=0$ theory there is a residual dependence on the choice of ultraviolet
${\rm Spin}^c$ structure. This dependence is just an overall sign. The sign can be canceled by a local counter-term, given in equation \eqref{phaseDW}, and this should be done to compare with the standard Donaldson invariants. The standard mathematical definition of the Donaldson invariants does \underline{not} rely on a choice of ${\rm Spin}^c$ structure. The price we then pay is that the  counterterm  \eqref{phaseDW}  can change sign under a change of lift $\bfmu$.  }
Let $\CM_{k,\nu,\frak{s}}$ denote the moduli space (stack) of solutions to the nonabelian monopole equations. There is a
Donaldson-type $\mu$ map:
\be
\mu: H_*(X) \to H^{4-*}(\CM_{k,\nu,\frak{s}})
\ee
and the mathematical definition of the generating function for the topological correlators is of the form
\be\label{eq:formalMathdef}
Z_\nu(x; \tau_{\rm uv}, c_{\rm uv}, t) := \sum_{k\geq 0} q_{\rm uv}^k \int_{\CM_{k,\nu,\frak{s}}} e^{\mu(x)}\, {\rm Eul}(\CE_{\fs}; t),
\ee
where $x\in H_*(X,\IC)$ and ${\rm Eul}(\CE_{\fs}; t) $ is the $U(1)$-equivariant Euler class of a certain
virtual bundle over the moduli space (stack) $\CM_{k,\nu,\frak{s}}$ (with equivariant parameter $t$). Here we are using a
natural $U(1)$ symmetry of the nonablian monopole equations that rotates the monopole field by a phase.
One arrives at the expression \eqref{eq:formalMathdef} by localization arguments, starting with the path integral of the twisted
theory. See Section \ref{VirDims} for some more details. The basic elements of the argument can be found in \cite{Witten:1988ze, LoNeSha, Atiyah:1990tm, Cordes:1994fc, Laba05, MooreNotes2017}. It must be said that giving a proper and careful mathematical definition of the intersection
numbers in \eqref{eq:formalMathdef} is beyond the scope of this paper, and is a task best left to experts on the theory of moduli spaces. Among other things, one needs to orient ${\cal M}_{k,\nu,\frak{s}}$. (For an ACS it will have a canonical orientation.)  Indeed, already for projective surfaces, a careful definition of the VW invariants (which, we have explained, are a
special case of our expressions) is a nontrivial and delicate mathematical problem \cite{Tanaka:2017jom, Gholampour:2017bxh}.

In this paper we will give very explicit formulae for \eqref{eq:formalMathdef}. For the case $b_2^+>1$, the final results
can be found in Section \ref{SWcontributions}. They are given as a sum of three expressions, each of which is a sum over (infrared)
${\rm Spin}^c$-structures. The three expressions are Equations \eqref{SW1Final}, \eqref{u2DPSfinal} and \eqref{u3DPSfinal} when $x=0$.
\footnote{The reader should not confuse the homology class $x$ with 
$\bfx$ appearing in those equations. The latter is an integral cohomology class. In most of the paper we will denote $x=p\oplus S \in H_0(X) \oplus H_2(X)$.} 
The rules for including the dependence on $x$ are summarized in Tables \ref{Tab:pObsCoupl} and \ref{Tab:SObsCoupl}.
 Upon specialization of the ${\rm Spin}^c$ structure, we find agreement with results in the physics
literature \cite{Vafa:1994tf, Dijkgraaf:1997ce, Labastida:1998sk}. In
the $m\to \infty$ limit, we reproduce the Witten conjecture \cite{Witten:1994cg}
from the perspective of $\CN=2^*$ theory. In addition we find a very interesting connection to the
recent mathematical works   \cite[Appendix
C]{Gottsche:2017vxs}\cite{Gottsche:2019vbi,Gottsche:2020ale}. These papers
 considered the addition of observables to the VW
partition function for smooth projective surfaces with $h^{2,0}>0$.
Our expressions for the path integral (derived in a completely independent way)  are in
precise agreement with those mathematical papers, as discussed in detail in Section \ref{SWcontributions}.
See especially Table \ref{TableDict}.

The key to the derivation of the results of Section \ref{SWcontributions}  is the so-called $u$-plane integral or
Coulomb branch (CB)  integral. This integral appears when discussing the invariants for the case
$b_2^+=1$. When $b_2^+=1$ the  full expression for the invariants derived from the path integral is of the form
\be\label{eq:SumCoulHiggs}
Z_\nu^J(x; \tau_{\rm uv}, c_{\rm uv}, t)  = \Phi_\nu^{J}  + \sum_{j=1}^{3} Z^J_{SW,\nu,j}
\ee
where $\Phi_\nu^{J}$ is the CB integral and $Z^J_{SW,\nu,j}$ is the ``Seiberg-Witten contribution'' arising from
singularities in the Coulomb branch measure at the point $u_j$ of the discriminant locus. We dropped the arguments of the functions on the 
right-hand side for readability and 
noted that in general the quantities that appear are $J$-dependent.

After discussing, in Section \ref{geometryN=2star}, several relevant aspects of the LEET of the $\CN=2^*$ theory, we
give an in depth discussion of the Coulomb branch integral in sections \ref{FormUplaneInt} and \ref{Analysis}.
The basic strategy for determining the measure on the Coulomb branch follows \cite{Moore:1997pc},
but there are several important new technical points that need to be overcome when $c_{\rm uv} \not=0$.
(The case when $c_{\rm uv} =0$ is discussed in \cite{Moore:1997pc,Labastida:1998sk}.) In one parametrization
(as the base of a Hitchin fibration) the Coulomb branch can be identified with the complex
plane punctured at three points $u_i \in \IC$, and it is often called the ``$u$-plane.''
The $u$-plane can also naturally be identified with the   modular
curve for the modular subgroup $\Gamma(2)$ of $SL(2,\IZ)$ with a puncture at the point $\tau_\uv\in \BH$. We will return to this modular
parametrization presently.

If one blindly uses the formulae for the measure from the 1990's one discovers that the measure is not single-valued.
\footnote{As was pointed out to one of us by Marcos Mari\~no 20 years ago.}
The problem is that one must take into account couplings to the background ${\rm Spin}^c$ connection.
According to   standard folkore, those couplings should be topological and
  holomorphic as functions of $u$ \cite{Witten:1995gf, Shapere:2008zf}. Somewhat surprisingly,
  it appears that no choice of such couplings will lead to a single-valued measure. Rather
  the coupling of the  $\CN=2^*$ theory to a ${\rm Spin}^c$ structure induces a new non-holomorphic
  coupling in the effective action on the Coulomb branch. This becomes clear if we view the background
  ${\rm Spin}^c$ connection as a connection for a weakly-coupled background $U(1)$ vectormultiplet in
  a theory with $SU(2) \times U(1)$ gauge symmetry, following a trick going back at least to the
  work of Nelson and Seiberg \cite{Nelson:1993nf}. See section \ref{SpincCoupling} below for a detailed
  description of the coupling. One important consequence of this coupling is to modify the ``theta function''
  $\Psi_\bfmu^J$ arising from the sum over fluxes of the low energy $U(1)$ photon. See equation \eqref{defPsi3} 
  where the coupling to $\fs_{\rm uv}$ is achieved by  
  the inclusion of the parameter denoted $z$ in that equation. While the non-holomorphic part of the
  coupling is  $Q$-exact, the global duality group of the theory requires
  the presence of this coupling. Explicit evaluation of the partition
  functions demonstrates a strong dependence on the characteristic class $c_{\rm uv}$ of the  ${\rm Spin}^c$ structure,
  giving rise  to an infinite  family of functions labelled by $\mathfrak{s}_{\rm uv} $.

As mentioned above, the Coulomb branch can be identified with the modular curve $\IH/\Gamma(2)$,
and the modular parametrization can be quite useful. The ingredients used in constructing the
Coulomb branch measure are modular objects in both the UV coupling $\tau_{\rm uv}$ and in the
parameter $\tau$ of the Seiberg-Witten curve. Checking that the measure is single-valued is a
delicate issue. For example, one factor in the measure is the topological coupling to the
signature $\sigma$ of $X$. It is given by $B^{\sigma/8}$ where $B$ is a holomorphic function of
$u$ with simple zeroes on the discriminant locus $\{ u_1, u_2, u_3 \}$. Thus, unless $\sigma=0~ \mod ~ 8$, this factor is multi-valued. Other factors, such as the
theta function $\Psi_\bfmu^J$, depend explicitly on the choice of duality frame, and 
there is no globally valid duality frame on the Coulomb branch. Only when all the factors are combined will we obtain a single-valued measure, as shown carefully in Section \ref{SingleValued}. 
The failure of the Coulomb branch measure should be viewed as a kind of anomaly. It would be nice to interpret it more directly in the modern formulation of the theory of anomalies. 
Once we have checked the measure is single-valued we must ask if the Coulomb branch 
integral is well-defined. In fact, an analysis of
the measure near the locus $\{ u_1, u_2, u_3 \}$ and at $u \to \infty $ in Section \ref{ABCboundaries} reveals singularities. Therefore, the definition of the Coulomb branch integral requires careful
definition. This is given in Section \ref{SecRegularize}.

In checking the single-valuedness and deriving the singularities of the measure we make use of
certain remarkable identities for topological couplings involving the UV ${\rm Spin}^c$ structure.
These identities should be of independent interest. These identities can be found in Equations (\ref{tauvtauUV}) and \eqref{CphijC}.  A partial derivation of these identities is given in section \ref{SecCouplingC} but further work is required to give a complete proof.

When $b_2^+=1$ the Coulomb branch integral exhibits interesting anomalies in the ``BRST'' $Q$-symmetry.
These anomalies are related to the boundary terms mentioned above. After localization of the path integral,
the relevant boundaries in field space become the boundaries near the discriminant locus and the circle at  $u \to \infty$.
Because of the anomalies there is non-holomorphic dependence on $\tau_{\rm uv}$ as well as metric dependence.
These are examined in Sections \ref{SecHolAnomaly} and \ref{metricwallcross}. The essential technical point is that the variation of the
measure with respect to $J$ can be written as an exact form which has good $q$-expansions near the cusps.
One finds familiar piecewise constant (``wall-crossing'') type behavior from the boundary contributions near $u_j$. (Indeed it is
exactly this behavior which allows us to derive the topological correlators for $b_2^+>1$ as explained in
Section \ref{SWcontributions}.) One also finds   behavior at $u \to \infty$ leading to \underline{continuous} metric dependence (along with a holomorphic anomaly).
See equations \eqref{eq:HoloAnom-Cond} and \eqref{CondCont} for conditions stating precisely which correlators do not have continuous metric dependence (nor holomorphic anomaly). 
(A similar phenomenon was noted for the $SU(2)$, $N_f=4$ theory in \cite{Moore:1997pc}.) 

It turns out that the Coulomb branch integral itself (not just the difference for two different period points) can
be evaluated using integration by parts. Performing such a computation makes essential use of the theory of
mock modular forms, and here we build on the results of reference \cite{Korpas:2019cwg}. Many works have noted
similar appearances of mock modular forms in the context of path
  integrals of two dimensional theories \cite{Manschot:2007ha, Troost:2010ud, Eguchi:2010cb, Murthy:2013mya,
    Gaiotto:2019gef, Dabholkar:2020fde} where the non-holomorphic
  contribution can be attributed to a non-compact direction in field space.
Since the different $\Phi^{J_1}_\nu - \Phi^{J_2}_\nu$ can be evaluated using integration
by parts it suffices to find a special period point where $\Phi^{J}_\nu$ can
conveniently be done by integration by parts. We identify such a point in Section \ref{sec:Evaluate}.
Of course, if we write the measure as a total derivative $\omega = d \Omega$ then the one form $\Omega$
should be single-valued and non-singular away from the singularities of $\omega$. The non-singularity of $\Omega$ turns out
to be a rather subtle condition that depends strongly on the choice of UV ${\rm Spin}^c$ structure.
 We give a thorough discussion when $\fs_{\rm uv}$ is determined by an ACS.
The Coulomb branch integral with $x=0$ is given rather explicitly in Equation \eqref{PhimuJOddFINAL},
where $\widehat g_\mu$ is a mock modular form defined in Equation \eqref{whG}.
When $\fs_{\rm uv}$ is not determined by an ACS we describe many partial results, but a complete picture must be left for future work. 
This includes finding a useful way to impose the non-singularity constraint on $\Omega$ and verifying the rather nontrivial modular properties of some of the expressions which appear. (See, for example equation \eqref{eq:funny-modular}.) 

Using the above results, and taking $\mathfrak{s}_{\rm uv}$ to be determined by an ACS,  we reproduce results from VW theory for
four-manifolds with $b_2^+=1$. For example for $X=\mathbb{CP}^2$ and other
  rational surfaces,
the (holomorphic part of the) partition functions of the $\CN=4$, $SU(2)$ VW
theory \cite{Vafa:1994tf} have long been known to be examples of mock modular forms \cite{Vafa:1994tf,Gottsche:1996, Minahan:1998vr, Alim:2010cf, Manschot:2011dj}. This was originally derived using techniques from algebraic geometry \cite{Yoshioka1994}, toric localization \cite{Klyachko1991} and analytic number theory \cite{Zagier:1975}.
We give in this paper a derivation of the partition functions for all four-manifolds with $b_2^+=1$ from the perspective of the $\CN=2^*$ theory. This further enhances the physical understanding of the holomorphic anomaly which has been of recent interest. In the context of the hypermultiplet geometry of Calabi-Yau compactifications of Type IIB string theory, a derivation of the mock modularity of the VW partition function was obtained  by Alexandrov et al. \cite{Alexandrov:2016tnf, Alexandrov:2019rth}.
Within the setting of the twisted $\CN=4$
  theory and the dual two-dimensional field theory, Dabholkar, Putrov and Witten \cite{Dabholkar:2020fde} have given a derivation of the
  non-holomorphic anomaly. Moreover, Bonelli
  et al. \cite{Bonelli:2020xps} have arrived at the holomorphic anomaly using toric localization in supersymmetric field theory.

In the example of $\mathbb{C}\mathbb{P}^2$ we also give very explicit results for the topological 
correlators when $x$ is the point class.
This can be found in Tables \ref{TablePointObs0} and \ref{TablePointObs1} for the ACS case. The results nicely interpolate
between non-vanishing Donaldson and Vafa-Witten invariants. Since $H^2(\IC \BP^2; \IZ) \cong \IZ$
we can identify $c_{\rm uv}$ with an odd integer. The ACS case is $c_{\rm uv} = 3$.
We also give results for the case $c_{\rm uv} = 1$ in Tables \ref{TablePointObs02} and \ref{TablePointObs12} and the data here 
exhibit some interesting features.

The $\CN=2^*$ theory is famously an example of an $S$-dual theory \cite{Seiberg:1994aj}. In Section \ref{Sdualrankone}, we review the action of $S$-duality on the relevant theories. As we 
show in Sections \ref{SecPhiSdual} and \ref{SWSdual}, our results are compatible with 
the expected action of $S$-duality.

Finally, in Section \ref{sec:Discussion} we list several open problems and avenues for
further research. In addition
Section \ref{SecClassS} discusses the CB  integral for  
general $d=4,\CN=2$ theories of class S. 
This entire section is more in the nature of a research proposal than a 
report on completed research. The results of the present paper
have some important implications for this research program. 
 
We have included four appendices with further details and background material. Appendix \ref{AModForms}  summarizes several useful facts about (mock) modular forms used in this paper. 
Appendix \ref{daduDelta} gives a derivation of some important identities in the modular parametrization 
of quantities used in the LEET of the $\CN=2^*$ theory. Appendix \ref{ACS4folds} reviews aspects of almost complex four-manifolds. Finally Appendix  \ref{AppSymbolList} gives a concordance 
for the notation used throughout the paper. We hope it will be useful to those brave souls 
attempting to read it. 

%\bigskip 
\bigskip
\noindent
\emph{Note on convention.} Our convention for the $u$ and $a$ variable
is that for large $a$, $u=a^2+$ subleading terms. The periodicity of the
complexified coupling constant $\tau$ is 2, and the duality group of the $N_f=0$ Seiberg-Witten theory is the
congruence subgroup $\Gamma(2)\in SL(2,\mathbb{Z})$ as in \cite{Seiberg:1994rs}. This $\tau$ is half the effective coupling used in the convention with
  duality group $\Gamma^0(4)$ in, for example, \cite{Seiberg:1994aj, Moore:1997pc}.

\section{The $\CN=2^*$, $SU(2)$ theory}
\label{geometryN=2star}
This Section discusses various aspects of $\CN=2^*$ gauge theory
with gauge group $SU(2)$ in $\mathbb{R}^4$. After reviewing the field content and the non-perturbative solution of the theory, we discuss the monodromies, effective couplings and the $S$-duality action for this theory.

\subsection{Field content}
\label{Sectwisting}
The $\CN=2^*$ theory consists of a vector multiplet and a
hypermultiplet, both in the adjoint representation of the gauge group. We 
introduce a mass $m$ for the hypermultiplet. The global symmetry group
of the theory is $SU(2)_{R}\times U(1)_B$, with $SU(2)_R$ the
${R}-$symmetry and $U(1)_B$ the baryon symmetry. The theory contains
furthermore an anomalous $U(1)_R$ symmetry. 

The field content of the vector multiplet is $(A_\mu,\phi, \psi^I_\alpha,
\bar \psi^I_{\dot \alpha})$, here $I=1,2$ and $\alpha$ and $\dot \alpha$
are spinor indices. The vector multiplet is invariant under $U(1)_B$.
The bosons of the vector multiplet form the following representation under the
spacetime rotation group $SO(4)=SU(2)_-\times SU(2)_+$, and $SU(2)_R$,
\be
({\bf 2},{\bf 2},{\bf 1})\oplus ({\bf 1}, {\bf 1}, {\bf 1})\oplus ({\bf 1}, {\bf 1}, {\bf 1}).
\ee
The fermions (and the supersymmetry generators) form the representation
\be
\label{Vfermireps}
({\bf 2},{\bf 1},{\bf 2})\oplus ({\bf 1}, {\bf 2}, {\bf 2}),
\ee
and are sections of the positive and negative spin bundles over
$\mathbb{R}^4$, $S^+$ and $S^-$. The $U(1)_R$-charge (also known as ghost number) of
$\phi$ is 2.
 
The hypermultiplet consists of two copies, $Q$ and $\tilde Q$, of the $\CN=1$ chiral
multiplet. $U(1)_B$ acts on $Q$ and $\tilde Q$ as $Q\to e^{i\varphi}
Q$ and $\tilde Q \to e^{-i\varphi} \tilde Q$. The bosons $q$ and $\tilde q$ of the hypermultiplet form the following representation under the
spacetime rotation group, and $SU(2)_R$-symmetry group,
\be
({\bf 1},{\bf 1},{\bf 2})\oplus ({\bf 1},{\bf 1},{\bf 2}),
\ee
where the first term is the doublet $(q, \tilde q^\dagger)$, and the
second term the doublet $( -\tilde q, q^\dagger)$. More precisely, the bosons form a quaternion
\be 
H=\left(\begin{array}{cc} q & \tilde q^\dagger \\ -\tilde q & q^\dagger \end{array} \right).
\ee 
The $SU(2)_R$ symmetry correspond to right multiplication of $H$ by unit quaternions. The fermions $\lambda_\alpha$ and $\chi_\alpha$ of the hypermultiplet form the representation
\be 
({\bf 2}, {\bf 1}, {\bf 1}) \oplus ({\bf 2}, {\bf 1}, {\bf 1}) \oplus
({\bf 1}, {\bf 2}, {\bf 1})\oplus ({\bf 1}, {\bf 2}, {\bf 1}). 
\ee
The mass $m$ of the hypermultiplet has $R$-charge 2. 

The $\CN=2^*$ superpotential reads  
\be
\label{superW}
W=m\mathrm{Tr}(\tilde Q Q)+\mathrm{Tr}(\tilde Q[\Phi,Q]),
\ee
where $Q$ and $\tilde Q$ are $\CN=1$ chiral superfields. 
The scalar potential is minimized by
\be
\label{MinPotential}
\begin{split}
&\left[\phi,\phi^\dagger\right]=0,\qquad  q\, q^\dagger- \tilde q^\dagger\,\tilde q=0,\\
& \left[q, \tilde q\right]=0,\\
&\phi \, q + m\, q=0,\qquad \tilde q\, \phi +m\,\tilde q=0.
\end{split}  
\ee
The Coulomb branch is the branch of solutions to \eqref{MinPotential} with $q=\tilde q=0$, and
$\phi,\phi^\dagger$ in a commutative subalgebra. Classically, a gauge
transformation can bring $\phi$ to $\phi=\frac{1}{\sqrt{2}}\,{\rm
  diag}(a,-a)$ with $a\in \mathbb{C}$. For generic $a$, the gauge group is partially broken to
$U(1)$. The $SU(2)$ bundle splits as $\CL\oplus \CL^{-1}$.

\subsection{The Seiberg-Witten solutions}
We review in this subsection the Seiberg-Witten solution of the
``pure'' $N_f=0$, $SU(2)$ theory followed by the $\CN=2^*$ theory with mass $m$.

\subsubsection*{The $\CN=2$, $SU(2)$ theory without hypermultiplets}
The Coulomb branch of the $N_f=0$, $SU(2)$ $\CN=2$ theory is parametrized by the order parameter $u_0=\frac{1}{16\pi^2}\left< {\rm Tr}[\phi^2]\right>_{\BR^4}$, and is topologically a
sphere $S^2$ with three points removed. The effective theories near the
three points are characterized by,
\begin{itemize}
\item $u_0\to\infty$, where $\tau\to i\infty$:\\ In this limit,
  two charged components of the adjoint vector multiplet become
  infinitely massive.
\item $u_0\to \Lambda_0^2$, where $\tau\to 0$:\\
In this limit, a hypermultiplet corresponding to a monopole becomes massless.
\item $u_0\to -\Lambda_0^2$, where $\tau\to 1$:\\
In this limit, a hypermultiplet corresponding to a dyon becomes massless.
\end{itemize}
Here and in the following, we label the quantities for the $N_f=0$ theory with a subscript 0.

The elliptic curve which provides the non-perturbative solution of  the
$N_f=0$, $\CN=2$ gauge theory with gauge group  $SU(2)$  is \cite{Seiberg:1994rs} 
\be
\label{N=2curve}
y^2 =  (x_0-u_0)(x_0^2-\Lambda_0^4). 
\ee
The effective coupling $\tau$ is identified with the complex structure of this curve. From (\ref{N=2curve}), one can derive that the order parameter $u_0$ is given as function of $\tau$ by,
\be 
\label{uSW}
\begin{split}
u_0(\tau)&=\left(2\frac{\vartheta_3(\tau)^4}{\vartheta_2(\tau)^4}-1\right)
\Lambda_0^2\\
&=\frac{\Lambda_0^2}{8}\,q^{-\frac{1}{2}}+\dots,
\end{split}
\ee
where $q=e^{2\pi i \tau}$ and $\vartheta_j$ are Jacobi theta series defined by Equation (\ref{Jacobitheta}) in the Appendix. Equation (\ref{uSW}) demonstrates that $u_0(\tau)$ is invariant under  $\Gamma(2)$ transformations of 
$\tau$. We
have moreover,
\be
\label{dauSW}
\begin{split} 
\left(\frac{da}{du}\right)_0&=\frac{1}{2\sqrt{2}\Lambda_0}\, \vartheta_2(\tau)^2\\
&=\frac{\sqrt{2}}{\Lambda _0}\, q^{\frac{1}{4}}+\dots,
\end{split}
\ee
and thus $a_0\sim \sqrt{u_0}$ for large $u_0$. Moreover, the discriminant $\Delta_{0,\rm phys}$ equals
\be
\frac{\Delta_{0,\rm phys}}{\Lambda_0^4}=(u_0^2/\Lambda_0^4-1)=64\,\left(\frac{\eta(\tau)}{\vartheta_2(\tau)}\right)^{12},
\ee
where $\eta$ is the Dedekind eta function defined by Equation (\ref{etafunction}) in the Appendix. For later reference, we state the form of the prepotential for $N_f=0$,
\be
\label{SWprepotential}
\CF_0=-\frac{1}{8\pi i }2(2a)^2\log(2a/\Lambda_0)+a^2\sum_{k=1}^\infty
d_k\,\left(\frac{\Lambda_0}{a} \right)^{4k},
\ee
 with $d_k\in \mathbb{Q}$. 
 
\subsubsection*{The $\CN=2^*$, $SU(2)$ theory with adjoint hypermultiplet}
The Coulomb branch of the $SU(2)$ $\CN=2^*$ theory is topologically a
sphere $S^2$ with four points removed. The effective theories near the
four points are characterized by,
\begin{itemize}
\item $u\to\infty$, where $\tau\to \tau_{\rm uv}$:\\ In this limit,
  two charged components of the adjoint vector multiplet become
  infinitely massive, while the effective mass of two components of the adjoint hypermultiplet also diverge. In this limit the mass parameter $m$ becomes negligible and the theory becomes a superconformal theory with conformal invariance only broken by the vev's of the scalars. 
\item $u\to  u_1=\frac{m^2}{4}e_1(\tau_{\rm uv})$, where $\tau\to
  i\infty$:\\
In this limit, a hypermultiplet corresponding to a quark becomes massless.
\item $u\to u_2=\frac{m^2}{4}e_2(\tau_{\rm uv})$, where $\tau\to 0$:\\
In this limit, a hypermultiplet corresponding to a monopole becomes massless.
\item $u\to u_3=\frac{m^2}{4}e_3(\tau_{\rm uv})$, where $\tau\to 1$:\\
In this limit, a hypermultiplet corresponding to a dyon becomes massless.
\end{itemize}

The $\CN=2^*$ Coulomb branch parametrizes a family of elliptic curves \cite[Section 16.2]{Seiberg:1994aj}
\begin{equation} \label{N2s2}
y^2 =  \prod_{j=1}^3 \left(x-e_j(\tau_{\rm uv})\, u - \frac{1}{4}\,e_j(\tau_{\rm uv})^2\,m^2\right),
\end{equation}
where $\tau_{\rm uv}$ is the UV coupling constant, and the half-periods $e_j$, $j=1,2,3,$ are
defined in Equation (\ref{halfperiods}). The curve is singular for
$u\in \{\infty,u_1,u_2,u_3\}$. Moreover, the curve is invariant under
SL$(2,\mathbb{Z})$ transformations of $\tau_{\rm uv}$,
since different half-periods are permuted by the modular transformations. The parameter $u$ differs from the trace of
$\phi^2$ by possible non-perturbative corrections \cite{Dorey:1996ez},
\begin{equation}
u= \frac{1}{16\pi^2} \langle \text{Tr}[\phi^2]\rangle + m^2 \sum_{n \in \mathbb{Z}_{\geq 0}} d_n\, q_{\rm uv}^{n},
\end{equation} 
where $q_{\rm uv} = \exp(2\pi i \tau_{\rm uv} )$. The physical discriminant is
defined as
\be
\Delta_{\rm phys}=(u-u_1)(u-u_2)(u-u_3).
\ee

The physical quantities of the $\CN=2^*$ theory are functions of the
UV coupling constant $\tau_{\rm uv}$ as well as the effective coupling
constant $\tau$. Since the modular group acts on each of these
variables, they are examples of bi-modular forms. Following
\cite{2020arXiv200302572W, yang2007}, we give a definition
of such function in Appendix \ref{AModForms}. In addition,
to the action of the modular group, they transform under the transformation where
$\tau$ circles around $\tau_{\rm uv}$ or vice versa.
Appendix \ref{daduDelta} provides
a derivation of $u$, $da/du$ and the physical discriminant $\Delta_{\rm phys}$. The final result reproduces the expression for $u$ derived in \cite{Huang:2011qx, Bonelli:2019boe}, while we find for $da/du$ and $\Delta_{\rm phys}$,
\be
\label{expdaduDelta*}
\begin{split}
\frac{da}{du}(\tau,\tau_{\rm uv})&=\frac{1}{4\,m} \frac{(\,\vartheta_4(\tau)^4\,\vartheta_3(\tau_{\rm uv})^4-\vartheta_3(\tau)^4\,\vartheta_4(\tau_{\rm uv})^4\,)^{\frac{1}{2}}}{\eta(\tau_{\rm uv})^{6}} ,\\
\Delta_{\rm phys}(\tau,\tau_{\rm uv})&=(2m)^6\,\frac{\eta(\tau_{\rm uv})^{24} \,\eta(\tau)^{12}}{\left(\vartheta_4(\tau)^4 \vartheta_3(\tau_{\rm uv})^4-\vartheta_3(\tau)^4 \vartheta_4(\tau_{\rm uv})^4\right)^{3}}.
\end{split}
\ee

The $N_f=0$ theory described above is obtained from the
$\CN=2^*$ theory in the double scaling limit, 
\be
\label{scaling_limit}
\begin{split}
&  \tau_{\rm uv}\to 1+i\infty, \\
& m\to \infty,\\
&  \sqrt{2}\,q_{\rm uv}^{1/4}\,m=i\,\Lambda_0\,\,{\rm fixed}, \\
& \Lambda \to \Lambda_0/\sqrt{2}.
\end{split}
\ee
We denote this limit by $\CN=2^*\to N_f=0$. The limit for $\tau_{\rm uv}$ is
chosen, such that (\ref{uxLimit}) and (\ref{dauDeltaLimit}) below hold. As before, the parameter $\Lambda_0$ is the dynamical scale of the $N_f=0$ Seiberg-Witten
theory. Moreover, $\Lambda$ is the natural scale of the massive $\CN=2^*$
theory, which will feature in the next Subsection.  

Together with the change of variables,
\be
\label{uxLimit}
\begin{split}
& u=u_0-\frac{1}{8} e_1(\tau_{\rm uv})m^2, \\
& x= x_0-\frac{1}{2} e_1(\tau_{\rm uv}) u-\frac{1}{8} e_1(\tau_{\rm uv})^2m^2,
\end{split}
\ee 
the SW curve of $\CN=2^*$ (\ref{N2s2}) reduces to the curve for $N_f=0$ (\ref{N=2curve}) in the limit (\ref{scaling_limit}). Moreover, the other $\CN=2^*$ quantities reduce in the limit to
\be
\begin{split}
\label{dauDeltaLimit}
\lim_{2^*\to N_f=0}\,\,
\frac{da}{du}=\left(\frac{da}{du}\right)_{0},\qquad \lim_{2^*\to
  N_f=0} -\frac{4}{m^2}\, \Delta_{\rm phys}=\Delta_{0, \rm phys}.
\end{split}
\ee

 \subsection{The prepotential}  
\label{N=2*prep}  

Key quantities of the $\CN=2^*$ theory are the central charge of the W-boson $a$, the central charge of the
magnetic monopole is $a_D$, and the effective coupling constant
$\tau$. The latter two can be derived from the prepotential $\CF$ \cite{Minahan:1997if, DHoker:1997hut, DHoker:1999hmo, Manschot:2019pog}.   
The perturbative part of $\CF$ reads 
\be         
\label{Fpert} 
\begin{split}
\CF_{\rm pert}(a,m)&=\frac{1}{2} \tau_{\rm uv} a^2 -\frac{1}{8\pi i}\left[\,  2(2a)^2 \log(2a/\Lambda) -(2a+m)^2\log((2a+m)/\Lambda)\right.\\
&\quad \left.-(2a-m)^2\log((2a-m)/\Lambda)-m^2\log(m/\Lambda) +\tfrac{9}{2}\,m^2\,\right]\\
&=\frac{1}{2} \tau_{\rm uv} a^2-\frac{1}{8\pi i}\left[\,  2(2a)^2 \log(2a/m)-(2a+m)^2\log(2a/m+1)\right.\\
&\quad \left.-(2a-m)^2\log(2a/m-1)-3m^2\log(m/\Lambda) +\tfrac{9}{2}\,m^2\right],
\end{split}
\ee
where $\Lambda$ is the dynamically generated scale. We recognize the
three contributions from the components of the adjoint hypermultiplet. The
dependence can also be understood by weakly gauging the $U(1)$ flavor
symmetry, and viewing $m$ as the scalar vev of the vector
multiplet \cite{Manschot:2019pog}. 

We introduce the variables $a_D$ and $m_D$, which are dual to $a$ and $m$
\be
a_D=\frac{\partial \CF}{\partial a},\qquad m_D=\frac{\partial
  \CF}{\partial m}.  
\ee
The perturbative parts of these couplings are
\be
\label{aDmDpert} 
\begin{split}
a_{D,{\rm pert}}&=\tau_{\rm uv}\,a -\frac{1}{2\pi i} \left(
  4a\,\log(2a/\Lambda)-(2a+m)\,\log((2a+m)/\Lambda)\right.\\
&\quad\left. -(2a-m)\,\log((2a-m)/\Lambda)+\dots\right),\\
m_{D,{\rm pert}}&=\frac{1}{4\pi i} \left(
  (2a+m)\,\log((2a+m)/\Lambda)-(2a-m)\,\log((2a-m)/\Lambda)\right.\\
&\quad\left. +m\,\log(m/\Lambda)-\tfrac{9}{2}\,m+\dots\right),
\end{split}
\ee
where the $\dots$ are polynomial in $a^{-1}$ and $m$.

The full prepotential receives further contributions from
instantons, which take the form
\be
\label{FPertplusInst}
\CF(a,m)=\CF_{\rm pert}(a,m) -\frac{1}{4\pi i} \sum_{n=2}^\infty  \frac{1}{(2n-2)}\frac{m^{2n}}{(2a)^{2n-2}}
\sum_{\ell=0}^\infty c_n(\lceil n/2\rceil +\ell)
\,q_{\rm uv}^{\lceil n/2\rceil+\ell},
\ee
where the $c_n(\ell)\in \mathbb{Q}$. The non-perturbative part
reproduces the non-perturbative prepotential of pure Seiberg-Witten theory in the limit 
$m\to \infty$, $q_{\rm uv}\to 0$, with $2|q_{\rm uv}^{1/2}m^2|=|\Lambda_0|$
fixed. 

Reference \cite{Minahan:1997if} demonstrated that the unknown
coefficients $c_n(\lceil n/2\rceil +\ell)$ can be determined by imposing that the
$q_{\rm uv}$-series are quasi-modular forms of $SL(2,\mathbb{Z})$. To this
end, one expands $\CF_{\rm pert}$ for $a\gg m$, which gives the
constant terms of the modular forms. Requiring that the next
non-vanishing coefficient is $q_{\rm uv}^{\lceil n/2\rceil}$ as in
(\ref{FPertplusInst}), allows to determine the expansion, since the
space of quasi-modular forms is finite dimensional. 

As a result, the full prepotential can be written as an expansion in
the ratio $m/(2a)$ \cite{Minahan:1997if, DHoker:1997hut, DHoker:1999hmo},
\be
\label{Fmassexp}
\begin{split}
\CF(a,m)&=\frac{1}{2}\,\tau_{\rm uv}\, a^2+\frac{1}{4\pi i}\, f_1(\tau_{\rm uv})\,
m^2\, \left(\log(2a/m)-\frac{3}{4} + \frac{3}{2}\,\log(m/\Lambda)\right)\\
&\quad -\frac{1}{4\pi i} \sum_{n=2}^\infty \frac{f_n(\tau_{\rm uv})}{(2n-2)}
\frac{m^{2n}}{(2a)^{2n-2}}. 
\end{split}
\ee
The $f_n$ are determined in \cite{Minahan:1997if} by requiring that $\CF$ reduces to the SW
prepotential (\ref{SWprepotential}). As discussed above, the coefficients of $f_n$ display a gap,
the constant part of $f_n$ is determined by the perturbative
prepotential, whereas the next non-vanishing coefficient is the one
of $q_{\rm uv}^{\lceil n/2\rceil}$. The first few $f_n$ are
\be
\label{fjs}
\begin{split} 
&f_1=1,\\
&f_2=\frac{1}{6}E_2,\\
&f_3=\frac{1}{18}E_2^2+\frac{1}{90} E_4,\\
&f_4=\frac{5}{216}E_2^3+\frac{1}{90}E_2E_4+\frac{11}{7560}E_6, \\
\end{split}
\ee
where the $E_k$ are the Eisenstein series of weight $k$ defined by Equation (\ref{Ek}) in the Appendix.
The prepotential (\ref{Fmassexp}) is homogeneous of degree 2 in $a$, $m$, and $\Lambda$,
\be
\label{Fhom}
a\frac{\partial \CF}{\partial a}+m\frac{\partial \CF}{\partial m}+\Lambda \frac{\partial \CF}{\partial \Lambda}=2\CF.
\ee
The derivative of $\Lambda \partial_\Lambda \CF$ is a monomial
in $m$, $\Lambda \partial_\Lambda \CF=-\frac{3}{8\pi i}m^2$. 
Let us also mention that the Coulomb branch parameter $u$ (\ref{expdaduDelta*}) can be expressed in terms of $\CF$ as
\be
\label{uCF}
u=2\,\frac{\partial \CF(a,m)}{\partial \tau_{\rm uv}}-\frac{m^2}{12}E_2(\tau_{\rm uv}). 
\ee

\subsection{$\CN=2^*$ couplings}
As mentioned before, our strategy to derive the topological couplings in the $u$-plane
integrand is to consider the mass $m$ as the vev of the scalar field of an additional
vector multiplet, which is ``frozen''. Then further couplings could be
derived from the prepotential,
\be
\label{2*couplings}
\tau=\frac{\partial^2 \CF}{\partial a^2},\qquad v=\frac{\partial^2 \CF}{\partial a \partial m}, \qquad \xi=\frac{\partial ^2
  \CF}{\partial m^2}. 
\ee 
Using the homogeneity of the prepotential (\ref{Fhom}), one can show that the coupling $v$ satisfies
\be
\label{vaDat}
v=\frac{1}{m}\left(a_D-a\, \tau\right).
\ee

From (\ref{Fmassexp}), we have the following expansions for $a\gg m$,
\be
\label{tauvxiexp}
\begin{split}
\tau&=\tau_{\rm uv}-\frac{1}{\pi i}
\sum_{n=1}^\infty (2n-1)\, f_n(\tau_{\rm uv})
\left(\frac{m}{2a}\right)^{2n},\\
v& = \frac{1}{\pi i}\sum_{n=1}^\infty n\, f_n(\tau_{\rm uv})
\left(\frac{m}{2a}\right)^{2n-1},\\
\xi &= \frac{1}{2\pi i}\ f_1(\tau_{\rm uv})\left( \log(2a/m)
  +\frac{3}{2}\log(m/\Lambda)\right)\\
&\quad -\frac{1}{2\pi i}\sum_{n=2}^\infty
\frac{n(2n-1)}{2n-2}\,f_n(\tau_{\rm uv})\,\left( \frac{m}{2a}\right)^{2n-2}.
\end{split} 
\ee 
Moreover, one can consider $\frac{d\tau}{da}=\frac{d^3\CF}{da^3}$. The
large $a$ expansions (\ref{tauvxiexp}) provide an independent check on
the expression for $da/d\tau$ in terms of $\Delta_{\rm phys}$ and $da/du$
(\ref{dadtau}).

For later calculations, it is useful to expand $a$ as function of $\tau-\tau_{\rm uv}$. To this end, we define the variable $s=-\pi i (\tau-\tau_{\rm uv})$, and assume
$\mathrm{Im}(\tau)>\mathrm{Im}(\tau_{\rm uv})$, with vanishing ${\rm
  Re}(\tau)={\rm Re}(\tau_{\rm uv})=0$. Then
\be 
\label{atautau0}
\begin{split}
a(s)&=\frac{m}{2\sqrt{s}}\left(1+\frac{3}{2}f_2\,s+\frac{1}{8}(20f_3-45f_2^2)\,s^2\right. \\
&\quad + \left.\frac{1}{16}(567\,f_2^3-420f_2f_3+56f_4)s^3+O(s^4)\right),
\end{split}
\ee  
with the $f_j$ as in (\ref{fjs}).

\subsection{Gravitational couplings and the $\Omega$-background}
In addition to the couplings discussed above, there are gravitational
couplings of the form $A^\chi B^\sigma$, where $A$ and $B$ are
\cite{Moore:1997pc, Witten:1995gf}
\be 
\label{ABdef}
\begin{split}
&A=\alpha\, \left(\frac{du}{da}\right)^\frac{1}{2}, \\
&B=\beta\,\Delta_{\rm phys}^{\frac{1}{8}}, 
\end{split} 
\ee   
where $\Delta_{\rm phys}$ is the physical discriminant, which vanishes
linearly at the singularities. Equation (\ref{expdaduDelta*})  expresses
$du/da$ and $\Delta_{\rm phys}$ as function of $\tau$, $\tau_{\rm uv}$
and $m$. We introduce furthermore,
\be
\label{Cexpxi}
C=\exp\!\left(-2\pi i \,\xi\right),
\ee
where $\xi=\frac{\partial^2 \CF}{\partial m^2}$ as in (\ref{tauvxiexp}).
These couplings can be derived from the $\Omega$-background
\cite{Nakajima:2003uh}, which we will briefly recall here.
Let $\mathcal{Z}$ be the Nekrasov partition function of the $\CN=2^*$
theory with gauge group $SU(2)$ in the
$\Omega$-background $\mathbb{C}^2_{\varepsilon_1,\varepsilon_2}$ with equivariant parameters $\varepsilon_1$ and
$\varepsilon_2$ \cite{Ne, Moore:1997dj}. Let also $\varepsilon_\pm=(\varepsilon_1\pm \varepsilon_2)/2$.
Note $\chi(\mathbb{C}^2_{\varepsilon_1,\varepsilon_2})=\varepsilon_1
\varepsilon_2$ and $\sigma(\mathbb{C}_{\varepsilon_1,\varepsilon_2}^2)=(\varepsilon_1^2+
\varepsilon_2^2)/3$ \cite{Nakajima:2003uh}. 

We define the refined free energy
\be
\CF(a,m,\varepsilon_1,\varepsilon_2)=\varepsilon_1\varepsilon_2 \log(\CZ).
\ee 
The free energy $\CF(a,m,\varepsilon_1,\varepsilon_2)$  has an
expansion \cite{Ne} 
\be 
\label{epsExp}
-4\pi i \CF(a,m)+(\varepsilon_1+\varepsilon_2)
\CF_1+\varepsilon_1\varepsilon_2  \CF_A +
\frac{1}{3}(\varepsilon_1^2+\varepsilon_2^2) \CF_B+\dots,
\ee
where the $\dots$ denote higher order terms in the $\varepsilon_i$,
and the prefactor in front of $\CF(a,m)$ is chosen such that
$\CF(a,m)$ agrees with (\ref{FPertplusInst}). The
term $\CF_A$ corresponds to the coupling $A=e^{\CF_A}$, and the term $\CF_B$ to
the coupling $B=e^{\CF_B}$ in the $u$-plane integral \cite{Nakajima:2003uh}. One finds \cite{Manschot:2019pog}
\be
\label{alphabetaOmega} 
\begin{split}
&\alpha=\Lambda^{-\frac{1}{2}} ,\\
&\beta= 2^{\frac{3}{4}} m^{\frac{1}{8}}\Lambda^{-\frac{7}{8}}. 
\end{split}
\ee
 
In the analysis of \cite{Manschot:2019pog}  a term $\CZ_{\rm
  extra}=\prod_{n=1}^\infty
(1-q_{\rm uv}^n)^{-\frac{2}{\varepsilon_1\varepsilon_2}(m^2-\varepsilon_-^2)}$
was divided out from the $U(2)$ partition function to determine the 
$SU(2)$ partition function \cite{Alday:2009aq, Nekrasov:2015wsu}. This also has the effect that $\CF_1$ vanishes.
On the other hand, $S$-duality and comparison with mathematical results lead us to keeping $\CZ_{\rm
  extra}$ for the analyses in this paper. In fact, rather than working with $\CZ_{\rm extra}$, let us include a
``classical'' term for $m$, and redefine $\CZ_{\rm
  extra}=\eta(\tau_{\rm uv})^{-\frac{2}{\varepsilon_1\varepsilon_2}(m^2-\varepsilon_-^2)}$
in terms of the $\eta$-function. The new refined prepotential $\widetilde \CF(a,m,\varepsilon_1,\varepsilon_2)$ is related to
the $\CF(a,m,\varepsilon_1,\varepsilon_2)$ by
\be
\label{tildeFeps}
\widetilde
\CF(a,m,\varepsilon_1,\varepsilon_2)=\CF(a,m,\varepsilon_1,\varepsilon_2)-2(m^2-\varepsilon_-^2)\,\log(\eta(\tau_{\rm
uv})), 
\ee
and the unrefined prepotential
\be
\label{tildeF}
\widetilde \CF(a,m)=\CF(a,m)+\frac{m^2}{2\pi i}\,\log(\eta(\tau_{\rm
uv})), 
\ee

Then (\ref{epsExp}) demonstrates that the couplings $\tilde A$, $\tilde
B$ and $\tilde C$ derived from $\widetilde \CF$ are related to  $A$,
$B$ and $C$ as
\be  
\label{ABCextra}
\begin{split}
&\tilde A=\eta(\tau_{\rm uv})^{-1}\, A,\\
&\tilde B=\eta(\tau_{\rm uv})^{\frac{3}{2}}\,B,\\
&\tilde C=\eta(\tau_{\rm uv})^{-2}\,C.
\end{split}
\ee 
These $\eta$-powers are in agreement with results in
the the literature. For example, they lead to the correct topological
couplings in the effective action \cite[Equation (3.6)]{Harvey:1996ir} (upto complex
conjugation). We will find further agreement in Sections
\ref{sec:Evaluate} and \ref{SWcontributions}.\footnote{The last paragraph of \cite[Section
  4.2]{Manschot:2019pog} determined a coupling to $K_X^2=2\chi+3\sigma$, denoted $\mu$ in \cite{Manschot:2019pog}, starting from $\CF$. This coupling was undetermined in
  \cite{Labastida:1998sk} and is important
  for comparison with the results of \cite{Labastida:1998sk} for spin manifolds. We have
  realized that the analysis of \cite[Section
  4.2]{Manschot:2019pog} must be in terms of $\tilde \CF$ rather than
  $\CF$. In particular, the $\tau_\uv$-dependence of $\mu$ is proportional to $\eta(\tau_\uv)$ for
  $\widetilde \CF$ instead of $\eta(\tau_\uv)^{-3/2}$ claimed in \cite{Manschot:2019pog}.\label{FootMu}}

We note that the relation for $u$ in terms of the
prepotential, Equation (\ref{uCF}), reads in terms of $\widetilde \CF$,
\be
\label{utCF}
u=2\frac{\partial \widetilde \CF}{\partial \tau_{\rm uv}}-\frac{m^2}{6}E_2(\tau_{\rm uv}).
\ee
For later reference, we also introduce the alternative coordinate $\tilde u$ on the $u$-plane
\be
\label{tutCF}
\tilde u=2\frac{\partial \widetilde \CF}{\partial \tau_{\rm uv}}.
\ee

We can understand the coupling to a ${\rm Spin}^c$ structure in the
$\Omega$-background as an $(\varepsilon_1,\varepsilon_2)$-dependent shift of the mass. If the ${\rm Spin}^c$ structure
corresponds to an almost complex structure, the shift $m\to m -
\varepsilon_+$ gives rise to the extra coupling $C^{\varepsilon_+^2/(\varepsilon_1\varepsilon_2)}$, since 
$C$ is the exponential of the second derivative of the prepotential with respect to $m$. 
Note that, equivariantly $\chi(\mathbb{C}_{\varepsilon_1,\varepsilon_2})= 
\varepsilon_1 \varepsilon_2$ and 
$\sigma(\mathbb{C}_{\varepsilon_1,\varepsilon_2})= \frac{1}{3}
p_1(\mathbb{C}_{\varepsilon_1,\varepsilon_2})= 
\frac{1}{3}(\varepsilon_1^2 + \varepsilon_2^2)$ so 
$\varepsilon_+^2=(2\chi(\mathbb{C}_{\varepsilon_1,\varepsilon_2})+3\sigma(\mathbb{C}_{\varepsilon_1,\varepsilon_2}))/4$. 
Thus one can view the Nekrasov partition function for the 
Vafa-Witten twist as the Nekrasov partition
function for the Donaldson-Witten twist with this modified mass. This is important for the application of toric localization to the topologically twisted $\CN=2^*$ theory  on non-spin toric manifolds \cite{Bershtein:2015xfa}.

\subsection{Monodromies}
\label{sec:monodromies}

As mentioned before, the $u$-plane is topologically a four punctured
sphere, with punctures at $u=\infty$ and the points $u_j$, $j=1,2,3$. The variables $(m_D, m, a_D, a)$
change under monodromies around the punctures by
matrix multiplication. To determine these monodromies, we choose a
base point $u_*$ as follows,
\be
u_*: \qquad \mathrm{Re}(\tau)=\mathrm{Re}(\tau_{\rm uv})=0,\qquad
\mathrm{Im}(\tau)>\mathrm{Im}(\tau_{\rm uv})\gg 0.
\ee
The monodromies of $u$ around $\infty$ and $u_1$ can be determined from the perturbative
prepotential (\ref{Fpert}). This approach can not be applied near
$u_2$ and $u_3$, since the theory is strongly coupled near these
singularities. We instead determine the monodromies around $u_2$ and $u_3$ by imposing
that they preserve the four-dimensional symplectic form
$\omega=da\wedge da_D+dm\wedge dm_D$, and are
conjugate to each other. This allows to solve for the elements of the
$4\times 4$ monodromy matrix. Where possible, we will confirm the
monodromies using the explicit expressions.

Under the monodromy $M_\infty$, $u$ follows a simple closed curve,
starting and ending at $u_*$, and moving counter-clockwise on the
$u$-plane. Since $u\sim (\tau-\tau_{\rm uv})^{-1}$ (\ref{u2star}), this is
equivalent with $\tau$ moving clockwise around $\tau_{\rm uv}$. From the perturbative
prepotential, we find that the vector $(m_D, m, a_D, a)$ changes under
the monodromy by,
\be
\begin{split}
&m_D\mapsto m_D+\tfrac{1}{2}\,m,\\
&m\mapsto m,\\  
&a_D\mapsto -a_D,\\
&a\mapsto -a,\\
\end{split}
\ee
or in matrix notation 
\be
M_\infty=\left( \begin{array}{rrrr} 1 & \frac{1}{2} & 0 & 0 \\ 0 & 1 & 0 & 0 \\ 0 & 0
    & -1 & 0 \\ 0 & 0 & 0 & -1    \end{array}\right).
\ee

We let $M_1$ be the monodromy for $u$ following a simple closed curve,
counter clockwise around $u_1$ and starting and ending at $u_*$. This
is the $T^2$-translation acting on $\tau$. We deduce
from the perturbative prepotential (\ref{Fpert}), that the variables change under this
monodromy as
\be
\label{aDaT2}
\begin{split}
&m_D\mapsto m_D+\tfrac{1}{2}m-a,\\
&m\mapsto m,\\ 
&a_D\mapsto a_D-m+2a,\\ 
&a\mapsto a,  \\
\end{split}
\ee   
or in matrix notation
\be
M_1=\left( \begin{array}{rrrr} 1 & \frac{1}{2} & 0 & -1 \\ 0 & 1 & 0 & 0 \\ 0 & -1
    & 1 & 2 \\ 0 & 0 & 0 & 1    \end{array}\right).
\ee

We let $M_2$ be the monodromy moving counter clockwise around
$u_2$. As an $SL(2,\mathbb{Z})$ transformation, it reads
 $ST^2S^{-1}=\left( \begin{array}{cc}  1 & 0\\ -2 &
     1  \end{array}\right)$. Using the approach mentioned at the
 beginning of this section, we find that the four variables transform as
\be
\label{aDaz2}  
\begin{split}
& m_D\mapsto m_D+\tfrac{1}{2}m-a_D,\\
& m\mapsto m,\\
&a_D\mapsto a_D,\\ 
&a\mapsto a-2a_D+m,
\end{split}
\ee
or as a matrix
\be
\label{M2} 
M_2=\left( \begin{array}{rrrr} 1 & \frac{1}{2} & -1 & 0 \\ 0 & 1 & 0 & 0 \\ 0 & 0
    & 1 & 0 \\ 0 & 1 & -2 & 1    \end{array}\right).
\ee

Finally, we let $M_3$ be the monodromy moving counter clockwise around
$u_3$. Its corresponding $SL(2,\mathbb{Z})$ matrix is $(TS)T^2(TS)^{-1}=\left( \begin{array}{cc}
    -1 & 2\\ -2 & 3  \end{array}\right)=\left( \begin{array}{cc}
    1 & 0\\ 2 & 1  \end{array}\right) \left( \begin{array}{cc}
    -1 & 2\\ 0 & -1  \end{array}\right)$. Again using the approach
from above, we arrive at 
\be
\begin{split}
\label{aDaz3}
& m_D\mapsto m_D+\tfrac{1}{2}m-a_D+a,\\
& m\mapsto m,\\ 
&a_D\mapsto m-a_D+2a,\\
&a\mapsto m-2a_D+3a,
\end{split}
\ee
with matrix form
\be
\label{M3}
M_3=\left( \begin{array}{rrrr} 1 & \frac{1}{2} & -1 & 1 \\ 0 & 1 & 0 & 0 \\ 0 & 1
    & -1 & 2 \\ 0 & 1 & -2 & 3    \end{array}\right).
\ee
The four matrices satisfy $M_1\,M_2\,M_3=M_\infty$. 

We see that the monodromy matrices $M_j$ generate a group in the set
of symplectic matrices with half-integer entries. The lower right $(2\times 2)$ blocks of the monodromy matrices
generate the congruence subgroup $\Gamma(2)\in SL(2,\mathbb{Z})$. If
we had chosen to work with the $\Gamma^0(4)$ convention for the $N_f=0$
 theory, the corresponding monodromy matrices would have
been elements in $Sp(2,\mathbb{Z})$.

\subsection{Action on the couplings $\tau$, $v$ and $\xi$}
\label{ActionTauVXI}
Since the monodromies modify the parameters $m$ and $a$, we have to be
careful determining the transformations of the couplings $\tau$, $v$
and $\xi$ (\ref{2*couplings}). We denote the transformed couplings by $\tilde m$ and
$\tilde a$. To determine the transformations of the coupling, it will
be important to determine the derivatives $\partial (\tilde m,\tilde
a)/\partial (m,a)$. We have
\be
\left(\begin{array}{cc} \partial m/\partial \tilde m &
    \partial m/\partial \tilde a\\ \partial a/\partial
      \tilde m & \partial a/\partial \tilde
      a\end{array}\right)= \left(\begin{array}{cc} \partial \tilde m/\partial  m &
    \partial \tilde m/\partial  a\\ \partial \tilde a/\partial
      m & \partial \tilde a/\partial  a\end{array}\right)^{-1}=
\frac{1}{\partial \tilde a/\partial a}\left(\begin{array}{cc}
    \partial \tilde a/\partial a & 0
    \\ -\partial \tilde a/\partial
     m & 1 \end{array}\right),
\ee
since $\partial \tilde m/\partial m=1$ and $\partial \tilde m/\partial
a=0$, for each of the monodromies. 

The action of the monodromies on the couplings $\tau$, $v$ and
$\xi$ (\ref{2*couplings}) are
\begin{itemize}
\item For $M_\infty$,
\be
\tau \mapsto \tau, \qquad v\mapsto -v, \qquad \xi\mapsto \xi+\frac{1}{2}.
\ee
\item For $M_1$,
\be
\label{M1trafos}
\tau \mapsto \tau+2, \qquad v\mapsto v-1, \qquad \xi\mapsto \xi+\frac{1}{2}.
\ee
\item For $M_2$,
\be
\label{M2trafos}
\begin{split}
&\tau \mapsto \frac{\tau}{-2\tau+1}, \qquad v\mapsto
\frac{v-\tau}{-2\tau+1},\\ & \xi\mapsto \xi+\frac{1}{2}-v
+\frac{(v-\tau)(2v-1)}{-2\tau+1}.
\end{split}
\ee
\item For $M_3$,
\be 
\label{M3trafos}
\begin{split}
&\tau \mapsto \frac{-\tau+2}{-2\tau+3},\qquad v\mapsto
\frac{v-\tau+1}{-2\tau+3},\\ & \xi\mapsto
\xi+\frac{1}{2}-v+\frac{(v-\tau+1)(2v-1)}{-2\tau+3}.
\end{split}
\ee
\end{itemize} 
We observe from these transformations that $v$ transforms as an
elliptic Jacobi variable. This is familiar for the off-diagonal
element of a genus 2 modular matrix transforming under
$Sp(2,\mathbb{Z})$. The action of the duality
group on $(\tau,v)$ can therefore be identified with subgroup of the Jacobi group
$\Gamma(2)\ltimes \mathbb{Z}^2$, where $\mathbb{Z}^2$ denote 
translations of the elliptic variable. Elements in $SL(2,\mathbb{Z})/\Gamma(2)$ map between different
theories, which we will discuss in more detail in Section \ref{Sdualrankone}.

\subsection{The coupling $C$ and Jacobi forms}
\label{SecCouplingC}
For the evaluation of the $u$-plane integral, we need to know
expressions for the couplings near the other singularities. An elegant
way to obtain these expansions is if we can determine expressions in
terms of modular functions in $\tau$ and $\tau_{\rm uv}$. We were able to
arrive at such relations for topological couplings $du/da$ and $\Delta_{\rm phys}$ in \eqref{expdaduDelta*}. We will
analyze here the coupling $C$, and demonstrate that it can be
expressed as a Jacobi form with arguments 
$\tau$ and $v$.

 Since $\CF$ is homogeneous
of degree $2$, $C$ is only a
function of $\tau$ and $\tau_{\rm uv}$.  
The monodromies discussed in Subsection \ref{ActionTauVXI}
demonstrate that $C(\tau,\tau_{\rm uv})$ transforms as a Jacobi form with $v$
as elliptic variable in the following sense: $C$ can be expressed as a
function of three variables, $\phi_{C}(\tau,v(\tau,\tau_{\rm uv});\tau_{\rm uv})$,
where $\phi_C(\tau,z;\tau_{\rm uv})$ transforms as a Jacobi form in $(\tau,z)$ for $\Gamma(2)$ with an independent elliptic variable $z$. It has index
1, and weight 0 in $\tau$, and is anti-symmetric in $z$, $\phi_C(\tau,-z;\tau_{\rm uv})=-\phi_C(\tau,z;\tau_{\rm uv})$. 
These properties are very restrictive, especially if we assume that $\phi_C(\tau,z;\tau_{\rm uv})$ is holomorphic in $z$ and weakly holomorphic in $\tau$ and $\tau_\uv$. The space of such Jacobi forms is finite dimensional \cite{Eichler:1985}, and $\phi_C$ must be a linear combination of $\vartheta_1(2\tau,2z)$, $\vartheta_1(\tau/2,z)$ and $\vartheta_1((\tau+1)/2,z)$. 

Indeed, if we substitute for $\tau$, $v$ and $\xi$ the series (\ref{tauvxiexp}),
we find by comparing expansions on each side of (\ref{Cexpxi}) that
$C(\tau,\tau_{\rm uv})$ equals $\phi_{1,C}(\tau,v;\tau_{\rm uv})$ with $\phi_{1,C}$ given by  
\be
\label{Cttv2}
\begin{split}
\phi_{1,C}(\tau,z;\tau_{\rm uv})&=\left( \frac{\Lambda}{m}\right)^{3/2} \frac{-i}{\vartheta_2(\tau_{\rm uv})^2}\,
\frac{\vartheta_1(2\tau,2z)}{\vartheta_4(2\tau)}.
\end{split}
\ee
This expression is checked up to order $O(\left(\frac{m}{2a}\right)^{7},q_{\rm uv}^5)$. 

Interestingly, this is not a unique expression for $C(\tau,\tau_{\rm uv})$ in terms of a Jacobi form. Using the series (\ref{tauvxiexp}), one can also verify
that $C(\tau,\tau_{\rm uv})$ equals $\phi_{2,C}(\tau,v;\tau_{\rm uv})$ with $C$ given by
\be
\label{Cttv}
\begin{split}
\phi_{2,C}(\tau,z;\tau_{\rm uv})&=\left( \frac{\Lambda}{m}\right)^{3/2} \frac{-i}{\vartheta_4(\tau_{\rm uv})^2}\,
\frac{\vartheta_1(\tau/2,z)}{\vartheta_2(\tau/2)}\\
& = \left( \frac{\Lambda}{m}\right)^{3/2} \frac{\vartheta_2(\tau_{\rm uv}/2)}{\vartheta_2(\tau/2)}\,
\frac{\vartheta_1(\tau/2,z)}{2i\,\eta(\tau_{\rm uv}/2)^3},
\end{split}
\ee 
where we used the identity
$\vartheta_4^2(\tau_{\rm uv})=2\eta(\tau_{\rm uv}/2)^3/\vartheta_2(\tau_{\rm uv}/2)$. Moreover,
the coupling $C$ must be invariant under the simultaneous
$T$-transformation $(\tau,\tau_{\rm uv})\to (\tau+1,\tau_{\rm uv}+1)$, leading to
the function
\be
\label{Cttv3}
\begin{split}
\phi_{3,C}(\tau,z;\tau_{\rm uv})&=\left( \frac{\Lambda}{m}\right)^{3/2} \frac{-i}{\vartheta_3(\tau_{\rm uv})^2}\,
\frac{\vartheta_1((\tau+1)/2,z)}{\vartheta_2((\tau+1)/2)}.
\end{split}
\ee

We stress that, for general values of $z$, 
\be
\phi_{1,C}(\tau,z;\tau_{\rm uv}) \neq \phi_{2,C}(\tau,z;\tau_{\rm uv}).
\ee
However, from the above remarks it is clear that when 
we substitute $z\to v(\tau,\tau_{\rm uv})$ then in fact 
we do have: 
\be
\label{C1C2}
\phi_{1,C}(\tau,v;\tau_{\rm uv})=\phi_{2,C}(\tau,v;\tau_{\rm uv}) \quad . 
\ee 
Moreover, the equality (\ref{C1C2}) implies an interesting
identity among the couplings $\tau_{\rm uv}$, $\tau$ and $v$. To express this
in a more useful form, we want to eliminate the dependence on
$\vartheta_2(\tau/2)$ and $\vartheta_4(2\tau)$. Note that
\be
\label{t4t3t0}
\left(\frac{\vartheta_4(\tau_{\rm uv})}{\vartheta_2(\tau_{\rm uv})}\right)^2=\frac{1}{2}\left(W(\tau_{\rm uv})^{-1}-W(\tau_{\rm uv})\right),
\ee
with 
\be
W(\tau_{\rm uv}):=\frac{\vartheta_2(2\tau_{\rm uv},0)}{\vartheta_3(2\tau_{\rm uv},0)}.
\ee
Moreover, the equality (\ref{C1C2}) demonstrates that
(\ref{t4t3t0}) equals
\be
\frac{\vartheta_1(\tau/2,v)}{\vartheta_1(2\tau,2v)}
\frac{\vartheta_4(2\tau)}{\vartheta_2(\tau/2)}=\frac{1}{2}\left( \frac{\vartheta_3(2\tau, v)}{\vartheta_2(2\tau,v)}-\frac{\vartheta_2(2\tau, v)}{\vartheta_3(2\tau,v)}\right).
\ee
The equality follows from identities for Jacobi theta series such as
(\ref{sumstheta}) and (\ref{tsproduct}).
 
Comparing with (\ref{t4t3t0}), we find the identity
\be
\label{tauvtauUV}
\frac{\vartheta_2(2\tau, v)}{\vartheta_3(2\tau,v)}=W(\tau_{\rm uv}),
\ee
We checked this identity up to order $O(\left(\frac{m}{2a}\right)^{12}, q_{\rm uv}^6)$
in the large $a$-expansion after substitution of $\tau=\tau(a,m,\tau_{\rm
  uv})$, $v=v(a,m,\tau_{\rm uv})$. Assuming that \eqref{tauvtauUV}
holds, we conclude that 
\be
\label{CphijC}
C(\tau,\tau_\uv)=\phi_{j,C}(\tau,v;\tau_\uv),\qquad j=1,2,3.
\ee
  
We will find \eqref{CphijC} useful in Section \ref{sec:Evaluate}, since it provides a way to deduce expressions for $v$
at the different singular points. Note that the rhs of (\ref{tauvtauUV}) RG flow independent. It is thus
a conserved quantity along the RG flow. Using the monodromies in
Section \ref{sec:monodromies}, one can verify that (\ref{tauvtauUV}) is invariant under
monodromies. The identity agrees moreover with the $\CN=2^*\to N_f=0$ limit detailed in Equation
(\ref{SWlimit}). 

The coupling $v(\tau,\tau_{\rm uv})$ is a complicated function of
$\tau$. A useful property for Section \ref{sec:Evaluate} is that the identity
(\ref{tauvtauUV}) implies that  $v$ can {\it not} take specific values for a
generic choice of $\tau_{\rm uv}$. Namely, if we consider the ratio
$\frac{\vartheta_2(2\tau, z)}{\vartheta_3(2\tau,z)}$ for arbitrary
$z\in \mathbb{C}$, then for 
\be
\label{vexcluded}
\begin{split}
&e^{2\pi i z}=-q^{n/2},\qquad n\in \mathbb{Z},\\
&e^{2\pi i z}=q^{n+\frac{1}{2}},\qquad n \in \mathbb{Z},
\end{split}
\ee
this ratio is independent of $\tau$ and equals $0,1,\infty$ or $\pm i$. Therefore given the
identity (\ref{tauvtauUV}), $v$ can never assume these values.

Another interesting consequence of the relation (\ref{C1C2}) is
$S$-duality. The analysis of Subsection \ref{ABCboundaries} demonstrates that $v$ transforms as
\be
v(-1/\tau,-1/\tau_{\rm uv})=-\frac{v(\tau,\tau_{\rm uv})}{\tau},
\ee
under simultaneous inversion of $\tau$ and $\tau_{\rm uv}$. The transformation $(\tau,\tau_{\rm uv})\mapsto (-1/\tau,-1/\tau_{\rm uv})$ applied to (\ref{tauvtauUV}) gives
the identity
\be
%\label{tauvtauUV}
\frac{\vartheta_4(\tau/2, v/2)}{\vartheta_3(\tau/2,v/2)}= \frac{\vartheta_4(\tau_{\rm uv}/2,0)}{\vartheta_3(\tau_{\rm uv}/2,0)}.
\ee
Using (\ref{sumstheta}) for Jacobi theta series, one can show that this identity is equivalent to (\ref{tauvtauUV}).

Using the transformations, in the appendix we find furthermore
\be
\begin{split} 
\varphi_{1,C}(-1/\tau,-1/\tau_{\rm uv},z/\tau)&=\tau_{\rm uv}^{-1} e^{2\pi i z^2/\tau } \varphi_{2,C}(\tau,\tau_{\rm uv},z/\tau).
\end{split}
\ee
We have therefore for the $S$-duality transformation of $C$,
\be
C(-1/\tau,-1/\tau_{\rm uv})=-\tau_{\rm uv}^{-1} e^{2\pi i v^2/\tau }\, C(\tau,\tau_{\rm uv}).
\ee

\subsection{S-duality action on the three non-abelian rank one theories}
\label{Sdualrankone}
\begin{figure}[ht]
 \begin{center} 
\begin{tikzpicture}[inner sep=2mm,scale=1]
 \node (1) at (0,6.9282) [circle,draw, fill=white, ultra thick] {$\,\CT_{SU(2)}\,$}; 
 \node (2) at (-4,0) [circle,draw, ultra thick] {$\CT_{SO(3)_+}$};
 \node (3) at (4,0) [circle,draw, ultra thick] {$\CT_{SO(3)_-}$};
\draw [dashed, opacity=.4] (0,-2) -- (0,9); 
\draw [dashed, opacity=.4] (5.732,-1) -- (-5.732,5.6188); 
\draw[bend right,<->,>=stealth]  (1) to node [left, pos=0.55] {$S\,\,$} (2);
\draw[bend right,<->,>=stealth]  (2) to node [below, pos=0.58] {$T$} (3);
\draw [-, ultra thick] (1) to node[auto] {} (3);
\draw [-, ultra thick] (2) to node[auto] {} (3);
\draw [-, ultra thick] (1) to node[left] {} (2);
\end{tikzpicture}
\end{center} 
\caption{Action of the generators $S$ and $T$ of $SL(2,\mathbb{Z})$ on
the theories $\CT_{SU(2)}$, $\CT_{SO(3)_+} $ and $\CT_{SO(3)_-}$. The
three nodes of the equilateral triangle represent the three
theories. The $S$
and $T$ transformations act as reflections on the triangle.}
\label{STTheories}  
\end{figure}
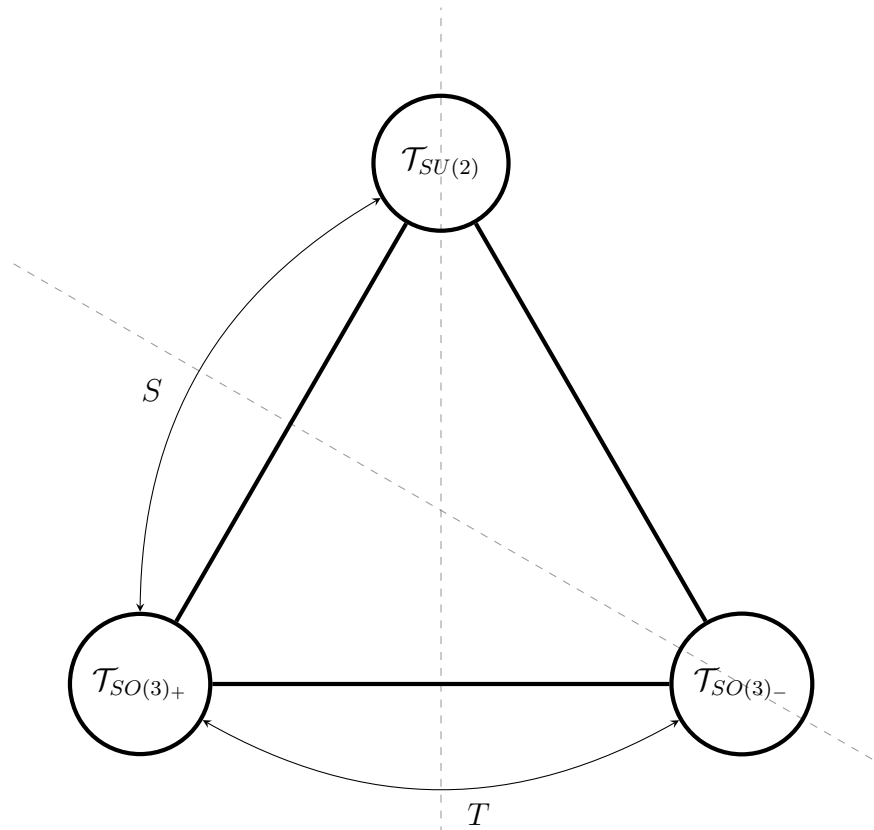

It is well-known that there are three inequivalent $\CN=2^*$ theories
with nonabelian rank one gauge group \cite{Gaiotto:2010be, Aharony:2013hda, Seiberg:2013}.
These theories are sometimes referred to as the $\CT_{SU(2)}$ and $\CT_{SO(3)_\pm}$ theories,
and can be distinguished by their (category of) line defects. The $S$-duality
group $SL(2,\mathbb{Z})$ acts through its quotient $S_3 \cong SL(2,\mathbb{Z})/\Gamma(2)$
via a permutation action with \cite{Gaiotto:2010be, Aharony:2013hda, Seiberg:2013}:
\begin{equation}
\begin{split}
&S:\quad \CT_{SU(2)} \leftrightarrow \CT_{SO(3)_+}, \qquad  \qquad \CT_{SO(3)_-} \leftrightarrow \CT_{SO(3)_-}, \\
&T: \quad \CT_{SO(3)_+} \leftrightarrow \CT_{SO(3)_-}, \qquad \qquad \CT_{SU(2)} \leftrightarrow \CT_{SU(2)}. \\
\end{split}
\end{equation} 
Figure \ref{STTheories} displays these relations between the theories.

The $SO(3)_-$ theory differs from the $SO(3)_+$ theory by 
an additional topological term in the action \cite{Aharony:2013hda}:
\begin{equation}
\exp\!\left( - \frac{i \pi}{2} \int P_2(w_2(E)) \right)
\end{equation}
where $w_2(E)\in H^2(X;\mathbb{Z}_2)$ is the Stiefel-Whitney class of the $SO(3)$ gauge bundle
and $P_2: H^2(X;\mathbb{Z}_2) \rightarrow H^4(X;\mathbb{Z}_4)$ is the Pontryagin square.
We will confirm this term in the action of the $\CT_{SO(3)_\pm}$ theory by
explicit computation of the partition function in Section \ref{sec:Evaluate} below.

Since $\exp\!\left( - \frac{i \pi}{2} \int P_2(w_2(E)) \right)$ is a fourth root of unity in general one might wonder if a new theory can be obtained by using instead the 
topological term $\exp\!\left(\frac{i \pi}{2} \int P_2(w_2(E)) \right)$. 
However,  
\begin{equation}
\exp\!\left(\frac{i \pi}{2} \int_X P_2(w_2(E)) \right)=
\exp\!\left(- \frac{i \pi}{2} \int_X P_2(w_2(E))  + i \pi \int_X w_2(E) w_2(X) \right)
\end{equation}
(where we have used $P_2(w_2(E)) = w_2(E)^2 \mod 4$ and the Wu formula).
We will interpret the insertion of $\exp( i \pi \int w_2(E) w_2(X) )$ as
defining two partition functions within the same theory $\CT_{SO(3)_\pm}$,
rather than as a definition of distinct theories.

If we consider the domain of definition of the theory to include non-spin manifolds and include the category of line defects in the definition of ``theory'' then the $S$-duality diagram will be more elaborate \cite{Ang:2019txy}.

\section{Formulation of the UV theory on a four-manifold} 
\label{Form4manifold}
This Section discusses aspects of the ultraviolet (UV) theory on a four-manifold. The first subsection starts by reviewing basic aspects of four-manifolds followed by a subsection on four-manifolds with $b_2^+$ odd. Section \ref{SpincStruct} discusses Spin and Spin$^{c}$ structures. After discussing topological twisting in Section \ref{TopTwisting}, we compare the $Q$-fixed equations of the Donaldson-Witten and Vafa-Witten twist in Section \ref{CompVWtwist}. Subsection \ref{VirDims} discusses index formula's for the virtual dimensions of various moduli spaces, namely moduli spaces of instantons and Seiberg-Witten equations. Subsection \ref{LocPathIntegral} discusses the supersymmetric localization of the topological path integral of the $\CN=2^*$ theory.

\subsection{Four-manifolds and lattices}
We consider a smooth, compact, orientable four-manifold $X$. We let $L$ be the image of the Abelian group  $H^2(X,\mathbb{Z})$ in $H^2(X,\mathbb{R})$. Since
$L$ is a lattice in a real vector space, we can divide its elements
without ambiguity. The intersection form on $H^2(X, \mathbb{Z})$ provides a natural non-degenerate bilinear form $B : (L \otimes \mathbb{R}) \times (L \otimes \mathbb{R}) \to \BR$ that pairs degree two co-cycles,
\be 
B(\bfk_1,\bfk_2):=\int_X \bfk_1\wedge \bfk_2,
\ee 
and whose restriction to $L \times L$ is an integral bilinear form. The bilinear form provides the quadratic form $Q(\bfk) := B(\bfk, \bfk) \equiv \bfk^2$, which is uni-modular and possibly indefinite. We let $H^{2,\pm}(X,\mathbb{R})$ be the positive and negative definite subspace of $H^2(X,\BZ)$ and set $b^\pm_2=\dim(H^{2,\pm}(X,\mathbb{R}))$ and $\sigma=b_2^+-b_2^-$. 

A vector ${\boldsymbol c}$ is a characteristic vector of $L$, if for all $\bfk\in L$,
\be
\label{KMcharacteristic} 
B(\bfc, \bfk)+\bfk^2\in 2\mathbb{Z}.
\ee
A characteristic vector ${\boldsymbol c}$ satisfies
\be 
\label{charL}
{\boldsymbol c}^2=\sigma \mod 8.
\ee
Any integral lift of the second Stiefel-Whitney class $w_2(T_X)$ to $L$ is a characteristic vector.

For manifolds with $b_2^+=1$, we let $J$ be the generator for the unique self-dual direction in $H^2(X,\mathbb{R})$.
We normalize it, such that it satisfies
\be
* J= J,\qquad Q(J)=\int_{X}J\wedge J=1,\qquad dJ=0.
\ee
It provides the projection $\bfk_+$ of a vector $\bfk \in L$ to the positive definite space $L^+\subset L\otimes \mathbb{R}$ spanned by $J$ and the orthogonal complement $L^-$. We define
\be 
\bfk_+=B(\bfk,J)\,J\in L^+,\qquad \bfk_-=\bfk-\bfk_+\in L^-.
\ee 

As in \cite{Korpas:2019cwg}, it will be important for us that a
lattice with signature $(1,b_2-1)$ can be brought to a simple standard
form \cite[Section 1.1.3]{Donaldson90}. The standard form depends on whether the lattice
is even or odd:
\begin{itemize}
\item If $Q$ is odd, an integral change of basis can bring
the quadratic form to the diagonal form
\be
\label{lattice_odd}
\left< 1 \right> \oplus m \left<-1\right>,
\ee
with $m=b_2-1$. This has an important consequence for characteristic
elements of such lattices. If $K$ is a characteristic element, $\bfk^2+B(K,\bfk)\in 2\mathbb{Z}$ for any $\bfk\in
L$. In the diagonal basis (\ref{lattice_odd}) this equivalent to
$\sum_{j=1}^{b_2} k_j^2 +K_j k_j\in 2\mathbb{Z}$ with
$K=(K_1,K_2,\dots, K_{b_2})$. This can only be true for all $\bfk\in L$
if $K_j$ is odd for all $j=1,\dots, b_2$.
\item If $Q$ is even,  the quadratic form
$Q$ can be brought to the form
\be
\label{lattice_even}
\mathbb{I}^{1,1} \oplus n\, L_{E_8},
\ee
where $\mathbb{I}^{1,1}$ and $L_{E_8}$ as defined above and $n=(b_2-2)/8$. The components $K_{j}$, $j=1,2$ must therefore be even in this basis.
\end{itemize}

\subsection{Simply connected four-manifolds with $b_2^+$ odd}
This paper is mostly concerned with simply connected four-manifolds with $b_2^+$ odd. 
We will discuss in Section \ref{VirDims} that for manifolds with $b_2^++b_1$ even, correlators of topological point and surface observables vanish. For a simply connected four-manifold $X$, $\pi_1(X)$ and $b_1=0$, and we thus naturally assume that $b_2^+$ odd.
Such manifolds are almost complex \cite{Scorpan}, which means that they admit an almost complex structure $\CJ$ on $TX$. We review a few aspects of almost-complex four-manifolds in Appendix \ref{ACS4folds}.

Since the tangent bundle $T_X$ of a manifold with almost complex
structure is complex, one can define its first Chern class $c_1(T_X)$.
Moreover, the first Chern class of the line bundle of
holomorphic 2-forms on $X$, $\Omega^{2,0}(X)$ 
is the canonical class $K_X=-c_1(T_X)$. The canonical class $K_X$ and
$c_1(T_X)$ are integral lifts of $w_2(T_X)$ to $H^2(X,\mathbb{Z})$ and are thus characteristic vectors of $L$. 

By the Hirzebruch signature theorem, we have
\be 
\label{K2chisigma}
K^2_X=c_1(T_X)^2=2\chi+3\sigma,
\ee 
which becomes for simply connected four-manifolds with $b_2^+=1$, 
\be
\label{KM2sigma}
K_X^2=\sigma+8.  
\ee
Any element $\bfk\in H^2(X,\mathbb{Z})$ satisfying
(\ref{K2chisigma}) gives rise to an almost complex structure. If $X$ is simply connected, such an $\bfk$ always exists if $b_2^+$ is odd. To shorten notation, we define 
\be
\label{lambda}
\lambda:=K_X^2.
\ee

The Riemann-Roch theorem gives for the holomorphic Euler characteristic 
\be
\label{holchi}
\chi_{\rm h}=\frac{\chi+\sigma}{4}.
\ee
If $X$ is complex, $\chi_{\rm h}$ equals $1-h^{0,1}+h^{0,2}$, where $h^{i,j}=\dim(H^{i,j}(X))$. In this case, combination of these two equations demonstrates that
$b_2^+=1+2\,h^{0,2}$, and there is a unique positive definite direction in $H^{1,1}(X,\mathbb{R})$ spanned a harmonic two-form. For generalizations 
of these statements to arbitrary almost complex manifolds see \cite{draghici2010}. 

We let $\omega$ be the fundamental (1,1)-form compatible with the metric. 
The bundle of self-dual two-forms $\Omega^{2+}$ therefore splits as,
\be
\label{Om2+}
\Omega^{2+} \cong R\oplus [ \Omega^{2,0}\oplus \Omega^{0,2}]_{\IR},
\ee
Here $R$ denotes the space of $(1,1)$ forms 
proportional to $\omega$. Note that $R$ is isomorphic to the space 
of real-valued functions on $X$.
The subscript $\IR$ in the second summand means we take the real 
points with respect to the obvious real structure that exchanges $\Omega^{2,0}$ 
with $\Omega^{0,2}$. 

\subsection{${\rm Spin}$ and ${\rm Spin}^c$ structures}
\label{SpincStruct}

As explained below in section \ref{TopTwisting}, the formulation 
of the topologically twisted UV theory makes use of spinor fields on $X$. 
In order to introduce spinors we must make a choice of a ${\rm Spin}^c$ 
structure. Let us recall what this means.   

The Lie group Spin$^c(4)$ can be defined as
\be 
\label{Spinc4}
{\rm Spin}^c(4)=\{(u_1,u_2)\,\vert\, {\rm det}(u_1)={\rm
  det}(u_2)\}\subset U(2)\times U(2). 
\ee
If we identify $\mathbb{R}^4$ with the quaternions 
then the action $q \mapsto u_1 q u_2^{-1}$ defines a 
homomorphism $\pi:{\rm Spin}^c(4)\to SO(4)$ so that 
we have an exact sequence
\be
\label{SpincES}
1\to U(1)\to {\rm Spin}^c(4)\to SO(4)\to 1 ~ . 
\ee
Since $X$ is oriented and Riemannian its tangent bundle has a 
reduction of structure group to $SO(4)$. A ${\rm Spin}^c$ structure, 
by definition, is a further ``reduction'' of structure group to 
${\rm Spin}^c(4)$ using the homomorphism $\pi$.
\footnote{Here is the precise definition of a ``reduction of 
structure group.'' Given two compact Lie groups $G_1, G_2$ and a homomorphism $\phi:G_1
\to G_2$ we can define a functor  from principal $G_1$
bundles on $X$ to principal $G_2$ bundles on $X$ by taking  a principal
$G_1$ bundle $ P \to X$ to a principal $G_2$ bundle $(P\times_{G_1} G_2) \to X$.
Recall that $(P\times_{G_1} G_2)$ is the set of pairs $(p,g)\in P
\times G_2$ with equivalence relation $(ph,g) = (p,\phi(h)g)$ for
$h\in G_1$, and this clearly admits a free right $G_2$ action.
Now,  if  $  P_2 \to X$ is a principal $G_2$
bundle, then a  \emph{reduction to $G_1$ under $\phi: G_1 \to G_2$} is a
principal $G_1$ bundle $ P_1 \to X$ \underline{together with}  an 
isomorphism of principal $G_2$ bundles: 
$\psi: (P_1\times_{G_1} G_2) \to  P_2$. 
If the homomorphism $\phi$ is the projection  $\pi: {\rm Spin}^c(4) \to SO(4)$ 
one might prefer to speak of a ``lift of structure group.'' }
What this means, 
in plain English, is that one can choose an atlas $\{ \CU_{\alpha} \}$ on $X$ with 
transition functions $g_{\alpha\beta}: {\cal{U}}_{\alpha\beta} \to {\rm Spin}^c(4)$ 
on patch overlaps ${\cal{U}}_{\alpha\beta}$  
that satisfy the cocycle condition and moreover 
are such that $\pi(g_{\alpha\beta})$ provide a set of transition functions 
for the oriented tangent bundle $TX$. Topological considerations allow one to prove that every oriented four-manifold 
admits a ${\rm Spin}^c$-structure \cite{Scorpan, Morgan, TeichnerSpinc}.
The set of ${\rm Spin}^c$-structures 
is a torsor for the group of line bundles since $U(1)$ is in the kernel of 
$\pi: {\rm Spin}^c \to SO(4)$.

The fundamental two-dimensional representation of $U(2)$ gives
two inequivalent representations  of Spin$^c(4)$ corresponding to the two projections of 
${\rm Spin}^c(4)$ to $U(2)$.
The associated bundles $W^{\pm}$ are known as the 
\emph{chiral spin bundles}. They satisfy
\be
\begin{split}
&{\rm Hom}(W^\pm,W^\pm) \cong (W^\pm)^\lor \otimes W^{\pm} \cong
\Lambda^{2,\pm} T^*X\otimes \mathbb{C},\\
&{\rm Hom}(W^+,W^-)\cong TX\otimes \mathbb{C}.
\end{split}
\ee
We define 
\be 
\CL := {\rm det}(W^\pm)\cong\Lambda^2(W^\pm) ~ . 
\ee
The first Chern class $c_1(\CL)$ is known as the \emph{characteristic class of the 
${\rm Spin}^c$-structure}. We will  denote it as   $c_1(\mathfrak{s})$ 
or simply $c(\mathfrak{s})$. One can show that its reduction modulo two 
is equal to $w_2(X)$. Thus $c(\mathfrak{s})$ is a characteristic vector in $H^2(X, \mathbb{Z})$.
Two different ${\rm Spin}^c$ structures 
define chiral spin bundles related by $W^+_1 \cong W^+_2 \otimes T$ where $T$ is a complex line bundle and hence $c(\mathfrak{s}_1) =c(\mathfrak{s}_2) + 2 c_1(T)$.

If $X$ is an almost complex manifold, there is a \underline{canonical} associated
Spin$^c$ structure, as we now explain:
If $X$ is an almost complex four-manifold, the real tangent bundle $TX$ has a complex structure and
one can choose a metric so that the structure group of $TX$ is reduced to $U(2)$. Let $s:U(2)\to SO(4)$ be the standard embedding,
\be 
s(u)=\left(\begin{array}{cc} A& B \\ -B & A \end{array}\right)\in SO(4).
\ee 
where $A,B$ are the real and imaginary parts of $u\in U(2)$, so $u= A+ i B$.
One embeds the structure group $U(2)$ in ${\rm Spin}^c(4)$ by
\be 
\phi(u)=\left( u, \begin{pmatrix} \det u & 0 \\ 0 & 1 \\ \end{pmatrix} \right)\in {\rm Spin}^c(4).
\ee 
Then we claim that the following diagram commutes,
\be  
\begin{tikzcd}
U(2) \arrow[r, "\phi"] \arrow[dr, "\tilde s" '] & [4em] {\rm Spin}^c(4) \arrow[d, "\pi"] \\[3em] & SO(4)
\label{U2embed}  
\end{tikzcd}
\ee 
Where $\tilde s$ is a suitable conjugate of the embedding $s$. 
Now, because the structure group can be reduced to $U(2)$ we can 
define transition functions for $TX$ on patch overlaps of the form 
$s(\hat g_{\alpha\beta})$ where 
$\hat g_{\alpha\beta}: {\cal U}_{\alpha\beta} \to U(2)$ satisfies the 
cocycle condition. But then, thanks to the above diagram 
$\phi(\hat g_{\alpha\beta})$ define ${\rm Spin}^c(4)$ transition 
functions which map correctly to $SO(4)$, thus defining the 
canonical ${\rm Spin}^c$ structure on $X$ associated to an ACS.   

An ACS determines a canonical class
$K_X$ and for the canonically associated ${\rm Spin}^c$-structure 
the line bundle $\CL$ above is isomorphic to $K_X^{-1}$. 
In this case we have 
\be 
c_1(\mathfrak{s})^2=2\chi+3\sigma ~ . 
\ee
Moreover, in this case there is an isomorphism  of vector bundles so that   
\be
\label{KSpinc}
\begin{split}
&W^+ \simeq\Lambda^{0,0}\oplus \Lambda^{0,2},\\
&W^- \simeq\Lambda^{0,1}.\\
\end{split}
\ee
A quick way to understand these decompositions is to note that, 
if $V$ is the fundamental two-dimensional representation of $U(2)$, 
then $W^+$ is the restriction of the representation $1\otimes V$ of $U(2)\times U(2)$
and $W^-$ is the restriction of the representation $V\otimes 1$ of $U(2)\times U(2)$. 
\footnote{Here we have a slight abuse of notation, letting $W^\pm$ denote 
both a representation and its associated bundle.}
Rather obviously we have 
\be 
\begin{split} 
\phi^*(W^+) & \cong \Lambda^2 V \oplus 1 \\
\phi^*(W^-) & \cong V \\ 
\end{split}
\ee
establishing the isomorphisms \eqref{KSpinc}. 
For another proof see  \cite[Corollary 3.4.6]{Morgan}.
We will discuss in the next subsection that this choice of ${\rm Spin}^c$
structure has interesting physical implications.

If $w_2(X)=0$ then $X$ admits a spin structure. In the language used 
above a spin structure is a ``reduction'' of 
structure group of $TX$ from $SO(4)$ to ${\rm Spin}(4)$ relative 
to the standard covering homomorphism $\pi:{\rm Spin}(4) \to SO(4)$. The pullback of this 
reduction is then a reduction of the 
${\rm Spin}^c(4)$-structure to a ${\rm Spin}(4) \times U(1)$ structure on $X$. 
In this case $c(\mathfrak{s})$ is an even class and $\CL$ has a square root. 
Then we can identify $W^+\cong S^+\otimes
\CL^{1/2}$, where a choice of square root of $\CL$ is determined by (and determines) a choice of ${\rm Spin}$-structure. It is important to recognize that, even if $X$ is spin, 
there is an infinite number of potential ${\rm Spin}^c$ structures. After all 
the set of ${\rm Spin}^c$-structures is a torsor for the group of line bundles. 
In the spin case, if 
we choose an almost complex structure we can identify the space of chiral 
spinors in the representation $(2,1)$ with $\Omega^{0,1}(X) \otimes K_X^{1/2}$ and anti-chiral spinors 
in the representation $(1,2)$ with $(\Omega^{0,0}(X) \oplus \Omega^{0,2}(X))\otimes K_X^{1/2}$.

\subsection{Topological twisting} 
\label{TopTwisting}
We consider first the case that the four-manifold $X$ is a
spin-manifold, such that the spinor bundles $S^+$ and $S^-$ are well-defined.
With the Donaldson-Witten twist \cite{Witten:1988ze}, we replace the $SU(2)_+$
representations by that of the diagonally embedded subgroup in
$SU(2)_+\times SU(2)_R$. The spacetime representations of the
bosonic fields in the vector multiplet are unchanged since their
$SU(2)_R$ representation is one-dimensional. The
representations for the vectormultiplet fermionic fields (\ref{Vfermireps}) become
$({\bf 2},{\bf 2})\oplus ({\bf 1},{\bf 2} \otimes {\bf 2})$, or
\be
\label{DtwistVfermions}
({\bf 2},{\bf 2})\oplus ({\bf 1},{\bf 1})\oplus ({\bf 1},{\bf 3}).
\ee
These terms corresponds geometrically to a 1-form $\psi$, a 0-form $\eta$ and a self-dual
2-form $\chi$.  This can be understood more precisely from the tensor
products $S^-\otimes S^+\simeq \Lambda^1$ and $S^+\otimes S^+ \simeq
\Lambda^0\oplus \Lambda^{2,+}$.

The spacetime representations of the bosons of the hypermultiplet
become
\be
\label{twistedhyperB}
({\bf 1},{\bf 2})\oplus ({\bf 1},{\bf 2}),
\ee 
where the first term corresponds to the $SU(2)_R$ doublet $(q,\tilde q^\dagger)$ of the
untwisted theory, and the second term to $(\tilde q,
-q^\dagger)$. This representation does thus not correspond to four
complex variables, but rather a real (quaternionic) subspace. We denote
the two terms in (\ref{twistedhyperB}) by $M_{\dot \alpha}$ and $\bar M_{\dot \alpha}$ respectively.
The representations of the fermionic fields become   
\be
\label{twistedhyperF}
({\bf 2},{\bf 1})\oplus ({\bf 2},{\bf 1})\oplus ({\bf 1},{\bf 2})\oplus ({\bf 1},{\bf 2}),
\ee 
which combine to a pair of massive Dirac spinors.

The spinor representations in (\ref{twistedhyperB}) and (\ref{twistedhyperF}) are problematic for
the formulation of the theory on a non-spin manifold, since if $w_2(X)\neq 0$ the spin bundles
$S^\pm$ are not
well-defined. The field $M$ is in that case a section of a Spin$^c$
bundle $W^+$. To implement this physically, we couple the massive hypermultiplet to a weakly coupled $U(1)$, whose
flux $\bfk_m$ equals the Chern class of a Spin$^c$ structure. The two terms with the same representation in
(\ref{twistedhyperB}) and (\ref{twistedhyperF}) have opposite charge
under this $U(1)$. After coupling to a Spin$^c$ bundle $\CL$,
the holomorphic bosons are sections of $S^+\otimes \CL^{1/2}$, and their
complex conjugates are sections of $S^+\otimes
\CL^{-1/2}$. Similarly, the fermions combine to sections of the
bundles 
$S^-\otimes \CL^{\pm 1/2}$ and $S^+\otimes \CL^{\pm 1/2}$. 
For a mathematical description of this procedure see equation \eqref{eq:keyhom} below. Since there is
an infinite family of Spin$^c$ structures $\bfk_m\in L+K_X/2$, we have an infinite family of
twisted theories. Put differently, the definition of the twisted theory in the UV 
requires the introduction of the extra datum of a ${\rm Spin}^c$ structure. Moreover,    
in order to write the Lagrangian one must also choose a ${\rm Spin}^c$ connection. Two different 
${\rm Spin}^c$ connections differ by a globally defined one-form.

Once we have chosen a ${\rm Spin}^c$-structure and formulated the 
topologically twisted theory the hypermultiplet fields become spinors in 
the chiral spin bundle $W^+$.   The $Q$-fixed point equations of this twist
are the non-Abelian (adjoint) monopole equations \cite{Witten:1994cg, Labastida:1998sk}
\be 
\label{MonopoleEqs}
\begin{split}
&F^+_{\mu\nu}+\frac{1}{2} \, (\bar \sigma_{\mu\nu})^{\dot \alpha
  \dot \beta} \left[\bar M_{(\dot \alpha}, M_{\dot \beta)}\right]=0,\\
&\slashed{D} M=\sum_{\mu=1}^4\sigma^\mu D_\mu M_{\dot \alpha}=0,
\end{split}
\ee   
Here $D_\mu$ is the chosen ${\rm Spin}^c$-connection on $W^+$. Moreover, 
in explicit computations we choose a representation of gamma matrices in an orthonormal 
frame in the form 
\be 
\gamma^a = \begin{pmatrix} 0 & \sigma^a \\  \bar\sigma^a & 0  \\ \end{pmatrix} 
\ee
where $\sigma^a_{\alpha\dot \beta}$ are the standard Pauli matrices in Euclidean space,
\be 
\sigma^1=\left( \begin{array}{cc} 0 & 1\\ 1 & 0 \end{array} \right), \qquad \sigma^2=\left( \begin{array}{cc} 0 & -i\\ i & 0 \end{array} \right), \qquad 
\sigma^3=\left( \begin{array}{cc} 1 & 0\\ 0 & -1 \end{array} \right), \qquad 
\sigma^4=\left( \begin{array}{cc} i & 0\\ 0 & i \end{array} \right).
\ee 
while $(\bar \sigma^a)^{\dot \alpha \beta}=\varepsilon^{\dot \alpha\dot \beta}\varepsilon^{\alpha\beta}\sigma^a_{\alpha\dot \beta}$ and 
\be 
\left(\bar \sigma^{ab} \right)^{\dot \beta}_{\,\,\dot \alpha}=\frac{1}{4}\left(\bar \sigma^a \sigma^b-\bar \sigma^b \sigma^a\right)^{\dot \beta}_{\,\,\dot \alpha} ~ .
\ee 
We will sometimes refer to \eqref{MonopoleEqs} as the adjoint-SW equations.

The equations \eqref{MonopoleEqs} are invariant under the transformation $M\to e^{i\varphi} M$, $\bar M\to
e^{-i\varphi} \bar M$ \cite{Hyun:1995hz, Labastida:1996tz}, which is the $U(1)_B$ symmetry in the
untwisted theory.  We deduce from (\ref{superW}) that the mass $m$
is neutral under this $U(1)$. The ghost number of $m$ is 2, the same
as the ghost number of $\phi$ \cite{Hyun:1995hz}. 
  
The fixed point locus of this equation consists of two components \cite{Hyun:1995hz, Labastida:1996tz}\footnote{A closely related set of equations, the so-called ``$SO(3)$ monopole equations,'' have been studied in the mathematical literature 
\cite{Pidstrigach:1995dg, Bradlow:1996alg, Teleman:1996, Okonek:1996, Feehan:1997gj}. In this case there are two sets of 
fixed point equations. The Abelian locus is a collection of Seiberg-Witten moduli spaces for the 
$U(1)$ case, and, although these spaces are smooth they can be non-smoothly embedded into 
the full moduli space.}:
\begin{itemize}
\item The instanton component: $M_{\dot \alpha}=0$ and $F^+=0$.
\item The Abelian component: This is the component where the  $U(1)_B$
  action is pure gauge. This can only be the case if the gauge group is broken to its diagonal
  subgroup, i.e. the $SU(2)$ bundle is a sum of line bundles
  $\CE=T\oplus T^{-1}$ for some line bundle $T$. Then $M_{\dot \alpha}$ must be of the form
  $\left(\begin{array}{cc} 0 & 0 \\ * & 0 \end{array} \right)$ or $\left(\begin{array}{cc} 0 & * \\ 0 & 0 \end{array} \right)$. The equations reduce to the Abelian Seiberg-Witten equations on this locus. 
\end{itemize}

If we choose the canonical ${\rm Spin}^c$ structure associated to an almost complex structure on $X$ then the adjoint-SW equations simplify. In this case the line bundle $\CL$ is identified with $K^{-1}_X$, and sections of $W^+ \cong S^+\otimes K_X^{-1/2}$ are identified with elements of $\Omega^{0,0}(X,\mathbb{C})\oplus \Omega^{0,2}(X,\mathbb{C})$ (\ref{KSpinc}). With $\alpha \in \Omega^{0,0}(X,\mathbb{C})$ and $\bar \beta\in \Omega^{0,2}(X,\mathbb{C})$, we let
\be 
M=\left(\begin{array}{c} \bar \beta \\ \alpha \end{array} \right).
\ee 
The curvature equation of (\ref{MonopoleEqs}) then reads,
\be 
\begin{split} 
& F^{2,0}+\frac{1}{2}[\alpha,\beta]=0,\\
& i\,\omega \wedge F^{1,1}+\frac{1}{2}\,\omega^2\,[\alpha,\bar \alpha]+[\beta,\bar \beta]=0,
\end{split}
\ee 
while the Spin$^c$ Dirac equation of the monopole equations reads \cite{Gauduchon},
\be 
\label{ACSDirac}
\slashed{D} M= \sqrt{2}(\bar \partial\alpha + \bar\partial^\dagger\bar \beta)+\frac{1}{4}\theta.M+\frac{1}{2}ia.M=0,
\ee 
where $\theta$ is the Lee form, defined through $d\omega=\theta\wedge \omega$. 
The $.$ stands for the Clifford action, acting on $W^+$ as wedge product on the 0-form, and contraction $\lrcorner$ on the 2-forms,
\be 
\theta.M=\sqrt{2}(\theta^{0,1}\wedge \alpha - \theta^{0,1} \lrcorner \bar \beta).
\ee 
Here $a\in \Omega^1(X)$ is such that the connection $\nabla_\CL$ for $\CL$ is $\nabla_\CL=\nabla'_\CL+ia$, with $\nabla'_\CL$ the Chern connection for $\CL$. We will in the following choose $a=0$. If $X$ is a spin manifold, $\slashed{D}$ is the Riemannian Dirac operator ($P$ in \cite[Eq. (3.5.1)]{Gauduchon}) twisted by $\nabla_{\CL}$. 

\subsection{Comparison with the Vafa-Witten twist}
\label{CompVWtwist}
In Section \ref{LocPathIntegral} below we are going to show 
that if we choose the canonical ${\rm Spin}^c$ structure associated 
to an almost complex structure then the $m \to 0$ limit of the partition function of the 
$\CN=2^*$ theory (without the insertion of observables) 
computes, at least formally, the Euler characteristic 
of the moduli space of instantons on $X$. On the other hand, we will 
also show later on that the partition function of the $N=2^*$ theory is 
$S$-duality covariant. These statements should remind one of the 
renowned Vafa-Witten (VW) topological quantum field theory obtained by a 
different topological twist of the $\CN=4$ SYM  \cite{Vafa:1994tf}. Indeed, the $S$-duality transformations of the $\CN=2^*$ partition function discussed in Sections \ref{sec:Evaluate} and \ref{SWcontributions} are identical to those of the VW partition functions. On the other hand, the VW twist and the twist used to define Donaldson theory are decidedly different. It is 
therefore interesting to compare the   $Q$-fixed equations in detail.

The fields appearing in the Vafa-Witten equations are a connection on a 
principal bundle $P\to X$ together with a real section $C \in \Omega^0(X;{\rm ad}P)$ 
and a real self-dual 2-form $B^+ \in \Omega^{2,+}(X;{\rm ad}P)$.   The $Q$-fixed equations of 
the twisted theory are \cite{Vafa:1994tf}:
\be
\label{VWQfixed}
\begin{split}
&F^+_{\mu\nu}+\frac{1}{2}[C,B^+_{\mu\nu}]+\frac{1}{4}[B^+_{\mu\rho},B^+_{\nu\sigma}]\,g^{\rho\sigma}=0,\\
&D_\mu C+D^\nu B^+_{\mu\nu}=0. 
\end{split}  
%See VW (2.57). (2.56) has a typo
\ee  
Refs \cite{MaresThesis, Tanaka:2014} discuss analytical aspects of these equations.

At first sight the matter fields in the adj-SW and VW equations appear to be very different. 
However, under the isomorphism \eqref{KSpinc} we can identify $M_{\dot 1} \in \Omega^{0,2}(X;{\rm ad}P)$ 
and $M_{\dot 2} \in \Omega^{0,0}(X;{\rm ad}P)$. Note that $\Omega^{0,0}(X;{\rm ad} P)$ is the space of \underline{complex}-valued sections of ${\rm ad}P$.
One the other hand, again using the almost complex structure we can identify  
\be
\Omega^{2,+}=R\oplus \left[\Omega^{2,0} \oplus \Omega^{0,2}\right]_{\BR},
\ee
where $R$ is the space of $(1,1)$ forms proportional to the almost K\"ahler form $\omega$. 
Put differently, we can decompose  
\be 
B^+=\kappa\,\omega+\beta +\bar \beta,
\ee 
with   $\beta\in \Omega^{2,0}(X;{\rm ad}P)$, $\bar \beta$ is its complex conjugate and $\kappa$ is a 
real section of ${\rm ad} P$. Under these isomorphisms 
can thus identify $M_{\dot 2}$ with $C-i\kappa$ and $M_{\dot 1}$ with $\bar \beta$. 

Indeed, one can check that on $\mathbb{R}^4$, if we set: 
\be 
\label{CBMbarM}
\begin{split} 
&C=\frac{1}{2}(M_{\dot 2}+\bar M_{\dot 2}),\\
&B^+_{12}=-\frac{i}{2}(M_{\dot 2}-\bar M_{\dot 2}),\\
&B^+_{13}=\frac{1}{2}(M_{\dot 1}+\bar M_{\dot 1}),\\
&B^+_{14}=\frac{i}{2}(M_{\dot 1}-\bar M_{\dot 1}),
\end{split}
\ee 
where the other components of $B^+_{\mu\nu}$ are determined by self-duality 
then the adjoint-SW equations are identical to the VW equations.  The ``first'' equation 
of each pair of equations, which determines the self-dual part of the curvature,  
becomes the same, while the four real equations of the second line of (\ref{VWQfixed}) in $\mathbb{R}^4$, are equivalent to the Dirac equation $\sigma^\mu \partial_\mu M=0$. If we choose a complex structure for $\mathbb{R}^4 \simeq \mathbb{C}^2$ by choosing complex coordinates $z_1=x_1+ix_2$ and $z_2=x_3+ix_4$, the $(1,1)$ component corresponds to $B^+_{12}\,\omega=B^+_{12}\,(\frac{i}{2}dz_1\wedge d\bar z_1+\frac{i}{2}dz_2\wedge d\bar z_2)\in \Omega^{1,1}(\mathbb{C}^2)$, while $\beta=\frac{1}{2}(B^+_{13}-i B_{14})dz_1\wedge dz_2\in \Omega^{2,0}(\mathbb{C}^2)$ and $\bar \beta=\frac{1}{2}(B^+_{13}+i B_{14})d\bar z_1\wedge d\bar z_2\in \Omega^{0,2}(\mathbb{C}^2)$. In particular in this case the VW equations have an $SO(2)$ symmetry: The $U(1)_B$ action on $M_{\dot \alpha}$ becomes an $SO(2)$ action on the separate pairs $(C,B^+_{12})$ and $(B^+_{13},-B^+_{14})$. 

On a general almost complex manifold the above identification of $\CN=2^*$ fields and VW 
fields again reveals that the first of the adjoint-SW equations is identical to the first 
of the VW equations. The relation between the second of the pairs of equations is a little 
more complicated. To understand this relation we begin by writing the   second equation of (\ref{VWQfixed}) in a coordinate independent way as 
\be 
\label{exteriorVW}
d_A C + d_A^\dagger B^+=0,
\ee
with $d_A^\dagger=-*d_A*$. 

%Note that $D\omega\neq 0$. It is more natural to consider a Hermitian connection $\nabla$, such that $\nabla\omega=0$.

The $(0,1)$ component of \eqref{exteriorVW} reads
\be 
\label{DiracCbeta}
\bar \partial_A C-i\bar \partial_A \kappa - i \kappa\,\pi_{0,1}\circ\theta +\bar \partial_A^\dagger\bar \beta+\pi_{0,1}\circ (d_A^\dagger\beta)=0,
\ee 
where we used that $*\,\omega\wedge \eta=i \eta$ for $\eta\in \Omega^{0,1}(X)$. 
The terms with $\beta$ and $\theta$   break the $U(1)_B$ symmetry, since $\alpha$ and $\bar \beta$ are multiplied by $e^{i\varphi}$, while $\beta$ is multiplied by $e^{-i\varphi}$. 

If $X$ is K\"ahler, $\pi_{0,1}\circ (d^\dagger\beta)=0$ and $\theta=0$, (\ref{DiracCbeta}) equals the Dirac equation (\ref{ACSDirac}) with $\alpha=C-i\kappa$. 
For a generic ACS manifold, the term $\pi_{0,1}\circ(d^\dagger_A \beta)$ in Eq. (\ref{DiracCbeta}) is non-vanishing, and expressed in terms of the Nijenhuis tensor defined in Eq. (\ref{Nijenhuis}). One has, in terms of components,
\be 
\pi_{0,1}\circ (d^\dagger\beta)=-\frac{1}{2}\,*  (N_\CJ)^a_{bc}\,\beta_{a\ell}\, e^{b}\wedge e^c\wedge e^\ell.
\ee 
Note well that the right hand side is tensorial in $\beta$.

While a non-vanishing Nijenhuis tensor and $\theta$ obstruct a direct equivalence between the adjoint-SW equations and VW equations, we can determine the deformation of the VW equations to which the adjoint-SW equations are equivalent. To this end, we consider the connection $\nabla_{A,\CJ} :\Omega^0 \oplus \Omega^{2+}\to \Omega^1$, defined by
\be 
\begin{split}
\nabla_{A,\CJ} C & = (d_A +\tfrac{1}{4}\theta)\,C, \\
\nabla_{A,\CJ} B^+ & = d^\dagger_A B^+ + \tfrac{1}{2}\,* (N_\CJ)^a_{bc}\, B^+_{a\ell} e^{b}\wedge e^c\wedge e^\ell +  \tfrac{1}{2}\, (*\theta\wedge \omega) (\omega \lrcorner B^+) - \tfrac{1}{4}\theta\lrcorner B^+\\
&=d^\dagger_A B^+ +\tfrac{1}{2}\,* (N_\CJ)^a_{bc}\, \beta_{a\ell} e^{b}\wedge e^c\wedge e^\ell
+\tfrac{1}{2}\,* (N_\CJ)^a_{bc}\,  \bar \beta_{a\ell} e^{b}\wedge e^c\wedge e^\ell\\
& \quad + \tfrac{3}{4}\, \kappa\,*\theta\wedge \omega - \tfrac{1}{4}\theta\lrcorner (\beta+\bar \beta).
\end{split}
\ee 
Using that $*\omega^2=2$, and $*\theta\wedge \omega=\theta\lrcorner \omega$, one verifies that the $(0,1)$ component of 
\be 
\label{VWnablaN}
 \nabla_{A,\CJ}C+\nabla_{A,\CJ} B^+=0,
\ee 
equals the Spin$^c$ Dirac equation (\ref{ACSDirac}) for an ACS manifold. Thus the $\CN=2^*$ equations are equivalent to the ``deformed" VW equations, (\ref{VWnablaN}) together with the first equation of (\ref{VWQfixed}).

The VW equations and the deformed VW equations \eqref{VWnablaN} differ by a 
compact operator on the space of fields. This leads one to conjecture that the 
moduli space of solutions will be topologically equivalent. Indeed we will 
confirm this with our computations. In the K\"ahler case the $Q$-fixed point equations are 
equivalent without deformation and a nice check on our computations is the agreement 
of our partition functions
(in the $m \to 0$ limit) with the known expressions for VW theory on  K\"ahler
manifolds given in \cite{Vafa:1994tf, Dijkgraaf:1997ce}. 
%For $b_2^+>1$, it can be
%expressed as a sum of Seiberg-Witten invariants multiplying functions
%of $\tau_{\rm uv}$, $\bfk_m$ and the topological data. 

It is interesting to compare the instanton and Abelian components in the adjoint-SW 
and VW equations. For $X$ a K\"ahler manifold, the VW equations are left invariant under the $U(1)_B$ symmetry, and the fixed point locus consists of two components \cite{Vafa:1994tf}. The instanton component corresponds to $B=C=0$ and $F^+=0$. For the abelian component\footnote{At other places in the literature, for example \cite{Dijkgraaf:1997ce, Tanaka:2017jom,  Gottsche:2017vxs} this component is referred to as the ``monopole branch''. Since the $Q$-fixed equations of low energy
    effective theory near each of the cusps $u_j$ are the SW monopole
    equations, we have opted to refer to this component as the abelian
  component.}, the gauge bundle is a sum of two line
  bundles $\CE=T\oplus T^{-1}$. Furthermore, $C=0$, and $B$ is a sum of a
  $(2,0)$-form $\beta$ and a $(0,2)$-form $\bar \beta$, which are of
  the form $\left(\begin{array}{cc} 0 & 0 \\ * & 0 \end{array}
  \right)$ or $\left(\begin{array}{cc} 0 & * \\ 0 & 0 \end{array}
  \right)$. The results for K\"ahler manifolds were earlier obtained using a mass deformation to
$\CN=1^*$ \cite{Vafa:1994tf}, which demonstrated that the abelian component of the theories
gives rise to the same abelian monopole equations \cite{Dijkgraaf:1997ce}. 
 
The vanishing theorem of \cite{Vafa:1994tf} states that the abelian component does not contribute for K\"ahler manifolds with curvature $R\geq 0$. These manifolds include K3 and the Fano surfaces, such as
$\mathbb{P}^2$, the Hirzebruch surfaces $\mathbb{F}_0$ and
$\mathbb{F}_1$. For more general algebraic surfaces with $b_2^+\geq 3$, a mathematical theory has recently developed for the
contribution from the abelian component 
\cite{Tanaka:2017jom, Gholampour:2017bxh, GHOLAMPOUR2020107046, Gholampour:2017dcp, Laarakker_2020}. It confirms and extends results by Vafa
and Witten \cite{Vafa:1994tf} for such four-manifolds. Analogously, explicit results for the contribution from the instanton component have been determined mathematically for such manifolds \cite{Gottsche:2017vxs, Gottsche:2019vbi, Gottsche:2020ale, Gottsche:2018meg}.

\subsection{Virtual dimensions of moduli spaces for instantons and monopoles}
\label{VirDims}
We recall in this subsection a number of aspects of bundles,
characteristic classes and moduli spaces.

\subsubsection*{Bundles and characteristic classes}
Let $G$ be a Lie group, and $P$ a principal $G$-bundle. We let $\CP_r$
be the complex vector bundle $r(P)\otimes \mathbb{C}$ associated to
the representation $r$. We are mostly interested in this article in
the groups $G=SU(2)$, $U(2)$ or $SO(3)$.  For the semi-simple groups
$SU(2)$ and $SO(3)$, we
let $\CA={\rm ad}(P)\otimes \mathbb{C}$ be
the complex vector bundle associated to their adjoint
representation, and $\CE={\rm
  sp}(P) \otimes \mathbb{C}$ the complex
bundle associated to the two-dimensional (spinorial) representation of the Lie group. 
  
Since the Lie algebra's of $SU(2)$ and
$SO(3)$ are isomorphic the principal bundle are closely related. If
the second Stiefel-Whitney class $w_2(P)$ vanishes for a principal $SO(3)$-bundle, it can be lifted to
a principal $SU(2)$ bundle, while it can be lifted to a principal $U(2)$-bundle
for any $w_2(P)\in H^2(X,\mathbb{Z}/2 \mathbb{Z})$. The first Chern
class of the complex associated bundle $\CE$ satisfies $c_1(\CE)=w_2(P)\mod 2H^2(X,\mathbb{Z})$. 

There are a number of relations between the Chern classes of the
different bundles. The adjoint representation gives an endomorphism of
the fundamental representation. As a result, the adjoint bundle $\CA$
for $G=U(2)$ can be understood as the tensor product, $\CA\cong
\CE^\lor \otimes \CE$. Using the familiar formulas for the Chern
character of a complex vector bundle $E$, ${\rm ch}(E)={\rm rank}(E) +c_1(E) +
c_1(E)^2/2-c_2(E)$, and its property under the tensor
product, one finds   
$$
c_1(\CA)=0,\qquad c_2(\CA)=4c_2(\CE)-c_1(\CE)^2. 
$$
Moreover, if $V$ is a complex bundle, its complexification $V\otimes_{\mathbb{R}}
\mathbb{C}\cong V\oplus iV$, is isomorphic to $E\oplus \bar E$. As a
result, the Chern classes of $V\otimes \mathbb{C}$ and $V$ are related as
$$
c_1(V\otimes \mathbb{C})=0, \qquad c_2(V\otimes \mathbb{C})=2c_2(V)-c_1(V)^2.
$$
Applying this to the tangent bundle of an almost complex manifold $X$,
we find that the first Pontryagin
class $p_1(X)=p_1(T_X)$ gives the signature of $X$,
\be
\int_X p_1(X)=-\int_X c_2(T_X\otimes \mathbb{C})= 3\sigma,
\ee 
where we used (\ref{K2chisigma}).

For a complex bundle $\CP_r$ with connection, the Chern character is expressed in
terms of the (real) field strength $F$ as 
\be
\label{chP}
{\rm ch}(\CP_r)={\rm Tr}_r[\,\exp(F/2\pi)\,], 
\ee 
where the trace is taken in the representation $r$. We define the instanton number $k(P)$ as
$$k(P)=-\frac{1}{8\pi^2} \int_X {\rm Tr}_{\bf 2}(F^2),$$ where the trace 
is taken in the two-dimensional spinorial representations of the Lie
algebras. In terms of Chern classes $c_j(\CE)$, $k(P)$ reads,
\be
\label{kPs}
\begin{split}
&SU(2):\quad  k(P)=c_2(\CE), \\
&U(2):  \quad  k(P)=c_2(\CE)-\frac{1}{2} c_1(\CE)^2, \\
&SO(3): \quad k(P)= c_2(\CE)-\frac{1}{4} c_1(\CE)^2.
\end{split} 
\ee 
Note the trace in the adjoint representation, ${\rm
  Tr}_{\bf 3}(F^2)$, is 4 times ${\rm
  Tr}_{\bf 2}(F^2)$. We have more generally for the instanton number $k(P)=-1/(16h^\lor\pi^2)  \int_X {\rm
  Tr}_{\dim(G)}(F^2)$, where ${\rm
  Tr}_{\dim(G)}$ is the trace in the adjoint representation of $G$,
and $h^\lor_G$ is its dual Coxeter number. (Recall $h^\lor_{SO(3)}=2$.)

\subsubsection*{Atiyah-Hitchin-Singer index formula}
The Atiyah-Hitchin-Singer index formula is an important tool for the
determination of the expected or virtual dimensions of the moduli
spaces. To state this formula, we consider a principal $G$-bundle $P$, and a
Dirac operator coupled to the complex associated $\CP_r$ in representation $r$. 
\be
{\rm \bf D}_A:\Gamma(S^+\otimes \CP_r)\to \Gamma(S^-\otimes \CP_r).
\ee 
If $X$ is not a spin-manifold, we consider $S^{\pm} \otimes \CP_r$ as
Spin$^c$ bundles. Its index is defined as
\be
{\rm Ind}({\bf D}_A)={\rm dim}({\rm Ker}({\bf D}_A))-{\rm dim}({\rm Ker}({\bf D}_A^\dagger)).
\ee 
The Atiyah-Hitchin-Singer index theorem expresses the index as
\be
\label{Dindex}
{\rm Ind}({\rm \bf D}_A)=\int_X {\rm ch}(\CP_r)\,\hat A(X),
\ee 
where for $X$ a four-manifold, $\hat A(X)=1-p_1(X)/24$.
 
\subsubsection*{Instantons}
Let $\CM_{\rm i}(k)$ be the moduli space of anti-self dual
connections with instanton number $k$. The complex dimension of $\CM_{\rm i}(k)$
is the index of the operator 
\be 
\delta_A=d_A^\dagger+d_A^+: \Omega^1(\CA) \to
\Omega^0(\CA)\oplus \Omega^{2+}(\CA).
\ee
To this end, we recall for ${\rm Index}(\delta_A^\dagger)=-{\rm
  Index}(\delta_A)$ \cite{Atiyah:1978}
\be 
\label{IndexSD}
{\rm Index}(\delta_A^\dagger)=-2\,h^\lor_G\, k(P) +{\rm dim}(G)\,(\chi+\sigma)/4.
\ee 
As a result, the virtual real dimension of the moduli space of $G=SO(3)$ instantons is
\be
\label{dimCMinst} 
\begin{split}
{\rm vdim}(\CM^{\rm i}_{k,\bfmu})&= 4\,h^\lor_G\, k(P) -{\rm dim}(G)\,(\chi+\sigma)/2\\
&=8\,k -\frac{3}{2}\,(\chi+\sigma).
\end{split} 
\ee

\subsubsection*{Abelian monopole equations (Seiberg-Witten equations)}
The monopole equations (\ref{MonopoleEqs}) consist of two equations,
and the deformation operator consists therefore of two operators
$\cD_A=(\delta_A, {\rm \bf D}_A)$. As discussed in Section \ref{SpincStruct}, the spinors are sections of a
Spin$^c$ bundle $W^+$, which can be informally written as $S^+\otimes
\CL^{1/2}$ with $c_1(\CL)=c_1(\mathfrak{s})$. The index of $\cD_A$ splits,
${\rm Index}(\cD_A)={\rm Index}(\delta_A)+{\rm Index}( {\rm \bf D}_A)$.
As before, we have  
\be
{\rm Index}(\delta_A)=-{\rm Index}(\delta_A^\dagger)=-\frac{1}{4}(\chi+\sigma),
\ee
since the Chern character of the associated bundle ${\rm ch}(\CE^\lor \otimes \CE)$ equals 1. Furthermore, for the index of the Dirac
operator coupled to the Spin$^c$ bundle, the Chern character is
${\rm ch}(\CA)=e^{c_1(\mathfrak{s})/2}$. As a result, the index is 
\be 
\label{IndexDA}
{\rm Index}({\rm \bf D}_A)=\frac{c_1(\mathfrak{s})^2-\sigma}{8}.
\ee
Adding the two contributions and multiplying by two, we find that the virtual
real (expected) dimension of the moduli space of solutions to the monopole equations is \cite{Witten:1994cg}
\be 
\label{dimSWmodspace}
{\rm vdim}(\CM^{\rm m}_\mathfrak{s})=\frac{1}{4}\,(c_1(\mathfrak{s})^2-(2\chi+3\sigma))\in 2\BZ.
\ee
Since $c_1(\mathfrak{s})$ is a characteristic vector for $L$,
$c_1(\mathfrak{s})^2=\sigma \mod 8$, and $(2\chi+2\sigma)/4=1-b_1+b_2^+$
rhs is indeed an integer if $X$ is smooth and compact.  
Taking $c_1(\mathfrak{s})=c_{\rm uv}$ we define 
\be
\label{defell}
\ell :=\frac{1}{8}\left(c_{\rm uv}^2 -2\chi-3\sigma \right)\in \mathbb{Z}.
\ee
The integer $\ell$ will play an important role in many formulae below. 
 
\subsubsection*{Non-abelian monopole equations}
The non-abelian monopole equations are the $Q$-fixed equations of the
$\CN=2^*$ theory, and the generalization of the
monopole equations (\ref{MonopoleEqs}) to an arbitrary gauge group $G$. We
are interested in $G=SO(3)$. As for the abelian monopole equations,
the deformation operator is a combination of two operators,
$\cD_A=(\delta_A, {\rm \bf D}_A)$. The contribution of $\delta_A$ is $-1$ times
(\ref{IndexSD}). For the index of ${\rm \bf D}_A$, ${\rm ch}(\CP_r)$
in the index formula (\ref{Dindex}) is given by $e^{c_1(\mathfrak{s})/2}\,{\rm
  ch}(\CA)$. This index then reads 
\be
\label{IndDA}
{\rm Index}({\rm \bf D}_A)=- 2\, h^\lor_G\, k(P)+{\rm dim}(G)\, \frac{c_1(\mathfrak{s})^2-\sigma}{8}.
\ee
Thus the two dimensions added together give for the virtual real dimension of
the moduli space of non-Abelian monopole equations $\CM^{Q}_{k,\bfmu,\mathfrak{s}}$,
\be
\label{dimCMm}
{\rm vdim}(\CM^{Q}_{k,\bfmu,\mathfrak{s}})=\dim(G)\, \frac{c_1(\mathfrak{s})^2-(2\chi+3\sigma)}{4}.
\ee 
Interestingly, this is independent of the instanton number $k$.
This is closely related to the fact that the correlation functions 
of any fixed topological operator $\CO(x)$ with $x\in H_*(X)$ 
are infinite series in $q_{\rm uv}$. If $b_1=0$,
$c_1(\mathfrak{s})^2$ needs to be sufficiently positive for the
${\dim}(\CM^{Q}_{k,\bfmu,\mathfrak{s}})\geq 0$. We deduce moreover that if $\mathfrak{s}$
corresponds to an almost complex structure, ${\dim}(\CM^{Q}_{k,\bfmu,\mathfrak{s}})= 0$, and the rank of the obstruction bundle equals the
dimension of the instanton moduli space (\ref{dimCMinst}). Our
main interest is in gauge groups with $\dim(G)=3={\rm odd}$, such that
${\dim}(\CM^{\rm i}_{k,\bfmu})= 1+b_2^+ \mod 2$. Correlation 
functions are therefore only non-vanishing for $b_2^+$ odd, 
since all observables have even ghost number for $b_1=0$.  

Finally, we can consider $N_f$ spinors in the fundamental
representation of  
$SO(3)$. Then ${\rm ch}(\CE)=2-k(P)$. The index of the Dirac
operator becomes in this case
\cite{Labastida:1995zj, Feehan:2001jc}
\be    
-k+\frac{1}{4}(c_1(\mathfrak{s})^2-\sigma).
\ee
The dimension of the moduli space of anti-self dual connections
coupled to $N_f$ spinors in the fundamental representation is thus
\be
{\rm vdim}(\CM^{{Q}, N_f}_{k,\bfmu,\mathfrak{s}})=2(4-N_f)k+\frac{1}{2}(N_f\,c_1(\mathfrak{s})^2-3\chi-(3+N_f)\sigma).
\ee  

\subsection{Localization of the $\CN=2^*$ path integral}
\label{LocPathIntegral}
This final subsection discusses the geometric content of the topological correlators of the $\CN=2^*$ theory. Before discussing the infinite dimensional path integral of the $\CN=2^*$ theory, we first discuss a finite dimensional model, namely the Mathai-Quillen model.

\subsubsection*{Mathai-Quillen formalism}
To understand the geometric content of the path integral, we review a few aspects of the Mathai-Quillen formalism
\cite{Atiyah:1990tm, Cordes:1994fc, MooreNotes2017, Mathai:1986tc}. Topologically twisted $\CN=2$
Yang-Mills theory is an infinite  
dimensional analogue of the Mathai-Quillen formalism. Let $M$ be a
compact oriented manifold of even dimension $d$, and $V$ a real
oriented vector bundle of even rank $r\leq d$ with $SO(r)$ connection
$A$, and
metric $g_{ab}$, $a,b=1,\dots,r$. Recall that the Euler class of $V$
is given by 
\be 
\label{EulerClass} 
{\rm Eul}(V)=\frac{1}{(2\pi)^{r/2}}\, {\rm Pf}(F),
\ee
where $F$ is the 2-form field strength of the connection $A$.

Let $s$ be a generic section of $V$. The connection $A$ defines a
linear operator for each point $p\in M$,
\be
\nabla_A s(p):\quad T_pM \to V_p.
\ee
Let $E\to M$ be the bundle with fiber ${\rm Ker}(\nabla_A s(p))$ over
the point $p\in M$. Similarly, let $W\to M$ be the bundle with fiber
${\rm Cok}(\nabla_A s(p))$ over $p$.
We then have the exact sequence
\be
\label{exseq}
0\longrightarrow E\longrightarrow TM \longrightarrow V \longrightarrow W \longrightarrow 0.
\ee

Let us recall a basic construction in supergeometry
\cite{Deligne:1999, Leites_1980, Witten:2012bg}. Given the
vector bundle $V\to M$. We denote by $\Pi V$ the superspace,
where the coordinates of the fiber of $V$ are considered odd.  For
$V=TM$, we let furthermore $\widehat M=\Pi TM$. We then have the
isomorphism 
\be
C^\infty(\widehat M)\cong \Omega^*(M),
\ee
since the odd
variables $\psi^i$, are linear functions on the fibers of $TM$,
$\psi^i(\partial_{x^j})=\delta^i_j$.

The MQ model has a topological symmetry $Q$, and a ghost number operator
$U$. We introduce local coordinates $x^i$, $i=1,\dots,d$ on $M$, and
fermionic partners $\psi^i$. The ghost numbers of $x^i$ and $\psi^i$ 
are respectively $0$ and $1$. We introduce in addition the anti-ghost
multiplets for $a=1,\dots, r$, consisting of a fermion $\chi^a$ (with ghost number -1) and
boson $H^a$ (with ghost number 0). 

 The action of $Q$ on the fields is
\be
\begin{split}
[Q,x^j]&=i\psi^j,\\
[Q,\psi^j]&=0,\\
[Q,\chi^a]&=H^a -i\psi^j A_{jb}^{a} \chi^b,\\
[Q,H^a]&=i\psi^j A_{jb}^{a} H^b-\tfrac{i}{2} \psi^i\psi^j F^a_{ijb} \chi^b.\\
\end{split}
\ee

Let
$s:M\to \mathbb{R}^r$ be a section of $V$. We consider the $Q$-exact ``Lagrangian''
\be
L=\frac{1}{2\hbar}g_{ab} H^a(H^b-is^b) +\frac{1}{\hbar} g_{ab}\chi^a
(\nabla_{A,i}x^b)\psi^i -\frac{1}{2\hbar} F_{ijab} \psi^i\psi^j\chi^a\chi^b.
\ee 
We define the partition function as
\be
Z=\frac{1}{(2\pi)^{(d+r)/2}}\int dx\, d\psi\,d\chi\,dH\, e^{-L}.
\ee
Since $L$ is $Q$-exact, $Z$ is independent of $s^a$ and $g_{ab}$; it depends only on topological properties of
$M$ and $V$. For $s^a=0$, we find
\be
\label{ZPfF}
Z= \int_{\Pi V} dx\,d\psi\, {\widehat {\rm Eul}}(V), 
\ee 
where
\be
{\widehat {\rm Eul}}(V)=\frac{1}{(2\pi)^{d/2}} {\rm
  Pf}(F_{ij}\psi^i\psi^j). 
\ee
Since the $\psi^i$ are odd, $Z$ vanishes for $r\neq d$. For $r=d$,
integration over the $\psi^i$ gives 
the Euler class (\ref{EulerClass}),
\be
Z=\int_{M} dx\, { {\rm Eul}}(V)=\chi(V),\qquad r=d. 
\ee
For a non-vanishing section $s^a$, $Z$ can also be evaluated and reduces to a weighted sum
of signs. Let $S(s^a)\in M$ be the collection of points where the
sections $s^a$ vanish. This is also the on-shell $Q-$fixed locus of
the fields, $M_Q$. 
Then,
\be 
\label{ZSsa}
Z=\sum_{x\in S(s)}  \mathrm{sgn}(\mathrm{det}(\nabla_{A,i}s^b)).
\ee
Equations (\ref{ZPfF}) and (\ref{ZSsa}) are well-known expressions for $\chi(V)$.
The equality is an example of the Thom isomorphism. 

If we have more variables than equations, $r<d$, one can insert observables with total ghost number $d-r$ to reach
a non-vanishing answer. Let $\iota: S(s) \hookrightarrow M$ be the
inclusion of $S(s)$. If $\nabla_As$ has maximal rank for each point on $S(s)$, the
dimension of $S(s)$ equals the dimension of ${\rm
  Ker}(\nabla_As)=d-r$. For an observable $\CO$, which corresponds
to the form $\omega_\CO$ on $M$ with degree $d-r$, we then arrive at
\be
\label{ZEulV}
\begin{split}
Z&=\int_M  \omega_{\CO}\wedge {\rm Eul}(V)\\
&=\int_{S(s)}  \iota^*(\omega_{\CO}).
\end{split}
\ee  

More generally, $\nabla_As(p)$ is not necessarily surjective, and can have a non-trivial
cokernel $V_p/{\rm Im}(\nabla_As(p))$. This gives rise to the Euler class ${\rm
  Eul}(W)$. For an observable $\CO$, which corresponds
to the form $\omega_\CO$ on $M$, (\ref{ZEulV}) is then modified to
\be
\label{ZEulVF}
\begin{split}
Z&=\int_M  \omega_{\CO}\wedge {\rm Eul}(V)\\
&=\int_{S(s)}  \iota^*(\omega_{\CO}) \wedge {\rm Eul}(W),
\end{split}
\ee
where $W$ is the cokernel bundle introduced above (\ref{exseq}). 
We can enhance this model by considering a model with the action of a
Lie group $G$. In that case, the Euler class becomes the equivariant
Euler class with respect to $G$.

\subsubsection*{$\CN=2^*$ path integral}
Let us first consider the path integral of the $N_f=0$ theory with only a vector multiplet. The topological twisted vector multiplet is an infinite dimensional
version of the MQ formalism, where
$d-r={\dim}_{\mathbb{R}}(\CM^{\rm i}_{k,\bfmu})$. The path integral reduces to an
integral over $\CM^{\rm i}_{k,\bfmu}$. The Donaldson observables are
\be
\label{Dobservables}
\begin{split}
u_0&=\frac{1}{16\pi^2}\mathrm{Tr}[\phi^2],\\
I_0(\bfx)&=\frac{1}{4\pi^2}\int_\bfx \mathrm{Tr}\left[\frac{1}{8} \psi\wedge \psi-\frac{1}{\sqrt{2}}\phi\,F\right].
\end{split}
\ee
where $\bfx\in H_2(X,\mathbb{Q})$. The first observable, $u_0$ has ghost
number four, and corresponds to a four-form $\omega_{u_0}$ on instanton
moduli $\CM^{\rm i}_{k,\bfmu}$. Similarly, the integrand of the surface observable
$I(\bfx)$ has ghost number 2 and corresponds to a two-form $\omega_{\bfx}$ on $\CM^{\rm i}_{k,\bfmu}$. It is
natural to include these in exponentiated form
\be
\left< e^{2p\,u_0/\Lambda_0^2+I_{0}(\bfx)/\Lambda_0}\right>.
\ee
Since $u_0$ has ghost number 4, the fugacity $p$ naturally has ghost
number $-4$. Similarly, we associate ghost number $-2$ to
$\bfx$. Donaldson polynomials of the pure $SU(2)$ theory are homogenous polynomials with
monomials of the form $p^s\,\bfx^{t}$, satisfying
$-4s-2t=-\mathrm{dim}(\CM_{k,\bfmu})$. 

We consider next the $\CN=2^*$ theory, with the massive
hypermultiplet. The topological observables (\ref{Dobservables}) are
closely related for this theory. While the surface operator is
essentially identical, Equation (\ref{SWlimit}) shows that the
$\CN=2^*$ analogue of Donaldson's point observable $\mu(p)$ is shifted, $u_{\rm D}=u+\frac{1}{8}e_1(\tau_{\rm uv})m^2$. We will explain
this in more detail in Section \ref{SecObservables}. 
From the perspective of the MQ formalism, the adjoint hypermultiplet is
corresponds to a bundle over $\CM^{\rm i}_{k,\bfmu}$ together
with anti-ghost multiplet, leading to the insertion of the Euler
class of this bundle \cite{LoNeSha}. More precisely, the bundle is the index bundle
of the Dirac operator in the adjoint representation of $SO(3)$. The
hypermultiplet fields, $M_{\dot \alpha}$, give rise to an infinite number of
additional variables and equations, whose difference is the rank of the index
bundle (\ref{IndDA}). 
 
The path integral of the UV theory, reduces to an integral over the
solution space of non-Abelian monopole equations
(\ref{MonopoleEqs}). From the perspective of the MQ formalism, these
equations correspond to the zero locus of sections over an ambient
space. It is most convenient to work to work with the fixed point locus of the $U(1)_B$ symmetry. The monopole equations then correspond to
sections over the union $\CM^{\rm i}_{k,\bfmu} \cup \CM^{\rm a}_{k,\bfmu,\mathfrak{s}}$ of the instanton
 component and abelian component respectively. For $\CM^{\rm i}_{k,\bfmu}$, the
hypermultiplet fields $M_{\dot \alpha}=0$ vanish, whereas for $\CM^{\rm a}_{k,\bfmu,\mathfrak{s}}$
the gauge connection is reducible. 

The equivariant parameter of the $U(1)_{B}$ symmetry is the mass $m$. Since $m$ has ghost number $2$,
correlation functions
will evaluate to sums of monomials of the form $m^r p^s\,\bfx^{t}$, 
where $r$, $s$ and $t$ satisfy 
\be
\label{SelectionRule}
2r-4s-2t=-\mathrm{vdim}(\CM^{Q}_{k,\bfmu,\mathfrak{s}}), 
\ee
with $\mathrm{vdim}(\CM^{Q}_{k,\bfmu,\mathfrak{s}})$ given in
(\ref{dimCMm}).
 
Let ${\rm Ind}({\rm \bf D}_A))$ be the
index bundle of ${\rm \bf D}_A$. This is a formal difference
of bundles; it is an element of K-theory or a virtual bundle,
with rank ${\rm Index}({\rm \bf D}_A)$ as in (\ref{IndDA}). We let
furthermore $W$ be the bundle with fibers ${\rm Cok}({\rm \bf
  D}_A^\dagger)$. This bundle $W$ is the analogue of the bundle $W$
introduced above (\ref{ZEulVF}) in the
Mathai-Quillen formalism. Correlation functions then evaluate to
\be
\label{O1Op0}
\begin{split}
\left< \CO_1\dots \CO_p\right>&=\sum_k  q_{\rm uv}^k\,m^{-{\rm Index}({\rm \bf D}_A)}
\int_{\CM^{Q}_{k,\bfmu,\mathfrak{s}}}\,c_\ell(W)\,\omega_1\dots \omega_p,
\end{split}
\ee
with ${\rm Index}({\rm \bf D}_A)$ as in (\ref{IndDA}).

We can express this as an integral over $\CM^{\rm i}_{k,\bfmu} \cup \CM^{\rm
  a}_{k,\bfmu,\mathfrak{s}}$. 
The equivariant Euler class with respect to $U(1)_B$ is given in terms of splitting classes
$x_i$ as $\prod_{i=1}^{{\rm rk}(E)} (x_i+m)$. We can express this in
terms of Chern classes $c_\ell$ as
\be
m^{-{\rm rk}(E)} \prod_{i=1} (x_i+m)=\sum_{\ell=0}^{{\rm rk}(E)} \frac{c_\ell}{m^\ell}.
\ee 
Note that if $\mathfrak{s}$ corresponds to an almost complex structure, the
index bundle ${\rm Ind}({\rm \bf D}_A^\dagger)$ (or obstruction bundle) is equal to the
tangent bundle of $\CM^{\rm i}_{k,\bfmu} \cup \CM_{k,\bfmu,\mathfrak{s}}^{\rm
  a}$. More generally, physical correlation functions evaluate to
intersection numbers of Chern classes and differential forms
corresponding to observables \cite[Equation (5.13)]{LoNeSha}. In the
present case of $\CN=2^*$, we arrive at
\be
\label{O1Op}
\begin{split}
\left< \CO_1\dots \CO_p\right>&=\sum_k  q_{\rm uv}^k\,m^{-{\rm Index}({\rm \bf D}_A)}
\int_{\CM^{\rm i}_{k,\bfmu} \cup \CM^{\rm
  a}_{k,\bfmu,\mathfrak{s}}} \sum_{\ell\geq 0} \frac{c_\ell}{m^\ell}
\,\omega_1\dots \omega_p\\
&= m^{-3\ell+D_\omega} \sum_k q_{\rm uv}^k\\
&\quad \times \left[\int_{\CM^{\rm i}_{k,\bfmu}} c_\ell\,\omega_1\dots
  \omega_p+\int_{\CM^{\rm a}_{k,\bfmu,\mathfrak{s}}} c_\ell\,\omega_1\dots \omega_p\right],
\end{split}
\ee 
where the $\omega_j$ are the differential forms on either $\CM^{\rm
  i}_{k,\bfmu}$ or $\CM^{\rm a}_{k,\bfmu,\mathfrak{s}}$,
corresponding to the topological observables $\CO_j$, and $D_\omega={\rm
  deg}(\omega_1\dots \omega_p)$. The latter can also be identified as
the $U(1)_R$-charge of $\CO_1\dots\CO_p$.  The factor
$m^{-{\rm Index}({\rm \bf D}_A)}$ is a consequence of integrating out
the massive hypermultiplet fields \cite{Hyun:1995hz}. We will 
find that the physical partition function of the $\CN=2^*$ theory 
is multiplied by a factor $\Lambda^{3\ell}$, where
$\Lambda$ is the scale. 

We can similarly write down the correlation function of exponentiated
observables. For example, if $\CO$ is the point observable with mass
dimension 2, then
\be
\label{O1Opv2} 
\begin{split}
\left< e^{\CO/\Lambda^2} \right>&=\sum_k \sum_{s=0}^\infty \sum_{\ell\geq 0} q_{\rm uv}^k\,m^{-{\rm Index}({\rm \bf D}_A)}\\
&\quad \times \left[  \int_{\CM^{\rm i}_{k,\bfmu}} 
\frac{1}{s!} \frac{c_\ell}{m^{\ell}\Lambda^{2s}}\,\omega_\CO^s+\int_{\CM^{\rm
  a}_{k,\bfmu,\mathfrak{s}}}
\frac{1}{s!} \frac{c_\ell}{m^{\ell}\Lambda^{2s}}\,\omega_\CO^s\right].
\end{split}
\ee 

When the UV ${\rm Spin}^c$ structure does not correspond to an almost complex structure
the obstruction bundle will be different and the cohomology classes
$c_\ell$ in (\ref{O1Op}) will not be the Chern classes of the tangent bundle, but rather
those of a different bundle (namely, the obstruction bundle).

It is interesting to consider the asymptotics of these formulae for $m \to 0$ and
$m \to \infty$, respectively. Physically, the limit $m\to 0$ corresponds to the
case of $\CN=4$ SYM. When we choose the UV ${\rm Spin}^c$ structure to be an almost
complex structure the leading terms in the $m \to 0$ limit are expected to be closely related to  Vafa-Witten theory as discussed in Section \ref{CompVWtwist}.
Indeed, for a fixed value of $k$ in the sum in (\ref{O1Op}) the leading term will be the
integral over $\CM^{\rm i}_{k}$ of the highest Chern class of
the tangent bundle thus, formally, producing the Euler character
$\chi(\CM^{\rm i}_{k,\bfmu})$ of instanton
moduli space. This leads to a conundrum, since the moduli space
$\CM^{\rm i}_{k,\bfmu}$
changes by a bordism under metric deformations of $X$, while the 
Euler character is not a bordism invariant. On the other hand, we
expect that the path integral is a topological invariant, at least for
four-manifolds with $b_2^+>1$. The envisioned resolution
is that the abelian solutions of (\ref{MonopoleEqs}) contribute
to the partition function, and that this contribution is also not
bordism invariant. However, the sum of the contributions is expected
to be a bordism invariant, such that a change of the Euler number
$\chi(\CM^{\rm i}_{k,\bfmu})$ due to a bordism, is exactly cancelled by a
change due to the abelian solutions.\footnote{We thank Richard Thomas for correspondence on
  these aspects.} 
 
In addition to the limit $m\to 0$, there is the limit $m \to \infty$
(with $q_{\rm uv} \to 0$ such that $4 m^4 q_{\rm uv} = \Lambda_0^4$ is held
fixed). This limit should physically reproduce pure $\CN=2$ SYM. In the twisted theory we therefore expect the leading
term in the $m \to \infty$ limit to reproduce
the Donaldson invariants. Indeed equation (\ref{O1Op}) makes this clear: The leading
term in the $m \to \infty$ asymptotics is given by the term with $\ell=0$: The
Chern-classes drop out and, for \underline{any} choice of UV ${\rm Spin}^c$ structure
we should reproduce the Donaldson invariants.
 
Equation (\ref{O1Op}) ignores the singularities in the moduli space
$\CM^{\rm i}_{k,\bfmu}$ due to reducible connections.
is that the obstruction ``bundle'' (really, a sheaf over a stack),
  i.e., the index ``bundle,'' cannot be identified
with $\ker({\rm \bf D}_A^\dagger)$.  Nevertheless, we would conjecture that when the
index in (\ref{IndDA}) is sufficiently negative that the obstruction bundle $E$ is
a well-defined vector bundle which can be identified with  the bundle of
fermion zeromodes with fiber $\ker({\rm \bf D}_A^\dagger)$.

\section{Path integral of the $\CN=2^*$ effective IR theory} 
\label{FormUplaneInt}
We formulate in the $u$-plane integral for the $\CN=2^*$ theory in this section. We start in Section \ref{sec:higher_rank} and \ref{SpincCoupling} with the effective Lagrangian and its coupling to the Spin$^c$ structure. Section \ref{Intmassive} deals with the analogue of an important non-trivial phase of Donaldson-Witten theory for the $\CN=2^*$ theory. After determining the sum over $U(1)$ fluxes in Section \ref{SFluxes}, we  determine the $u-$plane integrand in Section \ref{Formuplane}. We demonstrate that the integral is well-defined in Section \ref{SingleValued}. We discuss moreover the limit to the $N_f=0$ theory, inclusion of observables and regularization of the $u$-plane integral.

\subsection{General twisted, effective Lagrangian}
\label{sec:higher_rank}
Having discussed the formulation of the UV theory, we turn to the low
energy effective theory. To couple the hypermultiplet to a ${\rm Spin}^c$
structure, we consider first the low energy effective theory for a gauge group
with rank $r>1$, whose effective theory has gauge group
$U(1)^r$. For four-manifolds with $b_2^+=1$, the $u$-plane integral reduces to an integral over zero
modes \cite{Moore:1997pc}. To write this Lagrangian, we let the indices $I,J,K$ run
from 1 to $r$ for lower indices, and $(1)$ to $(r)$ for upper indices. The bosonic zero-modes of the twisted effective theory are the scalars $a^I$ and field strengths $F^I$ (with positive/negative definite projections $F^I_\pm$). When $b_1=0$, the fermionic zero modes are the 0-forms $\eta^I$ and self-dual 2-forms $\chi^I$. We have in addition the auxiliary fields $D^I$. Together with $\tau_{IJ}=\partial_{a^I}\partial_{a^J}\CF$, $y_{IJ}=\mathrm{Im}(\tau_{IJ})$ and $\CF_{IJK}=\partial_{a^I}\tau_{JK}$, the zero mode
Lagrangian reads \cite{Marino:1998bm}\footnote{This Lagrangian is
  multiplied by two compared to \cite{Marino:1998bm}, since we work
  here in the $\Gamma(2)$ convention for the duality group.}
\be
\label{Lhigherrank}
\begin{split}
\CL&=\frac{i}{8\pi} \left( \bar \tau_{IJ} F_+^I\wedge F_+^J + 
  \tau_{IJ} F_-^I\wedge F_-^J\right)-\frac{1}{4\pi} y_{IJ}D^I\wedge D^J \\
&\quad 
+\frac{i\sqrt{2}}{8\pi} \bar \CF_{IJK} \eta^{I}\chi^J\wedge (D+F_+)^K.
\end{split}
\ee 
The action of the BRST symmetry $Q$  on the zero modes is 
\be 
\label{higherranksusy}
\begin{split}
&[Q,A^I]=\psi^I=0,\qquad \qquad[Q,\psi^I]=4\sqrt{2}\,da^I,\\
&[Q,a^I]=0,\qquad \qquad \qquad[Q,\bar a^I]=\sqrt{2}i\,\eta^I,\\
&[Q,\eta^I]=0,\qquad \qquad \qquad[Q,\chi^I]=i(F_+^I-D_+^I),\\
&[Q,D^I]=(d\psi^I)_+=0.
\end{split}
\ee
We can express $Q$ in terms of the fields as
\be
\label{CQderivatives}
Q=\sqrt{2}i\eta^I\frac{\partial}{\partial \bar a^I}+i(F_+^I-D_+^I)\frac{\partial}{\partial \chi^I}.
\ee

The Lagrangian $\CL$ can be written as a topological, holomorphic term plus a
$Q$-exact term, 
\be
\label{CLtauQW}
\CL=\frac{i}{8\pi} \tau_{IJ} F^I\wedge F^J+\left\{Q,W\right\},
\ee
with 
\be
\label{WFD}
W=-\frac{i}{4\pi} y_{IJ} \chi^I (F_+^J+D^J).
\ee
This makes it manifest that $\{Q, \CL\}=0$.  
     
\subsection{Coupling to a ${\rm Spin}^c$ structure in the effective theory}
\label{SpincCoupling}
We model the effective theory of the $\CN=2^*$ theory coupled to a ${\rm Spin}^c$
structure as the effective theory of a theory  with gauge group
$SU(2)\times U(1)$. The hypermultiplet is charged under both factors
of the gauge group, and the fields of the vector multiplet associated
with the $U(1)$ are ``frozen'' in a specific way as discussed below.
The Lagrangian of this theory is that in (\ref{Lhigherrank}) with
$r=2$. We identify the index ``1''
with the unbroken $U(1)$ of the $SU(2)$ gauge group, and identify
the index ``2'' with the $U(1)$ factor of the gauge group.
The mass $m$ of the $\CN=2^*$ hypermultiplet is the vev
of the scalar field of the $U(1)$ vector multiplet,
$\left<\phi^{(2)}\right>=a^{(2)}=m$ \cite{Nelson:1993nf}. The field strength
$F^{(2)}$ provides the coupling of the $\CN=2^*$ hypermultiplet to the
${\rm Spin}^c$ structure $c_1(\mathfrak{s})$; we set $[F^{(2)}]=4\pi 
\bfk_m$ with $\bfk_m=c_1(\mathfrak{s})/2\in L/2$. We will later also frequently use
$c_{\rm uv}=2\bfk_m$.

 We determine the values
of the other fields from the BRST variations
(\ref{higherranksusy}). We set the fermionic fields to 0,
$\eta^{(2)}=\chi^{(2)}=0$, such that the BRST variation of $\bar a^{(2)}$ vanishes. Finally, to ensure that the 
BRST variation of $\chi^{(2)}$ vanishes, we set $D^{(2)}=F^{(2)}_+$. With this
prescription, the BRST operator $Q$ (\ref{CQderivatives}) of the $U(1)\times U(1)$ theory
reduces to the BRST operator of the $U(1)$ theory,
\be 
\label{reducedQ}
Q=\sqrt{2}i\eta^{(1)}\frac{\partial}{\partial \bar a^{(1)}}+i(F^{(1)}_+-D_+^{(1)})\frac{\partial}{\partial \chi^{(1)}}.
\ee

With the vanishing of the fermionic fields $\eta^{(2)}=\chi^{(2)}=0$, the Lagrangian (\ref{Lhigherrank}) evaluates to
\be 
\label{CLQW}
\begin{split} 
\CL&=\frac{i}{8\pi} \tau_{IJ} F^I\wedge F^J+\left\{Q,W\right\}\\
&=\frac{i}{8\pi} \tau_{IJ} F^I\wedge F^J+\frac{1}{4\pi} y_{11}
F_+^{(1)}\wedge F_+^{(1)}-\frac{1}{4\pi} y_{11}D^{(1)}\wedge D^{(1)}\\
&\quad+\frac{i\sqrt{2}}{8\pi} \bar \CF_{11J}\eta^{(1)}\chi^{(1)}\wedge
(D+F_+)^J+\frac{1}{4\pi}y_{12}(F_+^{(1)}-D^{(1)})\wedge (F_+^{(2)}+D^{(2)}),
\end{split}
\ee
where as before $y_{IJ}=\mathrm{Im}(\tau)_{IJ}$. Substitution of $D^{(2)}=F^{(2)}_+$ gives
\be
\label{CLD2isF2}
\begin{split}
\CL&=\frac{i}{8\pi} \tau_{IJ} F^I\wedge F^J+\frac{1}{4\pi} y_{11}
F_+^{(1)}\wedge F_+^{(1)}-\frac{1}{4\pi} y_{11}D^{(1)}\wedge D^{(1)}\\
&\quad+\frac{i\sqrt{2}}{8\pi} \bar \CF_{111}\eta^{(1)}\chi^{(1)}\wedge
(D^{(1)}+F_+^{(1)})+\frac{i\sqrt{2}}{4\pi} \bar \CF_{112}\eta^{(1)}\chi^{(1)}\wedge F_+^{(2)}\\
&\quad +\frac{1}{2\pi}y_{12}(F_+^{(1)}-D^{(1)})\wedge F_+^{(2)}.
\end{split}
\ee

The integral over $D^{(1)}$ is a standard Gaussian integral after analytic
continuation. The effective Lagrangian then becomes
\be
\label{LDF}
\begin{split}
\CL&=\frac{i}{8\pi} \tau_{IJ} F^I\wedge F^J+\frac{1}{4\pi} y_{11}
F_+^{(1)}\wedge
F_+^{(1)}+\frac{1}{4\pi}\frac{y_{12}^{2}}{y_{11}}\,F_+^{(2)}\wedge F_+^{(2)} +\frac{1}{2\pi}y_{12}F_+^{(1)}\wedge
F_+^{(2)}\\
&\quad +\frac{i\sqrt{2}}{8\pi} \bar \CF_{111}\eta^{(1)}\chi^{(1)}\wedge
(F_+^{(1)}-\frac{y_{12}}{y_{11}}F_+^{(2)})+\frac{i\sqrt{2}}{4\pi} \bar
  \CF_{112}\eta^{(1)}\chi^{(1)}\wedge
  F_+^{(2)}\\
  &\quad +\log\!\left(\frac{\sqrt{y_{11}}}{2\pi i}\right),
\end{split}
\ee
where the last term is due to the integral over $D^{(1)}$.
Integrating over the fermion zero modes, we arrive at
\be
\begin{split}
\label{intferm}
&\int d\eta^{(1)}\,d\chi^{(1)} \,e^{-\int_X \CL}\\
&=\left[\frac{\partial}{\partial \bar a_1}\left(
  \sqrt{2y_{11}}\,B(F^{(1)}+\frac{y_{12}}{y_{11}} F^{(2)}, J)
\right)\right] e^{-\int_X \CL_0}, 
\end{split}
\ee
where $\CL_0$ is now the first line of (\ref{LDF}), 
\be
\label{U12L0}
\begin{split}
\CL_0&=\frac{i}{8\pi} \tau_{IJ} F^I\wedge F^J+\frac{1}{4\pi} y_{11}
F_+^{(1)}\wedge
F_+^{(1)}+\frac{1}{4\pi}\frac{y_{12}^2}{y_{11}}\,(F_+^{(2)})^2+\frac{1}{2\pi}y_{12}F_+^{(1)}\wedge
F_+^{(2)}\\
&=\frac{i}{8\pi} \tau_{11} F^{(1)}_-\wedge F^{(1)}_-+\frac{i}{8\pi} \bar \tau_{11} F^{(1)}_+\wedge F^{(1)}_++\frac{i}{4\pi} \tau_{12} F^{(1)}_-\wedge F^{(2)}_-\\
&\quad+\frac{i}{4\pi}\bar \tau_{12} F^{(1)}_+\wedge F^{(2)}_++\frac{1}{4\pi}\frac{y_{12}^2}{y_{11}}\,(F_+^{(2)})^2+\frac{i}{8\pi} \tau_{22} F^{(2)}\wedge F^{(2)}.
\end{split} 
\ee 

An important point in the discussion we have just given is the
importance of the role of $Q$-exact terms. If we just drop 
the exact terms we will not produce convergent expressions, 
nor will we obtain a single-valued measure on the Coulomb branch. 
For example, dropping $Q$ exact terms we might expect that the 
action is holomorphic in $\tau_{11}$, but if we just keep that 
term the resulting theta function will not make sense.  
Similarly, the sum of the last two
terms in (\ref{CLD2isF2}) is $Q$-exact, $[Q, \frac{1}{2\pi
  i}y_{12}\chi^{(1)}\wedge F_+^{(2)}]$. 
  Without this term, the exponentiated action is holomorphic in $\tau_{12}$, and single-valued
around the large $u$ limit and around $u_1$. However, omitting
the term is incompatible with the monodromies around the
strong-coupling singularities $u_2$ and $u_3$. This will become more
clear when we discuss the modular transformations in Section
\ref{FormUplaneInt}. Similar remarks apply to the dependence on $\tau_{12}$ in \eqref{U12L0}. 
Dropping the $Q$-exact nonholomorphic dependence on $\tau_{12}$ turns out to 
lead to a measure which is not single-valued, at least when $F^{(2)}\neq 0$.  
Standard reasoning would lead us to believe that the coupling of the Coulomb branch measure to the background ${\rm Spin}^c$-connection should be holomorphic \cite{Shapere:2008zf}. 
Our example serves as a cautionary tale that this 
is not always the case. It is not completely clear that one can drop $Q$-exact terms
when the object that $Q$ acts on is not globally well-behaved. 
This important phenomenon deserves to be much better understood, since 
dropping $Q$-exact expressions is a very commonly used maneuver.

\subsection{Integrating out the massive modes}
\label{Intmassive}
The massive modes are irrelevant on four-manifolds
with $b_2^+=1$ \cite{Moore:1997pc}. Even so, they
do give rise to an important interaction in the $u$-plane integrand
for $N_f=0$, namely a flux dependent sign \cite{Witten:1995gf}, which
is crucial for the integrand to be single valued on the $N_f=0$ $u$-plane. 

We perform here the analogous analysis for the $\CN=2^*$
theory following closely Section 4.4 of \cite{Witten:1995gf}.
All fields of the $\CN=2^*$ theory are in the 3-dimensional, adjoint representation of
 $SU(2)$ or $SO(3)$. One component is neutral, while the other two 
components have charge $\pm 1$ under the unbroken $U(1)$ gauge group
on the $u$-plane. For a generic value of $u$, the charged components
become massive. In addition, the index of a two-dimensional $SU(2)_R$
representation labels after twisting and coupling to a ${\rm Spin}^c$ 
structure, the two components of $W^+\simeq S^+ \otimes K_X^{-1/2}$.  We can then label the fields of the vector
multiplet as $\psi^{\pm \pm}_\alpha$, where the first $\pm$ are the
indices for $W^+$, and the second $\pm$ labels the charge under the
unbroken $U(1)$ gauge group.

As explained in Section 4.4 of \cite{Witten:1995gf}, there 
are essentially two ways to write the measure for the massive
fermions, which differ by the sign. The
natural one in the UV theory is
\be\label{eq:NaturalFermiMeasure}
(\mathcal{D} \psi)_{\rm UV}=\prod_{I=1}^{N_{++}} d\psi^{++}_I d\psi^{--}_I \prod_{J=1}^{N_{+-}}  d\psi^{+-}_J
d\psi^{-+}_J \prod_{K=1}^{\bar N_{++}} d\bar \psi^{++}_K d\bar
\psi^{--}_K \prod_{L=1}^{\bar N_{+-}}
d\bar \psi^{+-}_L d\bar \psi^{-+}_L. 
\ee
In addition, there is a ``positive definite'' measure,
\be\label{eq:PositiveFermiMeasure}
(\mathcal{D}\psi)_{+}= \prod_{I=1}^{N_{++}}  d\psi^{++}_I d\psi^{--}_I \prod_{J=1}^{N_{+-}}  d\psi^{-+}_J
d\psi^{+-}_J \prod_{K=1}^{\bar N_{++}}  d\bar \psi^{++}_K d\bar
\psi^{--}_K  \prod_{L=1}^{\bar N_{+-}} d\bar \psi^{-+}_L d\bar \psi^{+-}_L.
\ee
These two measures differ by a sign,
\be
(\mathcal{D} \psi)_{\rm UV}=(-1)^{N_{+-}+\bar N_{+-}} (\mathcal{D} \psi)_{+}=(-1)^{N_{+-}-\bar N_{+-}} (\mathcal{D} \psi)_{+}.
\ee
While $N_{+\pm}$ and $\bar N_{+\pm}$ are individually infinite, $N_{+-}-\bar N_{+-}$ is
finite and equal to the index of the Dirac operator ${\rm \bf D}_{+-}$ acting on spinors
with charge $+-$. Thus, they are sections of $S^+\otimes
K^{1/2} \otimes E$ with $c_1(E) =\bar w_2(E) \mod 2H^2(X,\mathbb{Z})$. We
set $\bfk=c_1(E)/2= \bfmu  \mod  L$, with $\bfmu\in L/2$. With (\ref{Dindex}), we
find for the index 
\be
\label{MFSign}
\begin{split}
{\rm Index}({\rm \bf D}_{+-})&=\frac{1}{2}(K_X/2 +
2\bfk)^2-\sigma/8\\
&=\frac{1}{2}\left( B(K_X,2\bfk)+(2\bfk)^2\right) + \frac{1}{8}(K_X^2-\sigma)\in \mathbb{Z}.\\
\end{split}
\ee
Since $K_X$ is a lift $\bar w_2(X)$ of $w_2(X)$, $(-1)^{N_{+-}-N_{-+}}$ changes sign under a change of $\bfk$, $\bfk\to \bfk +\bfl$ with $\bfl\in L$,
if and only if the phase $(-1)^{B(\bar w_2(X),\bfk)}$ changes sign as well. The integration over massive fermionic modes  gives therefore rise to the insertion of $(-1)^{B(\bar w_2(X),\bfk)}$ in the sum over fluxes for the $N_f=0$ theory \cite{ Moore:1997pc},\cite{Witten:1995gf}.

For the $\CN=2^*$ theory, we must also take into account the massive modes of the fermions of the hypermultiplet, which decomposes in two $\CN=1$ copies of each chirality. We weakly couple this multiplet to the additional ``frozen" $U(1)$ gauge group with flux $c_1(\frak{s})/2$ for $\frak{s}$ a Spin$^c$ structure. The two $\CN=1$ copies
have opposite electric charge with respect to this $U(1)$. Thus we can write the fields as $\lambda^{\pm\pm}_\alpha$, where the
first $\pm$ now corresponds to the charge under the ``frozen"
$U(1)$, and the second the charge under the unbroken $U(1)$ of $SU(2)$ or $SO(3)$. Thus the measure for the hypermultiplet fermions $\cD(\lambda)_{\rm UV}$, which is completely analogous to equations \eqref{eq:NaturalFermiMeasure} and \eqref{eq:PositiveFermiMeasure}, with $\psi$ replaced by $\lambda$, gives rise to another sign $(-1)^{{\rm Index}({\rm \bf \tilde D}_{+-})}$, where ${\rm Index}({\rm \bf \tilde D}_{+-})$ is as in (\ref{MFSign}), but with $K_X$ replaced by $c_1(\frak{s})$. Multiplying the two signs for the two multiplets gives a $\bfk$-independent sign (for fixed $\bfmu$). If we choose $c_1(\frak{s})=K_X$, the vector and hypermultiplet give rise to the same sign, such that the combined sign is +1. 

Since the combined sign is $\bfk$-independent, there is no insertion of the type \linebreak[4] $(-1)^{B(\bar w_2(X),\bfk)}$ in the sum over fluxes for $\CN=2^*$ in Section \ref{SFluxes}. This is an important difference from the integrand for the $N_f=0$, $SU(2)$ theory. It suggests there is a canonical orientation for the integral in equation \eqref{eq:formalMathdef}. Interestingly, in the limit from 
$\CN=2^*$ to $N_f=0$ the sign $(-1)^{B(\bar w_2(X),\bfk)}$ appears in an
unexpected way, namely due to the limit of the coupling $v$ (\ref{SWlimit}).

\subsection{Sum over fluxes: $\Psi^J_\bfmu$}
\label{SFluxes}
We consider the $u$-plane integral for $\CN=2^*$ theory with gauge group $SU(2)$ and
fixed non-trivial 't Hooft flux $\bfmu\in L/2$. To evaluate the integrand, we
specialize (\ref{U12L0}) to  $\CN=2^*$ with the $U(1)$ flavor symmetry
weakly gauged. We will set $F^{(1)}=4\pi \bfk$ and $F^{(2)}=4\pi
\bfk_m$, with $\bfk\in L+\bfmu$ and $\bfk_m\in L/2$. We will find that
requiring that the integrand is single-valued will put further
constraints on $\bfk_m$. We identify furthermore the couplings
$\tau_{IJ}$ of (\ref{U12L0}) with the couplings $\tau$, $v$ and $\xi$
of Section \ref{N=2*prep}. In particular, $y_{11}=y$, $\tau_{12}=v$, and
$\tau_{22}=\xi=\frac{\partial^2\CF}{\partial m^2}$.
 
Summing the fluxes $\bfk$ in the rhs of (\ref{intferm}) over $L+\bfmu$ and multiplying by
$\partial\bar a/\partial \bar \tau$, we arrive at the sum over fluxes
$\Psi^J_\bfmu$,\footnote{This sum over fluxes is 
  very similar to those defined for $N_f=0$ SW theory
  \cite{Moore:1997pc, Korpas:2019cwg}. The arguments $\tau$ and $\bfz$
  differ by a factor of 2 to match the $\Gamma(2)$ convention, while the current $\Psi_\bfmu^J$ does not involve the sign $(-1)^{B(\bfk,K)}$. We will explain that this sign appears
  in the limit from the $\CN=2^*$ theory to the SW theory with $N_f=0$.}
\be
\label{defPsi3}
\begin{split} 
\Psi_\bfmu^J(\tau, \bar \tau, \bfz, \bar \bfz)& =e^{-4\pi y\, \bfb_+^2}\sum_{\bfk\in L+\bfmu} \partial_{\bar \tau}\left(
  \sqrt{4{y}}\,B(\bfk+\bfb, J)
\right) \,q^{-\bfk_-^2}\,\bar q^{\bfk_+^2}\\
&\quad \times \,e^{-4\pi i B(\bfk_-,\bfz)-4\pi i
  B(\bfk_+,\bar \bfz)},
\end{split}
\ee
where $\bfb=\mathrm{Im}(\bfz)/y$ and $\bfz=v\bfk_m$. When we include 
observables in Section \ref{SecObservables}, we will also use an argument $\bar \bfz$, which is not
the complex conjugate of $\bfz$. To include these cases, we define
\be
\label{Defb}
\bfb=\frac{\bfz-\bar \bfz}{2i\,y}. 
\ee

Using the (rescaled) error
function defined by Equation (\ref{Eerror}), we can write the summand of $\Psi_\bfmu^J$ as a total
derivative to $\bar \tau$,
\be
\label{defPsi4}
\begin{split} 
\Psi_\bfmu^J(\tau, \bar \tau, \bfz, \bar \bfz)& =\tfrac{1}{2}\sum_{\bfk\in
  L+\bfmu} \partial_{\bar \tau} E(\sqrt{4{y}}\,B(\bfk+\bfb, J)) \,q^{-\bfk^2}\,e^{-4\pi i B(\bfk,\bfz)}.
\end{split}
\ee
This expression does not directly allow to write the integrand as a
total derivative of a well-defined function, since the sum is
divergent if $\partial_{\bar \tau}$ is taken out. We will later
determine suitable anti-derivatives using mock modular forms in Section
\ref{sec:Evaluate}.  

We define more generally the theta series $\Psi_\bfmu^J[\CK]$ with kernel $\CK$,
\be
\label{defPsiK}
\begin{split} 
\Psi_\bfmu^J[\CK](\tau, \bar \tau, \bfz, \bar \bfz)& =e^{-4\pi y\,
  \bfb_+^2}\sum_{\bfk\in L+\bfmu} \CK(\bfk) \,q^{-\bfk_-^2}\,\bar q^{\bfk_+^2} \,e^{-4\pi i B(\bfk_-,\bfz)-4\pi i
  B(\bfk_+,\bar \bfz)}.
\end{split}
\ee
The function $\Psi_\bfmu^J(\tau,
\bar \tau,\bfz,\bar \bfz)$ is an example of a non-holomorphic Jacobi form. We list various of its
properties in the following with $\tau$ and $\bfz\in \mathbb{C}^{b_2}$
as independent variables.

From the definition follows for $\bfmu\in L/2$ and $\bfnu\in L$, that 
\be
\label{Psi0}
\begin{split}
\Psi_{\bfmu+\bfnu}^J(\tau,\bar \tau, \bfz, \bar \bfz)&=\Psi_{\bfmu}^J(\tau,\bar \tau, \bfz, \bar \bfz),\\
\Psi_\bfmu^J(\tau,\bar \tau, -\bfz, -\bar \bfz)&=
-\Psi_\bfmu^J(\tau,\bar \tau, \bfz, \bar \bfz),
\end{split}
\ee
and in particular $\Psi_\bfmu^J(\tau,\bar \tau, 0,0)=0$. 

As function of $\bfz$, $\Psi_\bfmu^J(\tau, \bfz)$ is quasi-periodic
under shifts by half-integer vectors in the lattice $L$. For $\bfnu\in L/2$, we have 
\be
\label{Psitrans}
\begin{split}
\Psi_\bfmu^J(\tau, \bar \tau, \bfz+\bfnu, \bar \bfz+\bfnu)&=e^{-4\pi i
  B(\bfnu,\bfmu)}\,\Psi_\bfmu^J(\tau,\bar \tau,\bfz,\bar \bfz),\\
\Psi_\bfmu^J(\tau, \bar \tau, \bfz+\bfnu\tau, \bfz+\bfnu\bar
\tau)&=q^{\bfnu^2}\,e^{4\pi i
  B(\bfnu,\bfz)}\,\Psi_{\bfmu+\bfnu}^J(\tau, \bar \tau,\bfz,\bar \bfz).
\end{split}
\ee

To determine the transformation properties of $\Psi_\bfmu^J$, it is
useful to introduce a characteristic vector $K$ of $L$. We find then
for the $\tau\mapsto \tau+1/2$ and $\tau\mapsto -1/\tau$ transformations: 
\be
\begin{split}
&\Psi^J_{\bfmu+K/2}(-1/\tau, -1/\bar \tau,\bfz/\tau, \bar \bfz/\bar \tau)=-i(-i\tau/2)^{b_2/2}(i\bar
\tau/2)^2\,\exp(-2 \pi i \bfz^2/\tau)\\
&\qquad \qquad \times
\Psi^J_0(\tau/4,\bar \tau/4,\bfz/2-\bfmu/2-K/4, \bar \bfz/2-\bfmu/2-K/4),\\
&\Psi^J_{\bfmu+K/2}(\tau+1/2,\bar \tau+1/2,\bfz,\bar \bfz)=e^{\pi i
  (\bfmu^2-K^2/4)}\,\Psi^J_{\bfmu+K/2}(\tau,\bar \tau,\bfz+\bfmu/2,
\bar \bfz+\bfmu/2),
\end{split}
\ee
where the shift of $\tau$ by $1/2$ on the last line is a consequence of the $\Gamma(2)$ convention. 
Removing the dependence on $K$ on the left-hand side by shifting $\bfmu$, we have
for the two transformations,
\be
\label{PsiTrafos}
\begin{split}
\Psi^J_\bfmu\!\left(-1/\tau,-1/\bar \tau, \bfz/\tau,\bar \bfz/\bar
  \tau \right) & =-i(-i\tau/2)^{b_2/2}(i\bar
\tau/2)^2 \, \exp\!\left( -2\pi i \bfz^2 /\tau \right) \\
& \quad \times \,\Psi^J_0\!\left(\tau/4,\bar
  \tau/4,\bfz/2-\bfmu/2,\bar \bfz/2-\bfmu/2 \right),\\
&=-i(-i\tau/2)^{b_2/2}(i\bar
\tau)^2 \, \exp\!\left( -2\pi i \bfz^2 /\tau \right) \\
& \quad \times \sum_{\bfnu\in (L/2)/L} e^{4\pi i
  B(\bfnu,\bfmu)}\,\Psi_\bfnu^J(\tau,\bar \tau,\bfz,\bar \bfz),\\
 \Psi^J_\bfmu(\tau+1/2,\bar \tau+1/2,\bfz,\bar \bfz) & =  e^{\pi i
   (\bfmu^2-B(\bfmu,K))}\\
& \quad \times \,\Psi_\bfmu^J(\tau,\bar
 \tau,\bfz+\bfmu/2-K/4,\bar \bfz+\bfmu/2-K/4).
\end{split} 
\ee
Furthermore, we have for the two generators of $\Gamma(2)$,
\be
\label{PsiST2S}
\begin{split}
&\Psi^J_{\bfmu}\!\left(\frac{\tau}{-2\tau+1},\frac{\bar \tau}{-2\bar \tau+1},\frac{\bfz}{-2
    \tau+1},\frac{\bar \bfz}{-2\bar 
    \tau+1}\right)=\\
&\qquad (-2\tau+1)^{b_2/2}(-2\bar
\tau+1)^2 \exp\!\left(\frac{4 \pi i \bfz^2}{-2\tau+1}
\right)\Psi^J_{\bfmu-K/2}(\tau,\bar \tau,\bfz,\bar \bfz)\\
&\Psi^J_{\bfmu}\!\left(\tau+2, \bar \tau+2,\bfz,\bar \bfz\right)=e^{2\pi i B(\bfmu,K)}
\Psi^J_{\bfmu}\!\left(\tau,\bar \tau,\bfz,\bar \bfz\right). 
\end{split} 
\ee

\subsection{Formulation of the $u$-plane integral}
\label{Formuplane}
We proceed by formulating the $u$-plane integral for the $\CN=2^*$ theory. This question has been addressed before in \cite{Moore:1997pc, Labastida:1998sk}. As for the
$N_f=0$ theory, the $\CN=2^*$ path
integral reduces for $b_2^+=1$ to an integral over the zero modes of
the vector multiplet. Besides the sum over fluxes $\Psi_\bfmu^J$ (\ref{defPsi3}) and the topological couplings $A$, $B$ (\ref{ABdef}), we also include the
coupling to a ${\rm Spin}^c$ structure, $C^{\bfk_m^2}$ \eqref{Cexpxi}, which
 was first put forward by Shapere and Tachikawa \cite{Shapere:2008zf}.
Equation (\ref{expdaduDelta*}) gives
non-perturbative, modular expressions
for $du/da$ and $\Delta_{\rm phys}$, and thus $A$ and $B$, while
Equation (\ref{Cttv}) gives a closed
expression for $C$ as function of $\tau$, $v$ and $\tau_{\rm uv}$.
 
Combining all ingredients, we find that the $u$-plane integral reads
\be 
\label{uintegral}
\Phi_\bfmu^J=K_u \,\eta(\tau_{\rm uv})^{-4\ell-2\chi}\,\int_{\mathcal{B}} da\wedge d\bar a\,\frac{d\bar\tau}{d\bar a}\,  A^\chi\, B^\sigma
\,C^{\bfk_m^2}\, \Psi^J_\bfmu(\tau, \bar \tau, v\bfk_m, \bar v \bfk_m),
\ee
where $\ell$ is as in (\ref{defell}) and we included the power of $\eta$ due to (\ref{ABCextra}). Note that if $\bfk_m$ corresponds to an
almost complex structure, $\bfk_m^2=(2\chi+3\sigma)/4$, the
power of $\eta$ simplifies to $\eta(\tau_{\rm uv})^{-2\chi}$. The
factor $d\bar \tau/d\bar a$ has arisen from the integration over
fermion zero modes of the vector multiplet for the unbroken $U(1)$
gauge group. It carries $R$-charge 2. The leading behavior of $A$, $B$
and $C$ can be understood from the local $R$ symmetry \cite{Witten:1995gf, Shapere:2008zf}. We
will comment on this in Section \ref{Analysis}.

The three parameters $(K_u, \alpha, \beta)$ in (\ref{uintegral}) are
normalization factors. We set 
\be
\label{Ku}
K_u=-\frac{i\,\kappa}{\pi\,\Lambda}
\ee
with $\kappa$ a numerical factor, which we fix in Section
\ref{Spinacs} by comparison with mathematical results for Euler numbers of instanton moduli spaces. Since
$\chi+\sigma=4$, there is an ambiguity \cite{Moore:2017cmm} 
$$
(K_u,\alpha,\beta)\sim (\zeta^{-4}K_u,\zeta\, \alpha,\zeta\,\beta).
$$

\begin{figure}[ht]\centering 
	\includegraphics[scale=1]{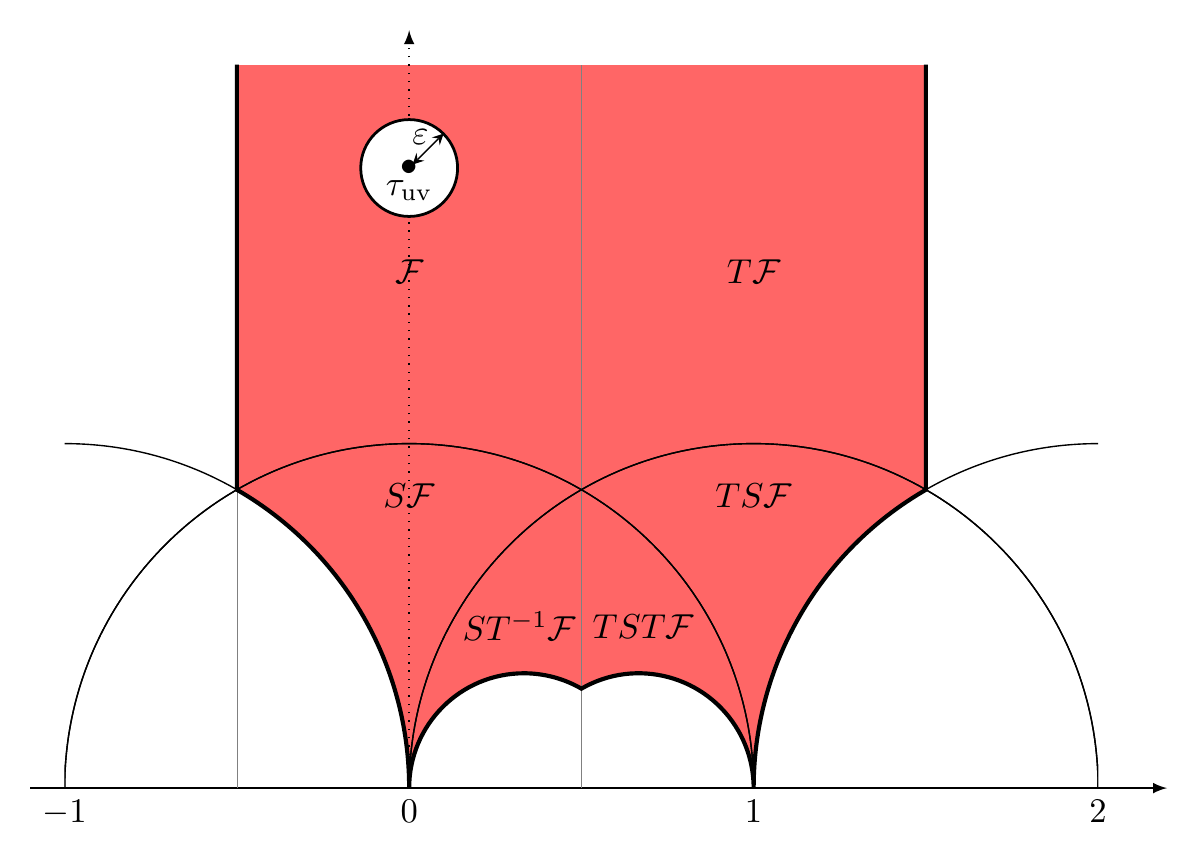} 
	\caption{Fundamental domain $\CU_\varepsilon$ for $\tau$: the fundamental domain
          $\mathbb{H}/\Gamma(2)$ with the
          $\varepsilon$-neighborhood of $\tau_{\rm uv}$, $B(\tau_{\rm uv},\varepsilon)$, excluded.}\label{fig:fund_dom_gamma2}
\end{figure}

The integration domain $\CB$ in (\ref{uintegral}) is the Coulomb branch for the variables $a$ and $\bar a$. 
A more rigorously defined integration domain is obtained by changing
coordinates to $\tau$. We let $\mathrm{Im}(\tau_{\rm uv})\gg 1$, and remove
an epsilon disk $B(\tau_{\rm uv},\varepsilon)$ around $\tau_{\rm uv}$ from the fundamental domain 
$\mathbb{H}/\Gamma(2)$. The Coulomb branch takes in these variables
the form of a ``punctured'' modular curve
\be 
\label{Ueps}
\CU_\varepsilon=(\mathbb{H}/\Gamma(2))\backslash B(\tau_{\rm uv},\varepsilon).
\ee 
The fundamental domain $\mathbb{H}/\Gamma(2)$ consists of six images
of elements of $SL(2,\mathbb{Z})/\Gamma(2)$
of the key hole fundamental domain $\CF$ for
$SL(2,\mathbb{Z})$. An example is the union of $\CF$,
$T\CF$, $S\CF$, $TST\CF$, $STS\CF$ and
$TS\CF$. We have displayed the domain and $\CU_\varepsilon$ in Figure \ref{fig:fund_dom_gamma2}.

The integral $\Phi_\bfmu^J$ now reads 
\be 
\label{UplanetvPsi}
\Phi_\bfmu^J=\lim_{\varepsilon\to 0}\int_{\CU_\varepsilon} d\tau\wedge d\bar
\tau\,\nu(\tau,\tau_{\rm uv},m)\, \Psi^J_\bfmu(\tau,\bar \tau,v\bfk_m,
\bar v\bfk_m),
\ee 
where we defined the measure $\nu$, 
\be
\nu(\tau,\tau_\uv,m)=K_u\,\eta(\tau_\uv)^{-4\ell-2\chi}\,\frac{da}{d\tau}\,A^\chi\, B^\sigma
\,C^{\bfk_m^2}.
\ee

We express $\nu$ more explicitly using \eqref{ABdef}, \eqref{alphabetaOmega}, \eqref{ABCextra} and Matone's relation \cite{Matone:1995rx} for $\CN=2^*$ theory,
\be
\label{Matone2*}
\frac{1}{2\pi i}\frac{du}{d\tau}= \frac{8}{m^2}\,\Delta_{\rm phys} \left( \frac{da}{du}\right)^2.
\ee
This gives,
\be 
\label{tnu}   
\begin{split}
\nu(\tau,\tau_{\rm uv},m)&=\frac{16\pi i\,K_u\,\alpha^\chi\, \beta^\sigma}{
  m^2}  \,\eta(\tau_{\rm uv})^{-4\ell - 2\chi}\,\left(\frac{du}{da}\right)^{\frac{\chi}{2}-3} \,\Delta_{\rm
  phys}^{\frac{\sigma}{8}+1}\,C^{\bfk_m^2}\\ 
&=\frac{\kappa\,m^{\frac{\sigma}{8}-2}}{4\,\eta(\tau_{\rm uv})^{2\chi+4\ell}}\left(
  \frac{4}{\Lambda}\right)^{\frac{3}{8}(2\chi+3\sigma)} \left(\frac{du}{da}\right)^{\frac{\chi}{2}-3} \,\Delta_{\rm
  phys}^{\frac{\sigma}{8}+1}\,C^{\bfk_m^2}.
\end{split}
\ee

\subsection{Single-valuedness of the $u$-plane integrand}
\label{SingleValued}

In the following subsections, we discuss the single-valuedness of the
integrand under monodromies around the four singularities of the Coulomb branch.

\subsubsection{$ \infty$}
For the monodromy around $u=\infty$, we have $a\to e^{\pi i}a$ and $v\to
e^{-\pi i} v$. The different ingredients of the integrand transform as follows:
\be
\begin{split} 
&\frac{du}{da}\mapsto e^{\pi i}\, \frac{du}{da},\\
&\Delta_{\rm phys}\mapsto e^{6\pi i}\,\Delta_{\rm phys},\\
&C\mapsto e^{-\pi i} \,C. 
\end{split}
\ee
The sum over fluxes changes by a sign (\ref{Psi0}).

The integrand of $\Phi^J_\bfmu$ therefore changes by a phase 
\be
e^{\pi i\, \chi/2+6\pi i\, \sigma/8-\pi i\, \bfk_m^2}.
\ee
For $b_2^+=1$, we have $\chi+\sigma=4$, such
that this phase equals
\be
\label{phaseInf}
e^{2\pi i\, \sigma/8-\pi i\, \bfk_m^2}.
\ee
Therefore, if we set $\bfk_m \in K/2 + L$, with $K$ a characteristic
vector of $L$ (\ref{KMcharacteristic}), then (\ref{phaseInf}) equals 1. We thus find that $\bfk_m$
must correspond to a ${\rm Spin}^c$ structure, for the $u$-plane integral to be single-valued with respect to monodromies around $\infty$. We will demonstrate that this requirement also implies single-valuedness with respect to the monodromies around $u_j$, $j=1,2,3$.

\subsubsection{$u_1$}
For the monodromy around $u_1=\frac{m^2}{4}e_1$ (or $\tau\to i\infty$), we have using
(\ref{aDaT2}) and (\ref{M1trafos})
\be  
\begin{split} 
& \tau \mapsto \tau+2,\\
& v\mapsto v-1,\\
&\frac{du}{da}\mapsto \frac{du}{da},\\
&\Delta_{\rm phys}\mapsto e^{2\pi i}\,\Delta_{\rm phys},\\
& C\mapsto e^{-\pi i} C.
\end{split}
\ee
For the sum over fluxes, we have using (\ref{PsiST2S}) and (\ref{Psitrans})
\be
\begin{split}
\Psi^J_\bfmu(\tau+2, \bar \tau+2,\bfz- \bfk_m, \bar \bfz- \bfk_m)&=
e^{2\pi i B(\bfmu,K)}\,\Psi^J_\bfmu(\tau,\bar \tau,\bfz-\bfk_m, \bar \bfz-\bfk_m)\\
&=  e^{2\pi i B(\bfmu,K+2\bfk_m)}\,\Psi^J_\bfmu(\tau,\bar \tau,\bfz,
\bar \bfz).
\end{split}
\ee
The integrand is therefore multiplied by the following phase
\be
e^{2\pi i\, \sigma/8-\pi i \,\bfk_m^2+ 2\pi i B(\bfmu,K+2\bfk_m)}.
\ee
Therefore for $\bfk_m\in K/2+L$, the integrand is single-valued for the monodromy around $u_1$.
This condition on $\bfk_m$ is compatible with the condition found for the monodromy around $\infty$. 

\subsubsection{$u_2$}
For the monodromy around $u_2=\frac{m^2}{4}e_2$, we have using
(\ref{aDaz2}) and (\ref{M2trafos})
\be
\begin{split} 
& \tau \mapsto \frac{\tau}{-2\tau+1},\\
&v \mapsto \frac{v-\tau}{-2\tau+1},\\
&\frac{du}{da}\mapsto (-2\tau+1)^{-1}\, \frac{du}{da},\\
&\Delta_{\rm phys}\mapsto e^{2\pi i}\,\Delta_{\rm phys},\\
& C\mapsto e^{-\pi i + 2\pi i v - 2\pi i(v-\tau)(2v-1)/(-2\tau+1)}\,C .
\end{split}
\ee

Finally we need to determine the transformation of
$\Psi^J_\bfmu$. 
We have
\be
\label{PsiM2}
\begin{split} 
&\Psi^J_\bfmu\!\left(\frac{\tau}{-2\tau+1}, \frac{\bar \tau}{-2\bar
    \tau+1},\frac{v-\tau}{-2\tau+1}\bfk_m, \frac{\bar v-\bar \tau}{-2\bar \tau+1}\bfk_m\right)\\
&= \exp\!\left( 2\pi i \left(
    \frac{\tau\,\bfk_m^2-2\,v\,\bfk_m^2}{-2\tau+1}  \right)\right)\, \Psi^J_{\bfmu-\bfk_m}\!\left(\frac{\tau}{-2\tau+1}, \frac{\bar \tau}{-2\bar
    \tau+1},\frac{v}{-2\tau+1}\bfk_m, \frac{\bar v}{-2\bar \tau+1}\bfk_m\right) \\
&=(-2\tau+1)^{b_2/2}\,(-2\bar \tau+1)^2\,  \exp\!\left( 2\pi i 
    \frac{(\tau-2\,v+2\,v^2)\,\bfk_m^2}{-2\tau+1}  \right) \\
&\quad \times  \Psi^J_{\bfmu-\bfk_m-K/2}\!\left(\tau, \bar \tau,
  v\,\bfk_m, \bar v\,\bfk_m\right),
\end{split}
\ee
where we used (\ref{Psitrans}) for the first equality sign, and
(\ref{PsiST2S}) for the second.

The weights of the various functions and
$d\tau\wedge d\bar \tau$ cancel among each other. Moreover, the terms
involving $v$ cancel beautifully between $C^{\bfk_m^2}$ and $\Psi^J_\bfmu$.  
The remaining phase is
\be
e^{2\pi i \sigma/8-\pi i \bfk_m^2}
\ee
which evaluates to $1$ for $\bfk_m\in K/2+L$. Finally, for this
choice of $\bfk_m$, $\Psi_{\bfmu-\bfk_m-K/2}^J=\Psi_{\bfmu}^J$. We
thus conclude that the integrand is single-valued under the monodromy
around $u_2$ if $\bfk_m$ represents a ${\rm Spin}^c$ structure.

\subsubsection{$u_3$}
This monodromy must be single-valued too, since
$M_1\,M_2\,M_3=M_\infty$. As a check, we determine the various
transformations using (\ref{aDaz3}) and (\ref{M3trafos})
\be
\begin{split}
&\tau\mapsto \frac{-\tau+2}{-2\tau+3},\\
&v\mapsto \frac{v-\tau+1}{-2\tau+3},\\
&\frac{du}{da}\mapsto (-2\tau+3)^{-1}\, \frac{du}{da},\\
&\Delta_{\rm phys}\mapsto e^{2\pi i}\,\Delta_{\rm phys},\\
& C\mapsto e^{-\pi i-2\pi i(\tau-1-2v+v^2)/(-2\tau+3)}C.
\end{split}
\ee 

\subsection{The limit to $N_f=0$}
\label{SecNf0limit}
Note the sum over fluxes $\Psi_\bfmu^J(\tau,\bar \tau,v\bfk_m, \bar
v\bfk_m)$ appears at first sight to differ from the theory with
$N_f=0$, since there is no insertion $(-1)^{B(\bfk,K)}$. 
However, (\ref{SWlimit}) demonstrates that this will appear in the limit.

The non-perturbative contributions to $v$ and $\xi$ vanish in
  the limit to the pure SW theory. Equation (\ref{aDmDpert}) shows that their
  perturbative parts evaluates to constants. From (\ref{Cttv}), one obtains for the three couplings
\be
\label{SWlimit}
\tau \mapsto \tau,\qquad v= -\frac{1}{2}+O(q_{\rm uv}),\qquad \xi\mapsto
-\frac{1}{4}-\frac{1}{2\pi i} \frac{3}{2} \log(\Lambda/m)+O(q_{\rm uv}),
\ee 
One of the striking effects of this limit is that the non-holomorphic, non-topological coupling
\be
e^{-4\pi i v B(\bfk_-,\bfk_m)-4\pi i \bar v B(\bfk_+,\bfk_m)}
\ee
in (\ref{defPsi3}) and (\ref{uintegral}), becomes holomorphic and
topological,
\be
e^{-2\pi i B(\bfk_m,\bfk)}.
\ee
This phase is crucial in the sum over fluxes of the effective theory
of pure SW theory.

The topological couplings for the $N_f=0$ theory, $A_0$ and $B_0$ are defined as in (\ref{ABdef}), with for $\alpha_0$ and $\beta_0$ \cite{Manschot:2019pog}
\be 
\alpha_0=\Lambda^{-1/2},\qquad \beta_0=\sqrt{2}\, \Lambda^{-1/2}.
\ee
From (\ref{ABdef}) and (\ref{alphabetaOmega}), we deduce that in the
limit to pure $SU(2)$ the gravitational couplings $A$, $B$ and $C$ become
\be
\label{AASW}
A \to A_0,\qquad  B\to  \left(\frac{m}{\Lambda}\right)^{3/8}
e^{-\pi i/8}  B_0,\qquad C\to
\left(\frac{\Lambda}{m}\right)^{3/2} e^{\pi i/2}.
\ee  
The measure $\nu_0(\tau)$ of the $u$-plane integral for pure SW-theory is 
\be
\label{nuSW}
\nu_0(\tau)=-\frac{i}{4} \frac{\vartheta_4(2\tau)^{13-b_2}}{\eta(2\tau)^9},
\ee
while the sum over fluxes reads\footnote{Both $\nu_0$
  and $\Psi^J_{{\rm SW},\bfmu}$ differ by an overall factor 2 and a
  factor 2 in the argument from the quantities in \cite{Korpas:2019cwg}
  (respectively Equation (4.6) and Equation (4.9/10)). These factors
  are due to the $\Gamma_0(4)$ convention in \cite{Korpas:2019cwg} and
the $\Gamma(2)$ convention in this paper. The overall factors of 2
take into account the change in the measure $d\tau\wedge d\bar \tau$.}
\be
\label{PsiSW}
\Psi^J_{0,\bfmu}(\tau,\bar \tau)=\sum_{\bfk\in
  L+\bfmu} \partial_{\bar \tau}\left( \sqrt{4y} B(\bfk,J)\right)
e^{\pi i B(\bfk,K_{0})}q^{-\bfk_-^2}\bar q^{\bfk_+^2},
\ee
with $K_{0}$ a characteristic vector of $L$, and therefore $K_{0}\in
\bar w_2(X) + 2L$. Two different choices
for the characteristic vector, $K_0$ and $\tilde K_0$, change
$\Psi^J_{0,\bfmu}$ by a sign, $(-1)^{B(\bfmu,K_0-\tilde K_0)}$. This
dependence on the lift of $w_2(X)$, can be compensated by an
additional sign in $\Psi^J_{0,\bfmu}$ \cite{Moore:1997pc},
\be
\label{phaseDW}
e^{\frac{i \pi }{2} ((2\bfmu)^2-B(2\bfmu,K_0))},
\ee
with $2\bfmu$ a lift of $w_2(E)$ to $L$ as before.
\footnote{Since $K_0$ is characteristic this factor is a sign. Note that under a shift $K_0 \to K_0 + 2\beta$ with $\beta$ an integral class in $L$, the sign changes by $(-1)^{w_2(E) \cdot \beta}$. On the other hand, under a shift $2\bfmu\to 2\bfmu + 2\beta$ the sign changes by $(-1)^{\beta^2}$. The phase 
\eqref{phaseDW} can be interpreted as a topological term in the action. One cannot define a partition function that depends on just $w_2(E)$ and $w_2(X)$. One must choose a lifting of one of the two classes. This can be interpreted in terms of mixed anomalies and five-dimensional anomaly inflow \cite{Cordova:2018acb, Wang:2018qoy}.}
With (\ref{phaseDW}), $\Psi^J_{0,\bfmu}$ is independent of $K_0$,
and the change of sign, due to a different choice of $\bfmu$,
corresponds to a change of orientation of
instanton moduli space discussed in \cite[Section 7.1.6]{Donaldson90}.

The $N_f=0$ limit of $\nu$ is
\be
\nu(\tau,\tau_{\rm uv},m) \sim i\,\kappa\,
\eta(\tau_{\rm uv})^{-4\ell-2\chi} \left(\frac{\Lambda}{m}\right)^{\frac{3}{2}
  (\bfk_m^2-\frac{\sigma}{4})}\,e^{\pi i (\bfk_m^2/2 - \sigma/8)}\,\nu_0(\tau).
\ee
Moreover, the $N_f=0$ limit of the $\CN=2^*$ sum over fluxes is
\be
\Psi^J_{\bfmu}(\tau,\bar \tau,v\bfk_m,\bar v\bfk_m)\to
(-1)^{B(\bfmu,K_{0}-2\bfk_m)}\,\Psi^J_{0,\bfmu}(\tau,\bar \tau).
\ee
Put differently, the $N_f=0$ limit leads to a canonical choice of
$K_0=2\bfk_m=c_{\rm uv}$. 
Combining the factors, we find that for the $N_f=0$ limit, $\Phi_\bfmu^J$ and $Z_\bfmu$
more generally, should be
multiplied by
\be
\label{Prefactor1} 
R_\bfmu^X(K_0,2\bfk_m)=e^{-\pi i (\chi+\sigma)/8}\,(-1)^{B(\bfmu,K_0- 2\bfk_m)+ (\bfk_m^2/2 - \sigma/8)}\,\eta(\tau_{\rm uv})^{2\chi+4\ell} \times
\left(\frac{m}{\Lambda}\right)^{\frac{3}{2}
  (\bfk_m^2-\frac{\sigma}{4})}, 
\ee
where we have replaced $-i$ by $e^{-\pi i (\chi+\sigma)/8}$, such that
this formula can be applied to study this limit for generic manifolds
of simple type in Section \ref{SWRedNf0}. We have also set $\kappa=1$,
which will follow from explicit calculation in Section \ref{tau0contribution}.

For later reference, we define the dimensionless partition function $\underline \Phi_\bfmu^J$,
\be 
\label{underlinePhi}
\underline \Phi_\bfmu^J(\tau_\uv,\bar \tau_\uv)=R_\bfmu^X(K_0,2\bfk_m)\,\Phi_\bfmu^J(\tau_\uv,\bar \tau_\uv).
\ee 
The leading term of $\underline \Phi_\bfmu^J$ is a constant $\sim q_\uv^0$, since 0 is the smallest instanton number.

\subsection{Observables}
\label{SecObservables}
The $\CN=2^*$ theory gives rise to correlation functions of
non-trivial observables. The observables are proportional to the mass
$m$, and vanish in the massless limit. 
\\
\\
{\it The point observable}\\
It is straightforward to include the point observable, which
corresponds to the order parameter $u$ on the Coulomb
branch. Mathematically, this observable corresponds to a 4-form
$\mu_D(p)$ on the instanton moduli space, where $\mu_D:
H_i(X,\mathbb{Q})\to H^{4-i}(\CM_{k,\bfmu},\mathbb{Q})$ is Donaldson's
$\mu$-map.
Since the limit to pure SW theory (\ref{SWlimit}) involves a shift,
the appropriate $\CN=2^*$ observable is
\be
\label{DonaldsonPointObs}
u_{\rm D}=u+\frac{1}{8}e_1(\tau_{\rm uv})m^2.
\ee 
We determine the normalization by the limit to the pure theory.
\be
\label{epu}
e^{p u_{\rm D} /\Lambda^2}.
\ee
If $u$ approaches the singularities $u_j$, $j=1,2,3$, $u_{\rm D}(\tau,\tau_{\rm uv})$ approaches
\be
\label{uDj}
\begin{split}
u_{{\rm D},1}&= \tfrac{3}{8}m^2 e_1(\tau_{\rm
  uv})=m^2\left(\tfrac{1}{4}+6\,q_{\rm uv}+O(q_{\rm uv}^2)\right),\\
u_{{\rm D},2}&= \tfrac{1}{4}m^2 e_2(\tau_{\rm uv})+\tfrac{1}{8}m^2 e_1(\tau_{\rm uv})=-2\,m^2\,q_{\rm uv}^{1/2}+O(q_{\rm uv})\\
&=\Lambda_0^2\,(1+O(q_{\rm uv}^{1/2})),\\
u_{{\rm D},3}&= \tfrac{1}{4}m^2 e_3(\tau_{\rm uv})+\tfrac{1}{8}m^2
e_1(\tau_{\rm uv})=2\,m^2\,q_{\rm uv}^{1/2}+O(q_{\rm uv})\\
&=-\Lambda_0^2\,(1+O(q_{\rm uv}^{1/2})).\\
\end{split}
\ee
Thus $u_{{\rm D},1}$ diverges in the limit to pure SW, while $u_{{\rm D},2}$ and
$u_{{\rm D},3}$ approach the strong coupling singularities of the pure theory.\\
\\
{\it The surface observable}\\
For $S\in H_2(X,\mathbb{Q})$, the surface observable reads
\be
\label{UVSobs}
I_-(S)=\frac{1}{4\pi^2} \int_{S} {\rm Tr}\left[ \frac{1}{8}
  \psi\wedge \psi-\frac{1}{\sqrt{2}}\phi F\right].
\ee
The coupling flows in the infra-red to:
\be
\label{tildeIMx}
\tilde I_-(S)=\frac{i}{\sqrt{2}\pi}\int_S \left ( \frac{1}{32}\frac{d^2u}{da^2}\psi\wedge \psi-\frac{\sqrt{2}}{4} \frac{du}{da}(F_-+D)\right).
\ee
Then we add to the effective action
\be
e^{\tilde I_-(S)/\Lambda}.
\ee
The overall normalization is such that the leading power of $q$ of the
Seiberg-Witten limit (\ref{SWlimit}) of $\tilde I_-$ is $ q^{-1/4}
B(S,F/4\pi)=q^{-1/4}
B(S,\bfk_m)$. 

After reduction to zero modes and integrating over $D$, the insertion
of the surface operator $e^{\tilde I_-(S)}$ in the effective action is
taken into account by adding an extra term to $v\bfk_m$ in (\ref{uintegral}), 
\be
\label{zSandv}
\bfz_S=\frac{S}{4 \pi\, \Lambda} \frac{du}{da},\qquad  v\bfk_m \mapsto v \bfk_m + \bfz_S,\qquad
\bar v \bfk_m \mapsto \bar v \bfk_m, 
\ee
In this way, we can easily take the surface observable into account in
the sum over fluxes. This becomes 
\be
\Psi_\bfmu^J(\tau,\bar \tau,v\bfk_m+\bfz_S, \bar v\bfk_m),
\ee
with $\Psi_\bfmu^J$ as in (\ref{defPsi3}). For the monodromy $M_2$, (\ref{PsiM2}) is modified to
\be
\label{PsiwithS}
\begin{split} 
&\Psi^J_\bfmu\!\left(\frac{\tau}{-2\tau+1}, \frac{\bar \tau}{-2\bar
    \tau+1},\frac{(v-\tau) \bfk_m+\bfz_S}{-2\tau+1}, \frac{(\bar v-\bar
    \tau) \bfk_m}{-2\bar \tau+1}\right)\\
&=  \exp\!\left( 2\pi i 
    \frac{(\tau-2\,v+2\,v^2)\,\bfk_m^2+(4v-2)B(\bfk_m,\bfz_S)+2\bfz_S^2}{-2\tau+1}  \right) \\
&\quad \times  (-2\tau+1)^{b_2/2}\,(-2\bar \tau+1)^2\,\Psi^J_{\bfmu}\!\left(\tau, \bar \tau,
  v\,\bfk_m+\bfz_S, \bar v\,\bfk_m\right).
\end{split}
\ee  
\\
\\  
{\it Contact term $T_2$}\\
Since $\bfz_S \sim S$ appears in the exponential prefactor, we deduce that
single valuedness of the integrand requires the presence of contact
terms in the effective action \cite{LoNeSha,
  Moore:1997pc, Marino:1998bm}. We denote the term for $S^2$ is by
$e^{T_2\, S^2}$, and for $B(\bfk_m,\bfz)$ by
$e^{T_3\,B(\bfk_m,\bfz)}$. We discuss first $T_2$, which is given by 
\be  
\label{T2u} 
T_2(\tau,\tau_\uv)=\frac{1}{2\pi i \Lambda^2}\frac{\partial^2 \widetilde \CF(a,m,\tau_{\rm uv})}{\partial \tau_{\rm uv}^2},
\ee 
where the derivative to $\tau_{\rm uv}$ is for fixed $a$ and $m$.
We can express this function in terms of $u$ and $du/da$ \cite{Marino:1997gj},
\be  
\begin{split}
T_2(\tau,\tau_{\rm uv})&=\frac{1}{24\Lambda^2}\left(-\frac{E_2(\tau)}{2}\left(
  \frac{du}{da}\right)^2+2 \, E_2(\tau_{\rm uv})\, u +\frac{m^2 }{6} E_4(\tau_{\rm uv})+\frac{m^2}{\pi i}\,\frac{dE_2(\tau_\uv)}{d\tau_\uv}\right)\\
&= \frac{1}{24\Lambda^2}\left(-\frac{E_2(\tau)}{2}\left(
  \frac{du}{da}\right)^2+2 \, E_2(\tau_{\rm uv})\, u +\frac{m^2}{6}\,E_2(\tau_\uv)^2\,\right),
\end{split}
\ee 
where we used (\ref{DE2}) for the second identity.

We can derive the modular properties also from (\ref{T2u}). Using
(\ref{utCF}), we can write $T_2$ as a derivative of $u$,
\be  
\frac{\partial^2 \widetilde \CF(a,m,\tau_{\rm uv})}{\partial
  \tau_{\rm uv}^2}=\left. \frac{\partial}{\partial \tau_{\rm uv}} \left( \tfrac{1}{2}
  u +\tfrac{m^2}{12}E_2(\tau_{\rm
    uv})\right)\right\vert_{a,m},
\ee
where we consider $u$ as a function of $a$, $m$ and $\tau_{\rm uv}$. We
next change variables from $a$ to $\tau$ for $u$, such that
$\partial^2_{\tau_{\rm uv}} \widetilde\CF$ becomes
\be
\begin{split}
\frac{\partial \widetilde \CF(a,m,\tau_{\rm uv})}{\partial
  \tau_{\rm uv}^2}=&\left.\frac{\partial}{\partial \tau_{\rm uv}} \left( \tfrac{1}{2}
  u(\tau,\tau_{\rm uv},m) +\tfrac{m^2}{12}E_2(\tau_{\rm
    uv})\right)\right\vert_{\tau,m} \\
&+\frac{1}{2}\left.\frac{\partial
  u(\tau,\tau_{\rm uv},m)}{\partial\tau}\right\vert_{\tau_{\rm uv},m}\,\left.\frac{\partial
  \tau(a,\tau_{\rm uv},m)}{\partial \tau_{\rm uv}}\right\vert_{a,m}. 
\end{split}
\ee
Using that the last term equals $\partial \tau/\partial
\tau_{\rm uv}=\frac{1}{2}\partial^2 u/\partial a^2$, we can determine the
transformations as in Section \ref{ActionTauVXI}. This gives
\be 
M_2:\qquad \frac{\partial^2 \widetilde \CF}{\partial \tau_{\rm uv}^2}\mapsto
\frac{\partial^2 \widetilde \CF}{\partial \tau_{\rm uv}^2}+\frac{1}{2(-2\tau+1)}\left( \frac{du}{da}\right)^2,
\ee
where we used $\partial_a\mapsto 1/(-2\tau+1) \partial_a$
(\ref{aDaz2}). 

We have moreover for $SL(2,\mathbb{Z})$ transformations of $\tau_{\rm
  uv}$,
\be
\label{T2Sdual}
\begin{split}
T_2\!\left(\tau,\frac{a\tau_{\rm uv}+b}{c\tau_{\rm uv}+d}\right)&=(c\tau_{\rm uv}+d)^4\,\left(T_2(\tau,\tau_{\rm
    uv})+\frac{1}{2\pi i\,\Lambda^2} \frac{c}{c\tau_{\rm uv}+d}\, \left(u+\frac{m^2}{6} E_2(\tau_\uv)\right)\right.\\
& \qquad \left. -\frac{m^2}{4} \frac{c^2}{(c\tau_\uv+d)^2}\right).
\end{split}
\ee
where we used (\ref{uCF}).

The last aspect we mention here is the limit of $T_2$ to the $N_f=0$ theory. Denoting this limit by  $T_{2,0}$, we find
$$
T_{2,0}(\tau)=\frac{1}{24\Lambda_0^2} \left( u_0 -E_2(\tau) \left( \frac{du}{da}\right)^2_0\right),
$$
where $u_0$ is as in Equation \eqref{uSW}.
\\
\\
{\it Contact term $T_3$}\\
Due to the coupling $B(\bfk_m,\bfz)$, there must be a third contact
term. We will demonstrate that this contact term is given by  
\be 
\label{T3u}
T_3(\tau,\tau_\uv)=-\frac{2i}{\Lambda}\frac{\partial^2 \widetilde \CF}{\partial \tau_{\rm uv}\partial m},
\ee
where the derivatives act for fixed $a$. Again using
(\ref{uCF}), we can write $T_3$ as a derivative of $u$ to $m$ for fixed $a$.
Changing variables from $a$ to $\tau$,
we can express $\partial^2 \widetilde \CF/\partial \tau_{\rm uv}\partial m$ as
\be 
\frac{\partial^2 \widetilde \CF(a,m,\tau_{\rm uv})}{\partial \tau_{\rm
    uv}\partial
  m}=\tfrac{1}{2} \partial_m(u+\tfrac{m^2}{6}E_2(\tau_{\rm
  uv}))\vert_{\tau,\tau_{\rm uv}}
+\tfrac{1}{2} \partial_\tau u\vert_{\tau_{\rm uv},m}\,\partial_m\tau\vert_{a,\tau_{\rm uv}}
\ee
Using that $\partial_m\tau=\partial^2 m_D/\partial a^2$, we can use
the transformations of $m_D$ and $a$ to derive the transformations of
$T_3$. We then find using $v=\partial_a m_D$,
\be
\begin{split}
&M_\infty, M_1:  \qquad \frac{\partial^2\widetilde \CF}{\partial \tau_{\rm uv}\partial m}
\mapsto \frac{\partial^2 \widetilde \CF}{\partial \tau_{\rm uv}\partial m},\\
&M_2 \qquad\,: \qquad \frac{\partial^2 \widetilde \CF}{\partial \tau_{\rm uv}\partial m}\mapsto
\frac{\partial^2 \widetilde \CF}{\partial \tau_{\rm uv}\partial
  m}+\frac{2v-1}{2(-2\tau+1)}\frac{du}{da}.
\end{split} 
\ee
These transformations are of the expected form to match the
transformation of the sum over fluxes (\ref{PsiwithS}). 

To write $T_3$ more explicitly, we use
\be
\left. \frac{\partial \tau}{\partial m} \right\vert_{a,\tau_{\rm
    uv}}=-\frac{a}{m} \left.\frac{\partial\tau}{\partial a}\right\vert_{m,\tau_{\rm uv}}.
\ee
Moreover, we determine from \eqref{vaDat} that
\be
a= - m\frac{d v}{d \tau}.
\ee
We then arrive at 
\be
\label{dFdtuvdm}
T_3(\tau,\tau_{\rm uv})=-\frac{2i}{\Lambda}\left(\frac{u}{m} +\frac{m}{6}\,E_2(\tau_{\rm
  uv}) + \frac{1}{2} \frac{d v}{d \tau}\,\frac{du}{da}\right).
\ee
Therefore, under an $SL(2,\mathbb{Z})$ transformation of $\tau_{\rm
  uv}$, we have
\be
\label{T3Sdual}
T_3\!\left(\tau,\frac{a\tau_{\rm uv}+b}{c\tau_{\rm
      uv}+d}\right)=(c\tau_\uv+d)^2 \left( T_{3}(\tau,\tau_\uv) -
  \frac{2m}{\pi\Lambda} \frac{c}{c\tau_\uv+d}  \right).
\ee
Finally, we consider the $N_f=0$ limit. This limit of $T_{3,0}$ in fact diverges, 
\be
\label{T3u0}
\lim_{\CN=2^*\to N_f=0} T_{3}=\infty. 
\ee 
On the other hand, the limit of $\partial^2\CF/\partial\tau_\uv\partial m$ vanishes.
Since the $N_f=0$ limit of $T_3$ is infinite, we define an
alternative contact term by replacing $u$ with $u_{\rm D}$ \eqref{DonaldsonPointObs} in
\eqref{dFdtuvdm}, and subtract the term involving $E_2$. We have then
\be
\label{TD3}
T_{{\rm D},3}=T_{3}+\frac{i\,m}{3\Lambda}E_2(\tau_\uv)-\frac{i\,m}{4\Lambda}e_1(\tau_\uv).
\ee

\subsection{Regularization of the $u$-plane integral}
\label{SecRegularize}
The $u$-plane integrand often diverges near the singularities, which will become more clear from the discussion in Section \ref{ABCboundaries}. Due to the divergences, it is necessary to reguralize the integral. To this end, we first compactify $\CU_\varepsilon$ by
introducing a cut-off $\mathrm{Im}(\tau)\leq Y$ for the cusp $\tau=\tau_1\to i\infty$, and similarly for the local couplings $\tau_2$ and $\tau_3$ near the cusps $\tau=0$ and $\tau=1$. We denote this space by
$\CU_{Y,\varepsilon}$. Then $\Phi_\bfmu^J$ is defined as
\be  
\label{PhiABC}
\begin{split}
\Phi_\bfmu^J&=\lim_{Y\to \infty,\varepsilon\to 0} K_u \,\eta(\tau_{\rm
  uv})^{-4\ell-2\chi}\,\alpha^{\chi}\beta^\sigma \\
&\qquad \times\int_{\CU_{Y,\varepsilon}} d\tau\wedge d\bar
\tau\,\left(\frac{du}{da}\right)^{\frac{\chi}{2}-1} \frac{du}{d\tau}
\,\Delta_{\rm phys}^\frac{\sigma}{8} \,C^{\bfk_m^2}
\,\Psi^J_\bfmu(\tau,\bar \tau, v \bfk_m, \bar v \bfk_m),
\end{split}
\ee
assuming that the limits exist. This integral requires careful regularization, especially if
observables are included. We will assume that the $\tau$-dependent of the
integrand is a linear combination of the following terms near the
boundaries $u_j$ and $\infty$,
\be
\label{IntegrandExp}
\begin{split}
&\text{For}\,\, u\to u_j:\qquad y^s q_j^m \bar q_j^n,\qquad n\geq 0, m-n\in
\mathbb{Z}/2,\\
&\text{For}\,\,
u\to \infty:\qquad  (\tau-\tau_\uv)^m (\bar \tau-\bar \tau_\uv)^n,
\qquad n\geq 0, m-n\in
\mathbb{Z}.
\end{split}
\ee
where $q_j=e^{2\pi i \tau_j}$ with $\tau_1=\tau$, $\tau_2=-1/\tau$ and
$\tau_3=-1/(\tau-1)$. The analysis in Section \ref{Analysis} will
confirm that the expansions of the $u$-plane integrand are of the form in \eqref{IntegrandExp}. For the cusps $u_j$, the "familiar" regularization is then applicable \cite{Dixon:1990pc, Harvey:1995fq, Borcherds:1996uda}. For example for $u_1$, we have
\be
\int^Y_{Y_0} dy \int_{-1/2}^{3/2} dx\,y^s q_j^m \bar
q_j^n=2 \delta_{m,n} \int^Y_{Y_0} y^s e^{-4\pi y m},
\ee
which is convergent for $Y\to \infty$ for $n>0$, and for $s<-1$ if $n=0$. Similarly for $u\to \infty$, we have
\be
\int^{r_0}_\varepsilon r\,dr \int_{0}^{2\pi} d\varphi\,(\tau-\tau_\uv)^m
(\bar \tau-\bar \tau_\uv)^n =2\pi\delta_{m,n}  \int^{r_0}_\varepsilon dr\,r^{2m+1},
\ee
with $\tau-\tau_\uv=r\,e^{i\varphi}$. This is also convergent for
$\varepsilon \to 0$ for $n\geq 0$. With this regularization, we thus have a well-defined $u$-plane integral.

It is possible to introduce observables to the path integral, which violate the condition $n\geq 0$ in
Equation \eqref{IntegrandExp} and which are therefore not included in Section \ref{SecObservables}. An example is the anti-holomorphic surface observable \cite{Korpas:2017qdo, Korpas:2019cwg}
$$\tilde I_+(S)=-\frac{i}{\sqrt{2}\pi}\int_S \frac{1}{2} \frac{d^2\bar u}{d\bar a^2}\eta\chi+\frac{\sqrt{2}}{4}\frac{d\bar u}{d\bar a}(F_+-D)$$
Such observables lead to a more severe divergence than the divergences
occuring if \eqref{IntegrandExp} is satisfied. We therefore had to
make the non-symmetric replacement of $v\bfk_m$ and $\bar v\bfk_m$ in
(\ref{zSandv}). For the $N_f=0$ theory,
such severe divergences were regularized in
\cite{Korpas:2019ava}. This regularization also
applies to the cusps $u_1$, $u_2$ and $u_3$ of the $\CN=2^*$
theory. We expect that such divergences for $u\to \infty$ can be treated in
a similar way, but will leave this for future work. 
 
\subsection{Evaluation using Stokes' Theorem}
Having established the regularization of the integral, we proceed with outlining the evaluation using Stokes' Theorem, which follows \cite{Korpas:2017qdo, Moore:2017cmm, Korpas:2019cwg}. The $u$-plane integral $\Phi_\bfmu^J$
is schematically of the form 
\be
\Phi^J_\bfmu=\int_{\CB} \omega.
\ee
We aim to express $\omega$ manifestly as an exact form,
$\omega=d\Omega$, with $\Omega$ being a smooth 1-form on $\CB$.  
We can then evaluate $\Phi^J_\bfmu$ on the boundary $\partial \CB$
\be
\label{PhiintdB}
\Phi^J_\bfmu=\int_{\partial \CB} \Omega.
\ee
There is an ambiguity in the choice of $\Omega$, since for a closed one-form
$\varphi$, replacing $\Omega$ by $\Omega+\varphi$ does not change the
integral. This is simply a choice of integration constant. $\varphi$
can be written as a linear combination of closed, non-exact
one-forms. Since the Coulomb branch of the $\CN=2^*$ is a four
punctured sphere, it has three independent one-cycles, and
consequently three independent closed, non-exact one-forms
$\varphi_j$, $j=1,2,3$. One can
choose the following representatives
\be
\begin{split}
\varphi_1&=\vartheta_3(\tau)^4\,d\tau,\\
\varphi_2&=\vartheta_4(\tau)^4\,d\tau,\\
\varphi_3&=\frac{\vartheta_3(\tau)^4\,\vartheta_4(\tau)^4}{\vartheta_3(\tau)^4\, \vartheta_4(\tau_{\rm uv})^4-\vartheta_4(\tau)^4\, \vartheta_3(\tau_{\rm uv})^4}\,d\tau.
\end{split}
\ee
The integrals of $\varphi_j$ along the boundary arcs in $\partial \CB$
vanish. The integral of $\varphi_1$ receives two non-vanishing,
opposite contributions from arcs around $u_1$ and $u_2$. Similarly,
the non-vanishing contributions for $\varphi_2$ are due to $u_1$ and $u_3$, and for
$\varphi_3$ due to $u_\infty$ and $u_1$. 
Using the ambiguity in $\Omega$, one can in principle ``fix'' the contribution from three out of
four cusps. On the other hand, we can require that the magnitude of
the anti-derivative near each of the boundaries of $\CB$ is of the same order
as that of the integrand. This fixes the anti-derivative uniquely, and
appears to be the most natural choice from the physical perspective.

Concretely, (\ref{UplanetvPsi}) demonstrates that $\omega$
reads
\be
d\tau\wedge d\bar \tau \,\nu\,\Psi_\bfmu^J.
\ee
To express the integrand as an exact form, we write the
integrand $\nu\,\Psi_\bfmu^J$ as a total deriviative using mock
modular forms,
\be
\nu(\tau,\tau_{\rm uv},m)\,\Psi_\bfmu^J(\tau,\bar \tau,v\bfk_m,\bar v\bfk_m)=\frac{d}{d\bar \tau} \widehat
\CH_\bfmu^J(\tau,\bar \tau, \tau_{\rm uv},m).
\ee
The 1-form $\Omega$ from (\ref{PhiintdB}) then reads
\be
\Omega=-d\tau\,\widehat \CH_\bfmu^J(\tau,\bar \tau, \tau_{\rm uv},m),
\ee
and the orientation of $\partial \CB$ is such that ``the interior of
$\CB$ is on the lhs''. The integral can now be written as a sum of contributions from the
various boundary components of the Coulomb branch. 
For $\CN=2^*$, the boundaries are a circle of radius $\varepsilon$ around
$\tau=\tau_{\rm uv}$, and the three arcs around the cusps of
$\mathbb{H}/\Gamma(2)$, which we take at
$\tau=0,2$ and $i\infty$. Reversing the orientation of $\partial \CB$,
(\ref{UplanetvPsi}) is expressed as  
\be
\begin{split}
\Phi_\bfmu^J&= 2[\widehat
\CH_\bfmu^J(\tau,\bar \tau, \tau_{\rm uv},m)]_{q^0}+2[\tau\mapsto -1/\tau]_{q^0}+2[\tau\mapsto
1-1/\tau]_{q^0}\\
&\quad + \lim_{\varepsilon\to 0} \oint_{|\tau-\tau_{\rm uv}|=\varepsilon}
d\tau\,\widehat \CH_\bfmu^J(\tau,\bar \tau, \tau_{\rm uv},m),
\end{split}
\ee 
with $\tau$ circling counter-clockwise around an $\varepsilon$-disc centered at
$\tau_{\rm uv}$.

\section{Analysis of the $u$-plane integral}   
\label{Analysis}
This section will analyze the $u$-plane integral. We start with determining the behavior of various quantities near the singularities in Section \ref{ABCboundaries}, followed by derivations of $S$-duality, holomorphic anomaly, metric dependence and the effect of blowing-up a point on $X$ on $\Phi_\bfmu^J$.

\subsection{Behavior near the boundaries}
\label{ABCboundaries}
Since the $u$-plane integral can be expressed as a sum of boundary
contributions, it is important to analyze the behavior of the
couplings $A$, $B$, $C$, $v$ and other quantities near
the boundaries. 

\subsubsection{$u\to\infty$}
We first consider the limit $u\to\infty$ or $\tau\to \tau_{\rm uv}$. 
To avoid ambiguities with the phases of square roots, we assume that
$\mathrm{Im}(\tau-\tau_{\rm uv})>0$ and $\mathrm{Re}(\tau-\tau_{\rm uv})=0$.

The first terms for large $a$ are
\be
\label{BehavioraInfty}
\begin{split}   
&\frac{da}{du} = \frac{1}{2a} +O(a^{-5}),\\
&u=a^2- \frac{m^2}{12}E_2(\tau_{\rm uv})+O(a^{-2}),\\
&\Delta_{\rm phys}=a^6-\frac{m^2}{4} E_2(\tau_{\rm uv})\,a^4+O(a^2),\\
&v =\frac{1}{2\pi i}\left(\frac{m}{a}+\frac{m^3}{12a^3}E_2(\tau_{\rm uv}) +O(a^{-5})\right),\\
&C =\left(\frac{\Lambda}{m}\right)^{3/2}\, \left(\frac{m}{2a} + \frac{m^3}{16a^3} E_2(\tau_{\rm uv})+O(a^{-5})\right).\\
\end{split}
\ee  

Substitution of (\ref{atautau0}), with $s=-\pi i (\tau-\tau_\uv)$, gives for the different ingredients of the $u$-plane integrand 
\be
\label{BehaviorInfty}
\begin{split}  
&\frac{da}{du} = \frac{1}{m}\, s^{1/2}\left(1 - \frac{E_2(\tau_{\rm uv})}{4}\,s +O(s^{2})\right),\\
&u=\frac{m^2}{4} s^{-1}+\frac{m^2}{24} E_2(\tau_{\rm uv})+O(s),\\
&\Delta_{\rm phys}=\left(\frac{m}{2}\right)^6 s^{-3}\left(1+\frac{E_2(\tau_{\rm uv})}{2}\,s+O(s^2)\right),\\
&v =\frac{1}{\pi i}\, s^{1/2}\left(1+\frac{E_2(\tau_{\rm uv})}{12}\,s+O\!\left( s^{2} \right)\right),\\
&C =  \left(\frac{\Lambda}{m}\right)^{3/2} s^{1/2}\left(1+\frac{E_2(\tau_{\rm uv})}{4}\,s+O(s^{2})\right).\\
\end{split}
\ee 

With these expansions, we are able to make a verification of the
selection rule (\ref{SelectionRule}). We find for the behavior of the
measure 
\be 
\label{leadingtau-tau0} 
\begin{split}
\nu(\tau,\tau_{\rm uv},m)&=\frac{\kappa}{4\,\eta(\tau_{\rm uv})^{2\chi+4\ell}} \left(\frac{\Lambda}{m}\right)^{3 \ell}
s^{\ell-\frac{3}{2}} \left(1 + N(c_{\rm uv})\, E_2(\tau_{\rm uv})\,s+ O(s^2) \right),
\end{split}
\ee
where we
used $\chi+\sigma=4$, and defined
\be
\label{ellandN}
\begin{split}
\ell&=\frac{\bfk_m^2}{2}-\frac{(2\chi+3\sigma)}{8}\in \mathbb{Z} \\
N(c_{\rm uv})
&=\frac{1}{4}\left(\frac{\chi}{2}-3\right)+\frac{1}{2}\left(\frac{\sigma}{8}+1\right)+\frac{\bfk_m^2}{4}\\
&=\frac{1}{16}(4-\sigma+c_{\rm uv}^2).
\end{split}  
\ee
For four-manifolds with $b_2=b_2^+=1$, $\ell\geq -1$, while if
$b_2>1$, $\ell$ is unbounded. Moreover,
since $c_{\rm uv}=2\bfk_m$, we have the identity
$2\dim(G)\,\ell={\rm vdim}(\CM^Q_{k,\bfmu,\mathfrak{s}})$ (\ref{dimCMm}),
such that we find agreement with the selection
  rule (\ref{SelectionRule}). If $\mathfrak{s}$ corresponds to an almost complex
  structure, $\ell=0$ and the dependence on $m/\Lambda$ disappears.  

The leading behavior $s^{\ell-3/2}$ of $\nu$ \eqref{leadingtau-tau0} can be understood from
the anomalous $U(1)_R$ symmetry. Integration over the fermion zero
modes in the $u$-multiplet contributes $R$-charge 2 to the
measure. Together with the change of
variables from $(a,\bar a)$ to $(\tau,\bar \tau)$, it gives rise to
the factor $da/d\tau\sim s^{-3/2}$. The masses of the two massive components of the $SU(2)$
vector multiplet diverge for $u\to \infty$. After integrating out,
these components contribute each $(\chi+\sigma)/2$ to the $R$-charge
of the measure \cite{Witten:1995gf}. In addition, the masses of two of the three components of the adjoint
hypermultiplet diverge for $u\to \infty$. These two components are charged under $U(1)_B$, such that their $R$-charge depends on $c_\uv$. Each component contributes $-2\,{\rm Index}({\rm \bf D})_{A}=(\sigma-c_\uv^2)/4$
\eqref{IndexDA} to the $R$-charge of the measure. The total $R$-charge of the measure is thus
$2-4\ell$. Since the $R$-charge of $u$ is 4, this gives for the leading behavior,
\be  
\nu\sim \frac{da}{d\tau}\, A^\chi\,B^\sigma\,C^{c_\uv^2/4}\sim s^{-3/2}\, u^{\frac{2\chi+3\sigma-c_\uv^2}{8}}\sim s^{\ell-3/2},
\ee 
in agreement with \eqref{leadingtau-tau0}.

Finally, we list the limits of the contact terms $T_{2,\infty}=\lim_{u\to \infty} T_2$ and $T_{3,\infty}=\lim_{u\to \infty} T_3$,
\be 
\begin{split}
&T_{2,\infty}(\tau_\uv)=\frac{m^2}{288\,\Lambda^2}\left(E_2(\tau_\uv)^2-E_4(\tau_\uv) \right),\\
&T_{3,\infty}(\tau_\uv)=-\frac{i\,m}{6\Lambda}\,E_2(\tau_\uv). 
\end{split}
\ee 

\subsubsection{$u\to u_1$}
This boundary corresponds to $\tau=\tau_1\to i\infty$. 
We start by listing the different quantities of the integrand in the
limit $\tau_1\to i\infty$. 

From (\ref{expdaduDelta*}), we have 
\be   
\label{duau1}
\frac{du}{da}=m\, \vartheta_3(\tau_{\rm uv})^2\,\vartheta_4(\tau_{\rm uv})^2+O(q_1^{1/2}),\\ 
\ee
where $q_1=e^{2\pi i \tau_1}$. For the leading term in the
$q_1$-expansion of the discriminant, we find
\be
\label{Deltau1}
\Delta_{\rm
  phys}=(2m)^6\frac{\eta(\tau_{\rm uv})^{24}}{\vartheta_2(\tau_{\rm uv})^{12}}\,q_1^{1/2}+O(q_1^{3/2}).
\ee
We find for the leading terms of the coupling $v$,
\be
\label{vtinfty}
v= -\frac{\tau_1}{2} + \frac{1}{2\pi i}\left(2\log(W)-2(W^{-2}-W^{2})\,q_1^{\frac{1}{2}}+O(q_1) \right).
\ee
Substition of this expansion in (\ref{Cttv}) demonstrates that the leading term in the $q_1$-expansion of $C$ is
\be
\label{Cu1}
\begin{split}
C(\tau_1,\tau_{\rm uv},v)&= \left(\frac{\Lambda}{m}\right)^{3/2}\,\frac{e^{2\pi i (v+\tau_1/4)}}{\vartheta_2(\tau_{\rm uv})^2}+O(1)\\
&=
\left(\frac{\Lambda}{m}\right)^{3/2}
\frac{\vartheta_2(2\tau_{\rm uv})}{2\vartheta_3(2\tau_{\rm uv})^3} \,q_1^{-1/4}
+O(1),
\end{split}
\ee 
where we used the identity
$\vartheta_2(\tau)^2=2\,\vartheta_2(2\tau)\,\vartheta_3(2\tau)$. 

With (\ref{duau1}), \eqref{Deltau1} and \eqref{Cu1}, we find for the leading behavior of the measure $\nu$ 
\be
\label{u1asympnu}
\begin{split}
\nu(\tau_1,\tau_\uv,m)&=\kappa\,2^{12+\sigma-2\ell}\left(
 \frac{\Lambda}{m}\right)^{3\ell}
\eta(\tau_\uv)^{4\ell+4\chi+9\sigma}\,\vartheta_2(\tau_\uv)^{-3\sigma/2-12}\\
&\qquad \times \vartheta_3(\tau_\uv)^{\chi-6}\,\vartheta_4(\tau_\uv)^{\chi-6}\left(
  \frac{\vartheta_2(2\tau_\uv)}{\vartheta_3(2\tau_\uv)^3}\right)^{\bfk_m^2}\,q_1^{-\ell/2}+\dots.
\end{split}
\ee

We can understand the leading behavior of $\nu$ from the
local $R$-symmetry  \cite{Witten:1995gf, Shapere:2008zf}. The local $R$-charges are normalized such
that the local coordinate $a_1$ has charge 2. Since $u-u_1$  vanishes linearly with the local
coordinate $a_1$ for $u$ near $u_1$, the local $R$-charge of $u-u_1$
is 2. The massless modes of the $u$-multiplet account for $(\chi+\sigma)/2=2$,
whereas the light quark contributes $-2$\,Index$({\rm \bf
  D}_A)=(\sigma-c_\uv^2)/4$ \eqref{IndexDA} to the
$R$-charge of the measure. The total $R$-charge of the measure is therefore $-2\ell$.
We therefore arrive at,  
\be
\nu\sim \frac{da}{d\tau}\, A^\chi\,B^\sigma\,C^{c_\uv^2/4}\sim (u-m^2\,e_1/4)^{-\ell} \sim q_1^{-\ell/2},
\ee
in agreement with (\ref{u1asympnu}). 

We finish this subsection by listing the limits of the contact terms, $T_{2,1}=\lim_{u\to u_1}T_2$ and $T_{3,1}=\lim_{u\to u_1}T_3$,
\be
\label{T2T3u1}
\begin{split}
T_{2,1}(\tau_\uv)&=\frac{m^2}{144\,\Lambda^2} \left( -3\, \vartheta_3^4(\tau_{\rm
    uv})\, \vartheta_4^4(\tau_{\rm uv}) \right. \\
& \left. \quad +\, E_2(\tau_{\rm uv})\,(\vartheta_3^4(\tau_{\rm
    uv})+\vartheta_4^4(\tau_{\rm uv}))+E_4(\tau_{\rm uv})+\frac{6}{\pi i}\,\frac{dE_2(\tau_\uv)}{d\tau_\uv}\right),\\
T_{3,1}(\tau_\uv)&=-\frac{i\,m}{6\,\Lambda} \left( \vartheta_3(\tau_{\rm
    uv})^4+\vartheta_4(\tau_{\rm uv})^4+2\,E_2(\tau_{\rm
    uv})\right)+\frac{i}{2\Lambda} \left(\frac{du}{da} \right)_1,
\end{split}
\ee
where $(du/da)_1=\lim_{u\to u_1} du/da$ is the leading term in \eqref{duau1}. 

\subsubsection{$u\to u_2$}
\label{ABCutou2}
We determine the behavior of the various quantities in the
strong-coupling limit $\tau\to 0$. We denote the dual weak coupling by
$\tau_2=-1/\tau$.\footnote{Note $\tau_2$ is {\it not} the imaginary
  part of $\tau$, which is $y$.}

We define the strongly coupled version of $du/da$,
$(du/da)_{2}(\tau_{2})=\tau_{2}(du/da)(-1/\tau_{2})$. In the limit $\tau \to
0$, this approaches
\be
\label{dadu2}
\begin{split}
\left(\frac{du}{da}\right)_{2}\!\!(\tau_{2})&=m
\,\vartheta_2(\tau_{\rm uv})^2\,\vartheta_3(\tau_{\rm uv})^2 + O(q_{2}^{1/2})\\
&=2\sqrt{2}i\,\Lambda_0 +O(q_{\rm uv}^{1/2}, q_2^{1/2}).
\end{split}
\ee
Similarly, the dual of the discriminant
$\Delta_{\rm phys}(\tau)=\Delta_{\rm phys}(-1/\tau_2)$. The leading term is
\be 
\label{Deltau2}
\Delta_{\rm phys,2}(\tau_{2}):=\Delta_{\rm phys}(-1/\tau_{2})=-(2m)^6
\frac{\eta(\tau_{\rm uv})^{24}}{\vartheta_4(\tau_{\rm uv})^{12}}\,q_{2}^{1/2}+O(q_{2}). 
\ee

To determine the dual version, $v_{2}$ of $v$, we deduce from the
monodromy (\ref{M2trafos}), that the leading term is
$\tau_{2}/2$. Using (\ref{tauvtauUV}), we deduce sub-leading terms,
\be
\label{tau2v2}
\begin{split}
&\tau_{2}= - \frac{1}{\tau}\mapsto i\infty, \qquad
v_{2}=-\frac{v}{\tau}\mapsto \frac{\tau_{2}}{2}+\frac{1}{\pi i}\log\!\left( \frac{\vartheta_3(\tau_{\rm uv}/2)}{\vartheta_4(\tau_{\rm uv}/2)}\right)+O\!\left(q_{2}^{1/2}\right).
\end{split}
\ee

We let $C_2$ be the coupling dual to $C$ near $u_2$, defined as 
\be 
C_2(\tau_2,\tau_{\rm uv},v_2)=\exp(-2\pi i
v_2^2/\tau_2)\,C(-1/\tau_2,\tau_{\rm uv},v_2/\tau_2).
\ee
Using (\ref{Cttv}), we find that it is given by
\be
\label{C2}
\begin{split}
C_{2}(\tau_{2},\tau_{\rm uv},v_{2})&=-\left( \frac{\Lambda}{m}\right)^{\frac{3}{2}}
\,\frac{1}{\vartheta_4(\tau_{\rm uv})^2}\,\frac{\vartheta_1(2\tau_{2},2v_{2})}{\vartheta_4(2\tau_{2})}\\
&=i\left( \frac{\Lambda}{m}\right)^{\frac{3}{2}}
\frac{1}{\vartheta_4(\tau_{\rm uv})^2}\, e^{-2\pi i v_2}\, q_{2}^{1/4} +O(q_{2}^{1/4})\\
&= i \left( \frac{\Lambda}{m}\right)^{\frac{3}{2}}  \frac{\vartheta_4(\tau_{\rm uv}/2)}{\vartheta_3(\tau_{\rm uv}/2)^3}\,q_2^{-1/4}+O(q_2^{1/4}).
\end{split}
\ee 
As expected, the coefficient of $q_{2}^{-1/4}$ is the UV $S$-dual of the
coefficient of $q_1^{-1/4}$ in (\ref{Cu1}).

With \eqref{dadu2}, \eqref{Deltau2} and \eqref{C2}, the leading
behavior of $\nu$ is
\be
\begin{split}
\nu(\tau,\tau_\uv,m)&=\kappa\,(-1)^{\ell}\,2^{10+3\sigma/2}\,\left(
 \frac{\Lambda}{m}\right)^{3\ell}
\eta(\tau_\uv)^{4\ell+4\chi+9\sigma}\,\vartheta_4(\tau_\uv)^{-3\sigma/2-12}\\
&\qquad \times  \vartheta_2(\tau_\uv)^{\chi-6}\,\vartheta_3(\tau_\uv)^{\chi-6}\left(
  \frac{\vartheta_4(\tau_\uv/2)}{\vartheta_3(\tau_\uv/2)^3}\right)^{\bfk_m^2}\,q_2^{-\ell/2}+\dots.
\end{split}
\ee
Analogously to $u_1$, we can understand the leading behavior of $\nu$ from the
local $R$-symmetry. The massless monopole contributes 
$(\sigma-c_\uv^2)/4$ to the $R$-charge of the measure.

We finish this subsection by listing the limits of the contact terms $T_2$ and $T_3$ for $u\to u_2$,
\be
\label{T2T3u2}
\begin{split}
T_{2,2}(\tau_\uv)&=\frac{m^2}{144\,\Lambda^2} \left( -3\, \vartheta_2^4(\tau_{\rm
    uv})\, \vartheta_3^4(\tau_{\rm uv}) \right. \\
& \left. \quad -E_2(\tau_{\rm uv})\,(\vartheta_2^4(\tau_{\rm
    uv})+\vartheta_3^4(\tau_{\rm uv}))+E_4(\tau_{\rm uv})+\frac{6}{\pi i}\,\frac{dE_2(\tau_\uv)}{d\tau_\uv}\right),\\
T_{3,2}(\tau_\uv)&=-\frac{i\,m}{6\,\Lambda} \left(-\vartheta_2(\tau_{\rm
    uv})^4-\vartheta_3(\tau_{\rm uv})^4+2\,E_2(\tau_{\rm uv})\right)-\frac{i}{2\Lambda} \left(\frac{du}{da} \right)_2.
\end{split}
\ee

\subsubsection{$u\to u_3$}
We define the dual coupling near this singularity by 
\be
\tau_{3}=-\frac{1}{\tau-1}.
\ee

We define the strong coupling version of $(du/da)_{3}=\tau_{3}
\,(du/da)(1-1/\tau_{3})$. Near $u_3$, this gives
\be
\label{duda3}
\begin{split}
\left( \frac{du}{da} \right)_{\!\! 3}(\tau_{3})&=-m\,i
\,\vartheta_2(\tau_{\rm uv})^2\,\vartheta_4(\tau_{\rm
  uv})^2+O(q_{3}^{1/2})\\
&=2\sqrt{2}\,\Lambda_0 + O(q_{\rm uv}^{1/2},q_{3}^{1/2})
\end{split}
\ee
We have resolved the ambiguity of the square root, by requiring that
the $N_f=0$ limit matches with the limit of $du/da$ of the $N_f=0$
theory.

The dual discriminant is $\Delta_{{\rm phys},3}(\tau_{3})=\Delta_{\rm
  phys}(1-1/\tau_{3})$, with leading term
\be
\label{Deltaphys3}
\Delta_{{\rm phys},3}(\tau_{3})=-(2m)^6\,\frac{\eta(\tau_{\rm uv})^{24}}{\vartheta_3(\tau_{\rm uv})^{12}}\,q_{3}^{1/2}+O(q_{3}).
\ee

The strong-coupling limit $\tau\to 1$,
\be 
\label{tau3v3}
\tau_{3} \mapsto i\infty,\qquad v_{3}=-\frac{v}{\tau-1}\mapsto \frac{\tau_{3}}{2}+\frac{1}{\pi i}\log\!\left(\frac{\vartheta_4((\tau_{\rm uv}+1)/2)}{\vartheta_3((\tau_{\rm uv}+1)/2)}\right)+O(q_{3}^{1/2}),
\ee
The dependence of $v_{3}$ on $\tau_{3}$ matches with the monodromy around $u_3$ (\ref{M3trafos}).
 
For $C_{3}(\tau_{3},\tau_{\rm uv},v_{3})=\exp(-2\pi i
v_{3}^2/\tau_{3})\,C(1-1/\tau_{3},\tau_{\rm uv},v_{3}/\tau_{3})$,
\be
\label{C3}
\begin{split} 
C_{3}(\tau_{3},\tau_{\rm uv},v_{3})&=\left(
  \frac{\Lambda}{m}\right)^{\frac{3}{2}}
\,\frac{-1}{\vartheta_3(\tau_{\rm uv})^2}\,\frac{\vartheta_1(2\tau_{3},2v_{3})}{\vartheta_4(2\tau_{3})}\\
&=i\left( \frac{\Lambda}{m}\right)^{\frac{3}{2}}
\frac{1}{\vartheta_3(\tau_{\rm uv})^2}\,e^{-2\pi i v_3}
q_3^{1/4}+O(q_{3}^{1/4})\\
&=i\left( \frac{\Lambda}{m}\right)^{\frac{3}{2}}
\frac{\vartheta_4((\tau_\uv+1)/2)}{\vartheta_4((\tau_\uv+1)/2)^3}\,q_3^{-1/4}+O(q_{3}^{1/4})
\end{split}
\ee

With \eqref{duda3}, \eqref{Deltaphys3} and \eqref{C3}, we find for the
leading behavior of $\nu$
\be
\begin{split}
\nu(\tau,\tau_\uv,m)&=\kappa\, e^{-\frac{\pi i}{2}
  (\frac{\chi}{2}-3)}\,(-1)^\ell\,2^{10+3\sigma/2}\,\left(
  \frac{\Lambda}{m}\right)^{3\ell}
\eta(\tau_\uv)^{4\ell+4\chi+9\sigma}\,\vartheta_3(\tau_\uv)^{-3\sigma/2-12}\\
&\qquad \times  \vartheta_2(\tau_\uv)^{\chi-6}\,\vartheta_4(\tau_\uv)^{\chi-6}\left(
  \frac{\vartheta_4((\tau_\uv+1)/2)}{\vartheta_3((\tau_\uv+1)/2)^3}\right)^{\bfk_m^2}\,q_3^{-\ell/2}+\dots.
\end{split}
\ee
Analogously to $u_1$, we can understand the leading behavior of $\nu$ from the
local $R$-symmetry. The massless dyon contributes
$(\sigma-c_\uv^2)/4$ to the $R$-charge of the measure.

We finish this subsection by listing the limits of the contact terms $T_2$ and $T_3$ for $u\to u_3$,
\be
\label{T2T3u3}
\begin{split}
T_{2,3}(\tau_\uv)&=\frac{m^2}{144\,\Lambda^2} \left( 3\, \vartheta_2^4(\tau_{\rm
    uv})\, \vartheta_4^4(\tau_{\rm uv}) \right. \\
& \left. \quad +E_2(\tau_{\rm uv})\,(\vartheta_2^4(\tau_{\rm
    uv})-\vartheta_4^4(\tau_{\rm uv}))+E_4(\tau_{\rm uv}) +\frac{6}{\pi i}\,\frac{dE_2(\tau_\uv)}{d\tau_\uv}\right),\\
T_{3,3}(\tau_\uv)&=-\frac{i\,m}{6\,\Lambda} \left(\vartheta_2(\tau_{\rm
    uv})^4-\vartheta_4(\tau_{\rm uv})^4+2\,E_2(\tau_{\rm uv})\right) -\frac{i}{2\Lambda} \left(\frac{du}{da} \right)_3.
\end{split}
\ee 

\subsection{$S$-duality of $\Phi^J_\bfmu$} 
\label{SecPhiSdual}
In this subsection, we will analyze relations and duality properties of
$\Phi^J_\bfmu=\Phi^J_\bfmu(\tau_{\rm uv},\bar \tau_{\rm uv};\bfk_m)$. Equation (\ref{Psi0})
implies a periodicity for $\bfmu$,
\be 
\Phi^J_{\bfmu+\bfnu}(\tau_{\rm uv},\bar \tau_{\rm uv};\bfk_m)=\Phi^J_{\bfmu}(\tau_{\rm uv},\bar \tau_{\rm uv};\bfk_m),\qquad \bfnu\in L.
\ee 
In other words, $\Phi^J_{\bfmu}$ is independent of the choice of lift $\bar w_2(E)=2\bfmu \in L$ of $w_2(E)$. For the dependence on $\bfk_m$, we have the relation,
\be
\Phi^J_\bfmu(\tau_{\rm uv},\bar \tau_{\rm uv};-\bfk_m)=-\Phi^J_\bfmu(\tau_{\rm uv},\bar \tau_{\rm uv};\bfk_m),
\ee
thus in particular $\Phi^J_\bfmu=0$ for $\bfk_m=0$.
(Nevertheless, the topological correlators can be nonzero 
when $\bfk_m=0$.) 

We first consider the $T$-transformation, $\tau_{\rm uv}\mapsto \tau_{\rm uv}+1$. To
determine the effect on  $\Phi^J_\bfmu$, we also shift $\tau\mapsto
\tau+1$. This changes the integration domain, but using a $\Gamma(2)$
transformation on some of the $\CF_\infty$ images, we can map this
back to the original domain. Since the couplings, $A$, $B$ and $C$ are
unchanged under this transformation, the only change is from $\eta(\tau_{\rm uv})^{-\chi+3/2\sigma-2\bfk_m^2}$ and 
$\Psi_\bfmu^J$ (\ref{PsiTrafos}). 
We find
\be
\label{PhiTtrafo}
\begin{split}
\Phi^J_\bfmu(\tau_{\rm uv}+1,\bar \tau_{\rm uv}+1)&=e^{2\pi i
  (\bfmu^2-B(\bfmu,K)-(\chi+2\ell)/12)}\,\Phi^J_\bfmu(\tau_{\rm uv},\bar \tau_{\rm uv})\\
&=e^{-2\pi i   (\bfmu^2+(\chi+2\ell)/12)}\,\Phi^J_\bfmu(\tau_{\rm uv},\bar \tau_{\rm uv}).
\end{split}
\ee
We note that while the integrand is invariant under
the shift of the effective coupling constant, $\tau\mapsto \tau+2$,
the integral $\Phi^J_\bfmu$ is in general multiplied by a phase for
$\tau_{\rm uv}\mapsto \tau_{\rm uv}+2$. 

We proceed similarly for the $S$-transformation, $\tau_{\rm uv}\mapsto
-1/\tau_{\rm uv}$. We change variables in the integrand, $\tau\mapsto
-1/\tau$. Then,
\be
\begin{split}
&\frac{da}{du}(-1/\tau,-1/\tau_{\rm uv})= -\frac{\tau}{\tau_{\rm uv}^2}\frac{da}{du}(\tau,\tau_{\rm uv}),\\
& \Delta_{\rm phys}(-1/\tau,-1/\tau_{\rm uv})= \tau_{\rm uv}^6\,\Delta_{\rm
  phys}(\tau,\tau_{\rm uv}),\\
&C(-1/\tau,-1/\tau_{\rm uv})=-\tau_{\rm uv}^{-1}\,e^{2\pi i \bfk_m^2 v^2/\tau}\,C(\tau,\tau_{\rm uv}). 
\end{split}
\ee

Therefore, $\nu$ transforms as
\be
\begin{split}
\nu(-1/\tau,-1/\tau_{\rm uv},m)&=i\,(-i\tau_{\rm
  uv})^{-\chi/2-4\ell} (-i\tau)^{3-\chi/2} (-1)^\ell\\
&\quad \times e^{2\pi i \bfk_m^2 v^2/\tau}\, \nu(\tau,\tau_{\rm uv},m).
\end{split} 
\ee
Using $\chi+\sigma=4$, and (\ref{Psi0}), (\ref{PsiTrafos}) for the transformation of
$\Psi^J_\bfmu$, we arrive at 
\be
\label{PhiSdual} 
\begin{split}
\Phi^J_\bfmu(-1/\tau_{\rm uv},-1/\bar \tau_{\rm uv})&=2^{-b_2/2}e^{\pi i (\bfk_m^2/2-\sigma/8)}\,(-i\tau_{\rm uv})^{-\chi/2-4\ell}\\
&\quad \times \sum_{\bfnu\in (L/2)/L}e^{4\pi i
  B(\bfmu,\bfnu)}\,\Phi^J_\bfnu(\tau_{\rm uv},\bar \tau_{\rm uv}),
\end{split} 
\ee
Thus $\Phi^J_\bfmu$ forms a vector-valued representation of
$SL(2,\mathbb{Z})$ for any choice of ${\rm Spin}^c$ structure $\bfk_m$. Note that for $\bfk_m^2$
an almost complex structure, $\ell=0$, the weight reduces to $-\chi/2$, which is
the familiar weight from VW theory \cite{Vafa:1994tf}. We make a more
thorough comparison in Section \ref{CompVWtheory}.

In Section \ref{SecNf0limit}, we introduced the ``renormalized" partition function $\underline \Phi^J_\bfmu$ (\ref{underlinePhi}). It transforms as \eqref{PhiSdual}, except that its weight is $\chi/2-2\ell$. 
\\
\\
{\it Observables and $S$-duality}\\ 
We collect all the couplings involving $p$ and $S$ discussed in
Section \ref{SecObservables}. These give rise to the factor 
\be
\label{pSObservables}
P(\tau,\tau_\uv, p,S)=e^{T_2\, S^2+T_3 B(S,\bfk_m)+p\,\tilde u/\Lambda^2-\frac{i}{\Lambda} \frac{du}{da}
  B(\bfk_-,S)}. 
\ee 
From the transformation properties for $u$, $du/da$, $T_2$ (\ref{T2Sdual})
and $T_3$ \eqref{T3Sdual}, we can deduce the $S$-duality action on $p$ and
$S$. For an element $\gamma=\left(\begin{array}{cc} a & b \\ c & d \end{array}\right)\in SL(2,\mathbb{Z})$, we set 
\be 
\begin{split}
&p\mapsto \gamma\cdot p:=\frac{1}{(c\tau_{\rm uv}+d)^2}\left( p -\frac{c(c\tau_{\rm
    uv}+d)}{2\pi i}\,S^2\right),\\
&S\mapsto \gamma\cdot S:=\frac{S}{(c\tau_{\rm uv}+d)^2}.
\end{split}
\ee
Similarly to before we determine the transformation of the simultaneous IR and UV S-duality. 
We find that $P$  (\ref{pSObservables}) transforms as
\be
\begin{split}
&P( \gamma\cdot \tau,\gamma\cdot \tau_\uv,\gamma\cdot p,\gamma\cdot S)=\\
&e^{- \frac{m^2\,S^2}{4\pi^2\,\Lambda^2} \frac{c^2
  }{(c\tau_\uv+d)^2}-\frac{2m}{\pi\Lambda}
  \frac{c}{c\tau_\uv+d}B(S,\bfk_m)}\,P\!\left(\tau,\tau_\uv, p, -\frac{S}{c\tau+d}\right).
\end{split}
\ee
The $-S/(c\tau+d)$ matches with the duality transformation of the elliptic variable
of $\Psi_\bfmu^J$.  

Then the $S$ and $T$ transformations of $\Phi^J_\bfmu$ including observables are
\be
\label{SdualObs}
\begin{split}
&\Phi^J_\bfmu\!\left(-1/\tau_\uv,-1/\bar \tau_\uv,(p-\tau_\uv\,S^2/2\pi
  i)/\tau_\uv^2,S/\tau_\uv^2\right)=\\
&\qquad e^{- \frac{m^2\,S^2}{4\pi^2\,\Lambda^2\,\tau_\uv^2}
  -\frac{2m}{\pi\Lambda\,\tau_\uv} B(S,\bfk_m)} \,2^{-b_2/2}e^{\pi i (\bfk_m^2/2-\sigma/8)}\,(-i\tau_{\rm uv})^{-\chi/2-4\ell}\\
&\qquad \times \sum_{\bfnu\in (L/2)/L}e^{4\pi i   B(\bfmu,\bfnu)}\,\Phi^J_\bfnu(\tau_{\rm uv},\bar \tau_{\rm
  uv},p,S),\\
&\Phi^J_\bfmu(\tau_{\rm uv}+1,\bar \tau_{\rm uv}+1,p,S)=e^{-2\pi i   (\bfmu^2+(\chi+2\ell)/12)}\,\Phi^J_\bfmu(\tau_{\rm uv},\bar \tau_{\rm uv},p,S).
\end{split}
\ee
{\it Remark} The transformation laws of $(p,S)$ can be understood from
the Lagrangian viewpoint by studying the standard Abelian S-duality
transformation at the path integral level, at least in a simplified
version of the low energy Abelian theory. See \cite[pages 41-43]{Labastida:1998sk}.
\\
\\
{\it $S$-Duality of the $SU(2)$ and $SO(3)_\pm$ theories}\\
Using the transformations under $S$ and $T$, we can analyze the
$S$-duality properties of the partition functions of the three theories $\CT_{SU(2)}$ and
$\CT_{SO(3)_\pm}$, which we discussed in Section \ref{Sdualrankone}. 
Leaving aside the Seiberg-Witten contributions, we define the following partition functions for these theories,
\be 
\begin{array}{ll}
Z_{\bfmu}[\CT_{SU(2)}]=\Phi_\bfmu^J, & \\
Z_{{\bfmu}}[\CT_{SO(3)_+}]= \sum_{\bfnu \in (L/2)/L} e^{4\pi i B(\bfmu,\bfnu)}\Phi_\bfnu^J, &\\
Z_{\bfmu}[\CT_{SO(3)_-}]=\sum_{\bfnu \in (L/2)/L} e^{4\pi i B(\bfmu,\bfnu)-2\pi i \bfnu^2}\Phi_\bfnu^J. & 
\end{array} 
\ee 
Thus for each of the $\CT_{SU(2)}$ and $\CT_{SO(3)_\pm}$ theories, we
define a $|L/(L/2)|$ partition functions, each with a fixed 't Hooft
flux $\bfmu$. 

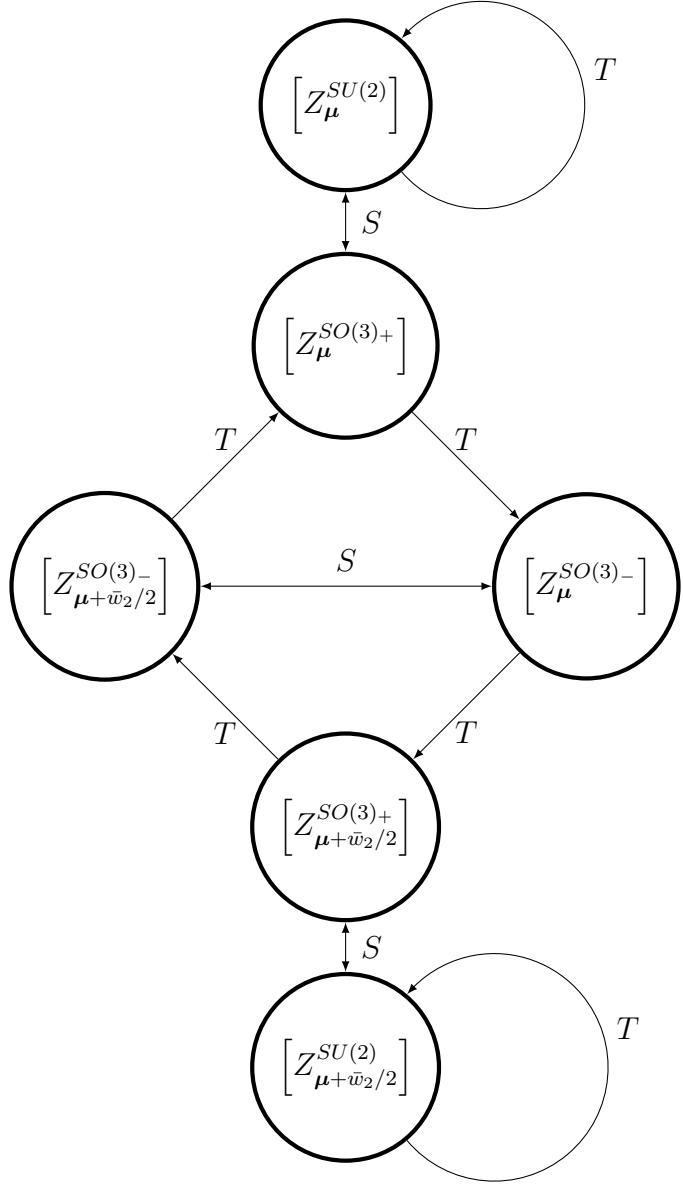
\begin{figure}[h!]
 \begin{center}
\begin{tikzpicture}[inner sep=2mm,scale=1.6]
 \node (1a) at (0,8) [circle,draw, ultra thick] {$\left[Z_{\bfmu}^{SU(2)}\right]$}; 
 \node (2a) at (0,6) [circle,draw, ultra thick] {$\left[Z_{\bfmu}^{SO(3)_+}\right]$}; 
 \node (3a) at (2,4) [circle,draw, ultra thick] {$\left[Z_{\bfmu}^{SO(3)_-}\right]$};
 \node (3b) at (-2,4) [circle,draw, ultra thick] {$\left[Z_{\bfmu+\bar w_2/2}^{SO(3)_-}\right]$};
 \node (2b) at (0,2) [circle,draw, ultra thick] {$\left[Z_{\bfmu+\bar w_2/2}^{SO(3)_+}\right]$}; 
 \node (1b) at (0,0) [circle,draw, ultra thick] {$\left[Z_{\bfmu+\bar w_2/2}^{SU(2)}\right]$}; 
\drawloop[->,stretch=1.0]{1a}{-50}{50} node[pos=-2,right]{$T$};
\drawloop[->,stretch=1.0]{1b}{-50}{50} node[pos=-2,right]{$T$};
\draw [<->] (1a) to node[auto] {$S$} (2a);
\draw [->] (2a) to node[above] {$T$} (3a);
\draw [->] (3a) to node[below] {$T$} (2b);
\draw [->] (2b) to node[below] {$T$} (3b);
\draw [->] (3b) to node[above] {$T$} (2a);
\draw [<->] (2b) to node[right] {$S$} (1b);
\draw [<->] (3a) to node[above] {$S$} (3b);
\end{tikzpicture}
\end{center} 
\caption{Action of the generators $S$ and $T$ of $SL(2,\mathbb{Z})$ on
  equivalence classes $[Z^G_\bfmu]$ of partition functions
  $Z^G_\bfmu:=Z_\bfmu[\CT(G)]$. $\bar w_2$ is an integral lift of
  $w_2(X)$ to $L$.}
\label{STPF}
\end{figure}

These partition functions are mapped to each other under $S,T\in
SL(2,\mathbb{Z})$, up to an overall factor due to the non-trivial
weight and phase. We identify partition functions, which differ by
such an overall factor and denote an equivalence class by $\left[
  Z_\bfmu(G)\right]$. The duality diagram for such equivalence classes
is given in Figure \ref{STPF}. This is in a sense a doubling of the
diagram of theories in Figure \ref{STTheories}, with three extra nodes
with partition functions for 't Hooft flux $\bfmu+\bar w_2(X)/2\in L/2$. We
consider these as partition functions for one of the three
theories, but with a different 't Hooft flux.  Note that if $X$ is a spin
manifold, then $\bar w_2(X)=0$, and the diagram in in Figure \ref{STPF}
reduces to one in Figure \ref{STTheories}. The diagram in Figure
\ref{STPF} with $\bfmu=0$ recently also appeared in \cite{Ang:2019txy,
  Gukov:2020btk}. It was suggested in these papers that the partition
function for $Z^{SU(2)}_{\bar w_2(X)}$, was that of a different theory,
namely spin-$SU(2)$, and similarly for $SO(3)_\pm$, which is slightly
different from our interpretation.

\subsection{Holomorphic anomaly}
\label{SecHolAnomaly}
An intriguing aspect is the dependence on the anti-holomorphic coupling
constant $\bar \tau_{\rm uv}$ of the Coulomb path integral $\Phi^J_\bfmu$. The $\bar \tau_{\rm uv}$
dependence of the Lagrangian (\ref{CLtauQW}) is contained in its
$Q$-exact part. The $\bar \tau_{\rm uv}$-dependence therefore evaluates to
the one-point function
\be
\frac{\partial}{\partial \bar \tau_{\rm uv}} \Phi^J_\bfmu=\left< [Q,G]
\right>, 
\ee
with $G=\int _X \partial_{\bar \tau_{\rm uv}} W$. Since $Q$ is a global
fermionic symmetry of the theory, a non-vanishing value of $\left< [Q,G]
\right>$ implies the
presence of a $Q$-anomaly. On the other hand, the path integral for such
correlation functions of a $Q$-exact observable can be 
expressed as a total derivative on field space, and therefore
relatively easily evaluated at the boundary of field space. We will demonstrate that the result is
non-vanishing. The result suggests that the $Q$-anomaly is due to the contribution from a class
of classical solutions which do not satisfy the $Q$-fixed point
equations. These are the reducible 
connections or abelian configurations, whose field strength in the two-dimensional
representation is of the form
\be
\left[ \frac{F}{2\pi} \right]=\left(\begin{array}{cc} \boldsymbol{c} &
    0 \\ 0 & -\boldsymbol{c}  \end{array} \right),
\ee
with $\bfc\in H^2(X,\mathbb{Z})$. These are solutions to the equations
of motion but are not anti-self-dual. The instanton
number of these solutions is $k=-\bfc^2$. Thus $k$ can be either
positive or negative, while $k\geq 0$
for proper $Q$-fixed solutions. Moreover, the
kinetic part evaluates to
\be
\frac{1}{8\pi^2} \int_X \mathrm{Tr}_{\bf 2}\left( F\wedge *F \right)=2\,B(\bfc,J)^2-\bfc^2,
\ee  
which for generic $J$ exceeds the lower bound $k=-\bfc^2$. For the 
$Q$-fixed anti-self-dual instanton solutions, the lower bound is
saturated. Solutions which exceed the lower bound give necessarily
rise to anti-holomorphic dependence of the path integral.
   
Solutions which do not saturate the $Q$-bound typically do not
contribute to the path integral due to cancellations between bosonic
and fermionic degrees of freedom. However, for four-manifolds with
$b_2^+=1$, the Coulomb branch is shown to contribute at tree
level \cite{Moore:1997pc}, and the Coulomb branch measure 
involves a sum over \underline{all} the harmonic connections 
on the low energy line bundle. If we view these as reducible 
connections, they are not self-dual.

A complementary viewpoint is given by string
dualities. This is most straightforward for the $\CN=4$ case, i.e. the massless theory with
$\bfk_m$ an ACS. Then the theory can be related to a two-dimensional
field theory with $(0,4)$ supersymmetry. The non-compact field space
of this theory can lead to a spectral asymmetry, which prevents the
precise cancellation among bosonic and fermionic states for non-BPS states
\cite{atiyah_patodi_singer_1975, Cecotti:1992qh, Bershadsky:1993ta, Dabholkar:2019nnc}. In our
context, $u\to \infty$ gives rise to a non-compact direction in field
space. One can also reduce further to a quantum-mechanical model of
BPS particles of an $\CN=2$ theory in four dimensions. From this perspective, the non-holomorphic contribution can be obtained as the contribution from multi-particle BPS states to the
Witten index in 
\cite{Alexandrov:2014wca, Pioline:2015wza}. \\
\\   
{\it Anomaly for $\bfk_m$ an almost complex structure}\\  
To determine the dependence of the partition function on $\bar
\tau_{\rm uv}$ explicitly, let us start with $\bfk_m$ an almost complex
structure. Changing variables to $\sigma=\tau-\tau_{\rm uv}$ in the
integrand, we arrive at
\be
\begin{split}
\partial_{\bar \tau_{\rm uv}} \Phi_\bfmu^J(\tau_{\rm uv},\bar \tau_{\rm uv})& =\partial_{\bar \tau_{\rm uv}} \int_{\CU_\varepsilon-\tau_{\rm uv}} d\sigma\wedge
d\bar \sigma\, \nu(\sigma+\tau_{\rm uv},\tau_{\rm uv},m)\,\\
&\quad \times \Psi_\bfmu^J(\sigma+\tau_{\rm uv},\bar \sigma+\bar \tau_{\rm uv},v\bfk_m,\bar v \bfk_m),
\end{split}
\ee
with $v=v(\sigma+\tau_{\rm uv},\tau_{\rm uv})$. Thus the integrand depends
explicitly on $\bar \tau_{\rm uv}$ through the shift of $\bar \sigma$, and
the explicit dependence in $\bar v$. The first dependence gives
naturally rise to a total $\bar \tau$-derivative. 

To understand the dependence through $\bar v$, we replace $v \bfk_m$
by a generic elliptic variable $\bfz\in H^2(X,\mathbb{C})$ in the sum over
fluxes $\Psi_\bfmu^J$ (\ref{defPsi3}). Then for fixed $\tau$, we have 
\be
\label{tau0Psi}
\partial_{\bar \tau_{\rm uv}}\Psi_\bfmu^J(\tau,\bar \tau,\bfz,\bar
\bfz)\vert_{\tau\,\,{\rm fixed}}=\partial_{\bar \tau} \left( 
  \frac{i}{\sqrt{y}} \left(\partial_{\bar \tau_{\rm uv}} \bar \bfz\vert_{\bar \tau\,\,{\rm fixed}}\right)\,\Psi_\bfmu^J[1](\tau,\bar \tau,\bfz,\bar
\bfz)\right).
\ee 
where $\Psi_\bfmu^J[1]$ is the series (\ref{defPsiK}) with kernel
$\CK=1$. In the $u$-plane integral $\bfz=v\,\bfk_m$. We therefore
arrive at
\be
\label{Dtau0Psi}
\begin{split}
&\partial_{\bar \tau_{\rm uv}}\Psi_\bfmu^J(\sigma+\tau_{\rm uv},\bar \sigma+\bar
\tau_{\rm uv},v\bfk_m,\bar v \bfk_m)=\partial_{\bar \sigma}\left( \Psi_\bfmu^J(\sigma+\tau_{\rm uv},\bar
  \sigma+\bar \tau_{\rm uv},v\bfk_m,\bar v \bfk_m) \right.\\
&\quad \left.+  \frac{i\,B(\bfk_m,J)}{\sqrt{y_\uv+\mathrm{Im}(\sigma)}} \left(\partial_{\bar \tau_{\rm uv}} \bar v
    \vert_{\bar \sigma+\bar \tau_{\rm uv}\,\,{\rm
        fixed}}\right)\,\Psi_\bfmu^J[1](\sigma+\tau_{\rm uv},\bar \sigma+\bar
  \tau_{\rm uv},v\bfk_m,\bar v 
  \bfk_m)\right).
\end{split}
\ee
We can express $\partial_{\bar \tau_{\rm uv}} \bar v \vert_{\bar \sigma+\bar \tau_{\rm uv}\,\,{\rm
        fixed}}$ as $(\partial_{\bar \tau_{\rm uv}} -\partial_{\bar
      \sigma})\overline{v(\tau_{\rm uv}+\sigma,\tau_{\rm uv})}$. The $\partial_{\bar
      \sigma}$-term of this derivative cancels against a term of the 
    kernel $\partial_{\bar \tau}(\sqrt{4y}B(\bfk+\bfb,J))$ of
    $\Psi_\bfmu^J$ (\ref{defPsi3}) after changing variables to $\tau=
    \sigma+\tau_{\rm uv}$. The theta series within the brackets in
    (\ref{Dtau0Psi}) can then be expressed as a theta
    series $\Psi_\bfmu^J[\CK_0]$ (\ref{defPsiK}),
\be
\partial_{\bar \tau_{\rm uv}}\Psi_\bfmu^J(\sigma+\tau_{\rm uv},\bar \sigma+\bar
\tau_{\rm uv},v\bfk_m,\bar v \bfk_m)=\partial_{\bar \sigma}\Psi_\bfmu^J[\CK_0](\sigma+\tau_{\rm uv},\bar \sigma+\bar
\tau_{\rm uv},v\bfk_m,\bar v \bfk_m),
\ee
with the kernel $\CK_0$ given by
\be   
\label{CK0} 
\CK_0=\frac{i}{2\sqrt{y_\uv+\mathrm{Im}(\sigma)}}B(\bfk-\bfb,J)+\frac{i}{\sqrt{y_\uv+\mathrm{Im}(\sigma)}}
 B(\bfk_m,J) \partial_{\bar
  \tau_{\rm uv}} \overline{v(\sigma+\tau_{\rm uv},\tau_{\rm uv})}.
\ee
 
Since the integrand is a total derivative in $\bar \sigma$, we need to determine the term proportional to
$\sigma^{1/2}=(\tau-\tau_{\rm uv})^{\frac{1}{2}}$ in the expansion near
$\tau=\tau_{\rm uv}$, or equivalently near $\sigma=0$. 
Given the behavior of $v$ for $\tau$ near $\tau_{\rm uv}$ (\ref{BehaviorInfty}),
we realize that the second term does not contribute. For $\tau$ near
$\tau_{\rm uv}$, the leading term of $\CK_0\,e^{-4\pi i v B(\bfk_-,\bfk_m)}$ therefore
reads
\be
\frac{1}{\pi i}(-\pi i(\tau-\tau_{\rm uv}))^{1/2}\left(
  -\frac{B(\bfk_m,J)}{4\,y_\uv^{3/2}}+\frac{2\pi}{\sqrt{y_\uv}} B(\bfk,J)\,B(\bfk_-,\bfk_m)   \right).
\ee
Computing the residue integral for $\tau$ around $\tau_{\rm uv}$, we arrive at
\be
\boxed{
\label{HolAnol}
\begin{split}
\partial_{\bar \tau_{\rm uv}}\Phi_\bfmu^J(\tau_{\rm uv},\bar \tau_{\rm uv})&=\frac{\kappa}{8\pi i y_\uv^{3/2}\,\eta(\tau_{\rm uv})^{2\chi}} \\
&\quad \times \sum_{\bfk\in L+\bfmu} \left[ B(\bfk_m,J)- 8\pi y_\uv
  B(\bfk,J)\,B(\bfk_-,\bfk_m)   \right] q_{\rm uv}^{-\bfk_{-}^2}\bar
q_{\rm uv}^{\bfk_+^2}.
\end{split}}
\ee

For $b_2=1$ and $\bfk_m=3/2$, the second term in the square bracket is necessarily 0,
such that we arrive at,
\be 
\label{HolAnolP2}
\begin{split} 
\partial_{\bar \tau_{\rm uv}}\Phi_\mu^{\mathbb{P}^2}(\tau_{\rm
  uv},\bar \tau_{\rm uv})&=-\frac{3i\kappa}{16\pi  y_\uv^{3/2}}\,\frac{
\Theta_\mu(-\bar \tau)}{\eta(\tau_{\rm uv})^{6}},\\
\end{split}
\ee 
where $\Theta_\mu$ is given by
\be
\label{Thetamu}
\Theta_\mu(\tau)=\sum_{k \in \mathbb{Z}+\mu}q^{k^2}.
\ee
This is in agreement with the known holomorphic anomaly for $\mathbb{P}^2$ \cite{Vafa:1994tf}. The comparison fixes the undetermined constant, $\kappa=1$, which we will use in the following. We deduce that $\eta^{6}\,\Phi_\mu^{\mathbb{P}^2}$ is annihilated by the
Laplacian $\partial_{\tau_\uv} y_\uv^{3/2} \partial_{\bar \tau_\uv}$, and is thus a harmonic Maass form.

Equation (\ref{HolAnol}) is also in agreement with the holomorphic anomaly derived for Hirzebruch surfaces on page 12 of
\cite{Manschot:2011dj} for $\kappa=1$ and $\bfk_m=-K/2$ with $K$ the
canonical class of the Hirzebruch surface. The second term in the brackets of the summand may be
absent in special cases. One case is if $J\sim \bfk_m$, and another case is if $J=(1,0,\dots,0)$. 
The rational elliptic surface is an interesting case \cite{Minahan:1998vr, Yoshioka_1999, Klemm:2012sx}. Since $K^2=0$, the period point $J=-K$ is at the boundary of the K\"ahler cone for this four-manifold. The holomorphic anomaly therefore changes from mock modular forms to that of quasi-modular forms. Moreover, T-duality relates the partition function to that of topological strings.\\
\\   
{\it Anomaly for $\bfk_m$ not an almost complex structure}\\  
Let us consider the case that $\bfk_m$ is not an ACS. If
$\bfk_m^2>(2\chi+3\sigma)/4$ or $\ell>0$, the leading term of $\nu$ goes as
$\sigma^{c}$, with $c\geq -1/2$. As a result, there are no poles in
$\sigma$, and the holomorphic anomaly vanishes. 
We discuss below that a non-vanishing holomorphic anomaly is possible for $\ell>0$ for correlation functions, since $u$ and  $du/da$ diverge for $\sigma\to 0$.

On the other hand, if $\bfk_m^2<(2\chi+3\sigma)/4$ or $\ell<0$, the leading term of $\nu$ goes as
$\sigma^{c}$, with $c\leq -5/2$, such that there is generically an
holomorphic anomaly.  To get an idea of the type of anomaly, let us also determine this for
$X=\mathbb{P}^2$ with $k_m=1/2$. Then $\nu$ diverges as
$\sigma^{-5/2}$, and we need to determine also the first subleading
term in the $\sigma$-expansion of $\nu$ and $\Psi_\bfmu^J$ using the
expansions in (\ref{leadingtau-tau0}). After collecting all the terms,
we find that the holomorphic anomaly becomes
\be 
\label{HolAnolkm12}
\partial_{\bar \tau_{\rm uv}}
\Phi_\mu^{\mathbb{P}^2}(\tau_{\rm uv},\bar\tau_{\rm
  uv})=-\frac{i}{48\pi\, y_{\rm uv}^{3/2}}\, \frac{\widehat
  E_2(\tau_{\rm uv},\bar\tau_{\rm uv})\,\Theta_\mu(-\bar \tau_{\rm uv})}{\,\eta(\tau_{\rm uv})^{2}},
\ee
where $\widehat E_2$ is the non-holomorphic Eisenstein series defined
in (\ref{whE2}) and $\Theta_\mu$ by (\ref{Thetamu}).
Since the theta series $\Theta_\mu$ is multiplied by $\widehat E_2$,
the numerator is more general than the class of harmonic Maass forms.
\\
\\    
{\it Anomaly for correlations functions}\\  
The holomorphic anomaly is generically also present for correlation
functions, even when absent for the partition function. There is an holomorphic anomaly for the coefficients of $S^rp^n$ for 
\be\label{eq:HoloAnom-Cond}
\left\{\begin{array}{ll}
\ell  \leq n+r/2, &\qquad r\,\, {\rm even}, \\
\ell  \leq n+(r+1)/2, &\qquad r\,\, {\rm odd}.
\end{array} \right. 
\ee
Thus only for $\ell>0$ can the holomorphic anomaly be absent,
and then it is only absent for a finite set of correlators. 

As an explicit example, we consider $X=\mathbb{P}^2$, $k_m=3/2$, and the
one-point function $\left< u\right>$. We find for the 
holomorphic anomaly
\be
\partial_{\bar \tau_{\rm uv}} \Phi_\mu^{\mathbb{P}^2}[u](\tau_\uv,\bar
\tau_\uv)= -\frac{3i\,m^2
}{64\pi\, y_{\rm uv}^{3/2}}\, \frac{\widehat E_2(\tau_{\rm uv},\bar\tau_{\rm uv})\,\,\Theta_\mu(-\bar \tau_{\rm uv})}{\,\eta(\tau_{\rm uv})^{6}}.
\ee
This equals (\ref{HolAnolkm12}) up to the power of $\eta$ and the
prefactor. We leave a more in depth analysis for future work.
\\ 
\\
Since the space of modular forms is finite dimensional for fixed
weight and multiplier system, combined knowledge of the holomorphic anomaly and
$S$-duality of Section \ref{SecPhiSdual} is very powerful. Let us recall that the weight of the "normalized" partition function $\underline \Phi_\bfmu^J$ (\ref{underlinePhi}) is $\chi/2-2\ell$ and that $\underline \Phi^J_\bfmu$ is holomorphic with leading term $q_\uv^0$ or smaller. Therefore for $\ell>\chi/4$, we have necessarily that $\underline \Phi^J_\bfmu$ and $\Phi_\bfmu^J$ vanish, since holomorphic vector-valued modular forms of negative weight necessarily vanish. Therefore for $X=\BP^2$, $\Phi_\mu^{\BP^2}$ is only non-vanishing for $k_m=1/2$ and $k_m=3/2$.

Similarly to the determination of partition functions for topological strings  \cite{Haghighat:2008gw}, a potential approach to determine $\Phi_\bfmu^J$ is to integrate the anomaly
\eqref{HolAnol} with respect to $\bar \tau_\uv$, and fix the holomorphic
ambiguity using independent information, . This avenue is taken in \cite{Alexandrov:2020bwg} for VW theory,\footnote{For VW theory and $X$ a rational and ruled surfaces, the $\Phi_\bfmu^J$ can also be determined using algebro-geometric techniques as reviewed in Section \ref{CompVWtheory}.}  However for generic $c_\uv$ and $\CN=2^*$ observables, we do not have enough information to fix the holomorphic ambiguity. The next Section will evaluate $\Phi_\bfmu^J$ by explicit evaluation of the $u$-plane integral.

\subsection{Metric dependence and wall-crossing} 
\label{metricwallcross}
The metric dependence of the Lagrangian is also contained in the
$Q$-exact part. However, the path integral and correlation functions
can still exhibit a metric dependence similarly to what found in the
previous subsection for the dependence on $\bar \tau_{\rm uv}$. To
determine the metric dependence, the aim of this subsection is to evaluate the difference
\be
\Phi_\bfmu^J-\Phi_\bfmu^{J'}
\ee 
for two period points $J$ and $J'$. To this end, we write the
difference $\Psi_\bfmu^J-\Psi_\bfmu^{J'}$ as a total derivative to $\bar \tau$,
\be
\begin{split}
&\Psi_\bfmu^J(\tau,\bar \tau,\bfz,\bar \bfz) -\Psi_\bfmu^{J'}(\tau,\bar \tau,\bfz,\bar \bfz)=\partial_{\bar \tau} \widehat \Theta_{\bfmu}^{JJ'}(\tau,\bar \tau,\bfz,\bar \bfz),
\end{split}
\ee
with
\be
\label{ThetaComplete}
\begin{split} 
\widehat \Theta_{\bfmu}^{JJ'}(\tau,\bar \tau,\bfz,\bar \bfz)& = \tfrac{1}{2}\sum_{\bfk\in L+\bfmu} \left[E(\sqrt{4y}B(\bfk+\bfb,J))-E(\sqrt{4y}B(\bfk+\bfb,J')) \right]\\
&\qquad \times q^{-\bfk^2}\,e^{-4\pi i B(\bfk,\bfz)},
\end{split}
\ee
with $E(t)$ is the rescaled error function defined in
(\ref{Eerror}). We also define the holomorphic part
\be
\label{IndefiniteTheta}
\begin{split} 
 \Theta_{\bfmu}^{JJ'}(\tau,\bfz)& = \tfrac{1}{2}\sum_{\bfk\in L+\bfmu} \left[\sgn(B(\bfk+\bfb,J))-\sgn(\sqrt{4y}B(\bfk+\bfb,J')) \right]\\
&\qquad \times q^{-\bfk^2}\,e^{-4\pi i B(\bfk,\bfz)}.
\end{split}
\ee
The transformations of $\widehat \Theta_{\bfmu}^{JJ'}$ are similar to (\ref{PsiTrafos}),
\be
\begin{split}
\widehat \Theta_{\bfmu}^{JJ'}(-1/\tau,-1/\bar \tau,\bfz/\tau,\bar
\bfz/\bar \tau)&=i (-i\tau/2)^{b_2/2}\,\exp(-2\pi i
\bfz^2/\tau)\\
&\quad \times \widehat \Theta_{0}^{JJ'}(\tau/4,\bar
\tau/4,\bfz/2-\bfmu/2,\bar \bfz/2-\bfmu/2)\\
&= i(-i\tau/2)^{b_2/2}\, \exp(-2\pi i
\bfz^2/\tau)\\
&\quad \times \sum_{\bfnu \in (L/2)/L} e^{4\pi i B(\bfmu,\bfnu)} \, \widehat
\Theta_{\bfnu}^{JJ'}(\tau,\bar \tau,\bfz,\bar \bfz),\\
\widehat \Theta^{JJ'}_\bfmu(\tau+1/2,\bar \tau+1/2,\bfz,\bar \bfz) & =  e^{\pi i
   (\bfmu^2-B(\bfmu,K))}\\
& \quad \times \widehat \Theta_\bfmu^{JJ'}(\tau,\bar
 \tau,\bfz+\bfmu/2-K/4,\bar \bfz+\bfmu/2-K/4).
\end{split} 
\ee 

The completed theta series $\widehat \Theta_{\bfmu}^{JJ'}$ is a smooth function of $J$ and 
$J'$. However, we will see below that the contributions to $\Phi_\bfmu^J-\Phi_\bfmu^{J'}$ from $u_1$,
$u_2$ and $u_3$ are discontinuous after taking the limit $Y\to
\infty$. \\
\\
{\it Contribution from $\infty$}\\
The contribution from $\infty$ is
\be
\label{DeltaPhiInf}
\left[\Phi^J_\bfmu - \Phi^{J'}_\bfmu\right]_\infty=\underset{\varepsilon\to
0}{\rm \lim} \oint_{|\tau-\tau_{\rm uv}|=\varepsilon} d\tau\,
\nu(\tau,\tau_{\rm uv},m)\,\widehat \Theta^{JJ'}_\bfmu(\tau,\bar \tau,v\bfk_m,\bar
v \bfk_m),
\ee
which is non-holomorphic in $\tau_\uv$. We deduce also that
the metric dependence is smooth at this boundary, since $\widehat \Theta^{JJ'}_\bfmu$ is smooth. If
$\bfk_m^2>(2\chi+3\sigma)/4$, the rhs of (\ref{DeltaPhiInf})
vanishes, since $\widehat \Theta^{JJ'}_\bfmu$ is a Taylor series in
$s=(-\pi i(\tau-\tau_{\rm uv}))^{1/2}$.

On the other hand, if we choose $\bfk_m$ to correspond to an almost complex structure, we find 
\be
\left[\Phi^J_\bfmu -
  \Phi^{J'}_\bfmu\right]_\infty=\frac{1}{\eta(\tau_{\rm uv})^{2\chi}}\,
\theta^{JJ'}_\bfmu(\tau_{\rm uv},\bar \tau_{\rm uv}),
\ee
with 
\be
\begin{split}
\widehat \theta^{JJ'}(\tau_{\rm uv},\bar \tau_{\rm uv})&=-\frac{1}{2\pi i} \partial_{v}
\widehat \Theta^{JJ'}(\tau,\bar \tau,v\bfk_m,\bar v \bfk_m)\vert_{\tau=\tau_{\rm uv}}\\
&=\sum_{\bfk\in
  L+\bfmu}
\left[E(\sqrt{4y_\uv}B(\bfk,J))-E(\sqrt{4y}B(\bfk,J'))
\right]B(\bfk,\bfk_m) \,q_{\rm uv}^{-\bfk^2}\\ 
&\quad +\frac{1}{2\pi y_\uv} \left(B(\bfk_m,J) \Psi^{J}_\bfmu[1](\tau_{\rm uv},\bar \tau_{\rm uv})- B(\bfk_m,J') \Psi^{J'}_\bfmu[1](\tau_{\rm uv},\bar \tau_{\rm uv})\right),
\end{split}
\ee
where the elliptic arguments in $\Psi^{J}_\bfmu[1]$ and
$\Psi^{J'}_\bfmu[1]$ on the last line are set to 0.

The wall-crossing of Seiberg-Witten invariants cancels the contributions to
$\Phi^J_\bfmu-\Phi^{J'}_{\bfmu}$ from $u_1$, $u_2$ and
$u_3$. Equation (\ref{DeltaPhiInf}) gives therefore the full
continuous metric dependence of $Z_\bfmu^J$. This metric dependence
is clearly only present for $b_2^+=1$, for which
$\ell=(c_\uv^2-\sigma)/8-1$.  Without observables, there is continuous metric dependence
for $\ell\leq 0$. Including observables, there is continuous metric
dependence for the coefficients of $S^rp^n$ for 
\be
\label{CondCont}
\left\{\begin{array}{ll}
\ell  \leq n+r/2, &\qquad r\,\, {\rm even}, \\
\ell  \leq n+(r+1)/2, &\qquad r\,\, {\rm odd}.
\end{array} \right. 
\ee
Thus only for $\ell>0$ can continuous metric dependence be absent,
and then it is only absent for a finite set of correlators.
\\
\\
{\it Contribution from $u_1$}\\
Let us start by determining the contribution from $u_1$ or
$\tau=\tau_1 \to i\infty$. Since $v\sim -\tau_1/2$ near $u_1$ (\ref{vtinfty}), 
it is useful shift $v$, such that $\widehat \Theta_{\bfmu}^{JJ'}$ is expressed as
\be
\label{u1HatTheta}
\begin{split} 
\widehat \Theta_{\bfmu}^{JJ'}(\tau,\bar \tau,v\bfk_m,\bar v\bfk_m) &=
e^{-2\pi i (v+\tau/4)\bfk_m^2}\,\widehat
\Theta_{\bfmu-\frac{1}{2}\bfk_m}^{JJ'}(\tau,\bar \tau,
(v+\tau/2)\bfk_m, (\bar v+\bar \tau/2)\bfk_m).
\end{split}
\ee 
Since the subscript on the rhs is $\bfmu-\frac{1}{2}\bfk_m$, the
summation variable $\bfk$ of the theta series is an element in $\bfmu-\frac{1}{2}\bfk_m + L$. We refer to $4\bfk$ as the infra-red
${\rm Spin}^c$ structure, thus $4\bfk=c_1(\mathfrak{s}_{\rm ir})=:c_{\rm ir}\in 
4\bfmu-2\bfk_m + 4L$.

With (\ref{u1HatTheta}), we find that the wall-crossing at $u_1$ is given by
\be 
\label{WCu1} 
\begin{split}
&\left[\Phi_\bfmu^{J}-\Phi_\bfmu^{J'}\right]_{1}=\frac{\kappa\,m^{\frac{\sigma}{8}-2}}{4\,\eta(\tau_{\rm uv})^{2\chi-4\ell}}
\left(\frac{4}{\Lambda}
\right)^{\frac{3}{8}(2\chi+3\sigma)} \\
&\quad \times \lim_{Y\to \infty} \int_{-\frac{1}{2}+iY}^{\frac{3}{2}+iY}
d\tau_1\,\left( \frac{du}{da}\right)^{\frac{\chi}{2}-3}\Delta_{\rm phys}^{\frac{\sigma}{8}+1}\left( C\, e^{-2\pi i
    (v_1+\tau_1/4)}\right)^{\bfk_m^2}\\
&\quad \times \widehat \Theta_{\bfmu-\frac{1}{2}\bfk_m}^{JJ'}(\tau_1, \bar \tau_1,(v_1+ \tau_1/2)\bfk_m, (\bar v_1+\bar \tau_1/2)\bfk_m).
\end{split}
\ee
The $Y\to \infty$ limit has the effect that the error functions in $\widehat \Theta_{\bfmu}^{JJ'}$ reduce to
$\sgn$'s in (\ref{ThetaComplete}).  
In other words, $\widehat \Theta_{\bfmu-\frac{1}{2}\bfk_m}^{JJ'}$ is  replaced by its holomorphic part $\Theta_{\bfmu-\frac{1}{2}\bfk_m}^{JJ'}$ (\ref{IndefiniteTheta}), and the $u_1$ contribution is thus holomorphic in $\tau_{\rm uv}$.
Given the behavior near $u_1$ of the various quantities, we see that SW wall-crossing
only occurs for $c_{\rm ir}$ and $\sigma$ satisfying
\be 
\sigma+8-c_{\rm ir}^2\leq 0. 
\ee
This is the same condition as for the theory without hypermultiplets \cite{Moore:1997pc}.
\\
\\
{\it Contribution from $u_2$}\\
For the contribution from $u_2$, we express the integrand in terms of
$\tau_2=-1/\tau$, and $v_2=-v/\tau$ (\ref{tau2v2}). The indefinite
theta series $\widehat
\Theta^{JJ'}_\bfmu$ reads in terms of these variables,
\be 
\begin{split}
&\widehat \Theta^{JJ'}_\bfmu(-1/\tau_2,-1/\bar \tau_2, v_2 \bfk_m /\tau_2,\bar
v_2 \bfk_m/\bar \tau_2)=i(-i\tau_2/2)^{b_2/2}\exp(-2\pi i
v_2^2/\tau_2)\\
&\quad \times  \widehat \Theta_\bfmu^{JJ'}(\tau_2/4,\bar \tau_2/4,v_2\bfk_m/2-\bfmu/2,\bar v_2\bfk_m/2-\bfmu/2)\\ 
&= i(-i\tau_2/2)^{b_2/2}\,e^{-2\pi i B(\bfk_m,\bfmu)} \,\exp(-2\pi i
v_2^2/\tau_2) \,e^{2\pi i (v_2-\tau_2/4)\bfk_m^2}\\
&\quad \times \,\widehat \Theta_{\bfk_m}^{JJ'}(\tau_2/4, \bar \tau_2/4,(v_2/2-\tau_2/4)\bfk_m-\bfmu/2,(\bar v_2/2-\bar \tau_2/4)\bfk_m-\bfmu/2).
\end{split}
\ee
The contribution from $u_2$ reads,
\be
\label{PhiWCu2}
\begin{split}
&\left[\Phi_\bfmu^{J^+}-\Phi_\bfmu^{J^-}\right]_{2}=\frac{\kappa\,m^{\frac{\sigma}{8}-2}}{4\,\eta(\tau_{\rm uv})^{2\chi-4\ell}}\left(\frac{4}{\Lambda}\right)^{\frac{3}{8}(2\chi+3\sigma)}\,2^{-b_2/2}\, 
e^{\pi i \sigma/4 -2\pi i B(\bfk_m,\bfmu)}\\
&\quad \times \lim_{Y\to \infty} 
\int_{-\frac{1}{2}+iY}^{\frac{3}{2}+iY} d\tau_2\,\left(
  \frac{du}{da}\right)_2^{\frac{\chi}{2}-3} \Delta_{\rm
  phys,2}^{\frac{\sigma}{8}+1}\,\left( C_2\, e^{2\pi i
    (v_2-\tau_2/4)}\right)^{\bfk_m^2}\\
& \quad \times \,\Theta_{\bfk_m}^{JJ'}(\tau_2/4,(v_2/2-\tau_2/4)\bfk_m-\bfmu/2),
\end{split}
\ee
where we have replaced $\widehat \Theta_{\bfk_m}^{JJ'}$ by $\Theta_{\bfk_m}^{JJ'}$ as result of the $Y\to \infty$ limit.
\\
\\
{\it Contribution from $u_3$}\\
For the contribution from $u_3$, we express the integrand in terms of
$\tau_3=-\frac{1}{\tau-1}$, and $v_3=-v/(\tau-1)$. We then have
\be
\begin{split}
&\widehat \Theta^{JJ'}_\bfmu(1-1/\tau_3,1-1/\bar
\tau_3,v_3\bfk_m/\tau_3,\bar v_3\bfk_m/\bar \tau_3)\\
&\quad = e^{-2\pi i \bfmu^2}\,\widehat \Theta^{JJ'}_\bfmu(-1/\tau_3,-1/\bar
\tau_3,v_3\bfk_m/\tau_3,\bar v_3\bfk_m/\bar \tau_3) \\
&\quad = i(-i\tau_3/2)^{b_2/2}\,e^{-2\pi i \bfmu^2-2\pi i B(\bfk_m,\bfmu)} \,\exp(-2\pi i
v_3^2/\tau_3) \,e^{2\pi i (v_3-\tau_3/4)\bfk_m^2}\\
&\qquad \times \,\widehat \Theta_{\bfk_m}^{JJ'}(\tau_3/4, \bar \tau_3/4,(v_3/2-\tau_3/4)\bfk_m-\bfmu/2,(\bar v_3/2-\bar \tau_3/4)\bfk_m-\bfmu/2).
\end{split}
\ee 
The contribution from $u_3$ reads,
\be
\label{PhiWCu3}
\begin{split}
&\left[\Phi_\bfmu^{J^+}-\Phi_\bfmu^{J^-}\right]_{3}=\frac{\kappa\,m^{\frac{\sigma}{8}-2}}{4\,\eta(\tau_{\rm uv})^{2\chi-4\ell}}\left(\frac{4}{\Lambda}\right)^{\frac{3}{8}(2\chi+3\sigma)}\,2^{-b_2/2}\, 
e^{\pi i \sigma/4 -2\pi i \bfmu^2 -2\pi i B(\bfk_m,\bfmu)}\\
&\quad \times \lim_{Y\to \infty}
\int_{-\frac{1}{2}+iY}^{\frac{3}{2}+iY} d\tau_3\,\left(
  \frac{du}{da}\right)_3^{\frac{\chi}{2}-3} \Delta_{\rm
  phys,3}^{\frac{\sigma}{8}+1}\,\left( C_3\, e^{2\pi i
    (v_3-\tau_3/4)}\right)^{\bfk_m^2}\\
& \quad \times \,\Theta_{\bfk_m}^{JJ'}(\tau_3/4,(v_3/2-\tau_3/4)\bfk_m-\bfmu/2).
\end{split}
\ee

\subsection{Blow-up formula}
The $u$-plane integral $\Phi_\bfmu^J(X)$ for a four-manifold $X$ and $\Phi_{\hat \bfmu}^{\hat J}(\hat X)$ for the blow-up $\hat X$ of $X$
are related by a simple formula. Blowing-up a point of $X$ adds another two-cycle
to $X$ with self-intersection $-1$, such that $\chi(\hat X)=\chi(X)+1$
and $\sigma(\hat X)=\sigma(X)-1$. 
For simplicity, we restrict to the case that $\bfk_m$ is an ACS, $\bfk_m^2=(2\chi+3\sigma)/4$. We set $\hat \bfk_m=(\bfk_m,1/2)$, such
$\hat \bfk_m$) corresponds to an ACS of $\hat X$. 

We consider the contribution from the cusp at $u=\infty$. We observe from
(\ref{leadingtau-tau0}) that the leading terms of the $s$-expansion of the measures $\nu_{X}$ and $\nu_{\hat X}$ are equal up to a power of $\eta$,
\be
\nu_{\hat X} = \frac{\nu_X}{\eta(\tau_{\rm uv})^2}+O(s^{1/2}). 
\ee
Let $\hat \bfmu=(\bfmu, \tilde \mu)$, where $\tilde \mu\in
\{0,\frac{1}{2} \}$ is the 't Hooft flux corresponding to the blow-up cycle.
Then, 
\be
\label{PsiBlowUp}
\Psi^{\hat J}_{\hat \bfmu}(\tau,\bar \tau,v\hat \bfk_m,\bar v\hat \bfk_m)=\Psi^J_{\bfmu}(\tau,\bar \tau,v\bfk_m,\bar v\bfk_m)\,\Theta_{\tilde \mu}(\tau,v/2),
\ee
where $\hat J=(J,0)$ and $\Theta_{0}(\tau,z)=\vartheta_3(2\tau,2z)$ and $\Theta_{1/2}(\tau,z)=\vartheta_2(2\tau,2z)$.
Evaluation of the $u$-plane integrals $\Phi_{\bfmu}^J(X)$ and $\Phi_{\hat \bfmu}^{\hat J}(\hat X)$ for $\ell=0$ detailed in Section \ref{sec:Evaluate}, gives
\be 
\frac{\Phi_{\hat \bfmu}^{\hat J}(\hat X)}{\Phi_{\bfmu}^J(X)}=\frac{\Theta_{\tilde \mu}(\tau_{\rm uv})}{\eta(\tau_{\rm uv})^2},
\ee 
which is a known relation for the VW partition function 
\cite{Vafa:1994tf, Yoshioka:1996, Li_1999, Gottsche:1999}. This
relation also implies a relation for generating functions of $\hat X$,
\be
\frac{\Phi_{(\bfmu,0)}^{\hat J}(\hat X)}{\Phi_{(\bfmu,\frac{1}{2})}^{\hat J}(\hat
  X)}=\frac{\theta_3(2\tau_{\rm uv})}{\theta_2(2\tau_{\rm uv})}.
\ee 
Remarkably due to (\ref{PsiBlowUp}) and the identity (\ref{tauvtauUV}), this relation holds in fact for the integrand, not just the integral. For more about the blow-up formula in the $\Omega$-background see \cite{Nakajima:2003uh}.

\section{Evaluation of $\Phi_\bfmu^J$ for a special period point $J$}
\label{sec:Evaluate}   
This section evaluates the $u$-plane integral $\Phi_\bfmu^J$ for a smooth, compact four-manifold $X$ with $(b_1,b_2^+)=(0,1)$, and a convenient choice of period point $J$. Together 
with the formula's for the metric dependence of $\Phi_\bfmu^J$ in Section \ref{metricwallcross}, this provides the result for arbitrary $J$.

\subsection{Factoring $\Psi_\bfmu^J$}
\label{FactPsi}
Our main technique is to choose $J$ such that the sum over fluxes
$\Psi_\bfmu^J$ factors in a non-holomorphic and holomorphic part.
This will make it possible to determine an anti-derivative for  $\Psi_\bfmu^J$ in terms of mock modular forms. The analysis is similar to
\cite{Korpas:2019cwg}, except for the small differences in
$\Psi_\bfmu^J$ and the sum over fluxes in \cite{Korpas:2019cwg}. We
will restrict here to the case of odd lattices. The case of even lattices will be treated in Section \ref{SecEvalEven}.

To factor $\Psi_\bfmu^J$, we choose for the period point, 
\be
J=(1,\boldsymbol{0})\in L\otimes \mathbb{R}.
\ee
With this choice, $\Psi_\bfmu^J$ can be expressed as
\be
\Psi_\bfmu^J(\tau,\bar \tau,\bfrho,\bar \bfrho)=  f_{\mu_1}(\tau,\bar
\tau, \rho_1, \bar \rho_1)\,\Theta_{L_-,\bfmu_-}(\tau,\bfrho_-),
\ee
where $\bfrho=v\bfk_m=(\rho_{m,1},\bfrho_{m,-})$, $\bfmu=(\mu_1,\bfmu_-)$, with $\bfmu_-\in (\mathbb{Z}/2)^{b_2-1}$,
and 
\be
\label{fmutbart}
f_{\mu}(\tau,\bar \tau,\rho,\bar \rho)=\exp(-4\pi y\,b^2)\,
\sum_{k\in \mathbb{Z}+\mu} \tfrac{d}{d\bar \tau} (\sqrt{4y}\,(k+b))\,\bar q^{k^2}\,e^{-4\pi i\,\bar \rho\, k},
\ee
and
\be
\Theta_{L_-,\bfmu_-}(\tau,\bfrho_-)=\sum_{\bfk_-\in L_-+\bfmu_-}
q^{-\bfk_-^2}\,e^{-4\pi i B(\bfk_-,\bfrho_-)}.
\ee

For the evaluation of the $u$-plane integral using Stokes' theorem, we
aim to determine a family of functions $\widehat
G_\mu(\tau,\bar \tau,v,\bar v; k_{m,1})$, with $k_{m,1}\in
\mathbb{Z}+\frac{1}{2}$, which satisfy the following conditions:
\begin{enumerate}
\item $f_{\mu}$ is the anti-holomorphic derivative of $\widehat G_\mu$, 
\be
f_{\mu}(\tau,\bar \tau, v k_{m,1},\bar v k_{m,1})=\frac{d \widehat G_\mu(\tau,\bar \tau,v,\bar v; k_{m,1})}{d\bar \tau},
\ee
where $d/d\bar \tau$ also differentiates $\overline{v(\tau)}$. 
\item The 1-form  
\be 
\label{1formeta}
\Omega=d\tau\, \nu(\tau,\tau_{\rm uv})\,\widehat G_\mu(\tau,\bar \tau,v,\bar
v;k_{m,1})\,\Theta_{L_-,\bfmu_-}(\tau, v \bfk_{m,-}),
\ee
 is single-valued on $\CU_\varepsilon$.
\item Since $\omega =d\Omega$ on $\CU_\varepsilon$, the derivative to
  $\bar \tau$ of $\nu\,\widehat G_\mu\Theta_{L_-,\bfmu_-}$ exists at
  each point of $\CU_\varepsilon$. In particular, $\widehat G_\mu$ has no
  singularities on $\CU_\varepsilon$. 
\end{enumerate} 

We can arrive at such functions using the Appell-Lerch sum $M(\tau,u,v)$
(\ref{Mtuv}) and its completion (\ref{mucomplete})
\cite{ZwegersThesis, MR2605321}. More generally, the $\widehat G_\mu$ are part of the class of (Heisenberg harmonic) Jacobi-Maass forms
\cite{2012arXiv1207.5600B}.\footnote{Reference \cite{Dabholkar:2012nd} introduced the notion of mock Jacobi form, which is slightly different from the Jacobi-Maass forms used here. One difference is that the completion of a mock Jacobi form does not depend on the anti-holomorphic elliptic variable, whereas this is the case for a Jacobi-Maass form.} To determine $\widehat G_\mu$, we first
determine a family of auxiliary functions $\widehat F_\mu$, which satisfy Condition 1. and
2., namely
\be
\label{whF}
\begin{split}
\widehat F_\mu(\tau,\bar \tau, \rho,\bar \rho) & = -  \,q^{-\nu^2} e^{-4\pi i \rho \nu}
\left( \, M(2\tau,2\rho+2\mu\tau+1/2, \tau) \right.\\ 
&\qquad \left. +\tfrac{i}{2}R(2\tau,2\bar \tau,
  2\rho+2\nu \tau+1/2, 2\bar \rho+2\nu \bar \tau+1/2 )\,\right)
\end{split}
\ee 
with $\nu=\mu-\frac{1}{2}$. The holomorphic parts $F_{\mu}$ of $\widehat F_{\mu}$ read 
\be 
\label{F0andF1}
\begin{split}
F_0(\tau,\rho)&=\frac{1}{2}-\frac{1}{\vartheta_4(2\tau)}\sum_{n\in
  \mathbb{Z}} \frac{(-1)^n\,q^{n^2}}{1+e^{4\pi i \rho}\,q^{2n}},\\
F_{\frac{1}{2}}(\tau,\rho)&=-\frac{e^{2\pi i\rho}}{\vartheta_4(2\tau)}
\sum_{n\in\mathbb{Z}}
\frac{(-1)^n\,q^{n^2-\frac{1}{4}}}{1+e^{4\pi i \rho} \,q^{2n-1}},
\end{split}
\ee
and
\be
\label{hatFE}
\widehat F_\mu(\tau,\bar \tau,\rho,\bar \rho)=F_\mu(\tau,\rho)
-\frac{1}{2}\sum_{n\in \mathbb{Z}+\mu} \left(\,
  \sgn(n)-E(2(n+b)\sqrt{y})\, \right) e^{-4\pi i \rho n}\,q^{-n^2}, 
\ee
where $b=\mathrm{Im}(\rho)/y$. 
We then have the desired relation 
\be
f_\mu(\tau,\bar \tau,\rho,\bar \rho)=\frac{d\widehat F_\mu(\tau,\bar \tau,\rho,\bar \rho)}{d\bar \tau},
\ee
where $d/d\bar \tau$ also differentiates the elliptic variable $\bar \rho=\overline{v(\tau)}\,k_{m,1}$. In the following, we abbreviate the first component $k_{m,1}$ to $k_m$.

To verify that $\widehat F_\mu$ will lead to a well-defined
anti-derivative on the $u$-plane, we determine its transformations under the monodromies
using (\ref{M-u-v}) and (\ref{Mqutr}). With $k_m\in 1/2+ \mathbb{Z}$, we have 
\begin{itemize}
\item 
For $M_\infty$,
\be
\widehat F_\mu(\tau,\bar \tau,-v k_m,-\bar v k_m)=-\widehat
F_\mu(\tau,\bar \tau,v k_m,\bar v k_m),
\ee
where we abbreviated the first component $k_{m,1}$ to $k_m$. We will use this in the remainder of this subsection.
\item For $M_1$,
\be
\widehat F_\mu(\tau+2,\bar \tau+2,(v-1) k_m,(\bar v-1) k_m)=\widehat
F_\mu(\tau,\bar \tau,v k_m,\bar v k_m).
\ee 
\item For $M_2$,
\be
\begin{split}
&\widehat F_\mu\!\left(\frac{\tau}{-2\tau+1},\frac{\bar \tau}{-2\bar
    \tau+1}, \frac{v-\tau}{-2\tau+1} k_m,\frac{\bar v - \bar
    \tau}{-2\bar \tau+1} k_m\right)=\\ 
&\qquad (-2\tau+1)^{1/2}\exp\!\left(2\pi i \frac{(\tau-2v
    +2v^2)\,k_m^2}{-2\tau+1}\right) \widehat F_\mu(\tau,\bar
\tau,v k_m,\bar v k_m).
\end{split} 
\ee 
\end{itemize}
This is precisely the required behavior under monodromies. The function $\widehat F_\mu$ is furthermore quasi-periodic with index $-1$,
\be
\begin{split}
&\widehat F_\mu(\tau,\bar \tau,\rho+\tau,\bar \rho+\bar \tau)=e^{4\pi i \rho} q\,\widehat F_\mu(\tau,\bar \tau,\rho,\bar \rho),\\
&\widehat F_\mu(\tau,\bar \tau,\rho+1,\bar \rho+1)=\widehat F_\mu(\tau,\bar \tau,\rho,\bar \rho).
\end{split}
\ee
$F_\mu$ has poles for $\rho=1/4$ and $\rho=3/4 \mod
\mathbb{Z}+\tau\mathbb{Z}$, with residue 
\be
\begin{split}
&\mu=0:\qquad -\frac{1}{4\pi i\,\vartheta_4(2\tau)},\\
&\mu=1/2:\qquad \mp \frac{q^{-1/4}}{4\pi \,\vartheta_4(2\tau)}.\\
\end{split}
\ee
If we substitute $\rho=z k_m$ in $\widehat F_\mu$,
there will be $8k_m^2$ poles as function of $z$ in the parallelogram spanned by 2 and
$2\tau$. 

The third condition is the absence of singularities on
$\CU_\varepsilon$. The functions $F_\mu$ do not satisfy this
requirement in general. To see this, note that if we replace
$\rho=vk_m$ with $zk_m$, for an arbitrary $z\in \mathbb{C}$, $F_\mu$
has singularities in $z$ for
\be 
\label{polesFmu}
e^{4\pi i z k_m}=-q^{n},
\ee
with $n$ even for $F_0$, and $n$ odd for $F_{1/2}$. While many of these
poles can not be assumed by the coupling $v$, due to the identity
(\ref{tauvtauUV}), not all poles are contained in the ``excluded''' set (\ref{vexcluded}). The non-singular functions $\widehat G_{\mu}(\tau,\bar \tau,v,\bar v;k_m)$ have the same non-holomorphic part as $\widehat F_\mu(\tau,\bar
\tau,vk_m,\bar vk_m)$ but a different holomorphic part to avoid the poles.

Starting with $k_m=\pm 1/2$, we deduce all poles (\ref{polesFmu})
are excluded by the identity (\ref{tauvtauUV}). Therefore, the
$\widehat F_\mu$ are non-singular for these $k_m$, and we define
\be
\widehat G_{\mu}(\tau,\bar \tau,v,\bar v;1/2)=\widehat F_\mu(\tau,\bar
\tau,v/2,\bar v/2).
\ee 

Moving on to $k_m=\pm 3/2$. Now the $F_\mu$ have singularities in $v$,
which do also lead to singularities on $\CU_\varepsilon$ after
substitution $v=v(\tau,\tau_{\rm uv})$. To remove the undesirable
singularities, we add meromorphic functions $D_\mu$,
\be
\label{GFD}
\begin{split}
&G_0(\tau,z;3/2)= F_0(\tau,3z/2)+D_0(\tau,z),\\
&G_{\frac{1}{2}}(\tau,z;3/2)=
F_{\frac{1}{2}}(\tau,3z/2)+D_\frac{1}{2}(\tau,z). 
\end{split}
\ee
with 
\be
\label{D0half}
\begin{split} 
D_0(\tau,z)&=\frac{i\,\eta(2\tau)^3\,\vartheta_1(2\tau,2z)\,\vartheta_3(2\tau,z)}{\vartheta_2(2\tau,3z)\,\vartheta_4(2\tau)\,\vartheta_4(2\tau,2z)\,\vartheta_2(2\tau,z)},\\   
D_\frac{1}{2}(\tau,z)&=\frac{i\,\eta(2\tau)^3\,\vartheta_1(2\tau,2z)\,\vartheta_2(2\tau,z)}{\vartheta_3(2\tau,3z)\,\vartheta_4(2\tau)\,\vartheta_4(2\tau,2z)\,\vartheta_3(2\tau,z)}.  
\end{split}
\ee
The holomorphic part can be written as
\be       
\begin{split}  
G_{0}(\tau,z;3/2)&=-\frac{1}{2}+q^{-\frac{1}{4}} e^{3\pi i z}\,M(2\tau, 2z-\tau,\tfrac{1}{2}-z)  \\
& =\frac{1}{2}+\frac{q^{-\frac{3}{4}} e^{-5\pi i z}}{\vartheta_2(2\tau,z)}\sum_{n\in
\mathbb{Z}} \frac{q^{n^2+n}e^{2\pi i n z}}{1-e^{-4\pi i z}q^{2n-1}},\\
G_{\frac{1}{2}}(\tau,z;3/2)&= M(2\tau, 2z-\tau,\tfrac{1}{2}-\tau-z)\\
&= \frac{q^{-\frac{1}{4}}e^{-3\pi i z}}{\vartheta_3(2\tau,z)}\sum_{n\in
\mathbb{Z}} \frac{q^{n^2}e^{2\pi i nz}}{1-e^{-4\pi i z}q^{2n-1}}.
\end{split}
\ee 
These functions have singularities where $\vartheta_2(2\tau,z)=0$,
$\vartheta_3(2\tau,z)=0$, or for $z_{m,n}=(n-\frac{1}{2})\tau+m/2$, with
$m,n\in\mathbb{Z}$. All these values are excluded for the coupling $v$
by the identity (\ref{tauvtauUV}). Therefore, the
 functions $\widehat G_{\mu}(\tau,v;3/2)$ provide non-singular
 functions on $\CU_\varepsilon$ suitable for the evaluation of $\Phi_\bfmu^J$. 
The functions $\widehat G_{\mu}(\tau,\bar \tau,z,\bar z;3/2)$
transform as a vector-valued modular form \cite{Bringmann:2010sd},
\be
\begin{split} 
\label{whGmutrafo}
\widehat G_\mu(-1/\tau,z/\tau;3/2)&=\frac{i}{\sqrt{2}}
(-i\tau)^{1/2}\,\exp(-18 \pi i z^2/4)\\
&\quad \times \sum_{\nu=0,\frac{1}{2}} e^{4\pi i \mu \nu} \widehat G_\nu(\tau,z).
\end{split}
\ee
On the other hand, the $\widehat G_{\mu}(\tau,\bar \tau,z,\bar z;1/2)$
do not transform as a 2-dimensional vector-valued modular form, but are part of a longer of a higher-dimensional vector. Nevertheless, we will confirm in the next subsection that the integral $\Phi^J_\bfmu$ does transform as (\ref{PhiSdual}).
\\
\\
{\it Generic $k_m\in \mathbb{Z}+1/2$}\\
We briefly comment on more general $k_m\in \mathbb{Z}+1/2$. We note
that a suitable anti-derivative can be found, since we are free to
chose functions with singularities in the ``allowed'' set
\eqref{vexcluded}. For sufficiently high degree of the poles such functions do
exist. We leave finding suitable examples for future work.

\subsection{Contribution from $\infty$ for odd intersection form}
 \label{tau0contribution}
This section determines the contribution to $\Phi_\bfmu^J$ from
$u=\infty$, which confirms the anomaly \eqref{HolAnol}. This contribution is given by
\be
\Phi_{\bfmu,\infty}^J=\lim_{\varepsilon\to 0}
\oint_{|\tau-\tau_{\rm uv}|=\varepsilon}d\tau\,\nu(\tau,\tau_{\rm uv},m)\,\widehat G_{\mu_1}(\tau,\bar \tau,v,\bar v; k_{m,1})\,\Theta_{L_-,\bfmu_-}(\tau,\bfrho_-).
\ee

\subsubsection*{${\rm Spin}^c$ structure equal to an almost complex structure}
\label{Spinacs}
We first consider the case that the ${\rm Spin}^c$ structure corresponds to an
almost complex structure such that $\bfk_m^2=(2\chi+3\sigma)/4$, and
in particular $k_{m,1}=3/2$. We 
deduce from (\ref{leadingtau-tau0}) that the $u$-plane integral is determined
by the term proportional to $\sqrt{-\pi i(\tau-\tau_{\rm uv})}$ in the expansion of
$\widehat G_\mu(\tau,\bar \tau,v, \bar v; 3/2)\,\Theta_{L_-,\bfmu_-}(\tau,\bfrho_-)$ around
$\tau_{\rm uv}$. Given the behavior of $v$ in this limit
(\ref{BehaviorInfty}), and since $\widehat G_\mu$ is odd as function of
$v$ while $\Theta_{L_-,\bfmu_-}$ is even, we realize that this is determined by the
$v^1$ term of the Taylor expansion in $v$ of $\widehat
G_\mu(\tau,\bar \tau,v,\bar v;3/2)$. Let $\widehat g_\mu(\tau,\bar \tau)=-\frac{1}{2\pi
  i}\partial_z \widehat G_\mu(\tau,\bar \tau,z,\bar
z;3/2)\vert_{z=0}$. The holomorphic part, $g_\mu(\tau)=-\frac{1}{2\pi i}\partial_z
G_\mu(\tau,z;3/2)\vert_{z=0}$ is
\be 
\label{qseriesg} 
\begin{split}
g_0(\tau)&=-\frac{3}{\vartheta_4(2\tau)}\sum_{n\in
  \mathbb{Z}} \frac{(-1)^n\,q^{n^2+2n}}{(1+q^{2n})^2}+\frac{1}{2}\vartheta_3(2\tau)^3\\
&=-\frac{1}{4}+\frac{3}{2}\,q+3\,q^2+4\,q^3+\frac{9}{2}\,q^4+6\,q^5+O(q^6),\\
g_{1/2}(\tau)&=-\frac{3}{\vartheta_4(2\tau)}  
\sum_{n\in\mathbb{Z}}
\frac{(-1)^n\,q^{n^2+2n-\frac{5}{4}}}{(1+q^{2n-1})^2}+\frac{1}{2}\vartheta_2(2\tau)^3\\
&= q^{\frac{3}{4}}(1+3\,q+3\,q^2+6\,q^3+3\,q^4+O(q^5)).
\end{split}
\ee
These functions contain interesting arithmetic information; they are generating functions of the Hurwitz class numbers $H(n)$,
\be
\label{genClassNum}
g_{\mu}(\tau)=3\sum_{n\geq 0} H(4n-2\mu)\,q^{n-\mu/2}.
\ee
We express $\widehat g_\mu$ as a holomorphic $q$-series plus a
non-holomorphic part
\be
\label{whG}
\widehat g_\mu(\tau,\bar \tau)=g_\mu(\tau)-\frac{3i}{4\sqrt{2}\pi}\int_{-\bar \tau}^{i\infty} \frac{\Theta_{\mu}(u)}{(-i(\tau+u))^{3/2}}\,du,
\ee
with $\Theta_\mu$ defined in (\ref{Thetamu}).

The behavior of $\widehat G_\mu(\tau,\bar \tau,v,\bar v;3/2)$ for $\tau$ near $\tau_{\rm uv}$ is therefore
\be
\widehat G_\mu(\tau,\bar \tau,v,\bar v;3/2)= -2\sqrt{-\pi
  i(\tau-\tau_{\rm uv})}\,\widehat g_\mu(\tau_{\rm uv},\bar \tau_{\rm uv})+ O((\tau-\tau_{\rm uv})^{3/2}, (\bar
\tau-\bar \tau_{\rm uv})^{1/2}).
\ee
The subleading terms in the $(\tau - \tau_{\rm uv},\bar\tau -\bar \tau_{\rm uv})$-expansion do not contribute to the contour integral of $\tau$ around
$\tau_{\rm uv}$. We therefore find that the contribution to $\Phi_\bfmu^J$
from this boundary is 
\be
\label{PhimuJOddFINAL}
\Phi_{\bfmu,\infty}^J(\tau_{\rm uv},\bar \tau_{\rm uv})= \frac{\widehat g_\mu(\tau_{\rm uv},\bar \tau_{\rm uv})\,\Theta_{L_-,\bfmu_-}(\tau_{\rm uv})}{\eta(\tau_{\rm uv})^{2\chi}},
\ee
with $\bfmu=(\mu,\bfmu_-)$. This is compatible with the appropriate
specialization of the holomorphic anomaly equations \eqref{HolAnol}
and \eqref{HolAnolP2} derived earlier. 

For manifolds with $b_2=1$, such as $\mathbb{P}^2$, $\Theta_{L_-,\bfmu_-}=1$, and we find agreement with the partition function (\ref{VWP2}) of Vafa-Witten theory \cite{Vafa:1994tf}. 
The $q_\uv$-series perfectly agree, which is
expected from our discussion in Section \ref{TopTwisting} for $\bfk_m$ an almost
complex structure. Note that for $\mathbb{P}^2$, the class numbers have an interpretation as a count of the number of toric fixed points of rank 2 bundles \cite{Klyachko1991} We will expand more on this agreement in Subsection \ref{CompVWtheory}. 

\subsubsection*{${\rm Spin}^c$ structure different from an almost complex structure}
We next proceed with a different ${\rm Spin}^c$ structure, namely
$k_{m,1}=1/2$, which provides a non-trivial test for $S$-duality transformations of Section \ref{SecPhiSdual}. 

To express the partition functions, we define $\widehat g^{(n)}_\mu(\tau,k_m)$ by
\be
\widehat g^{(n)}_\mu(\tau,\bar
\tau;k_m)=(-2\pi i)^{n} \partial^n_z \widehat
G_\mu(\tau,\bar \tau,z,\bar z;k_m)\vert_{z=\bar z=0},
\ee
and similarly for $g^{(n)}_\mu$ in terms of $G_\mu$. The functions
vanish for $n$ even. For $n=1$, we have
\be
\begin{split}
g^{(1)}_0(\tau;1/2)&=-\frac{1}{\vartheta_4(2\tau)}\sum_{n\in
  \mathbb{Z}} \frac{(-1)^n\,q^{n^2+2n}}{(1+q^{2n})^2}\\
&=-\frac{1}{4}-\frac{1}{2}q-q^2+\frac{1}{2}\,q^4-2\,q^5-2\,q^6+O(q^7),\\
g^{(1)}_{1/2}(\tau;1/2)&=-\frac{1}{\vartheta_4(2\tau)}  
\sum_{n\in\mathbb{Z}}
\frac{(-1)^n\,q^{n^2+2n-\frac{5}{4}}}{(1+q^{2n-1})^2}\\
&= q^{\frac{3}{4}}(-1+q-3\,q^2+2\,q^3-3\,q^4+3\,q^5+O(q^6)),
\end{split}
\ee
The function $g^{(1)}_{1/2}(\tau)$ appears in Ramanujan's Lost Notebook
\cite{Ramanujan:1988}. The coefficients $c_\mu(m)$ of
$g^{(1)}_{\mu}$ can be expressed as
\be  
\label{coeffGmu}
\begin{split}
c_0(m)&=4\,H(m)- H(4m)\\
&=H(4m)-\frac{1}{6}r_{3}(m),\\
c_{1/2}(m-1/4)&=  H(4m-1)-\frac{1}{6}r_3(4m-1),
\end{split}
\ee 
where $H(m)$ is the Hurwitz class number and $r_{3}(m)$ is the number
of representations of $n$ as the sum of three squares. 

We note that the coefficients of $f_{2\mu}$ differ from those of
$G_\mu$ (\ref{coeffGmu}) by the term with $r_3(4n-2\mu)$. More
precisely, we have for $k_m=3/2$
\be
g^{(1)}_\mu(\tau;1/2)=\frac{1}{3} g_\mu(\tau)-\frac{1}{6}\, \Theta_\mu(\tau)^3,
\ee
with $\Theta_\mu$ as in (\ref{Thetamu}). The transformations (\ref{JacobiSdual}) and (\ref{JacobiTdual}) demonstrate that $\widehat g^{(1)}_\mu(\tau;1/2)$ does not transform as a 2-dimensional vector-valued modular form, but of a higher dimensional vector. 

Similarly, $G_\mu(\tau,\bar \tau,z,\bar z;1/2)$ does not
transform as a modular form.
For $n=3$, the first coefficients of $g^{(3)}_\mu$ are
\be
\begin{split}
g^{(3)}_0(\tau;1/2)&= \tfrac{1}{8}(1+2\,q+4\,q^2+24\,q^3+46\,q^4-40\,q^5+O(q^6)),\\
g^{(3)}_{1/2}(\tau;1/2)&=\tfrac{1}{4}q^{\frac{3}{4}}(-1+25\,q-75\,q^2+196\,q^3-339\,q^4+O(q^5))\\
g^{(3)}_0(\tau;3/2)&=\tfrac{1}{8}( 1+18\,q+324\,q^2+1208\,q^3+2862\,q^4+5400\,q^5+O(q^6)),\\
g^{(3)}_{1/2}(\tau;3/2)&=\tfrac{1}{4}q^{\frac{3}{4}}(1+99\,q+459\,q^2+1206\,q^3+2259\,q^4+O(q^5)).
\end{split}
\ee

For simplicity, we specialize to $X=\mathbb{P}^2$ and $k_{m}=1/2$. Together with $k_m=3/2$, this is the only value of $k_m$ for which $\Phi_\mu^{\BP^2}$ is non-vanishing. For $k_m=1/2$, we find that $\Phi_\mu^{\BP^2}$ takes the form,
\be
\Phi_\mu^{\mathbb{P}^2}(\tau_{\rm uv},\bar \tau_{\rm uv})=\frac{\widehat g_{\mu}(\tau_{\rm uv},\bar \tau_{\rm uv};1/2)}{\eta(\tau_{\rm uv})^2},
\ee
where the non-holomorphic part of $\widehat g_{1/2}$ follows from (\ref{HolAnolkm12}), and
the holomorphic part is given by 
\be
g_{\mu}(\tau_{\rm uv};1/2)=- \tfrac{2}{2\pi i}\partial_{\tau_{\rm uv}}
  g^{(1)}_{\mu}(\tau_{\rm uv};1/2)+\frac{2}{3}\,g^{(3)}_{\mu}(\tau_{\rm uv};1/2)+\frac{4}{3}\,
E_2(\tau_{\rm uv})\,g^{(1)}_{\mu}(\tau_{\rm uv};1/2),
\ee
which gives the following first few coefficients
\be 
\begin{split}
g_{0}(\tau_{\rm uv};1/2)&= 3\,q_{\rm uv} +14\,q_{\rm uv}^2+30\,q_{\rm uv}^3+54\,q_{\rm uv}^4+84\,q_{\rm uv}^5+124\,q_{\rm uv}^6+\dots,\\
g_{1/2}(\tau_{\rm uv};1/2)&=q_{\rm uv}^{3/4}(1+9\,q_{\rm uv}+19\,q_{\rm uv}^2+50\,q_{\rm uv}^3+51\,q_{\rm uv}^4+131\,q_{\rm uv}^5+\dots).
\end{split}
\ee
Since $g_{\mu}(\tau_{\rm uv};1/2)$ is expressed in terms of quantities for $k_m=1/2$, it is a priori unclear whether
the modular completion of this expression satisfies the expected $S$-duality diagram. However, comparison of the first $q_\uv$-expansion demonstrates that $g_\mu(\tau_\uv;1/2)$ can be expressed in terms of the functions for
$k_m=3/2$, which do transform as a 2-dimensional vector-valued modular forms after completion. This reads
\be
\label{eq:funny-modular}
\begin{split}
g_{\mu}(\tau_{\rm uv};1/2)&=\frac{10}{9 }\frac{1}{2\pi i}\partial_{\tau_{\rm uv}}
  g^{(1)}_{\mu}(\tau_{\rm uv};3/2)+\frac{2}{9}g^{(3)}_{\mu}(\tau_{\rm uv};3/2)\\
&\quad   +\frac{1}{9}E_2(\tau_{\rm uv})g^{(1)}_{\mu}(\tau_{\rm uv};3/2),
\end{split}
\ee
and confirms the duality diagram \ref{STPF}. We
leave an in depth analysis of the modular properties for future work.
  
\subsection{Including point observables for $\mathbb{P}^2$}

We continue by studying correlation functions for point
observables for $\mathbb{P}^2$. For comparison with $N_f=0$, we consider the correlation
functions $\Phi_\mu^{\mathbb{P}^2}[u_{\rm   D}^n]=\left< u_{\rm   D}^n\right>^{\mathbb{P}^2}_\mu$ with $u_{\rm D}$ as in
(\ref{DonaldsonPointObs}). In terms of the exponentiated observable
$e^{pu_{\rm D}/\Lambda^2}$, we have
\be 
\Phi_\mu^{\mathbb{P}^2}[u_{\rm   D}^n/\Lambda^{2n}]=\frac{\partial^n}{\partial p^n}\left.\,
\Phi_\mu^{\mathbb{P}^2}\!\left[e^{p\,u_{\rm D}/\Lambda^2}\right]\right
\vert_{p=0}. 
\ee
The correlation functions of the $N_f=0$
theory equal Donaldson invariants. For an easy comparison, we
introduce the dimensionless correlation functions of the $\CN=2^*$ theory,
\be
\label{underPhiP2}
\underline{\Phi}^{\mathbb{P}^2}_\mu[u_{\rm
  D}^n/(2\Lambda^2)^{n}]= R_\bfmu^{\mathbb{P}^2}(3,2k_m)\, \,\Phi^{\mathbb{P}^2}_\mu[u_{\rm
  D}^n/(2\Lambda^2)^{n}],
\ee 
where $R_\bfmu^{\mathbb{P}^2}$ is the factor defined in (\ref{Prefactor1}) relating
$\CN=2^*$ to $N_f=0$. In the $N_f=0$ limit, $(2\Lambda)^{2n}$ becomes $\Lambda_0^{2n}$, such that we arrive at
the appropriate point observable in Donaldson theory. 
 
We list the first terms of the holomorphic parts of $\underline{\Phi}^{\mathbb{P}^2}_\mu$ (\ref{underPhiP2}) for $k_m=3/2, 1/2$ and $0\leq n\leq 4$.  
Tables \ref{TablePointObs0} and \ref{TablePointObs1} give the first
few coefficients for $\mu=0$, $\mu=1/2$ and $k_m=3/2$. For $k_m=1/2$, we list observables in Tables \ref{TablePointObs02} and 
\ref{TablePointObs12}. Up to a factor of $\Lambda^{3\ell}$, the series match with
the formal series (\ref{O1Op}), which demonstrates the geometric
meaning of the coefficients in the $q_{\rm uv}$-series. 
  
 Let us consider the limit to pure SW theory.
 For $k_m=3/2$ and $\mu=1/2$, the $N_f=0$ limit of the series in Table \ref{TablePointObs1} reproduces exactly the
known expressions for Donaldson invariants
\cite{ellingsrud1995wall, Korpas:2019cwg}. 
On the other hand, the expressions for $\mu=0$ in Table \ref{TablePointObs0} diverge for all $n$, while the Donaldson invariant vanishes. We expect that this mismatch is due to 
contributions from reducible connections, which are strictly semi-stable. The situation is understood for $n=0$, which is discussed in detail in Section \ref{CompVWtheory}: The $q_\uv$-series for $n=0$ is the generating series of weighted Euler characteristics,  which receive a contribution from reducible connections. Due to this contribution, the coefficients of the $q_\uv$-series starts with the constant term $-\frac{1}{4}q_\uv^0$, causing the divergence. Equation (\ref{int_invariants}) is the $q_\uv$-series after correcting for the contribution from reducible connections. For this series, the limit to $N_f=0$ clearly vanishes. We do not have a similar quantitative understanding for the $q_\uv$-series with $n>0$ at this time, but we expect that this can developed. 

For the case of $k_m=1/2$, we find that the $N_f=0$ limit of the expressions in Tables \ref{TablePointObs02} and 
\ref{TablePointObs12} do reproduce the Donaldson invariants for both $\mu=0$ and $\mu=1/2$.

\begin{center}
\begin{table}[h!]  
\centering
\renewcommand{\arraystretch}{1.8}
\begin{tabular}{| l | r |}   \hline
$n$ & Hol. part of $\underline{\Phi}_0^{\mathbb{P}^2}[u_{\rm D}^n/(2\Lambda^2)^n]$  \\ \hline
0 &  $i \,t^3\left(-\frac{1}{4}+\frac{3}{2}\,q_{\rm uv}+3\,q^2+4\,q_{\rm uv}^3+\frac{9}{2}\,q_{\rm uv}^4+\dots \right)$ \\
1 & $ i\, t^5 \left(  -\frac{1}{64}+\frac{9}{32}\,q_{\rm uv}+\frac{11}{16}\,q_{\rm uv}^2-\frac{23}{8}\,q_{\rm uv}^3-\frac{207}{32}\,q_{\rm uv}^4+\dots \right)$ \\
2  & $i\, t^7\left(-\frac{1}{256}-\frac{3}{128}\,q_{\rm uv}-\frac{27}{64}\,q_{\rm uv}^2-\frac{39}{32}\,q_{\rm uv}^3+\frac{1001}{128}\,q_{\rm uv}^4  +\dots \right)$ \\
3 & $i\,t^9\left(-\frac{1}{2048}-\frac{15}{1024}\,q_{\rm uv}+\frac{87}{512}\,q_{\rm uv}^2+\frac{757}{256}\,q_{\rm uv}^3+\frac{14697}{1024}\,q_{\rm uv}^4  +\dots\right)$ \\
4 & $i\,t^{11}\left( -\frac{1}{16384}-\frac{27}{8192}\,q_{\rm uv}-\frac{99}{4096}\,q_{\rm uv}^2-\frac{403}{1024}\,q_{\rm uv}^3-\frac{37887}{8192}\,q_{\rm uv}^4+\dots \right)$\\
\hline 
\end{tabular}
\caption{For $\mathbb{P}^2$, $k_m=3/2$ and $\mu=0$, first terms of the
  holomorphic part of $\underline{\Phi}_0^{\mathbb{P}^2}[u_{\rm D}^n/(2\Lambda^2)^n]$ with $u_{\rm D}$ and $0 \leq
  n\leq 4$. }
\label{TablePointObs0}
\end{table}
\end{center}

\begin{center}
\begin{table}[h!]  
\centering
\renewcommand{\arraystretch}{1.8}
\begin{tabular}{| l | r |}   \hline
$n$ & Hol. part of
$\underline{\Phi}_{\frac{1}{2}}^{\mathbb{P}^2}[u_{\rm D}^n/(2\Lambda^2)^n]$  \\
\hline 
0 &  $i\,t^3\left(q_{\rm uv}^{3/4}+3\,q_{\rm uv}^{7/4}+3\,q_{\rm uv}^{11/4}+6\,q_{\rm uv}^{15/4}+\dots \right)$\newline \\
1 & $-i\,t^5\left(\frac{3}{4}\, q_{\rm uv}^{7/4}+6\,q_{\rm uv}^{11/4}+ \frac{35}{2}\,q_{\rm uv}^{15/4}+\dots \right)$ \\
2  & $i\,t^7\left(\frac{19}{64}\, q_{\rm uv}^{7/4}+ \frac{31}{8}\,q_{\rm uv}^{11/4}+\frac{89}{4}\,q_{\rm uv}^{15/4}+\dots \right)$ \\
3 & $- i\,t^9\left( \frac{15}{32}\,q_{\rm uv}^{11/4} + \frac{971}{128}\,q_{\rm uv}^{15/4}+\dots\right)$ \\
4 & $i\,t^{11}\left( \frac{85}{512}\,
  q_{\rm uv}^{\frac{11}{4}}+\frac{15151}{4096}\,q_{\rm uv}^{\frac{15}{4}} +\dots \right)$\\
\hline 
\end{tabular}
\caption{For $\mathbb{P}^2$, $k_m=3/2$ and $\mu=1/2$, first terms of the
  holomorphic part of $\underline{\Phi}_{1/2}^{\mathbb{P}^2}[u_{\rm
    D}^n/(2\Lambda^2)^{n}]$ with $t=m/\Lambda$, $u_{\rm D}$ (\ref{DonaldsonPointObs}) and $0 \leq
  n\leq 4$. In the $N_f=0$ limit,
  $t\,q_\uv^{1/4}\to i$, such that the non-vanishing terms match with the
  results on Donaldson invariants in the literature, for example
  \cite[Theorem 4.4]{ellingsrud1995wall}, \cite[Table 2]{Korpas:2019cwg}.}
\label{TablePointObs1}
\end{table}
\end{center}

\begin{center} 
\begin{table}[h!]
\centering 
\renewcommand{\arraystretch}{1.8}
\begin{tabular}{| l | r |}   \hline
$n$ & Hol. part of $\underline{\Phi}^{\mathbb{P}^2}_0[u_{\rm D}^n/(2\Lambda)^n]$  \\ \hline
0 &  $-i\,t^3\left( 3\,q_{\rm uv}+14\,
  q_{\rm uv}^{2}+ 30\, q_{\rm uv}^{3}+54\,q_{\rm uv}^{4}+\dots \right)$ \\
1 & $i\,t^5\left( \frac{1}{2}\, q_{\rm uv}^{2}+\frac{13}{2}\, q_{\rm uv}^{3}+\frac{97}{4}\, q_{\rm uv}^{4}+\dots\right) $ \\
2  & $-i\,t^7\left(\frac{13}{16}\,
  q_{\rm uv}^{2}+\frac{173}{16}\,q_{\rm uv}^{3}+\frac{2045}{32}\,q_{\rm uv}^{4}+\dots \right)$ \\
3 & $i\,t^9\left( \frac{83}{256}\,
  q_{\rm uv}^{3}+\frac{1545}{256}\,q_{\rm uv}^{4}+\dots \right)$ \\
4 & $-i\,t^{11}\left(\frac{879}{2048}\,q_{\rm uv}^{3}+\frac{18997}{1024}\,q_{\rm uv}^{4}+\dots\right)$\\
\hline 
\end{tabular} 
\caption{For $\mathbb{P}^2$, $k_m=1/2$ and $\mu=0$, the first terms
  of the holomorphic part of   $\underline{\Phi}_0^{\mathbb{P}^2}[u_{\rm D}^n/(2\Lambda)^n]$, with $u_{\rm D}$ and $0 \leq
  n\leq 4$.  In the $N_f=0$ limit, these series vanish in agreement
  with the vanishing of Donaldson invariants for this case.}
\label{TablePointObs02}
\end{table}
\end{center}

\begin{center} 
\begin{table}[h!]
\centering 
\renewcommand{\arraystretch}{1.8}
\begin{tabular}{| l | r |}   \hline
$n$ & $\underline{\Phi}^{\mathbb{P}^2}_\frac{1}{2}[u_{\rm D}^n/(2\Lambda)^n]$  \\ \hline
0 &  $i\,t^3\left( q_{\rm uv}^{3/4}+9\,
  q_{\rm uv}^{7/4}+ 19\, q_{\rm uv}^{11/4}+50\,q_{\rm uv}^{15/4}+\dots \right)$ \\
1 & $-i\, t^5\left(\frac{5}{8}\, q_{\rm uv}^{7/4}+3\, q_{\rm uv}^{11/4}+ \frac{43}{2}\, q_{\rm uv}^{15/4}+\dots\right) $ \\
2  & $i\,t^7\left(\frac{19}{64}\,  q_{\rm uv}^{7/4}+\frac{19}{4}\,q_{\rm uv}^{11/4}+\frac{581}{16}\,q_{\rm uv}^{15/4}+\dots \right)$ \\
3 & $-i\,t^9\left( \frac{23}{64}
  q_{\rm uv}^{11/4}+\frac{2599}{512}q_{\rm uv}^{15/4}+\dots \right)$ \\
4 & $i\,t^{11}\left(\frac{85}{512}\,q_{\rm uv}^{11/4}+\frac{16025}{4096}\,q_{\rm uv}^{15/4}+\dots\right)$\\
\hline 
\end{tabular} 
\caption{For $\mathbb{P}^2$, $k_m=1/2$ and $\mu=1/2$, the first terms
  of the holomorphic part of   $\underline{\Phi}^{\mathbb{P}^2}_{1/2}[u_{\rm D}^n/(2\Lambda)^n]$, with $u_{\rm D}$ and $0 \leq
  n\leq 4$. In the $N_f=0$ limit,
  $t\,q_\uv^{1/4}\to i$, such that the non-vanishing terms match with those
  in Table \ref{TablePointObs1}, and thus with the
  results on Donaldson invariants in the literature, for example
  \cite[Theorem 4.4]{ellingsrud1995wall}, \cite[Table 2]{Korpas:2019cwg}. }
\label{TablePointObs12}
\end{table}
\end{center}

\subsection{Contribution from $u_i, i=1,2,3$ for odd intersection form}
Besides the contribution from $u=\infty$ discussed in the previous
subsections, $\Phi_\bfmu^J$ can receive contributions from the other
three singularities $u \to u_i$, $i=1,2,3$. We will determine these
contributions in this Subsection for the integrands of Section
\ref{FactPsi} for $k_{m,1}=1/2$ and $3/2$ and $L$
an odd lattice. The analysis is similar to
Section \ref{metricwallcross}, but with the indefinite theta series
$\widehat \Theta_\bfmu^J$ replaced by $\widehat
G_{\mu_1}\Theta_{L_-,\bfmu_-}$. The analysis shows that these cusps
only contribute if $\sigma\leq -7$. In particular, the $u$-plane
integral for $\mathbb{P}^2$ is completely determined by the
contribution from $\infty$. 
\\
\\
{\it Contribution from $u_1$}\\
For this contribution, we need to determine the leading behavior of $\widehat G_{\mu}(\tau_1,\bar
\tau_1,v_1,\bar v_1;k_{m,1})$ for $\tau_1\to
i\infty$. To this end, we need to be
careful with contributions from the non-holomorphic part of $\widehat
G_\mu$, which can contribute to the holomorphic part since $v_1$
approaches $-\tau_1/2$ for this cusp. We set $\tilde
v_1=v_1+\tau_1/2$, which remains finite near the cusp.
To follow the structure we found in Section \ref{metricwallcross}, we
define 
\be
\nonumber
\begin{split}
& G_{\mu,1}(\tau_1, \tilde v_1; k_m)= e^{2\pi i ( v_1+\tau_1/4)
   k_{m,1}^2}\\
&\qquad \times \left( G_{\mu}(\tau_1, \tilde v_1-\tau/2; k_m)  -\tfrac{1}{2}\sum_{n\in
  \mathbb{Z}+\mu} (\sgn(n)-\sgn(n-k_m/2)) e^{-4\pi i v_1
k_m n} q^{-n^2+n k_m} \right),
\end{split}
\ee
where the sum over $n$ are holomorphic contributions arising from the
non-holomorphic part (\ref{hatFE}). We find for the leading terms of
$G_{1,\mu}$ in the $q_1$-expansion
\be
\begin{split}
 G_{0,1}(\tau_1,\tilde v_1;1/2)&=-e^{-3\pi i \tilde v_1/2}\,q^{7/16}+\dots,\\
 G_{1/2,1}(\tau_1,\tilde v_1;1/2)&=e^{3\pi i \tilde v_1/2}\,q^{7/16}+\dots,\\
 G_{0,1}(\tau_1, \tilde v_1;3/2)&=\frac{q^{-1/16}}{e^{5\pi i \tilde v_1/2}(-1+e^{4\pi i \tilde v_1})}+\dots,\\
 G_{1/2,1}(\tau_1, \tilde  v_1;3/2)&=\frac{q^{-1/16}}{e^{3\pi i \tilde v_1/2}(-1+e^{4\pi i \tilde v_1})}+\dots.
\end{split}
\ee 
Moreover, $\Theta_{L_-,\bfmu_-}(\tau_1,v_1\bfk_{m,-})=e^{-2\pi
  i(v_1+\tau_1/4)\bfk_{m,-}^2}\Theta_{L_-,\bfmu_-}(\tau_1,(v_1+\tau_1/2)\bfk_{m,-})$. Thus
the contribution from $u_1$ reads 
\be
\begin{split}
\Phi_{\bfmu,1}^J&=\frac{m^{\frac{\sigma}{8}-2}}{4\,\eta(\tau_{\rm uv})^{\chi-\frac{3}{2}\sigma+2\bfk_m^2}}
\left(\frac{4}{\Lambda} \right)^{\frac{3}{8}(2\chi+3\sigma)}\\
&\quad \times \lim_{Y\to \infty} \int_{-\frac{1}{2}+iY}^{\frac{3}{2}+iY}
d\tau_1\,\left( \frac{du}{da}\right)^{\frac{\chi}{2}-3}\Delta_{\rm phys}^{\frac{\sigma}{8}+1}\left( C\, e^{-2\pi i
    (v_1+\tau_1/4)}\right)^{\bfk_m^2}\\
&\quad \times G_{\mu,1}(\tau_1, v_1+\tau_1/2;k_{m,1})\,\Theta_{L_-,\bfmu_-}(\tau_1,(v_1+\tau_1/2)\bfk_{m,-}).
\end{split}
\ee
From the $\tau_1$ behavior of $du/da$, $\Delta_{\rm phys}$ and $C$
near $u_1$, it follows that
$\Phi_\bfmu^J$ receives a contribution from $u_1$ for
$\sigma\leq -15$ for $k_{m,1}=1/2$ or $\leq -7$ for $k_{m,1}=3/2$. 
\\
\\ 
{\it Contribution from $u_2$}\\
For the contributions from $u_2$, we determine the expansion of the
``dual'' function $$\widehat
F_{\mu,2}(\tau_2,\bar \tau_2,\rho_2,\bar \rho_2)=-i(-i\tau_2/2)^{-1/2}\,e^{2\pi i
  \rho_2^2/\tau_2}\,\widehat F_{\mu}(-1/\tau_2,-1/\bar
\tau_2,\rho_2/\tau_2, \bar \rho_2/\bar \tau_2).$$ We find using the
transformation of the Appell-Lerch sum (\ref{whMtrafo}) that $\widehat
F_{\mu,2}$ equals 
\be
\label{FDmu}  
\begin{split}
&\widehat F_{\mu,2}(\tau_{2},\bar \tau_2 ,\rho_{2},\bar \rho_2)=e^{\pi i
  \mu}\,q_2^{-1/16}w_2^{-1/2} \left( \,
  M(\tau_2/2,\rho_2-\mu+\tau_2/4,1/2)\right. \\
& \left. \quad  + \frac{i}{2} R(\tau_2/2,\bar \tau_2/2,\rho_2-\mu +
  \tau_2/4-1/2,\bar \rho_2-\mu +\bar \tau_2/4-1/2) \right)\\
&=F_{2,\mu}(\tau_2,\rho_2)\\
&\quad -\frac{1}{2} \sum_{n\in \mathbb{Z}} \left( \sgn(n)
  -E((n+2b_2)\sqrt{y_2})\right)\,(-1)^{2\mu n} q_2^{-n^2/4} w_2^{-n},
\end{split}
\ee
with 
\be
\begin{split}
F_{\mu,2}(\tau_2,\rho_2)&=\frac{1}{2}-\frac{1}{\vartheta_{2}(\tau_2/2)} \sum_{n\in
  \mathbb{Z}+\frac{1}{2}} \frac{q_2^{n^2/4}}{1-(-1)^{2\mu}
  w_2\,q_2^{n/2}}.
\end{split}
\ee
This function is anti-symmetric in $\rho_2$. By (\ref{GFD}) and (\ref{whGmutrafo}), this also equals
\be
F_{\mu,2}(\tau_2,3z_2/2)= G_{0}(\tau,z;3/2)+e^{2\pi i\mu}\,G_{1/2}(\tau,z;3/2)+D_{\mu,2}(\tau_2,z_2),
\ee 
where $D_{\mu,2}$ are the dual of the functions $D_{\mu}$ (\ref{D0half}),
\be
\begin{split}
D_{0,2}(\tau,z)&=-\frac{i\,\eta(\tau/2)^3\,\vartheta_1(\tau/2,z)\,\vartheta_3(\tau/2,z/2)}{\vartheta_4(\tau/2,3z/2)\,\vartheta_2(\tau/2)\,\vartheta_2(\tau/2,z)\,\vartheta_4(\tau/2,z/2)}, \\
D_{\frac{1}{2},2}(\tau,z)&=-\frac{i\,\eta(\tau/2)^3\,\vartheta_1(\tau/2,z)\,\vartheta_4(\tau/2,z/2)}{\vartheta_3(\tau/2,3z/2)\,\vartheta_2(\tau/2)\,\vartheta_2(\tau/2,z)\,\vartheta_3(\tau/2,z/2)}.
\end{split}
\ee

To determine the leading term near $u_2$, we set $v_2=\tau_2/2 +\tilde
v_2$, and introduce
\be
\begin{split}
& \widehat G_{\mu,2}(\tau_2, \tilde v_2; k_{m,1})= -i (-i\tau_2/2)^{-1/2}
e^{2\pi i v_2^2 k_{m,1}^2} e^{2\pi i (
  -v_2+\tau_2/4) 
   k_{m,1}^2}\\
&\qquad \times  \widehat G_{\mu}(-1/\tau_2, (\tilde v_2+\tau_2/2)/\tau_2; k_{m,1}).
\end{split}
\ee
Using the explicit expressions, we can determine the leading term of the
holomorphic part of $\widehat G_{\mu,2}$. We find
\be
\begin{split}
 G_{0,2}(\tau_2,\tilde v_2;1/2)&=\frac{e^{\pi i \tilde
     v_2/2}}{2(-1+e^{\pi i \tilde v_2})}\,q^{-1/16}+\dots,\\
 G_{1/2,2}(\tau_2,\tilde v_2;1/2)&=\frac{e^{\pi i \tilde
     v_2/2}}{2(1+e^{\pi i \tilde v_2})}\,q^{-1/16}+\dots,\\
 G_{0,2}(\tau_2, \tilde v_2;3/2)&=\frac{e^{3\pi i \tilde
     v_2/2}}{-1+e^{\pi i \tilde
     v_2/2}-e^{2\pi i \tilde
     v_2/2}+e^{3\pi i \tilde
     v_2}}\,q^{-1/16} +\dots,\\
 G_{1/2,2}(\tau_2, \tilde  v_2;3/2)&=-\frac{e^{3\pi i \tilde
     v_2/2}}{1+e^{\pi i \tilde
     v_2/2}+e^{2\pi i \tilde
     v_2/2}+e^{3\pi i \tilde
     v_2}}\,q^{-1/16}+\dots.
\end{split}
\ee 
The contribution from $u_2$ reads 
\be
\begin{split}
\Phi_{\bfmu,2}^J&=\frac{m^{\frac{\sigma}{8}-2}}{4\,\eta(\tau_{\rm uv})^{\chi-\frac{3}{2}\sigma+2\bfk_m^2}}
\left(\frac{4}{\Lambda}
\right)^{\frac{3}{8}(2\chi+3\sigma)}2^{-b_2/2}e^{\pi i \sigma/4-B(\bfk_{m,-},\bfmu_-)}\\
&\quad \times \lim_{Y\to \infty} \int_{-\frac{1}{2}+iY}^{\frac{3}{2}+iY}
d\tau_2\,\left( \frac{du}{da}\right)_2^{\frac{\chi}{2}-3}\Delta_{\rm phys,2}^{\frac{\sigma}{8}+1}\left( C_2\, e^{2\pi i
    (v_2-\tau_2/4)}\right)^{\bfk_m^2}\\
&\quad \times G_{\mu,2}(\tau_2, v_2-\tau_2/2;k_{m,1})\,\Theta_{L_-,\bfmu_-}(\tau_2/4,(v_2/2-\tau_1/4)\bfk_{m,-}-\bfmu_-/2).
\end{split}
\ee
Therefore the cusp at $u_2$ only contributes for four-manifolds with
$\sigma\leq -7$. 
\\
\\ 
{\it Contribution from $u_3$}\\ 
One can show similarly that the cusp at $u_3$ only contributes for
$\sigma\leq -7$.

\subsection{Evaluation for a special period point $J$ and even
  intersection form}
  \label{SecEvalEven}
This subsection evaluates $\Phi_\bfmu^J$ for $X$ spin and $\bfk_m$
an almost complex structure. If $X$ is a spin four-manifold,
$w_2(X)=0$ and $\bfk_m\in L$. Moreover, the lattice $L$ is even, and since $b_2^+=1$, it can be brought to the
form
\be
L=\mathbb{I}^{1,1}\oplus nL_{E_8},
\ee 
where $\mathbb{I}^{1,1}$ is the two-dimensional
lattice with quadratic form ${\scriptsize \left(\begin{array}{cc}  0 & 1 \\ 1 &
    0 \end{array}\right)}$, and $L$ is minus the $E_8$ root
lattice. We choose for the period point
\be
J=\frac{1}{\sqrt{2}} (1,1,{\bf 0}),\qquad C=\frac{1}{\sqrt{2}} (1,-1,{\bf 0})
\ee
where the first two components correspond to $\mathbb{I}^{1,1}\subset
L$, and ${\bf 0}$ is the $(b_2-2)$-dimensional 0-vector. The squares
of the positive and negative definite components of $\bfk\in L$ are
\be
\bfk_+^2=\frac{1}{2}(k_1+k_2)^2,\qquad \bfk_-^2=-\frac{1}{2}(k_1-k_2)^2+\bfk_n^2,
\ee
where $\bfk_n\in n L_{E_8}$. We set furthermore 
\be
\mu_+=\mu_1+\mu_2,\qquad \mu_-=\mu_1-\mu_2.
\ee
As before $\bfrho=v\bfk_m$. We introduce
\be
\rho_+=\sqrt{2}B(\bfrho,J),\qquad \rho_+=\sqrt{2}B(\bfrho,C).
\ee

The sum over fluxes $\Psi_\bfmu^J$ now factors for this choice of $J$ as
\be
\Psi^J_\bfmu(\tau,\bar \tau,\bfrho,\bar
\bfrho)=\Psi_{\mathbb{I},(\mu_+,\mu_-)}(\tau,\bar
\tau,\rho_-,\rho_+,\bar \rho_+)\,\Theta_{nE_8,\bfmu_n}(\tau,\bfrho_n),
\ee
with 
\be
\begin{split}
&\Psi_{\mathbb{I},(\mu_+,\mu_-)}(\tau,\bar
\tau,\rho_-,\rho_+,\bar \rho_+)=\exp(-2\pi y b_+^2) \sum_{\bfk\in
  \mathbb{I}^{1,1}+(\mu_+,\mu_-)} \partial_{\bar
  \tau}(\sqrt{2y}(k_1+k_2+b_+))\\
&\qquad \times q^{(k_1-k_2)^2/2}\bar q^{(k_1+k_2)^2/2}e^{2\pi i
  \rho_-(k_1-k_2)-2\pi i \bar \rho_+(k_1+k_2)}
\end{split}
\ee
where $b_+=\mathrm{Im}(\rho)/y$, and
\be
\Theta_{nE_8,\bfmu_n}(\tau,\bfrho_n)=\sum_{\bfk_n\in
  nL_{E_8}+\bfmu_n}q^{-\bfk_n^2}e^{-4\pi i B(\bfrho_n,\bfk_n)}. 
\ee 

Our aim is to determine suitable anti-derivatives for
$\Psi_{\mathbb{I},(\mu_+,\mu_-)}$,
\be
\partial_{\bar \tau} \widehat G_{\mathbb{I},(\mu_+,\mu_-)}(\tau,\bar
\tau,v,\bar v;\bfk_m)=\Psi_{\mathbb{I},(\mu_+,\mu_-)}(\tau,\bar
\tau,v\bfk_m,\bar v \bfk_m).
\ee
To this end, we decompose $\Psi_{\mathbb{I},(\mu_+,\mu_-)}$ further by setting
$n_\pm=k_1\pm k_2$. Then,
\be
\Psi_{\mathbb{I},(\mu_+,\mu_-)}(\tau,\bar
\tau,\rho_-,\rho_+,\bar \rho_+)= \sum_{j=0,1} s_{\mu_++j}(\tau,\bar
\tau,\rho_+,\bar \rho_+)\,t_{\mu_-+j}(\tau,\rho_-),
\ee
with 
\be
\begin{split}
&s_{\nu}(\tau,\bar \tau,\rho,\bar \rho)=\exp(-2\pi y
b_+^2)\sum_{n\in 2\mathbb{Z}+\nu} \partial_{\bar \tau}
(\sqrt{2y}(n+b)) \bar q^{n^2/2}e^{-2\pi i\bar \rho n},\\
&t_{\nu}(\tau,\rho)=\sum_{n\in 2\mathbb{Z}+\nu} q^{n^2/2}e^{2\pi i
  \rho n}. 
\end{split}
\ee
We define furthermore,
\be
\theta^\pm_{\nu}(\tau,\rho)=t_{\nu}(\tau,\rho)\pm t_{\nu+1}(\tau,\rho).
\ee
These can be expressed as 
\be
\begin{split}
&\theta_\nu^{+}(\tau,z)=\left\{   \begin{array}{ll}
    \vartheta_3(\tau,z),\qquad  & \nu =0 \mod \mathbb{Z}, \\ \vartheta_2(\tau,z),\qquad  & \nu =\frac{1}{2} \mod \mathbb{Z},  \end{array}                    \right.\\
&\theta_\nu^{-}(\tau,z)=\left\{   \begin{array}{ll}
    (-1)^\nu \vartheta_4(\tau,z),\qquad  & \nu =0 \mod \mathbb{Z}, \\
    -e^{\pi i \nu}\,\vartheta_1(\tau,z),\qquad  & \nu =\frac{1}{2} \mod \mathbb{Z}.  \end{array}                    \right.
\end{split}
\ee

We define new functions $f_\mu^\pm$ as linear combinations of the $s_\nu$,
\be
\begin{split}
&f^\pm_\mu(\tau,\bar \tau,z,\bar z) =s_{\nu}(\tau,\bar \tau,\rho,\bar
\rho)\pm s_{\nu+1}(\tau,\bar \tau,\rho,\bar \rho).
\end{split} 
\ee
These are simply related to the functions $f_\mu$ (\ref{fmutbart}) from the discussion for odd lattices . We have
\be
\begin{split}
&f^+_\mu(\tau,\bar \tau,z,\bar z)=f_{\mu}(\tau/2,\bar \tau/2,z/2,\bar z/2),\\
&f^-_\mu(\tau,\bar \tau,z,\bar z)=e^{\pi i \mu}\, f_{\mu}(\tau/2,\bar \tau/2,(z+1/2)/2,(\bar z+1/2)/2).
\end{split}
\ee

We arrive finally at,
\be
\label{PsiII}
\begin{split}
&\Psi_{\mathbb{I},(\mu_+,\mu_-)}(\tau,\bar
\tau,\rho_-,\rho_+,\bar \rho_+)=\tfrac{1}{2}f^+_{\mu_+}(\tau/2,\bar \tau/2,\rho_+/2,\bar \rho_+/2)\,\theta^+_{\mu_-}(\tau,\rho_-)\\
&\qquad  +\tfrac{1}{2}\,f^-_{\mu_+}(\tau/2,\bar \tau/2,(\rho_++1/2)/2,(\bar \rho_++1/2)/2)\,\theta^-_{\mu_-}(\tau,\rho_-).
\end{split}
\ee 

We proceed with $X$ such that $b_2=2$, and $\bfk_m$ being an almost complex
structure implies that $k_{m,1}=k_{m,2}=\pm 1$; we take $+1$ without lack
of generality. 
We have then $\rho_+=2v$ and $\rho_-=0$. Therefore $\theta^-_\nu(\tau,\rho_-)$ vanishes, and it is not
necessary to find an anti-derivative for $f^-_{1/2}$.

We will determine a set of anti-derivatives for $\Psi_{\mathbb{I},((\mu_+,\mu_-))}$  in terms of the
function $\widehat M$ (\ref{Mtuv}), adapted for the present
context. We choose the arguments of $\widehat M$ in such a way that
the theta series $\theta^\pm_\nu$ in (\ref{PsiII}) cancel against the
theta series in front of the sum of $\widehat M$ (\ref{Mtuv}). To this
end, we let $\partial_{\bar \tau} \widehat
A_{(\mu_+,\mu_-)}^\pm(\tau,\bar \tau, z,\bar z)= f^\pm_{\mu_+}(\tau,\bar \tau,z,\bar
z)\,\theta^\pm_{\mu_-}(\tau,0)$, and define $\widehat
A_{(\mu_+,\mu_-)}^\pm$ as,
\be
\begin{split}
&\widehat A^+_{(\mu_+,\mu_-)}(\tau,\bar \tau,z,\bar z)=e^{-2\pi i
  (\mu_+-1/2)z} q^{-(\mu_+-1/2)^2/2} \,\theta^+_{\mu_-}(\tau,0)\\
&\qquad \qquad \times \widehat M(\tau,\bar
\tau,z,\bar z, (1/2-\mu_+)\tau+1/2, (1/2-\mu_+)\bar \tau+1/2), \\
&\widehat A^-_{(\mu_+,\mu_-)}(\tau,\bar \tau,z,\bar z)=ie^{-2\pi i
  (\mu_+-1/2)z} q^{-(\mu_+-1/2)^2/2} \,\theta^-_{\mu_-}(\tau,0)\\
&\qquad \qquad \times  \,\widehat M(\tau,\bar \tau,z,\bar z, (1/2-\mu_+)\tau, (1/2-\mu_+)\bar \tau).
\end{split}
\ee
For the second line, $\mu_-\neq \pm \frac{1}{2}$, since $\theta^-_{\pm
  1/2}(\tau,0)=0$. For the four choices of $\mu_{1,2}\in \{0,\frac{1}{2}\}$, 
we arrive then at the following holomorphic parts
\be
\begin{split} 
&A^+_{(0,0)}(\tau,z)=A^+_{(1,0)}(\tau,z)=\frac{1}{2}\vartheta_3(\tau)-
\sum_{n\in \mathbb{Z}} \frac{q^{n^2/2}}{1-e^{2\pi i z}q^n},\\
&A^-_{(0,0)}(\tau,z)=-A^-_{(1,0)}(\tau,z)= \frac{1}{2}\vartheta_4(\tau)-
\sum_{n\in \mathbb{Z}} \frac{(-1)^nq^{n^2/2}}{1-e^{2\pi i z}q^n},\\
&A^+_{(1/2,\pm 1/2)}(\tau,z)= -e^{\pi i z} \sum_{n\in \mathbb{Z}}
\frac{q^{n(n+1)/2}}{1-e^{2\pi i z} q^n}.
\end{split}
\ee

Using these functions, we can determine suitable anti-derivatives of
$\Psi_{\mathbb{I},(\mu_+,\mu_-)}$. Let us first consider
$(\mu_+,\mu_-)=(0,0)$. Equation (\ref{PsiII}) demonstrates
that the holomorphic part of a potential anti-derivative is
\be
\frac{1}{2} A^+_{(0,0)}(\tau,\rho_+)+ \frac{1}{2}A^-_{(0,0)}(\tau,\rho_+)=
\frac{1}{2}\sum_{n \,\,{\rm even}} q^{n^2/2}-\sum_{n\,\,{\rm even}}
\frac{q^{n^2/2}}{1-e^{2\pi i \rho_+}q^n }.
\ee 
Substitution of $\rho_+=2v$, we find that this function has poles at
$e^{4\pi i v} q^{2n}=1$ for $n\in \mathbb{Z}$. Among these poles, the
ones at $e^{2\pi i v}q^n=1$ are not excluded by (\ref{vexcluded}). We
will substract these by subtracting the $SL(2,\mathbb{Z})$ modular
form,  $i\eta(\tau)^3/\vartheta_1(\tau,z)$ (\ref{InvThetaId}). The
holomorphic part of the anti-derivative for $\bfk_m=(1,1)$ is therefore
\be
G_{\mathbb{I},(0,0)}(\tau,v;\bfk_m)=\frac{1}{2}\sum_{n \,\,{\rm even}} q^{n^2/2}-\sum_{n\,\,{\rm even}}
\frac{q^{n^2/2}}{1-e^{4\pi i v}q^n}-\frac{i\eta(\tau)^3}{\vartheta_1(\tau,2v)}.
\ee
For $(\mu_+,\mu_-)=(\frac{1}{2},\pm \frac{1}{2})$, we arrive using a
similar analysis at
\be
\begin{split}
G_{\mathbb{I},(\frac{1}{2},\pm \frac{1}{2})}(\tau,v;\bfk_m)&=\frac{1}{2}A^+_{(1/2,1/2)}(\tau,2v)+\frac{i}{2} \frac{\eta(\tau)^3}{\vartheta_1(\tau,2v)}\\
&=-e^{2\pi iv}\sum_{n\,\,{\rm odd}}
\frac{q^{n(n+1)/2}}{1-e^{4\pi i v}q^n}.
\end{split}
\ee
Finally, for $(\mu_+,\mu_-)=(1,0)$, 
\be
\begin{split}
G_{\mathbb{I},(1,0)}(\tau,v;\bfk_m)&=\frac{1}{2} A^+_{(1,0)}(\tau,2v)+\frac{1}{2} A^-_{(1,0)}(\tau,2v)\\
&=\frac{1}{2}\sum_{n \,\,{\rm odd}} q^{n^2/2}-\sum_{n\,\,{\rm odd}}
\frac{q^{n^2/2}}{1-e^{4\pi i v}q^n}.
\end{split}
\ee
These functions are closely related to the refined VW partition function
for $\mathbb{P}^1\times \mathbb{P}^1$ \cite[(4.36)]{Haghighat:2012bm}. We comment more
on this relation in Section \ref{CompVWtheory}.

Similarly to the case of odd intersection form, we define
\be
g_{\mathbb{I},(\mu_+,\mu_-)}(\tau)=-\frac{1}{2\pi i} \partial_{z} G_{\mathbb{I},(\mu_+,\mu_-)}(\tau,z;\bfk_m)\vert_{z=0},
\ee
and similarly for the non-holomorphic $\widehat g_{\mathbb{I},(\mu_+,\mu_-)}$.
This gives explicitly
\be
\begin{split}
&g_{\mathbb{I},(0,0)}(\tau)=-\sum_{n\,\,{\rm
    even}}\frac{q^{n(n+1)/2}}{(1+q^{n/2})^2}+\sum_{n\,\,{\rm odd}}
\frac{q^{n(n+1)/2}(1+q^n)}{(1-q^n)^2},\\
&g_{\mathbb{I},(\frac{1}{2},\pm \frac{1}{2})}(\tau)=\sum_{n\,\,{\rm
    odd}} \frac{q^{n(n+1)/2}(1+q^n)}{(1-q^n)^2},\\
&g_{\mathbb{I},(\frac{1}{2},\pm \frac{1}{2})}(\tau)=2\sum_{n\,\,{\rm
    odd}} \frac{q^{n^2/2+n}}{(1-q^n)^2}.
\end{split}
\ee

Using these expressions, we can evaluate the $\Phi_\bfmu^J$ for all
$b_2>2$, and $\bfk_m$ of the form $\bfk_m=(1,1,\bfk_n)\in (L/2)/L$, with $\bfk_n^2=-2n
$ such that $\bfk_m$ is an almost complex structure.
Determination of $\Phi^J_\bfmu$ then follows by taking the residue at
$\tau-\tau_{\rm uv}=0$ as in the case of odd lattices
(\ref{PhimuJOddFINAL}). This gives our final formula
for $X$ with an even intersection form $\bfk_m=(1,1,\bfk_n)$
\be
\Phi_\bfmu^J(\tau_\uv,\bar \tau_\uv)=\frac{\widehat g_{\mathbb{I},(\mu_+,\mu_-)}(\tau_{\rm uv})}{\eta(\tau_{\rm uv})^{2\chi}}\,\Theta_{nE_8,\bfmu_n}(\tau_{\rm uv},0),
\ee
where
$\bfmu=(\frac{1}{2}(\mu_++\mu_-),\frac{1}{2}(\mu_+-\mu_-),\bfmu_n)$. This
matches with known partition functions for VW theory if $X$ is Fano.

\subsection{Review of Vafa-Witten partition functions for rational surfaces}
\label{CompVWtheory}
We recall a few aspects of partition functions of Vafa-Witten theory \cite{Vafa:1994tf} for rational surfaces. Throughout this subsection, $\tau=\tau_\uv$ and $X$ one of the rational surfaces. 
Let us introduce the Chern character $\gamma=(N,c_1,k)$ of a $U(N)$
bundle with 't Hooft flux (or first Chern class) $c_1=N\bfmu$, and
instanton number $k$. The holomorphic part of the $U(N)$ Vafa-Witten partitition function
with 't Hooft flux $\bfmu\in \frac{1}{N}L/L$ is a generating function
of weighted Euler characteristics reads 
\be
 h^X_{N,\bfmu}(\tau)=\sum_{k} \bar \chi(\gamma)\,q^{k-\frac{N\chi(X)}{24}}.
\ee  
On a Fano surface with $b_2^+=1$, the $\bar \chi(\gamma)$ are
explicitly given by
\be
\label{rational_inv}
\bar \chi(\gamma)=(-1)^{\mathrm{dim}_\mathbb{C} \CM_\gamma}
\sum_{m\geq 1, m\vert \gamma} (-1)^{\dim_\mathbb{C} \CM_{\gamma/m}}
\frac{\chi(\gamma/m)}{m^2},
\ee
where the $\chi(\gamma)\in \mathbb{Z}$ are Euler numbers for
intersection cohomology \cite{Yoshioka:1995, Manschot:2016gsx}. Many explicit results for the generating series $h^X_{N,\bfmu}$ are available for all $N$ \cite{Gottsche1990, Vafa:1994tf, Yoshioka1994, Yoshioka:1995, Gottsche:1999, Manschot:2010nc, Manschot:2011dj, Manschot:2014cca, Mozgovoy:2013zqx, Beaujard:2020sgs, Alexandrov:2020bwg}.

For manifolds with $b_2^+=1$, the VW partition functions are known to involve a non-holomorphic contribution \cite{Vafa:1994tf, Minahan:1998vr, Alim:2010cf, Manschot:2011dj, Manschot:2017xcr, Alexandrov:2019rth, Dabholkar:2020fde, Alexandrov:2020bwg}. With the appropriate non-holomorphic term, they transform as \cite{Vafa:1994tf}
\be 
\begin{split} 
&S:\quad  \widehat
h^X_{N,\bfmu}(-1/\tau, -1/\bar \tau)=(-1)^{(N-1)(\chi+\sigma)/4} N^{-\frac{b_2(X)}{2}}
(-i\tau)^{-\frac{\chi(X)}{2}}\\
&\qquad \qquad\qquad \qquad \qquad \qquad \times \sum_{\bfnu \in \frac{1}{N}L/L} e^{-2\pi i NB(\bfmu,\bfnu)}\,\widehat h^X_{N,\bfnu}(\tau,\bar \tau), \\
&T:\quad \widehat h^X_{N,\bfmu}(\tau+1,\bar \tau+1)=\exp\left(-\pi i
  N(N-1)\bfmu^2-2\pi i N\chi(X)/24\right)\,\widehat h^X_{N,\bfmu}(\tau, \bar \tau).
\end{split}
\ee
We have furthermore $h^X_{N,\bfmu}=h^X_{N,-\bfmu}=h^X_{N,\bfmu+\bfk}$ with $\bfk\in L$. 

For $N=1$, we have the explicit result for any $X$ \cite{Gottsche1990, Vafa:1994tf},
\be
h^X_{1,0}(\tau)=\frac{1}{\eta(\tau)^{\chi(X)}}.
\ee
In the following, we briefly review explicit expressions for $N=2$ and $X=\mathbb{P}^2$ and $X$ an Hirzebruch surface.

\subsubsection*{The complex projective plane $\mathbb{P}^2$}
We fix $X=\mathbb{P}^2$ and $N=2$ and suppress these labels from the
generating functions.  The holomorphic partition function $h^{\mathbb{P}^2}_{2,\mu}=h_\mu$ takes the following
form \cite{Vafa:1994tf}
\be
\label{VWP2}
h_{\mu}(\tau)=\frac{g_{\mu}(\tau)}{\eta(\tau)^6},
\ee  
Here, $g_\mu(\tau)$ is the generating function of class numbers $H(4n-2\mu)$ for $\mu=0,1/2$ defined earlier in Equation (\ref{genClassNum}), These are rational numbers for $\mu=0$. From \eqref{rational_inv}, one can deduce that the generating series for the integer invariants is $\eta^{-6}$ times
\be 
\label{int_invariants}
\begin{split}
g_{0}(\tau)+\frac{1}{4}\,\vartheta_4(2\tau)^3=6\,q^2+2\,q^3+6\,q^4+12\,q^6+6\,q^7+\dots
\end{split}
\ee
 
The functions $g_\mu(\tau)$ are examples of mock modular forms, whose non-holomorphic completions read \cite{Vafa:1994tf, Zagier:1975}
\be
\label{whfj}
\widehat g_\mu(\tau,\bar \tau)=g_\mu(\tau)-\frac{3i}{4\sqrt{2}\pi}\int_{-\bar
  \tau}^{i\infty} \frac{\Theta_{\mu}(v)}{(-i(v+\tau))^\frac{3}{2}}dv,
\ee 
with $\Theta_\mu$ as in \eqref{Thetamu}. The $g_\mu$ transform as a modular vector under $SL(2,\mathbb{Z})$. The transformations under the $S$ and $T$ generators are 
\be 
\begin{split} 
&S:\quad  \widehat
g_{\mu}(-1/\tau,-1/\bar \tau)=-\frac{1}{\sqrt{2}}(-i\tau)^\frac{3}{2}
\left(\widehat g_{0}(\tau)+(-1)^{2\mu}\, \widehat g_{1/2}(\tau,\bar \tau)\right), \\
&T:\quad \widehat g_{\mu}(\tau+1,\bar \tau+1)=(-i)^{2\mu}\,\widehat g_{\mu}(\tau,\bar \tau).
\end{split}
\ee

The $U(2)$ VW partition function for $\mathbb{P}^2$ with 't Hooft flux $\mu$ is \cite{Vafa:1994tf, Dabholkar:2020fde}
\be
\widehat h_\mu(\tau,\bar \tau)=\frac{\widehat g_{2\mu}(\tau,\bar \tau)}{\eta(\tau)^6}.
\ee
As mentioned in Section \ref{tau0contribution}, this is in perfect agreement with the holomorphic anomaly derived from the $\CN=2^*$ $u$-plane integral (\ref{HolAnol}), as well as the complete expression in Equation (\ref{whG}). The partition functions for the three gauge groups are 
\be 
\begin{split}
&\widehat h^{SU(2)}_0= \tfrac{1}{2} \widehat h_{0},\qquad \qquad \quad \widehat h^{SU(2)}_{1/2}= \tfrac{1}{2} h_{1/2}  \\
&\widehat h^{SO(3)_+}_0= \widehat h_{0}+ \widehat h_{1/2}, \qquad \quad \widehat h^{SO(3)_+}_{1/2}= \widehat h_{0}-\widehat h_{1/2},\\ 
&\widehat h_0^{SO(3)_-}= \widehat h_{0}-i\,\widehat h_{1/2},\qquad \widehat h_{1/2}^{SO(3)_-}= \widehat h_{0}+i\,\widehat h_{1/2}.
\end{split}
\ee
These functions satisfy the duality diagram in Figure \ref{STPF}.

Let us also consider the refined partition function for $\mathbb{CP}^2$ 
\cite{Yoshioka1994, Yoshioka:1995},
\be  
h_{\mu}(\tau,z)=-\frac{f_{2\mu}(\tau,z)}{\vartheta_1(\tau,z)^2},
\ee
which is the generating function of (intersection) Poincaré polynomials of moduli spaces of semi-stable sheaves. With $w=e^{2\pi i z}$, these are expressed in terms of the Appell-Lerch sum \eqref{Mtuv} as \cite{Bringmann:2010sd}:
\be       
\label{g01}
\begin{split}  
f_{0}(\tau,z)&=\frac{1}{2}-q^{-\frac{1}{4}}w^3\,M(2\tau, 4z-\tau,\tfrac{1}{2}-2z)  \\
& =\frac{1}{2}+\frac{q^{-\frac{3}{4}}w^5}{\vartheta_2(2\tau,2z)}\sum_{n\in
\mathbb{Z}} \frac{q^{n^2+n}w^{-2n}}{1-w^4q^{2n-1}},\\
f_{1}(\tau,z)&= -M(2\tau, 4z-\tau,\tfrac{1}{2}-\tau-2z)\\
&= \frac{q^{-\frac{1}{4}}w^3}{\vartheta_3(2\tau,2z)}\sum_{n\in
\mathbb{Z}} \frac{q^{n^2}w^{-2n}}{1-w^4q^{2n-1}}.
\end{split}
\ee 
They form a two-dimensional representation of $SL(2,\mathbb{Z})$. 
The functions $f_j(\tau,z)$ are very closely related to the functions $G_\mu$ (\ref{GFD}). 
\be 
\label{gmuFmu}
G_\mu(\tau,z;3/2)=f_{2\mu}(\tau,-z/2).
\ee

The physical significance for VW theory of the structure of the poles as function of $z$ of the
refined partition functions (\ref{g01}) is not well understood. Interestingly, the
$\CN=2^*$ $u$-plane integral gives some partial insight into this: The only allowed poles
are those given by (\ref{vexcluded}). The poles of $G_\mu(\tau,z;3/2)$ in $z$ are thus not
``physical'' singularities upon replacing $(\tau_{\rm uv},z)$ by
$(\tau,v/2)$.

\subsubsection*{The Hirzebruch surfaces}
The holomorphic part of the $U(2)$ VW partition functions can be evaluated for
all rational and ruled surfaces \cite{Yoshioka1994, Gottsche:1999}. Using the theory of mock modular
forms, one can determine the non-holomorphic terms expected for
a partition function compatible with S-duality. Let $\Sigma_n$ be a Hirzebruch surface, with self-intersection of the base curve equal to $n$, and canonical class $K_n$.
For a generic period point $J$, we express the partition function of $U(2)$ VW theory with 't Hooft flux $\bfmu$ as
\be
\widehat h_{2,\bfmu}(\tau,\bar \tau;{\Sigma_n},J)=\frac{\widehat g_{2,\bfmu}(\tau,\bar \tau;\Sigma_n,J)}{\eta(\tau)^8}.
\ee
Then, $\widehat g_{2,\bfmu}(\tau,\bar \tau; \Sigma_n,J)$ is given by \cite[Section
3.2]{Manschot:2011dj}\footnote{The expression differs by a sign from \cite{Manschot:2011dj}, since \cite{Manschot:2011dj} defined generating functions in terms of ``signed" Euler characteristics.}
\be
\label{fHirzSurface}
\begin{split}
&\widehat g_{2,\bfmu}(\tau,\bar \tau;\Sigma_n,J)=g_{2,\bfmu}(\tau;
\Sigma_n,J)\\
& \qquad +\sum_{\bfk\in L+\bfmu} \left(\frac{B(K_n,J)\,|B(\bfk,J)|}{4\pi}
\beta_{\frac{3}{2}}(4 y \bfk_+^2)\right.\\
&\qquad\left.-\frac{1}{2}\,
B(K_n,\bfk_-)\,\sgn(B(\bfk,J))\,\beta_{\frac{1}{2}}(4
y\bfk_+^2)\right) q^{-\bfk^2},
\end{split}
\ee
where
\be 
\beta_\nu(x)=\int_x^\infty u^{-\nu} e^{-\pi u}\,du.
\ee 
If the period point $J$ is chosen proportionally to $-K_n$, then $B(K_n,\bfk_-)$ on the last line vanishes and the holomorphic anomaly is similar in form to (\ref{whfj}). The non-holomorphic dependence in \eqref{fHirzSurface} matches with that found from the $\CN=2^*$ $u$-plane integral \eqref{HolAnol}.

\section{Seiberg-Witten contributions for manifolds with $b_2^+>1$}
\label{SWcontributions}

Besides the contribution from the $u$-plane $\Phi^J_\bfmu$, the partition function
receives in general localized contributions from the cusps $u_j$,
$j=1,2,3$. These contributions are known as the Seiberg-Witten
contributions, and are a sum of (infra-red) ${\rm Spin}^c$ structures $\mathfrak{s}_{\rm ir}$,
\be
\label{ZSWsir}
Z_{SW,j,\bfmu}^J=\sum_{\mathfrak{s}_{\rm ir}} Z^J_{SW,j,\bfmu}(\mathfrak{s}_{\rm ir}),
\ee 
where $Z^J_{SW,j,\bfmu}(\mathfrak{s}_{\rm ir})$ is the contribution from the cusp
$u_j$ for a fixed 't Hooft flux $\bfmu$. The full partition function
thus reads
\be
\label{SW123}
Z^J_\bfmu=\Phi_\bfmu^J+\sum_{j=1}^3 Z_{SW,j,\bfmu}^J.
\ee
While $Z^J_\bfmu$ depends on the metric through $J$ for manifolds with $b_2^+(X)=1$, this dependence is absent for manifolds with $b_2^+>1$. In the latter case, $\Phi_\bfmu^J=0$, and $J$ can be omitted from the notation. 

For $b_2^+=1$, the SW contributions $Z_{SW,j,\bfmu}^J$ are not independent from the $u$-plane
contribution $\Phi^J_\bfmu$. Since the metric dependence of $Z_\bfmu^J$ can only come
from the non-compact region $u\to \infty$, wall-crossing of $\Phi_\bfmu^J$ at the
cusps $u_j$, $j=1,2,3$ is cancelled by the wall-crossing of the
$Z_{SW,j,\bfmu}^J$. Therefore, the change of $\Phi_\bfmu^J$ at cusp $j$ across a wall equals
\be
\label{DPhiSW}
\left[\Phi^{J^+}_{\bfmu}-\Phi^{J^-}_{\bfmu}\right]_j=Z_{SW,j,\bfmu}^{J^-}-Z_{SW,j,\bfmu}^{J^+}.
\ee
The $u$-plane provides in this way a route to
determine the $Z_{SW,j,\bfmu}^J$ for manifolds with $b_2^+=1$. Then using the topological nature of the partition function, this can be generalized to four-manifolds with $b_2^+>1$ \cite{Moore:1997pc}. In the remainder of this Section, we will determine the $Z_{SW,j,\bfmu}^J$ by following this approach.

\subsection{Seiberg-Witten invariants and wall-crossing} 
The Seiberg-Witten contribution from the cusp $u_j$ is expressed as a residue in the local coordinate $a_j$. While $a$ changes sign under the monodromy around $u=\infty$, the
local coordinate $a_j$ near the cusp $u_j$ is single-valued under
the monodromy. A key idea is that in the  MQ formalism $a_j$ should be identified with the first
Chern class of the Seiberg-Witten moduli space (for each component). The justification
is that it generates the $Q$-cohomology and has the right R-charge, which translates
into ghost number to be a degree two class. The SW moduli  space is a union of projective spaces, and the contribution $Z^J_{{SW},j,\bfmu}$ from each cusp takes the form  
\be
\label{ZjSW}
Z^J_{{SW},j,\bfmu}= \sum_{c_{\rm ir}} {\rm SW}(c_{\rm ir};J)\,\underset{a_j=0}{\mathrm{Res}}\!\left[
a_j^{-1-n(c_{\rm ir})} e^{-S_{{SW},j}}\right], 
\ee
where ${\rm SW}(c;J)$ is the SW-invariant, and
$n(c_{\rm ir})\in \mathbb{Z}$ is half the real dimension of the SW moduli space (\ref{dimSWmodspace}). The term $e^{-S_{{SW},j}}$ is the exponentiated action near the singularity, which is the product of the couplings $\CA_j,\dots,\CF_{\bfmu,j}$ and a normalization factor $\kappa_{j}$,
\be
\label{eSWj}
e^{-S_{{SW},j}}=\kappa_{j}\,\CA_j^\chi\,\CB_j^\sigma\,\CC_j^{c_{\rm uv}^2}\,\cD_j^{B(c_{\rm
  uv}, c_{\rm
  ir})}\,\CE_j^{c_{\rm ir}^2}\,\CF_{\bfmu,j},   
\ee
where $c_{\rm uv}=c_1(\mathfrak{s}_{\rm uv})$ and $c_{\rm
  ir}=c_1(\mathfrak{s}_{\rm ir})$. The factor $\kappa_j=\kappa_j(\tau_\uv,m,\Lambda)$ is independent of $a_j$, but can in principle depend on $\tau_\uv$, $m$ and $\Lambda$. On the other hand, the coupling $\CA_j=\CA_j(a_j,\tau_{\rm uv},m,\Lambda)$
can depend on $a_j$, and similarly for the couplings $\CB_j,\dots,\CF_{\bfmu,j}$.
The coupling $\CF_{\bfmu,j}$ is a shorthand for the three couplings
depending on $\bfmu$,
\be
\label{CFj}
\CF_{\bfmu,j}=f_{1,j}^{\bfmu^2}\,f_{2,j}^{B(c_{\rm uv},\bfmu)}\,f_{3,j}^{B(c_{\rm ir},\bfmu)}.
\ee
Since the lhs of (\ref{DSW}) changes only by a phase under shifts of
$c_{\rm uv/ir}$ and $\bfmu$, the $f_{i,j}$ are roots of unity,
independent of $a_j$, $\tau_{\rm uv}$, $m$ and $\Lambda$. In total, we thus have eight couplings and a normalization to determine.

The wall-crossing of SW invariants cancels
the wall-crossing of the $u-$plane integral, such that the
wall-crossing of the full partition function is only due to
$u=\infty$. The wall-crossing formula for Seiberg-Witten invariants
reads \cite{Li:1995, Morgan, park_2004}
\be
\label{SW-WC}
{\rm SW}(c;J^+)-{\rm SW}(c;J^-)=\Delta{\rm SW}(c)=-(-1)^{n(c)}.
\ee
The rhs of (\ref{DPhiSW}) must therefore be of the form
\be  
\label{DSW0} 
\begin{split}
\left[\Phi^{J^+}_{\bfmu}-\Phi^{J^-}_{\bfmu}\right]_j&=\kappa_{j} \sum_{c_{\rm ir}} (-1)^{n(c_{\rm ir})}\tfrac{1}{2} (\sgn(B(c_{\rm ir},J^+))-\sgn(B(c_{\rm ir},J^-)))\\
&\times \underset{a_j=0}{\mathrm{Res}}\!\left[
a_j^{-1-n(c_{\rm ir})} \, \CA_j^\chi\,\CB_j^\sigma\,\CC_j^{c_{\rm uv}^2}\,\cD_j^{B(c_{\rm
  uv}, c_{\rm 
  ir})}\,\CE_j^{c_{\rm ir}^2}\,\CF_{\bfmu,j}\right], 
\end{split} 
\ee
where the precise sum over $c_{\rm ir}$ depends on $j$, and follows
from the indefinite theta series (\ref{WCu1}), (\ref{PhiWCu2}) and
(\ref{PhiWCu3}). Substituting
$\chi=4-\sigma$ in (\ref{DSW}) gives
\be  
\label{DSW} 
\begin{split}
\left[\Phi^{J^+}_{\bfmu}-\Phi^{J^-}_{\bfmu}\right]_j&=-\kappa_{j}
\sum_{c_{\rm ir}} (-1)^{(c_{\rm ir}^2-\sigma)/8}\,\tfrac{1}{2} (\sgn(B(c_{\rm ir},J^+))-\sgn(B(c_{\rm ir},J^-)))\\
&\times \underset{a_j=0}{\mathrm{Res}}\!\left[
a_j^{-(c_{\rm ir}^2-\sigma)/8} \, \CA_j^4\,(\CB_j/\CA_j)^\sigma\,\CC_j^{c_{\rm uv}^2}\,\cD_j^{B(c_{\rm
  uv}, c_{\rm 
  ir})}\,\CE_j^{c_{\rm ir}^2}\,\CF_{\bfmu,j}\right], 
\end{split} 
\ee
To determine the couplings from the $u$-plane integral, we have to
carefully compare (\ref{ZjSW}) with wall-crossings determined in Section \ref{metricwallcross}.

A useful relation for the SW-invariants is 
\be
\label{SW-c}
{\rm SW}(-c;J)=(-1)^{(\chi+\sigma)/4}\, {\rm SW}(c;J). 
\ee  
For the sums over $c_{\rm ir}$ as in Equation (\ref{DSW}), this implies the following identity
\be
\label{cto-c}
\begin{split}
&\sum_{\bfx\in L} {\rm SW}(c_{\rm ir}) \,\cD^{B(\bfx,\bfy)}\,\CE^{\bfx^2}\\
&\qquad =(-1)^{(\chi+\sigma)/4}\,\cD^{-B(c_{\rm uv},\bfy)}\CE^{2c_{\rm uv}^2}\sum_{\bfx\in L} {\rm SW}(c_{\rm ir}) \,\cD^{-B(\bfx,\bfy)}\,\CE^{2c_{\rm ir}^2-\bfx^2}.
\end{split}
\ee
where $\bfx=c_{\rm ir}-c_{\rm uv}$ and $\bfy\in H^2(X)$.

 A four-manifold is said to be of {\it Seiberg-Witten simple type} if ${\rm SW}(c_{\rm ir};J)$ is only non-vanishing for $c_{\rm ir}$, which correspond to an almost complex structure or equivalently $n(c_{\rm ir})=0$. While manifolds with $b_2^+=1$ exihibit wall-crossing and are never of simple type, all known four-manifolds with $b_2^+>1$ are of this type. An important example of a four-manifold of SW simple type is the K3 surface, whose only non-vanishing SW-invariant is ${\rm SW}(0)=1$. For later reference, we list the VW partition functions for K3 \cite{Vafa:1994tf}
\be
\label{K3VW}
\begin{split}
Z_{0}(\tau_{\rm uv})&=\frac{1}{4}\, G(2\tau_{\rm uv})+\frac{1}{2}\, G(\tau_{\rm uv}/2)+\frac{1}{2}\, G((\tau_{\rm uv}+1)/2),\\
Z_{\bfmu}(\tau_{\rm uv})&=\frac{1}{2}\, G(\tau_{\rm uv}/2)+\frac{1}{2}\, G((\tau_{\rm uv}+1)/2),\qquad \bfmu^2\in \mathbb{Z},\,\,\bfmu\neq 0,\\
Z_{\bfmu}(\tau_{\rm uv})&=\frac{1}{2}\, G(\tau_{\rm uv}/2)-\frac{1}{2}\, G((\tau_{\rm uv}+1)/2),\qquad \bfmu^2\in \mathbb{Z}+1/2,
\end{split}
\ee
where $G(\tau)=\eta(\tau)^{-24}$. These partition functions are of the form \eqref{ZSWsir} and (\ref{SW123})  with $\Phi_\bfmu^J=0$. Moreover, the term $\frac{1}{4}\,
G(2\tau_{\rm uv})$ for $\bfmu=0$ can be viewed as the $m=2$ contribution to the rational invariant (\ref{rational_inv}).

\subsection{Determination of Seiberg-Witten contributions}
In this subsection, we express the wall-crossing of $\Phi_\bfmu^J$ at
$u_j$ in a form similar to (\ref{DSW}), which makes it straightforward
to read off the couplings $\CA_j,\dots,\CF_{\bfmu,j}$. We determine the $\kappa_j$ by comparison with the result for the K3 surface (\ref{K3VW}).\\
\\
{\it Contribution from $u_1$}\\
For $j=1$, we will find that the couplings read,
\be
\label{CACE}
\begin{split}  
%&\kappa_{\bfmu,1}=1,\\
&\CA_1=  2^{1/4} \kappa_1^{-1/4}\,\Lambda^{-3/4} e^{\pi i/4}
\frac{1}{\eta(\tau_{\rm uv})}\left(\frac{du}{da} \right)^\frac{1}{2},\\
&\CB_1=  2\,\kappa_1^{-1/4}\,e^{3\pi i/8}\Lambda^{-9/8} m^{1/8} \,
\eta(\tau_{\rm uv})^\frac{3}{2} \left(\frac{\Delta_{\rm phys}}{a_1}\right)^\frac{1}{8},\\
&\CC_1= \frac{C^{1/4}}{\eta(\tau_{\rm uv})^{1/2}} \,e^{-\pi i v_1/2}\,q_1^{-1/16},\\
&\cD_1=e^{-\pi i v_1/2}\,q_1^{-1/8},\\
&\CE_1=e^{-\pi i/8} \,a_1^{1/8} q_1^{-1/16},\\
&\CF_{\bfmu,1}=1.
\end{split}
\ee

For fixed $c_{\rm ir}=4\bfmu-2\bfk_m\mod 4L$ with $\sgn(B(c_{\rm ir},J^+))=-\sgn(B(c_{\rm ir},J^-))>0$, Equation (\ref{WCu1}) shows that the wall-crossing at $u_1$ is given by
\be
\label{DPhi+-}
\begin{split} 
\left[\Phi_\bfmu^{J^+}-\Phi_\bfmu^{J^-}\right]_{1,c_{\rm
    ir}}&=\frac{1}{64\pi i }
\int_{-\frac{1}{2}+iY}^{\frac{3}{2}+iY}
d\tau_1\,\frac{da}{d\tau_1}\,\left((2^{3/2}\Lambda^{-3/4} \eta(\tau_{\rm uv})^{-1}
\left( du/da \right)^{1/2} \right)^\chi \\
&\times  \left( 2^{9/4}\,m^{1/8} \Lambda^{-9/8}
  \eta(\tau_{\rm uv})^{3/2}\,\Delta_{\rm phys}^{\frac{1}{8}}\right)^\sigma \\
& \times \left(
C^{1/4} e^{-\pi i(v_1/2+\tau_1/8)}\,\eta(\tau_{\rm uv})^{-1/2}\right)^{c_{\rm
    uv}^2}\,\\
&\times \left(e^{-\pi i v_1/2}\,q_1^{-1/8}\right)^{B(c_{\rm ir},c_{\rm uv})}\,
q_1^{-c_{\rm ir}^2/16},
\end{split}
\ee
where we substituted (\ref{dadtau}) for $da/d\tau$. Next, we change
the integration variable to the local coordinate $a_1=2a-m$. Since
$da/d\tau$ (\ref{dadtau}) is a power series in $q^{1/2}$ for 
large ${\rm Im}(\tau)$, we learn that $a_1$ is well-defined under the monodromy $M_1$. This is also
expected physically since a single hypermultiplet becomes
massless. Comparing with (\ref{DSW}), we can read off the couplings
$\CC_1$, $\cD_1$, $\CE_1$ and $\CF_{\bfmu,1}$ in agreement with (\ref{CACE}), such that  (\ref{DPhi+-})
becomes
\be
\label{DPhi+-2}
\begin{split} 
\left[\Phi_\bfmu^{J^+}-\Phi_\bfmu^{J^-}\right]_{1,c_{\rm
    ir}}&=\frac{1}{32} \,\underset{a_1=0}{\mathrm{Res}}\!\left[
  \left((2^{3/2}\Lambda^{-3/4} \eta(\tau_{\rm uv})^{-1}
\left( du/da \right)^{1/2} \right)^\chi \right.\\
&\times  \left( 2^{9/4}\,m^{1/8} \Lambda^{-9/8}
  \eta(\tau_{\rm uv})^{3/2}\,\Delta_{\rm phys}^{\frac{1}{8}}\right)^\sigma \\
&\times \left. \CC_1^{c_{\rm
    uv}^2}\, \cD_1^{B(c_{\rm ir},c_{\rm uv})}\, \CE_1^{c_{\rm ir}^2}\right].
\end{split}
\ee
Substituting $\chi=4-\sigma$, we determine $\CA_1$ and $\CB_1$ in agreement with \eqref{CACE}.

Using (\ref{dadtau}), we find for the behavior of the local coordinate $a_1$ as function of $\tau_1$
near the singularity,
\be 
a_1= 16\, m    
\frac{\eta(\tau_{\rm uv})^6}{\vartheta_2(\tau_{\rm uv})^6}\,q_1^{\frac{1}{2}} + O(q_1^1).
\ee  
The leading constants in the $q_1$-expansion of the couplings are,
\be   
\label{u1CoupExp}
\begin{split} 
\CA_1 &= \,2^{5/4}\,\kappa_1^{-1/4}\, \Lambda^{-3/4}
m^{\frac{1}{2}}\,e^{2\pi i/8}\,\frac{\eta(\tau_{\rm uv})^2}{\vartheta_2(\tau_{\rm uv})}+\dots,\\
&= 2^{1/4} \Lambda^{-3/4}
m^{\frac{1}{2}}\,\,e^{2\pi i/8}\,\frac{\eta(\tau_{\rm uv})^3}{\eta(2\tau_{\rm uv})^2} +\dots,\\ 
\CB _1&= 2^{5/4}\kappa_1^{-1/4}\,m^{\frac{3}{4}}\, e^{3\pi i/8} 
\Lambda^{-9/8}
\left(\frac{\eta(\tau_{\rm uv})^5}{\vartheta_2(\tau_{\rm uv})}\right)^{3/4}+\dots,\\
&= 2^{1/2}\,m^{\frac{3}{4}}\, e^{3\pi i/8}
\Lambda^{-9/8}
\left(\frac{\eta(\tau_{\rm uv})^3}{\eta(2\tau_{\rm uv})}\right)^{3/2}+\dots,\\
\CC_1 & =\left(
  \frac{\Lambda}{m}\right)^\frac{3}{8}\frac{1}{\eta(\tau_{\rm uv})^{1/2}\,\vartheta_2(\tau_{\rm uv})^{1/2}}+\dots,\\
&=2^{-1/2}\left( \frac{\Lambda}{m}\right)^\frac{3}{8}\frac{1}{\eta(2\tau_{\rm uv})}+\dots,\\
\cD_1&=\left(\frac{\vartheta_3(2\tau_{\rm uv})}{\vartheta_2(2\tau_{\rm uv})}\right)^{1/2}+\dots,\\
\CE_1&= 2^{1/2}\,m^{1/8} e^{-\pi i/8} \left(\frac{\eta(\tau_{\rm uv})}{\vartheta_2(\tau_{\rm uv})}\right)^{3/4}+\dots\\
&=2^{-1/4}\,m^{1/8} e^{-\pi i/8} \left(\frac{\eta(\tau_{\rm uv})}{\eta(2\tau_{\rm uv})}\right)^{3/2}+\dots. 
\end{split}
\ee 
where we used $\vartheta_2(\tau_{\rm uv})\eta(\tau_{\rm uv})=2\,\eta(2\tau_{\rm uv})^2$ to simplify expressions.

For simplicity, we assume in the following that $X$ has $b_2^+\geq 3$, and is of Seiberg-Witten simple
type. Then $J$ can be omitted from the notation, $c_{\rm ir}^2=2\chi+3\sigma$, and the leading terms in the expansions \eqref{u1CoupExp} suffice. Combining all the terms, we find that the contribution from $u_1$ is
\be
\label{ZSW1mu1} 
\begin{split}
Z_{SW,1,\bfmu}(\tau_{\rm uv}) &=\kappa_1^{1-(\chi+\sigma)/4} \, 2^{-(\chi+\sigma)/4-c_{\rm uv}^2/2}\,\left(\frac{\Lambda}{m} \right)^{\frac{3}{8}(c_{\rm uv}^2-2\chi-3\sigma)}\eta(\tau_{\rm uv})^{6\chi+9\sigma}\eta(2\tau_{\rm uv})^{-5\chi-6\sigma-c_{\rm uv}^2} \\
&\quad \times \sum_{c_{\rm ir}= 4\bfmu-c_{\rm uv} \mod 4L} {\rm SW}(c_{\rm ir})\left( \frac{\vartheta_3(2\tau_{\rm uv})}{\vartheta_2(2\tau_{\rm uv})}\right)^{B(c_{\rm ir}, c_{\rm uv})/2}.
\end{split}
\ee 

We can rewrite this in a more convenient form to compare to
\cite{Vafa:1994tf, Dijkgraaf:1997ce}. To this end, we replace $c_{\rm
  ir}$ by $-c_{\rm ir}$ in \eqref{ZSW1mu1} and substitute (\ref{SW-c}) for
${\rm SW}(-c_{\rm ir};J)$. Since now $c_1\in -4\bfmu+c_\uv \mod
4L=4\bfmu+c_\uv \mod 4L$, we introduce $\bfx \in 2\bfmu + 2L$
defined by
\be
c_{\rm ir}=2\bfx+c_{\rm uv}=2\bfx+2\bfk_m.
\ee 
Using moreover the identity
(\ref{tsproduct}) and
$\vartheta_4(2\tau_{\rm uv})=\eta(\tau_{\rm uv})^2/\eta(2\tau_{\rm
  uv})$, we write $Z_{SW,1,\bfmu}$ as the sum,
\be
\label{ZSWSumu1}
Z_{SW,1,\bfmu}(\tau_{\rm uv})=\sum_{ \bfx = 2\bfmu \mod 2L} Z_{SW,1,\bfmu}(\tau_{\rm uv};\bfx),
\ee
with summand,
\be  
\label{u1DPS}
\begin{split}
Z_{SW,1,\bfmu}(\tau_{\rm uv};\bfx) &=\kappa_1\,\, {\rm SW}(c_{\rm
  ir})
\left( -2\,\kappa_1\,\eta(2\tau_{\rm uv})^{12} \right)^{-\chi_{\rm h}} \,\left(
  \frac{(\Lambda/m)^3}{4\,\eta(\tau_{\rm uv})^4\,\vartheta_3(2\tau_{\rm uv})}
\right)^{\ell} 
\\  
&\quad \times \left( \frac{\eta(\tau_{\rm
      uv})^2}{\vartheta_3(2\tau_{\rm uv}) }\right)^{\lambda} \left( \frac{\vartheta_3(2\tau_{\rm
      uv})}{\vartheta_2(2\tau_{\rm uv})}\right)^{\bfx^2}, 
\end{split}
\ee 
where we used $\ell$ (\ref{defell}) to substitute for $c_\uv$, and also $\lambda$ \eqref{lambda} and $\chi_{\rm h}$ \eqref{holchi} to shorten notation.
 
Let us specialize to $X=K3$ to fix $\kappa_1$. The only non-vanishing SW-invariant is for $c_{\rm ir}=0$. With
$(\chi,\sigma)=(24,-16)$, we find for the contribution at $u_1$,
\be 
\label{K3cuv}
Z^{K3}_{SW,1,\bfmu}=\left\{ \begin{array}{rl}\kappa_1^{-1}\,\left( 2^{4}t^3 \right)^{-\ell} \frac{1}{4\,\eta(2\tau_{\rm uv})^{24+8\ell}}, & \qquad c_\uv=4\bfmu \mod 4L, \\ 0, & \qquad {\rm otherwise.} \end{array} \right.
\ee 
We can compare this result for $c_\uv=0$ with the Vafa-Witten result for $K3$ (\ref{K3VW}). It is clear that we should compare (\ref{K3cuv}) with
the first term in (\ref{K3VW}), which demonstrates that
$\kappa_1=1$.  Our final formula for the $u_1$ contribution is then,
\be
\label{SW1Final}
\boxed{
\begin{split}
Z_{SW,1,\bfmu}(\tau_{\rm uv})&=\left( -2\,\eta(2\tau_{\rm uv})^{12} \right)^{-\chi_{\rm h}} \,\left(
  4\,t^3\,\eta(\tau_{\rm uv})^4\,\vartheta_3(2\tau_{\rm uv})^4
\right)^{-\ell} \,\left( \frac{\eta(\tau_{\rm
      uv})^2}{\vartheta_3(2\tau_{\rm uv}) }\right)^{\lambda} \\   
&\quad \times \sum_{ \bfx = 2\bfmu \mod 2L} {\rm SW}(c_{\rm
  ir})\left( \frac{\vartheta_3(2\tau_{\rm
      uv})}{\vartheta_2(2\tau_{\rm uv})}\right)^{\bfx^2}.
\end{split}}
\ee
\\
\\ 
{\it Contribution from $u_2$}\\
For the contribution from $u_2$, we proceed similarly by determining
the couplings $\CA_2,\dots,\CF_{\bfmu,2}$ from (\ref{PhiWCu2}). The local
coordinate near $u_2$ is $a_2=-2a_D+m$. The wall-crossing contribution from $u_2$ reads
\be
\begin{split}
\left[\Phi_\bfmu^{J}-\Phi_\bfmu^{J'}\right]_{2} &=\kappa_2\,(-1)^{n(c_{\rm ir})}\, \sum_{c_{\rm ir}= c_{\rm uv}\mod 2L} \tfrac{1}{2} (\sgn(B(c_{\rm ir},J))-\sgn(B(c_{\rm ir},J')))\\
&\quad \times \, \underset{a_2=0}{\mathrm{Res}}\left[ a_2^{-1-n(c_{\rm ir})} \CA_2^\chi\,\CB_2^\sigma\,\CC_2^{c_{\rm uv}^2}\, \cD_2^{B(c_{\rm ir},c_{\rm
      uv})}\,\CE_2^{c_{\rm ir}^2}\CF_{\bfmu,2}\right],
\end{split}
\ee
 with 
\be
\begin{split}
\label{u2Couplings}
\CA_2&=\kappa_2^{-1/4}\Lambda^{-3/4} e^{2\pi i/8} \frac{1}{\eta(\tau_{\rm uv})}\,\left(
  \frac{du}{da}\right)_2^{1/2},\\
\CB_2&= 2^{5/4}\,\kappa_2^{-1/4}\,e^{5\pi i/8} \Lambda^{-9/8}\,m^{1/8}\,\eta(\tau_{\rm uv})^{3/2} \left(\frac{\Delta_{\rm phys,2}}{a_2}\right)^{1/8},\\
\CC_2&= \frac{C_2^{1/4}}{\eta(\tau_{\rm uv})^{1/2}}\,e^{\pi i v_2/2}q_2^{-1/16},\\
\cD_2&=e^{-i\pi v_2/2} q_2^{1/8}, \\
\CE_2&=e^{-\pi i /8}\, a_2^{1/8}\,q_2^{-1/16},\\
\CF_{\bfmu,2}&=e^{\pi i B(c_{\rm ir}-c_{\rm uv},\bfmu)}.
\end{split}
\ee
The behavior of the local coordinate $a_2$ near $u_2$ follows from (\ref{dadtau}),
\be
a_2=-16\,m\,\frac{\eta(\tau_{\rm uv})^6}{\vartheta_4(\tau_{\rm uv})^6}\,q_2^{1/2}+\dots.
\ee
The leading terms are  
\be
\begin{split}
\CA_2&=2\,\kappa_2^{-1/4}\Lambda^{-3/4}m^{1/2} e^{2\pi i/8} \frac{\eta(\tau_{\rm uv})^3}{\eta(\tau_{\rm uv}/2)^2}+\dots,\\
\CB_2&= 2^{3/2}\kappa_2^{-1/4} \Lambda^{-9/8} e^{\pi i/8}m^{3/4}\left( \frac{\eta(\tau_{\rm uv})^3}{\eta(\tau_{\rm uv}/2)}\right)^{3/2}+\dots,\\
\CC_2&=\left(\frac{\Lambda}{m}\right)^{3/8}\, \frac{e^{\pi i/8}}{\eta(\tau_{\rm uv}/2)} +\dots,\\
\cD_2&=\left(\frac{\vartheta_4(\tau_{\rm uv}/2)}{\vartheta_3(\tau_{\rm uv}/2)}\right)^{1/2}+O(q_2^{1/2}),\\
\CE_2&= 2^{1/2}\,m^{1/8}\,\left(\frac{\eta(\tau_{\rm uv})}{\eta(\tau_{\rm uv}/2)}\right)^{3/2}+\dots,
\end{split}
\ee
using that $\eta(\tau)\,\vartheta_4(\tau)=\eta(\tau/2)^2$. 

Combining all the contributions and assuming that $X$ has $b_2^+\geq
3$ and is of SW simple type, we arrive for $Z_{SW,2,\bfmu}$ at
\be 
\label{u2ZSW}
\begin{split}
Z_{SW,2,\bfmu}(\tau_{\rm uv})&=\kappa_2^{1-\chi/4-\sigma/4}\,2^{2\chi+3\sigma}\,\left( \frac{\Lambda}{m}\right)^{\frac{3}{8}(c_{\rm uv}^2-2\chi-3\sigma)}\,(-1)^{(c_{\rm uv}^2-2\chi-3\sigma)/8}\\
&\quad \times \eta(\tau_{\rm uv})^{6\chi+9\sigma}\eta(\tau_{\rm uv}/2)^{-5\chi-6\sigma-c_{\rm uv}^2}\\
&\quad \times \sum_{c_{\rm ir}= \bar w_2(X) \mod 2L} {\rm SW}(c_{\rm ir})\, (-1)^{B(c_{\rm ir}-c_{\rm uv},\bfmu)} \left( \frac{\vartheta_4(\tau_{\rm uv}/2)}{\vartheta_3(\tau_{\rm uv}/2)}\right)^{B(c_{\rm ir},c_{\rm uv})/2}.
\end{split}
\ee  
With $c_{\rm ir}=2\bfx+c_{\rm uv}$, we can express $Z_{SW,2,\bfmu}$ as a sum over $\bfx\in L$, 
\be 
\label{ZSWSumu2}
Z_{SW,2,\bfmu}(\tau_\uv)=\sum_{\bfx\in L} Z_{SW,2,\bfmu}(\tau_\uv;\bfx), 
\ee 
with summand,
\be
\label{u2DPS}
\begin{split}
Z_{SW,2,\bfmu}(\tau_{\rm uv};\bfx)&=\kappa_2^{1-\chi/4-\sigma/4}\,(-1)^{(c_{\rm uv}^2-2\chi-3\sigma)/8}\left( \frac{\Lambda}{2^{4/3}\,\eta(\tau_{\rm uv})^4\,m} \right)^{\frac{3}{8}(c_{\rm uv}^2-2\chi-3\sigma)}\\
& \quad \times \eta(\tau_{\rm uv}/2)^{-3(\chi+\sigma)}\left(\frac{\vartheta_3(\tau_{\rm uv}/2)}{2\eta(\tau_{\rm uv})^2}\right)^{-c_{\rm uv}^2/2-\chi-3\sigma/2}\\
&\quad \times {\rm SW}(c_{\rm ir})\, (-1)^{2B(\bfx,\bfmu)} \left( \frac{\vartheta_3(\tau_{\rm uv}/2)}{\vartheta_4(\tau_{\rm uv}/2)}\right)^{\bfx^2}.
\end{split}
\ee 
We determine again $\kappa_2$ by comparing to the Vafa-Witten result for $K3$
(\ref{K3VW}). Specializing (\ref{u2ZSW}) to $K3$, we find
\be 
Z_{SW,2,\bfmu}^{K3}(\tau_\uv)=\kappa_2^{-1} t^{-3\ell} (-1)^{\ell+B(c_\uv,\bfmu)}\,\eta(\tau_\uv/2)^{-24-8\ell}.
\ee 
Comparing with the second term on the first line of (\ref{K3VW}) for $c_\uv=0$, gives $\kappa_2=2$. Using \eqref{defell}, we find for our final result for $Z_{SW,2,\bfmu}$,
\be
\label{u2DPSfinal}
\boxed{
\begin{split}
Z_{SW,2,\bfmu}(\tau_{\rm uv})&=2\,\left(2\, \eta(\tau_{\rm uv}/2)^{12}\right)^{-\chi_{\rm
  h}}\left(-t^3\,\eta(\tau_{\rm uv})^{4}\,\vartheta_3(\tau_{\rm uv}/2)^4 \right)^{-\ell}\left(\frac{2\eta(\tau_{\rm uv})^2}{\vartheta_3(\tau_{\rm uv}/2)}\right)^{\lambda}\\
&\quad \times \sum_{\bfx\in L} {\rm SW}(c_{\rm ir})\, (-1)^{2B(\bfx,\bfmu)} \left( \frac{\vartheta_3(\tau_{\rm uv}/2)}{\vartheta_4(\tau_{\rm uv}/2)}\right)^{\bfx^2}.
\end{split}}
\ee 
\\
\\ 
{\it Contribution from $u_3$}\\
For the contribution from $u_2$, we proceed similarly by determining
the couplings $\CA_3,\dots,\CF_{\bfmu,3}$ from (\ref{PhiWCu3}). The local
coordinate near the singularity is $a_3=-2a_D+2a+m$. The wall-crossing due
to $u_3$ reads
\be
\begin{split}
\left[\Phi_\bfmu^{J}-\Phi_\bfmu^{J'}\right]_{3} &=\kappa_3\,(-1)^{n(c_{\rm ir})}\, \sum_{c_{\rm ir}= c_{\rm uv}\mod 2L} \tfrac{1}{2} (\sgn(B(c_{\rm ir},J))-\sgn(B(c_{\rm ir},J')))\\
&\quad \times \, \underset{a_3=0}{\mathrm{Res}}\left[ a_3^{-1-n(c_{\rm ir})} \CA_3^\chi\,\CB_3^\sigma\,\CC_3^{c_{\rm uv}^2}\, \cD_3^{B(c_{\rm ir},c_{\rm
      uv})}\,\CE_3^{c_{\rm ir}^2}\,\CF_{\bfmu,3}\right],
\end{split}
\ee
with the couplings $\CA_3,\dots$ the same as in (\ref{u2Couplings}),
but with $2$ replaced by $3$ on the rhs. For the behavior of the local
coordinate we have near $u_3$,
\be
a_3=16\,i\,m\,\frac{\eta(\tau_{\rm uv})^6}{\vartheta_3(\tau_{\rm uv})^6}\,q_3^{1/2}+\dots.
\ee
Using $e^{\pi i/12}
\eta(\tau)\,\vartheta_3(\tau)=\eta((\tau+1)/2)^2$, this gives for the leading terms,
\be
\begin{split} 
\CA_3&=2\,\kappa_2^{-1/4}\,\Lambda^{-3/4}m^{1/2} e^{7\pi i/12} 
\frac{\eta(\tau_{\rm uv})^3}{\eta((\tau_{\rm uv}+1)/2)^2}+\dots,\\
\CB_3&=2^{3/2}\,\kappa_2^{-1/4}\,\Lambda^{-9/8}e^{3\pi i/4}m^{3/4} \left( 
  \frac{\eta(\tau_{\rm uv})^3}{\eta((\tau_{\rm uv}+1)/2)} \right)^{3/2}+\dots,\\
\CC_3&=\left(\frac{\Lambda}{m}\right)^{3/8}\, \frac{e^{\pi
    i/6}}{\eta((\tau_{\rm uv}+1)/2)} +\dots,\\
\cD_3&=\left(
  \frac{\vartheta_3((\tau_{\rm uv}+1)/2)}{\vartheta_4((\tau_{\rm uv}+1)/2)}
\right)^{1/2}+\dots, \\
\CE_3&=2^{1/2}\,m^{1/8}\left( \frac{\eta(\tau_{\rm uv})}{\eta((\tau_{\rm uv}+1)/2)}\right)^{3/2}+\dots,\\
\CF_{\bfmu,3}&= e^{-2\pi i\bfmu^2+\pi i B(c_{\rm ir}-c_{\rm uv},\bfmu)}.
\end{split}
\ee
 Combining all the contributions and assuming that $X$ has $b_2^+\geq
 3$ and is of SW simple type, we arrive at
\be
\label{ZSW3}
\begin{split}
Z_{SW,3,\bfmu}(\tau_{\rm uv})&=\kappa_3^{1-\chi/4-\sigma/4}\, 2^{2\chi+3\sigma}\,\left( \frac{\Lambda}{m}\right)^{\frac{3}{8}(c_{\rm uv}^2-2\chi-3\sigma)}\,e^{-2\pi i\bfmu^2}\,(-1)^{(c_{\rm uv}^2-\sigma)/8}\\
&\quad \times e^{\frac{2\pi i}{24}(c_{\rm uv}^2/2+\chi+3\sigma/2)}\,\eta(\tau_{\rm uv})^{6\chi+9\sigma}\eta((\tau_{\rm uv}+1)/2)^{-5\chi-6\sigma-c_{\rm uv}^2}\\
&\quad \times \sum_{c_{\rm ir}= \bar w_2(X) \mod 2L} {\rm SW}(c_{\rm
  ir})\, (-1)^{B(c_{\rm ir}-c_{\rm uv},\bfmu)} \left( \frac{\vartheta_3((\tau_{\rm uv}+1)/2)}{\vartheta_4((\tau_{\rm uv}+1)/2)}\right)^{B(c_{\rm ir},c_{\rm uv})/2}.
\end{split}
\ee
 
We again bring this to a more convenient form using (\ref{SW-c}), and
various manipulations. We arrive at the sum,
\be 
\label{ZSWSumu3}
Z_{SW,3,\bfmu}(\tau_{\rm uv})=\sum_{\bfx\in L} Z_{SW,3,\bfmu}(\tau_{\rm uv};\bfx),
\ee 
with summand, 
\be 
\label{ZSWkappa}
\begin{split}
Z_{SW,3,\bfmu}(\tau_{\rm uv};\bfx)&= \kappa_3^{1-\chi/4-\sigma/4}\, e^{2\pi i\bfmu^2} (-1)^{(c_{\rm uv}^2-2\chi-3\sigma)/8}
\left(\frac{\Lambda}{2^{4/3}\eta(\tau_{\rm uv})^4m}
\right)^{\frac{3}{8}(c_{\rm uv}^2-2\chi-3\sigma)} \\
&\quad \times \eta((\tau_{\rm uv}+1)/2)^{-3(\chi+\sigma)}\left(
  \frac{\vartheta_3((\tau_{\rm uv}+1)/2)}{2\eta(\tau_{\rm uv})^2}\right)^{-c_{\rm uv}^2/2-\chi-3\sigma/2}\\
&\quad  \times  {\rm SW}(c_{\rm ir})\, (-1)^{2B(\bfx,\bfmu)} \left( \frac{\vartheta_3((\tau_{\rm uv}+1)/2)}{\vartheta_4((\tau_{\rm uv}+1)/2)} \right)^{\bfx^2}.
\end{split}
\ee 
Finally, we fix $\kappa_3$ by comparing with $K3$. Specialization of \eqref{ZSW3} to $K3$ gives
\be 
Z_{SW,3,\bfmu}^{K3}(\tau_\uv)=\kappa_3^{-1} t^{-3\ell}(-1)^{\ell+B(c_\uv,\bfmu)}\,e^{-2\pi i\bfmu^2+2\pi i \ell/6}\,\eta((\tau_\uv+1)/2)^{-\chi-8\ell}.
\ee 
Clearly, $Z_{SW,3,\bfmu}$ corresponds to the third term on the first line of (\ref{K3VW}). This gives $\kappa_3=2$. With (\ref{defell}), we find
that $Z_{SW,3,\bfmu}(\tau_{\rm uv})$ reads
\be 
\boxed{
\begin{split}
\label{u3DPSfinal}
Z_{SW,3,\bfmu}(\tau_{\rm uv})&= 2\,e^{2\pi i\bfmu^2} 
\left(-t^3\,\eta(\tau_{\rm uv})^4\,\vartheta_3((\tau_{\rm uv}+1)/2)^4
\right)^{-\ell} \\
&\quad \times \left(2\,\eta((\tau_{\rm
    uv}+1)/2)^{12}\right)^{-\chi_{\rm h}}\,\left(
  \frac{2\,\eta(\tau_{\rm
      uv})^2}{\vartheta_3((\tau_{\rm uv}+1)/2)}\right)^{\lambda}\\
&\quad  \times \sum_{\bfx\in L}  {\rm SW}(c_{\rm ir})\,
(-1)^{2B(\bfx,\bfmu)} \left( \frac{\vartheta_3((\tau_{\rm
      uv}+1)/2)}{\vartheta_4((\tau_{\rm uv}+1)/2)} \right)^{\bfx^2}. 
\end{split}}
\ee  
\\ 
\\
{\it Remark}\\
As a special case of our formula's, we can set $-c_{\rm uv}$ equal to
the canonical divisor of $X$, such that $c_{\rm uv}^2=2\chi+3\sigma$ and
$\ell=0$. We find then precise agreement of (\ref{u1DPS}),
(\ref{u2DPSfinal}) and (\ref{u3DPSfinal}) with the respective terms in \cite[Equation (5.38), (5.45) and
(5.50)]{Vafa:1994tf}, the three terms of \cite[Equation
(6.1)]{Dijkgraaf:1997ce}, and the unrefined ($y=1$) specialization of
\cite[Conjecture 1.3]{Gottsche:2019vbi} ($u_1$ contribution) and
\cite[Conjecture 5.7]{Gottsche:2017vxs} ($u_2$ and $u_3$ contributions) for
algebraic surfaces of simple type. We make a precise comparison in
Section \ref{CompGKNW}.

As another special case, we can set $c_{\rm
  uv}=0$ if $X$ is a spin manifold. This corresponds to the
Donaldson-Witten twist studied by Labastida and Lozano \cite{Labastida:1998sk}. After multiplying \cite[Equation (4.34) and (4.36)]{Labastida:1998sk}  with
$\eta(\tau_\uv)^{\lambda}$ (See Footnote \ref{FootMu} for more details on this factor.) and a
$\Lambda$-dependent factor, we find agreement. Specifically, we have the relation
\be
Z_{SW,\bfmu}(\tau_\uv)\vert_{c_\uv=0}= (-1)^{\chi_{\rm h}} \Lambda^{-3\lambda/8}\,\eta(\tau_\uv)^{\lambda} \left( Z_\xi(\tau_0) \right)_{\rm LL},
\ee
where $Z_\xi(\tau_0)$ is given in \cite[Equation
(4.36)]{Labastida:1998sk}, and we identify $\tau_0=\tau_\uv$ and $\xi=2\bfmu$.

In addition to the seven couplings of \cite{Vafa:1994tf,
  Dijkgraaf:1997ce, Labastida:1998sk, Gottsche:2017vxs, Gottsche:2019vbi},
our expressions include an eighth coupling $\CN_j^{\ell}$ with
\be
\begin{split} 
&\CN_1= \left(4\,t^3\,\eta(\tau_\uv)^4\,\vartheta_3(2\tau_\uv)^4\right)^{-1},\\
&\CN_2= \left(-t^3\,\eta(\tau_\uv)^4\,\vartheta_3(\tau_\uv/2)^4\right)^{-1},\\
&\CN_3= \left(-t^3\,\eta(\tau_\uv)^4\,\vartheta_3((\tau_\uv+1)/2)^4\right)^{-1}.
\end{split}
\ee

\subsection{Observables}
The coupling to the point observable enters the Coulomb branch integral in the form
$\exp( p\,\hat u/\Lambda^2)$   where $\hat u$ can take several different forms depending on how,
exactly, we define the partition function. There are three basic options for $\hat u$,
each of which has advantages and disadvantages.

\begin{enumerate} 
\item We can take $\hat u$ to be the function $u(\tau, \tau_\uv;m)$ which appears in the Seiberg-Witten curve
and is the bi-modular form given in Equation \eqref{u2star}. This is the definition which is most convenient
for giving simple S-duality covariance of the partition function.
\item We could instead take $\hat u$ to be $u_D$ defined in equation \eqref{DonaldsonPointObs}. This is the best choice for
defining a partition function with a smooth $N_f=0$ limit (after subtracting the contribution of the cusp
at $u=u_1$). We obtain the partition function from the previous choice by multiplying by
$\exp( \frac{m^2}{8} e_1(\tau_\uv) p )$. Physically, this corresponds to a redefinition of the coordinate on the $u$-plane.
\item A similar redefinition of the partition function results in the integral defined by taking $\hat u$ to be
$\tilde u := 2 \p_{\tau_{\uv}} \tilde \CF$. This is the best definition to matching to the results in the mathematics
literature and therefore corresponds to the proper generating function of intersection numbers.
\end{enumerate}
In each of these evaluations the variable $p$ enters the contribution of the effective theory via a factor of
$e^{ \frac{\hat u_j}{\Lambda^2} p}$. where $\hat u_j$ is the value of $\hat u$ at $u=u_j$. In Table \ref{Tab:pObsCoupl}
we give the values of $\frac{\hat u_j}{\Lambda^2}$ at the three cusps for the three natural choices of $\hat u$.

\vskip1in  
\begin{table}
\renewcommand{\arraystretch}{1.8}
\begin{center}
\begin{tabular}{|c|c|c|c| }
\hline
  $j$ & $u$ & $u_{\rm D}$ & $\tilde u$   \\
\hline
$1$ & $\frac{t^2}{12} \left(\vartheta_3^4 + \vartheta_4^4\right)$  &
$\frac{t^2}{8} \left(\vartheta_3^4 + \vartheta_4^4\right) $  &
$\frac{t^2}{12} \left(\vartheta_3^4 + \vartheta_4^4 +2E_2 \right)$ \\ \hline
$2$ & $-\frac{t^2}{12} \left(\vartheta_2^4 + \vartheta_3^4\right)$ &
$-\frac{t^2}{8} \left(\vartheta_3^4 - \vartheta_4^4\right) $   & $
-\frac{t^2}{12} \left(\vartheta_2^4 + \vartheta_3^4-2E_2\right)$ \\ \hline
$3$ & $\frac{t^2}{12} \left(\vartheta_2^4 - \vartheta_4^4\right)$ & $\frac{t^2}{8} \left(\vartheta_3^4 - \vartheta_4^4\right) $
& $\frac{t^2}{12} \left(\vartheta_2^4 - \vartheta_4^4 +2E_2\right)$ \\ \hline
\end{tabular} 
\end{center}
\caption{Table with the leading point observables $\hat u\in \{ u,
  u_{\rm D}, \tilde u\}$ at the
  cusps $u_j$, $j=1,2,3$.}\label{Tab:pObsCoupl} 
\end{table}
\vskip1in

Upon including the surface observable
(\ref{tildeIMx}) for a two-cycle $S \in H_2(X,\mathbb{C})$, new couplings appear in the effective action \eqref{eSWj},
namely for the four combinations $S^2$, $B(S,c_{\rm ir})$, $B(S,c_{\rm
  uv})$ and $B(S,\bfmu)$, for $j=1,2,3$,
\be 
\CS_j(\tau_\uv,\tau_j)^{B(S,c_{\rm ir})}, \qquad \CT_j(\tau_\uv,\tau_j)^{S^2},
\qquad 
\CU_j(\tau_\uv,\tau_j)^{B(S,c_{\rm uv})}, \qquad \CV_j(\tau_\uv,\tau_j)^{B(S,\bfmu)}.
\ee
We first consider the couplings $\CV_j$. Since $S$ is a
formal expansion parameter, and the partition function should be
independent of the choice of lift of $\bfmu$ to $L/2$, we have 
\be
\CV_j(\tau_\uv,\tau_j)=1, \qquad j=1,2,3.
\ee
The limit of the contact terms $e^{T_2}$ and $e^{T_3/2}$ contribute to $\CT_j$
and $\CU_j$. There is a further contribution $\CU_j$. This arises similarly to
$\cD_j^{B(c_{\rm uv},c_{\rm ir})}$, and follows most easily from the
substitution (\ref{zSandv})
\be
v\,c_{\rm uv} \mapsto v\,c_{\rm uv}+\frac{S}{2\pi\,\Lambda}\frac{du}{da},
\ee 
in $\CC_j^{c_{\rm uv}^2}$ and $\cD_j^{B(c_{\rm uv},c_{ir})}$. 
Assuming that $X$ is of SW simple type, subleading terms in the $q_j$
expansion of $\CS_j$, $\CT_j$ and $\CU_j$ do not contribute, and it suffices to determine the
leading constant term. We define $\CS_j(\tau_\uv)=
\lim_{\tau_j\to\infty} \CS_j(\tau_\uv,\tau_j)$ and similarly for
$\CT_j$ and $\CU_j$. One finds for these couplings, 
\be 
\label{STUj}
\begin{split}
\CS_j(\tau_\uv)&= e^{-\frac{i}{4\Lambda}\left(du/da\right)_j},\\
\CT_j(\tau_\uv)&=e^{T_{2,j}},\\
\CU_j(\tau_\uv)&=\left\{ \begin{array}{ll} e^{T_{3,1}/2
      -\frac{i}{4\Lambda}\left(du/da\right)_1}, & \qquad u=u_1, \\ e^{T_{3,j}/2
      +\frac{i}{4\Lambda}\left(du/da\right)_j}, & \qquad u=u_j\,\,{\rm
    for}\,\,j=2,3. \end{array} \right.
\end{split}
\ee
The $T_{2,j}$ and $T_{3,j}$ are the limits of the contact terms
$T_2$ (\ref{T2u}) and $T_3$ (\ref{T3u}) at the cusps. These are given
for $T_2$ and $T_3$ in \eqref{T2T3u1}, (\ref{T2T3u2}), (\ref{T2T3u3}). Note that the shift by
$(du/da)_j$ in the exponent of $\CU_j$ precisely cancels the last
term of $T_{3,j}$ in (\ref{T2T3u1}), (\ref{T2T3u2}) and (\ref{T2T3u3}). Table \ref{Tab:SObsCoupl} lists the different couplings in terms of modular forms.

Having discussed the couplings including observables, the partition functions with observables are included by the following replacements,
\begin{itemize}
\item $Z_{SW,1,\bfmu}(\tau_\uv;\bfx)$ in Eq. (\ref{ZSWSumu1}) is replaced by,  
\be
\begin{split}
Z_{SW,1,\bfmu}(\tau_\uv;\bfx,p,S)&=e^{p\tilde u_1/\Lambda^2}
\CS_1(\tau_\uv)^{-B(S,c_{\rm
    uv}+2\bfx)}\,\CT_1(\tau_\uv)^{S^2}\,\CU_1(\tau_\uv)^{B(S,c_{\rm
    uv})}\\
&\quad \times \,Z_{SW,1,\bfmu}(\tau_\uv;\bfx).
\end{split}
\ee
\item $Z_{SW,2,\bfmu}(\tau_\uv;\bfx)$ in Eq. (\ref{ZSWSumu2}) is replaced by,  
\be
\begin{split} 
Z_{SW,2,\bfmu}(\tau_\uv;\bfx,p,S)&=e^{p\tilde u_2/\Lambda^2}
\CS_2(\tau_\uv)^{B(S,c_{\rm
    uv}+2\bfx)}\,\CT_2(\tau_\uv)^{S^2}\,\CU_2(\tau_\uv)^{B(S,c_{\rm
    uv})}\\
&\quad \times \,Z_{SW,2,\bfmu}(\tau_\uv;\bfx).
\end{split}
\ee
\item $Z_{SW,3,\bfmu}(\tau_\uv;\bfx)$ in Eq. (\ref{ZSWSumu3}) is replaced by,
\be
\begin{split} 
Z_{SW,3,\bfmu}(\tau_\uv;\bfx,p,S)&=e^{p\tilde u_3/\Lambda^2}
\CS_3(\tau_\uv)^{-B(S,c_{\rm
    uv}+2\bfx)}\,\CT_3(\tau_\uv)^{S^2}\,\CU_3(\tau_\uv)^{B(S,c_{\rm
    uv})}\\
&\quad \times \,Z_{SW,3,\bfmu}(\tau_\uv;\bfx).
\end{split}
\ee
\end{itemize}

Finally, for a proper limit to the $N_f=0$ theory, we introduce
\be 
\label{UDj}
\begin{split}
\CU_{{\rm D},j}&=\left\{ \begin{array}{ll} e^{T_{{\rm D}, 3,1}/2
      -\frac{i}{4\Lambda}\left(du/da\right)_1}, & \qquad u=u_1, \\ e^{T_{{\rm D},3,j}/2
      +\frac{i}{4\Lambda}\left(du/da\right)_j}, & \qquad u=u_j\,\,{\rm
    for}\,\,j=2,3, \end{array} \right.
\end{split}
\ee
with $T_{{\rm D}, 3}$ as in (\ref{TD3}). 

\begin{table} 
\renewcommand{\arraystretch}{1.8}
\begin{center}
\begin{tabular}{|c|c|c|c| }
\hline
  $j$  & $B(S, c_{\rm ir})$: $\log(\CS_j)$ & $S^2$: $\log(\CT_j)$ &
  $B(S, c_{\rm uv})$: $\log(\CU_j)$   \\
\hline 
$1$ &  $-\frac{it}{4}\vartheta_3^2\vartheta_4^2$   & 
$\frac{t^2}{144}\left[ -3 \vartheta_3^4 \vartheta_4^4 + E_2
  (\vartheta_3^4 + \vartheta_4^4)+E_2^2 \right]$  & $\frac{t}{12i}[\vartheta_3^4+\vartheta_4^4+2E_2]$   \\ \hline
$2$ &
$-\frac{it}{4}\vartheta_2^2\vartheta_3^2$   &
$\frac{t^2}{144} \left[ -3 \vartheta_2^4 \vartheta_3^4 - E_2 (\vartheta_2^4 + \vartheta_3^4)+E_2^2\right]$    &  $\frac{t}{12i}[-\vartheta_2^4-\vartheta_3^4+2E_2]$\\ \hline
$3$ &
$-\frac{t}{4}\vartheta_2^2\vartheta_4^2$  &
$\frac{t^2}{144}\left[ 3 \vartheta_2^4 \vartheta_4^4 + E_2 (\vartheta_2^4 - \vartheta_4^4)+E_2^2 \right]$  &   $\frac{t}{12i}[\vartheta_2^4-\vartheta_4^4+2E_2]$\\ \hline
\end{tabular}\end{center}
\caption{Table with the leading couplings to $S$ at the
  cusps $u_j$, $j=1,2,3$.} \label{Tab:SObsCoupl}
\end{table} 
\vskip1in

\subsection{Verification of $S$-duality}
\label{SWSdual}
With the explicit expressions, we can verify the $S$-duality
transformation. This confirms that the generators $S$ and $T$ act as
\be
\label{ZSWSdual}
\begin{split}
S:\qquad Z_{SW,\bfmu}(-1/\tau_{\rm uv})&=2^{-b_2/2} (-1)^{ \ell -(\chi+\sigma)/4} (-i\tau_{\rm uv})^{-\chi/2-4\ell}\\
&\times \sum_{\bfnu\in (L/2)/L} e^{4\pi i B(\bfmu,\bfnu)}
Z^J_{SW,\bfnu}(\tau_{\rm uv}),\\
T:\qquad Z_{SW,\bfmu}(\tau_{\rm uv}+1)&=e^{-2\pi i(\bfmu^2+(\chi+2\ell)/12)} Z_{SW,\bfmu}(\tau_{\rm uv}).
\end{split}
\ee

The generators exchange the contributions $Z_{SW,j,\bfmu}$ from the different cusps.
We have for the $S$-transformations
\be
\begin{split}
Z_{SW,1,\bfmu}(-1/\tau_{\rm uv})&=2^{-b_2/2} (-1)^{ \ell -(\chi+\sigma)/4} (-i\tau_{\rm uv})^{-\chi/2-4\ell} \\
&\quad \times \sum_{\bfnu\in (L/2)/L} e^{4\pi i B(\bfmu,\bfnu)}
Z^J_{SW,2,\bfnu}(\tau_{\rm uv}), \\
Z_{SW,2,\bfmu}(-1/\tau_{\rm uv})&=2^{-b_2/2} (-1)^{\ell -(\chi+\sigma)/4} (-i\tau_{\rm uv})^{-\chi/2-4\ell} \\
&\quad \times \sum_{\bfnu\in (L/2)/L} e^{4\pi i B(\bfmu,\bfnu)} Z^J_{SW,1,\bfnu}(\tau_{\rm uv}),\\
Z_{SW,3,\bfmu}(-1/\tau_{\rm uv})&=2^{-b_2/2} (-1)^{\ell -(\chi+\sigma)/4} (-i\tau_{\rm uv})^{-\chi/2-4\ell} \\
&\quad \sum_{\bfnu\in (L/2)/L} e^{4\pi i B(\bfmu,\bfnu)} Z^J_{SW,3,\bfnu}(\tau_{\rm uv}).\\
\end{split}
\ee 

The modular transformation of $Z_{SW,3,\bfmu}$ is most easily derived
starting from (\ref{ZSW3}). Using the modular transformations of
$\eta$, $\vartheta_j$ and the relation (\ref{SW-c}) for the
SW-invariant, we find for the
left hand side
\be
\label{ZSW3lhs}
Z_{SW,3,\bfmu}(-1/\tau_{\rm uv})=(-i\tau_{\rm uv})^{-\chi/2-4\ell}
(-1)^{\frac{\chi+\sigma}{4}} e^{-\pi i B(c_{\rm ir},c_{\rm uv})/4+4\pi i \bfmu^2}\,Z_{SW,3,\bfmu}(\tau_{\rm uv}).
\ee
For the sum on the right hand side, we find using the Gauss sum \cite[Section 3.3]{Vafa:1994tf}
\be
\sum_{\bfnu\in (L/2)/L} e^{2\pi i \bfnu^2}=2^{b_2/2}\,e^{\pi i \sigma/4},
\ee
that it equals 
\be
\begin{split}
&\sum_{\bfnu\in (L/2)/L} e^{4\pi i B(\bfmu,\bfnu)}
Z_{SW,3,\bfnu}(\tau_{\rm uv})=\\
&\qquad\qquad  2^{b_2/2} (-1)^{\ell}e^{-\pi i B(c_{\rm ir},c_{\rm uv})/4+4\pi
  i\bfmu^2}Z_{SW,3,\bfmu}(\tau_{\rm uv}).
\end{split}
\ee
Combining with the other factors on the rhs, this reproduces precisely (\ref{ZSW3lhs}).

We have for the $T$-transformations  
\be
\begin{split}
Z_{SW,1,\bfmu}(\tau_{\rm uv}+1)&=e^{-2\pi i(\bfmu^2+(\chi+2\ell)/12)} Z_{SW,1,\bfmu}(\tau_{\rm uv}),\\
Z_{SW,2,\bfmu}(\tau_{\rm uv}+1)&= e^{-2\pi i(\bfmu^2+(\chi+2\ell)/12)} Z_{SW,3,\bfmu}(\tau_{\rm uv}),\\
Z_{SW,3,\bfmu}(\tau_{\rm uv}+1)&=e^{-2\pi i(\bfmu^2+(\chi+2\ell)/12)} Z_{SW,2,\bfmu}(\tau_{\rm uv}).
\end{split}
\ee

We deduce from these transformations that the sum $Z_{SW,\bfmu}=\sum_{j=1}^3Z_{SW,j,\bfmu}$ transforms
identically to $u$-plane contribution $\Phi_\bfmu^J$
(\ref{PhiSdual}). Moreover, the sum transforms
as (\ref{SdualObs}) if we include observables. We thus conclude that
the full partition function $Z_{SW,\bfmu}$ transforms in this way.

\subsection{Reduction to $N_f=0$}
\label{SWRedNf0}
We discuss the reduction of the three contributions to the $N_f=0$
theory. We start with the $u_2$ and $u_3$ contribution, which
reproduce the Witten conjecture \cite{Witten:1994cg}. We then discuss the $u_1$
contribution which is generically divergent.\\
\\
{\it Limit of $u_2$ contribution}\\
The  leading term in $q_{\rm uv}$ of the $u_2$ contribution (\ref{u2DPSfinal}) is 
\be
\label{LeadZSW2}
\begin{split}
Z_{SW,2,\bfmu}(\tau_{\rm uv})&=2^{1+7\chi/4+11\sigma/4}\,(-1)^{(c_{\rm uv}^2-2\chi-3\sigma)/8} \left(
  \frac{\Lambda}{m}\right)^{\frac{3}{8}(c_{\rm uv}^2-2\chi-3\sigma)} \\
&\qquad \times
q_{\rm uv}^{-c_{\rm uv}^2/48+7\chi/48+\sigma/4} \sum_{\bfx\in L} {\rm
  SW}(c_{\rm ir})\,(-1)^{2B(\bfx,\bfmu)}+\dots. 
\end{split}
\ee
Multiplying by the factor (\ref{Prefactor1}) gives
\be
\begin{split}
&2^{1+7\chi/4+11\sigma/4}\,(-1)^{(\chi+\sigma)/4} e^{\frac{\pi
    i}{8}(\chi+\sigma)}\,(-1)^{B(\bfmu,c_{\rm uv}+K_0)}\left(
  \frac{m}{\Lambda}\right)^{\frac{3}{4}(\chi+\sigma)} \\
&\qquad \times
q_{\rm uv}^{\frac{3}{16}(\chi+\sigma)} \sum_{\bfx\in L} {\rm
  SW}(c_{\rm ir})\,(-1)^{2B(\bfx,\bfmu)}. 
\end{split}
\ee
Taking the limit to $N_f=0$ gives 
\be
\label{WittenC2}
\begin{split}
Z_{2,\bfmu}&=2^{1+\lambda-\chi_{\rm h}}\, \sum_{\bfx_0\in L} {\rm
  SW}(c_{\rm ir})\,(-1)^{2B(\bfx_0,\bfmu)},
\end{split}
\ee
with $2\bfx_0=c_{\rm ir}-K_{0}$. This agrees
with the first term of the Witten conjecture \cite[Equation
(2.17)]{Witten:1994cg} without observables and if we identify $-K_0$ with the lift of
$w_2(X)$ in \cite[Equation (2.17)]{Witten:1994cg}.

Finally to include observables, we multiply the summand by 
\be
\begin{split}
&e^{p\,u_{\rm D,2}/\Lambda^2}\,\CS_2^{B(c_{\rm ir},S)}\,
\CT_{2}^{S^2}\,\,\CU_{\rm D,2}^{B(S,c_\uv)} \\
&= e^{-2p\,t^2\,q_{\uv}^{1/2}-it\,q_\uv^{1/4} B(c_{\rm ir},S)-
  t^2\,q_\uv^{1/2}\,S^2/2+2i\,t\,q_\uv^{1/2}B(c_{\rm uv},S)+O(q_\uv^{3/4})}. 
\end{split}
\ee
Therefore we find after taking the limit,
\be
\label{WittenC2pS} 
\begin{split}
Z_{2,\bfmu}(p,S)=2^{1+\lambda-\chi_{\rm h}}\, e^{2p+S^2/2}\,\sum_{\bfx_0\in L} {\rm
  SW}(c_{\rm ir})\,(-1)^{2B(\bfx_0,\bfmu)}e^{B(c_{\rm ir}, S)},
\end{split}
\ee
confirming the first term of the Witten conjecture \cite[Equation
(2.17)]{Witten:1994cg}.
\\
\\
{\it Limit of $u_3$ contribution}\\  
The  leading term in $q_{\rm uv}$ of the $u_3$ contribution (\ref{u3DPSfinal}) 
agrees with the leading term of $u_2$ (\ref{LeadZSW2}) up to a phase
\be
\label{LeadZSW3}
Z_{SW,3,\bfmu}(\tau_{\rm uv})= e^{2\pi i \bfmu^2-\pi i (\chi+\sigma)/8}\,Z_{SW,2,\bfmu}(\tau_{\rm uv})+\dots.
\ee
Multiplying by the factor (\ref{Prefactor1}), taking the $N_f=0$ limit and using (\ref{SW-c}) gives
\be 
\label{WittenC3} 
\begin{split}
Z_{3,\bfmu}&=2^{1+\lambda-\chi_{\rm h}}\, e^{\pi i \chi_{\rm h}/2-2\pi i \bfmu^2} \sum_{\bfx_0\in L} {\rm
  SW}(c_{\rm ir})\,(-1)^{2B(\bfx_0,\bfmu)},
\end{split}
\ee 
wit $2\bfx_0=c_{\rm ir}-K_{0}$. This agrees exactly with the second term of the Witten conjecture
\cite[Equation (2.17)]{Witten:1994cg} in the absence of observables
and if we identify $-K_0$ with the lift of $w_2(X)$.

To include observables, we multiply the summand by 
\be
\begin{split}
&e^{p\,u_{\rm D,3}/\Lambda^2}\,\CS_3^{B(c_{\rm ir},S)}\,
\CT_3^{S^2}\,\,\CU_{\rm D,3}^{B(S,c_\uv)} \\
&=  e^{2p\,t^2\,q_{\uv}^{1/2}-t\,q_\uv^{1/4} B(c_{\rm ir},S)+
  t^2\,q_\uv^{1/2}\,S^2/2-2i\,t\,q_\uv^{1/2}B(c_{\rm uv},S)+O(q_\uv^{3/4})}. 
\end{split}
\ee
The full contribution from $u_3$ now becomes
\be
\label{WittenC3pS}  
\begin{split}
Z_{3,\bfmu}(p,S)&=2^{1+\lambda-\chi_{\rm h}}\, e^{\pi
  i\chi_{\rm h}/2-2\pi i \bfmu^2} \, e^{-2p-S^2/2}\\
&\quad \times \sum_{\bfx_0\in L} {\rm
  SW}(c_{\rm ir})\,(-1)^{2B(\bfx_0,\bfmu)}e^{-iB(c_{\rm ir}, S)},
\end{split}
\ee
confirming the second term of the Witten conjecture \cite[Equation
(2.17)]{Witten:1994cg}.
\\
\\
{\it Limit of $u_1$ contribution}\\
The  leading term in $q_{\rm uv}$ of the $u_1$ contribution (\ref{u1DPS}) is 
\be
\begin{split}
Z_{SW,1,\bfmu}(\tau_{\rm uv})&=(-1)^{\frac{\chi+\sigma}{4}} 2^{-c_{\rm
  uv}^2+\chi/4+\sigma/2}\,\left(\frac{\Lambda}{m}
\right)^{\frac{3}{8}(c_{\rm uv}^2-2\chi-3\sigma)}\,q_{\rm uv}^{-c_{\rm
  uv}^2/48-\chi/24+3\sigma/48} \\
&\quad \times \sum_{c_{\rm ir}=4\bfmu-c_{\rm uv} \atop\mod 4L} {\rm
  SW}(c_{\rm ir})\,q_{\rm uv}^{-\bfx^2/4}+\dots,
\end{split}
\ee 
with $c_{\rm ir}=2\bfx-c_{\rm uv}$ as before. After multiplying with (\ref{Prefactor1}), the leading term of the
$u_1$ contribution is 
\be
e^{-\pi i (\chi+\sigma)/8}(-1)^{B(\bfmu,K_0-c_\uv)+\ell}
2^{-2\ell-(\chi+\sigma)/4}\left( \frac{m}{\Lambda}\right)^{\frac{3}{4}(\chi+\sigma)}\sum_{\bfx=2\bfmu\mod 2L} {\rm SW}(c_{\rm ir})\,q_{\rm uv}^{-\bfx^2/4}
\ee
The $N_f=0$ limit of the terms in the sum depend crucially on
$\bfx^2$. The terms either vanish or diverge, except for those with
$\bfx^2=0$. Therefore, there is no smooth $N_f=0$ limit of the partition function of the $\CN=2^*$
theory even after allowing for the multiplicative renormalization 4.53   (and a similar one to adjust to $\hat u = u_D$.
Rather, one must subtract by hand the contribution of the cusp $u=u_1$.

\subsection{Comparison with G\"ottsche-Kool-Nakajima-Williams}
\label{CompGKNW}
Similar formulas for observables to those in this section have been determined in a
parallel development in algebraic geometry \cite{Gottsche:2010ig}, \cite[Conjecture
C.1]{Gottsche:2017vxs}, \cite{Gottsche:2019vbi, Gottsche:2020ale},
which build on the approach of Mochizuki \cite{Mochizuki}. These
papers develop the generating functions for algebraic surfaces with
$b_2^+>1$ for small rank, and including a refinement parameter $y$.
Our formulas should be compared with the unrefined expressions for
rank $2$ in those papers, that is to say the limit $y\to 1$. They
deduce twelve couplings $A_j$, $j=1,\dots,12$ for each of the cusps,
which agree with our couplings $\CA_j,\dots,\CF_j,\CS_j,\CT_j, \CU_j, \CV_j$
and $e^{p\,u/\Lambda^2}$. Note $\CF_j$ contains three couplings
(\ref{CFj}), and we have an extra coupling $\CC_j$ for generic $c_\uv^2$.

By comparison with the mathematical results we learn an interesting
fact which is not a priori evident from our derivation: The UV evaluation
of the partition function is naturally a sum of two terms corresponding
to the contribution of the instanton locus and the abelian locus of the
U(1)-fixed points. These are naturally identified with the sum of the
contributions of the IR cusps $u_2, u_3$ for the instanton locus and the
cusp $u_1$ for the abelian locus, respectively. Of course, there is no S-duality invariant
distinction between the three cusps. Here we are making use  of the
distinction that emerges when $t$ is large and $q_{\uv}$ is small but
$t^4 q_{\uv}$ is order one.

\begin{center}
\begin{table}
\renewcommand{\arraystretch}{1.8}
\begin{tabular}{|p{3cm}|p{5cm}|p{5cm}|}
\hline
Concept & This paper & GKNW \\ 
\hline
Geometry & Smooth, compact four-manifold $X$ with $b_1=0$ of SW simple
type & Projective
complex surface $S$
with $b_2^+>1$, $b_1=0$ of SW simple
type \\
\hline
Mass/Scale & $m/\Lambda=t$ & $t$\\
\hline
Modular parameter & $q_\uv$ & $x^4$ \\
\hline
Refinement & $1$ (unrefined) & $y$ \\
\hline
UV ${\rm Spin}^c$ structure & $c_{\rm uv} \in \bar w_2(X)+H_2(X,2\mathbb{Z})$ & Canonical class $K_S$ \\
\hline
IR ${\rm Spin}^c$ structure & $c_{\rm ir} \in \bar w_2(X)+H_2(X,2\mathbb{Z})$
& SW basic class $K_S-2a_i$ \\
\hline
Integral lift of 't Hooft flux & 2$\bfmu\in H^2(X,\mathbb{Z})$ & first Chern class $c_1$\\
\hline
0-observable & $p$ & $-u$ \\ 
\hline
2-observable & $S$ &  $i\alpha z$ \\
\hline
\end{tabular}
\caption{Dictionary between some of the concepts in this paper and in \cite{Gottsche:2017vxs,
  Gottsche:2019vbi, Gottsche:2020ale}}\label{TableDict}
\end{table}
\end{center}

To facilitate comparison, we give a detailed dictionary in this Subsection. 
Following \cite{Gottsche:2017vxs}, we set $\bar G_2(x)=(1-E_2(\sigma))/24$ with
$x=e^{2\pi i \sigma}$, and $\bar \eta(\tau)=q^{-1/24}\,\eta(\tau)$. \\
\\
{\it Point observable}\\
The coupling for the point observable in \cite[Appendix C]{Gottsche:2017vxs} reads $G_p(x)=2\bar
G_2(x^2)=2x^2+6x^4+\dots$. On the other hand, we have the expansion
$\tilde u_2=u_2+m^2E_2(\tau_\uv)/6=2\partial_{\tau_{\rm uv}}\widetilde
\CF\vert_{u=u_2}=(1/12-2q_\uv^{1/2}-6q_\uv-8q_\uv^{3/2}+\dots)\,t^2$. With $x=q_\uv^{1/4}$, we then have the identification,
\be
\left( e^{G_p(x)\,u\,t^2}\right)_{\rm
  GKNW}=e^{-p\,t^2/12}\,e^{p\,\tilde u_2/\Lambda^2}.
\ee
\\  
\\
{\it Surface observable}\\
The coupling for the surface observable in \cite[Appendix C]{Gottsche:2017vxs} reads
$G_S(x)=(\bar G_2(x)-\bar G_2(-x))/2$. This matches with $(du/da)_2$ (\ref{dadu2}). We have the identification,
\be
\left( e^{G_S(x)\left< K_S-2a_i,\alpha\right>zt}  \right)_{\rm
  GKNW}=\CS_2(\tau_\uv)^{B(S,c_{\rm ir})},
\ee
with $\CS_2$ given in \eqref{STUj}.
\\ 
\\ 
{\it Contact term $T_2$}\\
The analog of the contact term $T_2$ in \cite[Appendix C]{Gottsche:2017vxs} is $G_Q(x)=1/2 (DG_2)(x^2)=x^2/2+3x^4+6x^6+\dots$, where $D=x\,d/dx$. We have the identification,
\be
\left( e^{G_Q(x)Q(\alpha)z^2t^2}  \right)_{\rm
  GKNW}=\CT_2(\tau_\uv)^{S^2},
\ee
with $\CT_2$ as in \eqref{STUj}.
\\
\\
{\it Contact term $T_3$}\\
The analog of $\log(\CU_2)$ is $G_p$ in
\cite[Appendix C]{Gottsche:2017vxs}. This is in agreement up to the
constant term. We have the
identification
\be
\left( e^{-G_p(x) \left<K_S,\alpha\right>zt}  \right)_{\rm GKNW}=\left(e^{it/12}\,\CU_2(\tau_\uv)\right)^{B(S,c_\uv)}.
\ee
\\
{\it Generating functions}\\
With these identifications, $(\widetilde Z_{c_1}^S)_{GKNW}$ of
\cite[Appendix C]{Gottsche:2017vxs} is upto a prefactor equal to the contribution from $u_2$
in our notation,
\be
(\widetilde Z_{c_1}^S(\alpha z+pu,t,x))_{GKNW}=2\, q_\uv^{\chi_{\rm h}/4-\lambda/12} e^{(-p\,t^2+itB(S,c_\uv))/12} Z_{SW,2,\bfmu}(\tau_\uv;p,S),
\ee  
with $\ell=0$. Moreover, the generating function of \cite[Conjecture
C.1]{Gottsche:2017vxs} for the instanton component equals the sum of the contributions from $u_2$ and $u_3$,
\be
(Z_{c_1}^S(\alpha z+pu,t,x))_{GKNW}=q_\uv^{\chi_{\rm h}/4-\lambda/12}
e^{(-p\,t^2+itB(S,c_\uv))/12} \sum_{j=2,3}Z_{SW,j,\bfmu}(\tau_\uv;p,S)
\ee  
The observables (or $\mu$-classes) can be obtained from a (modified) $y\to 1$
limit of the K-theoretic $\chi_{-y}$-genera \cite[Proposition 4.8]{Gottsche:2019vbi}.

Finally, the $u_1$ contribution corresponds to the Abelian
component. As mentioned in the {\it Remark}, we find agreement with the unrefined
($y=1$) version of \cite[Conjecture 1.3]{Gottsche:2019vbi} without observables. 

\section{A broad-brush sketch of the generalization to class S theories}
\label{SecClassS}

The topological twist of Donaldson-Witten theory \cite{Witten:1988ze}  is based purely
on the structure of the $d=4, \CN=2$ superalgebra and as such can
be implemented in any theory with $d=4, \CN=2$ supersymmetry. Given the
spectacular applications of $SU(2)$ supersymmetric Yang-Mills theory (without
matter hypermultiplets) to the theory of four-manifolds it is natural to ask
if anything else can be learned from
topological twists of other $d=4,\CN=2$ field theories.

We should distinguish between ``old invariants'' and ``new invariants.''
 By ``old invariants'' we mean both the
the cohomology ring $H^*(X,\IZ)$ and the Donaldson/Seiberg-Witten
invariants. The cohomology ring characterizes the homotopy
type of $X$ \cite{Milnor}. Thanks to the Witten conjecture
(see section 7 above) the Donaldson invariants can be
expressed in terms of the Seiberg-Witten invariants and hence
do not carry new information. The Seiberg-Witten invariants
can distinguish between four-manifolds which are homotopy equivalent
but not diffeomorphic. In 1994 they were new and revolutionary invariants.
Nevertheless, it is clear that there are
many aspects of smooth four-dimensional topology which are not
accessible via the Seiberg-Witten invariants. By ``new invariants''
we mean smooth invariants of $X$ which cannot be computed
just knowing the cohomology ring and the Seiberg-Witten invariants.
\footnote{There is an interesting
refinement of the Seiberg-Witten invariants known as
Bauer-Furuta invariants. But we would include these in the
list  of ``old invariants.'' }

There are two kinds of new things one might learn from the study of other
topologically twisted $d=4, \CN=2$ theories. First, one might learn new
facts about the old invariants. This has in fact happened
via the study of $SU(2)$ with one fundamental quark hypermultiplet \cite{Marino:1998uy,Marino:1998tb,Feehan:1997rt},
where the existence of a superconformal theory with compact moduli space
led to sum rules on Seiberg-Witten invariants.
It is entirely possible that there are other new things to be learned
from other $d=4, \CN=2$ theories.
Second, one might discover new four-manifold invariants
in the sense described above.
\footnote{This question was perhaps first raised in \cite{Moore:1997pc} and was the central motivation for \cite{Marino:1998bm,Marino:1998uy,Marino:1998tb,Moore:2017cmm}
It was also addressed in \cite{Dedushenko:2017tdw,Gadde:2013sca,Gukov:2017zao,Gukov:2018iiq}.}

One way to search for the new invariants is to try to write the partition function
(or generator of topological correlators) in terms of $H^*(X;\IZ)$ and the Seiberg-Witten
invariants. As explained at length in the present paper, the correlators of twisted $\CN=2^*$
 are all constructed out of the old invariants.
The results of \cite{kronheimer2005, Marino:1998bm} show that for $SU(N)$ theories  without hypermultiplets
the topological correlators can again be written in terms of the old invariants. This result
suggests that in fact   all Lagrangian theories will
ultimately be expressed in terms of the old invariants.
We stress that this is \emph{suggested} by the results \cite{Marino:1998bm} but
certainly there is no proof: At the moment the idea that the topological correlators
of Lagrangian theories are always expressible in terms of the old invariants is just folklore.
 We also stress again that even if this conjecture proves to be correct the
 topological correlation functions of these
theories might teach us interesting lessons about physics, and might also lead to the
discovery of interesting new facts about the known invariants.

The above considerations give a very natural motivation to try to investigate the topological correlation functions of class S theories. Since class S theories are in general not Lagrangian theories, there is
no obvious ``UV formulation'' of the invariants in terms of intersection theory on moduli spaces of solutions to PDE's
generalizing the instanton equation. That certainly leaves open the possibility for something really new.
This in turn motivates the study of the Coulomb branch integral of these theories. Focusing on the
Coulomb branch integral makes sense because, as we will see, on the one hand it is something well within reach
using present technology, and yet the method of Section \ref{SWcontributions} of this paper
\footnote{also used in \cite{Moore:1997pc,Marino:1998bm,Labastida:1998sk,Moore:2017cmm}
}
asserts that the full topological correlation functions can be derived from the Coulomb branch integral alone.
The Coulomb branch integral, while subtle and complicated, is expected to depend only on the classical four-manifold invariants.  The main question is whether the analog of the condition
\eqref{DPhiSW}  for cancellation of wall-crossing at finite loci within the Coulomb branch can be implemented using only
Seiberg-Witten invariants. If that is the case, then we expect a generalization of the Witten conjecture  
to hold even in these general class S cases. If it is not the case, then there is room for something
really new. Thus the wall-crossing behavior of the Coulomb branch integral is of central importance. If the wall-crossing in strong coupling regimes 
always happens at walls corresponding to ${\rm Spin}^c$ structures then a 
complete invariant can be formed using just the Seiberg-Witten invariants. 

The result of \cite{Moore:2017cmm} for the simplest of the Argyres-Douglas theories (sometimes called the $(A_1, A_2)$ theory)
casts some doubt on the hope that twisted class S topological correlation functions will provide new invariants, but it
remains an open possibility that this Argyres-Douglas theory is an exceptional case.  Therefore, we believe
that the formulation of the Coulomb branch integral for general class $S$ theories is an interesting, and
potentially important, open problem. The results of the present paper, along with \cite{Marino:1998bm,Moore:1997pc},
constitute important benchmarks for general results for this problem.

In what follows we will give a broad-brush sketch of some of the most obvious
issues that arise in trying to formulate the Coulomb branch measure of
class S theories. In contrast to the rest of this paper, where we have taken
pains to be meticulous about every detail of our computations, the discussion
in this section is broad-brush and programmatic, and we will skate over many
subtleties and details - some of which might be quite important in the final
formulation.

\subsection{Setting notation for class S theories}

First let us recall the definition of a class S theory  \cite{Gaiotto:2009we,Gaiotto:2009hg,Klemm:1996bj,Witten:1997sc}.
For more extended discussions see \cite{Moore-FelixKlein,Tachikawa:2013kta}.
Given an ADE Lie algebra $\fg$, a Riemann surface $\CC$, without
boundary or punctures, but equipped with ``puncture data'' $D$,  one can define a four-dimensional
$d=4$, $\CN=2$ supersymmetric QFT denoted $S[\fg,\CC,D]$. One considers the six-dimensional $(2,0)$ theory
\cite{Seiberg:1996vs,Seiberg:1996qx,Witten:1995zh} on $\CC \times X$
 with ``defect data'' $D$. (In some cases, these represent transverse five-branes.)
 The data $D$ consists of a divisor $Supp(D) = \sum_\alpha p_\alpha$ on $\CC$
and choices of orbits in $\fg^\vee\otimes \IC$ at each  point $p_\alpha$. The data of the divisor can be thought of as boundary
conditions for a $\fg_{\IC}$-Hitchin system on $\CC$ with singularities at $Supp(D)$. See
\cite{Balasubramanian:2018pbp,Chacaltana:2012zy,Gaiotto:2009we,Gaiotto:2009hg} and references therein for detailed discussions.
One then performs a partial topological twist known as the class S twist \cite{Gaiotto:2009hg}. Then
certain physical quantities are formally independent of the K\"ahler structure on $\CC$ and one can take the length scales on $\CC$ to zero.
\footnote{In a small set of exceptional cases this limit will not be well-defined \cite{Gaiotto:2011xs}.}
The Coulomb branch of the resulting four-dimensional theory is well-studied. It is the base of a Hitchin moduli space and
can be identified with the affine space
\be
\CB = \oplus_i  H^0(\CC,  K_{\CC}^{\otimes d_i} \otimes \Delta_i )
\ee
where $d_i$ are the exponents of $\fg$ and $\Delta_i$ is a sheaf describing the pole structures of symmetric differentials at the support
$p_\alpha$ of the defects.
The spectral curve $\Sigma \subset T^*\CC$ is identified with the Seiberg-Witten curve and the Seiberg-Witten differential is then just
the pullback of the Liouville form.  The physics of the theory on $X$ can be described
by the decoupling limit of a single M-theory fivebrane wrapped on $\Sigma \times X$  \cite{Witten:1997sc}.

\subsection{Topological twisting of general $d=4,\CN=2$ theories}

One central aspect of the present paper is that in the presence of matter extra
data must be provided even to define the topological twisting of the theory.
The Donaldson-Witten twist itself
in general will not suffice and we will need to introduce extra data generalizing the
ultraviolet ${\rm Spin}^c$ structure $c_{\uv}$
that plays a significant role in the present paper.

To illustrate this general point let us first consider the topological twisting of the general Lagrangian $d=4,\CN=2$ theory.
The basic data of such a theory is a choice of compact semisimple gauge group $G$ (defining the
vectormultiplets) together with a quaternionic representation $\CR$ of $G$ (defining the hypermultiplets).
The representation is equipped with a   hyperk\"ahler metric such that $G$ is a subgroup of the
group of hyperk\"ahler isometries.  A choice of $\CN=1$ sub-algebra
determines a choice of complex structure on $\CR$. In that case the $\CN=1$ D-term is  written in terms of the
real moment map. The holomorphic symplectic form defines
a moment map which, when contracted with the $\CN=1$ chiral multiplet in the $\CN=2$ vectormultiplet gives the $\CN=1$ superpotential.
The ``flavor group,'' a.k.a. the group of global symmetries
$G^f$ is the group of hyperk\"ahler isometries commuting with $G$. It can be gauged
using external vectormultiplets. The vev's of scalars in these vectormultiplets are mass terms.
All together, $\CR$ is a representation of
\be
\rho: G \times G^f \times SU(2)_R
\ee
If we choose an isomorphism $\CR \cong \IH^n$ then $SU(2)_R$ is right action by
unit quaternions while $\rho(G\times G^f)$ acts on the left.

When we couple the theory to background fields we choose

\begin{enumerate}

\item  A principal $G/Z(G)$-bundle with connection, denoted $(P\to X,\nabla)$. Here $Z(G)$ is the
center of $G$.

\item  A principal $SO(3)_R$ ``R-symmetry bundle'' with connection $\nabla_R$.  The Donaldson-Witten twist
means we choose an isomorphism with the structure group of the bundle of self-dual
two forms such that $\nabla_R$ is identified with the Levi-Civita connection.

\item  A principal bundle with connection for global symmetries. We will be more precise about this below.

\end{enumerate}

For simplicity, we henceforth assume $G$ is simply connected.
In the $G$-gauge theory all of these fields are external, nondynamical fields except for the connection $\nabla$ on $P$.

Topological twisting identifies, locally, $SU(2)_R$ with a factor $SU(2)_+$
of the spin lift of the local frame group $Spin(4) \cong SU(2)_+ \times SU(2)_-$. On non-spin manifolds we cannot define a spin $SU(2)_+$ principal bundle globally. However, there does exist a principal bundle for $SO(3)_+ = SU(2)_+/\IZ_2$. As we have said, it can be taken  to be the structure group of the bundle of self-dual 2-forms on $X$.
Choosing a complex structure, $\CR$ is of the form $\CR' \otimes \textbf{2}$
where $\CR'$ is a representation of $G\times G^f $ and $\textbf{2}$ is the fundamental
of $SU(2)_R$.  Accordingly, when we make a topological twist on a non-spin manifold
we might not be able to define the associated bundle with fiber $\CR$.
We can define an element of a twisted $K$-theory of $X$, but not an honest bundle.
By the same token, in the presence of nonzero 't Hooft flux the gauge bundle $P$ is a
$G_{\rm adj} = G/Z(G)$-principal bundle  that will not lift to a principal $G$-bundle.
If $\CR$ transforms nontrivially under $Z(G)$ then the associated bundle will not
exist, but it will define an element of a twisted $K$-theory of $X$.

In both of these cases, the twisting is torsion and its isomorphism class
is characterized by an element of $H^2(X,U(1))$. The basic criterion for a
sensible theory  is that
the total twisting of $\CR$ should vanish, so that the associated
bundle exists as an honest bundle.

To put this in more concrete terms, suppose we choose an atlas $\{ \CU_\alpha \}$ for $X$
and a trivialization of our putative $G,G^f$ and $SU(2)_+$ bundles relative to this atlas.
There are   transition functions $g_{\alpha\beta}^{\rm gauge}$ for the gauge bundle, while there
will be analogous transition functions $g_{\alpha\beta}^{\rm flavor}$ for the global symmetry bundle,
and finally $g_{\alpha\beta}^s$ for the spin bundle. In all three cases the
bundle will exist locally, but not globally: On triple overlaps of patches $\CU_{\alpha\beta\gamma}$
the cocycle condition might fail. But it only fails by a phase:
\be
\begin{split}
g_{\alpha\beta}^{\rm gauge} g_{\beta\gamma}^{\rm gauge} g_{\gamma\alpha}^{\rm gauge} & = \zeta_{\alpha\beta\gamma}^{\rm gauge} 1_{G} \\
g_{\alpha\beta}^{\rm flavor} g_{\beta\gamma}^{\rm flavor} g_{\gamma\alpha}^{\rm flavor} & = \zeta_{\alpha\beta\gamma}^{\rm flavor} 1_{G^f} \\
g_{\alpha\beta}^{s} g_{\beta\gamma}^{s} g_{\gamma\alpha}^{s} & = \zeta_{\alpha\beta\gamma}^{s} 1_{SU(2)} \\
\end{split}
\ee
In all three cases $\zeta_{\alpha\beta\gamma}^{x}$ is a function from $\CU_{\alpha\beta\gamma}$ to $U(1)$.
(Note that $\zeta_{\alpha\beta\gamma}^{s}$ is valued in the subgroup $\{\pm 1\} \subset U(1)$.)
In global terms $\zeta_{\alpha\beta\gamma}^{\rm gauge}$ is a cocycle representing the 't Hooft flux
$w_2(P) \in H^2(X,Z(G))$ of the gauge bundle, while $\zeta_{\alpha\beta\gamma}^{s}$ is a cocycle
representing $w_2(X)$.  If $\CR_a$ is an irreducible representation of $G\times G^f \times SU(2)_R$ of
the form $\rho_a \otimes \rho_a^f \otimes \rho_{fund} $ then
the phases act on $\CR_a$ by the function $\CU_{\alpha\beta \gamma} \to U(1)$ given by the phase multiplying the identity operator on $\CR_a$ in: 
\be
\rho_a(\zeta_{\alpha\beta\gamma}^{\rm gauge}1_{G}) \rho_a^f( \zeta_{\alpha\beta\gamma}^{\rm flavor}1_{G^f}  ) 
\zeta_{\alpha\beta\gamma}^{s}  ~ .
\ee
The criterion we must impose is that for every irreducible component of $\CR$ this cocyle is
trivializable in cochains with values in $U(1)$.

We illustrate these general considerations with two examples:

\begin{enumerate}

\item
In $SU(2)$ theory , with $N_f$ fundamental flavors the twisting of the bundle associated
to $\CR$ from the doublet of $SU(2)_R$ is $w_2(X)$. Therefore, we must choose
$w_2(P) = w_2(X)$ to cancel the twisting \cite{Moore:1997pc}. This means the
cocycle $\zeta_{\alpha\beta\gamma}^{\rm gauge}  \zeta_{\alpha\beta\gamma}^{s} $ is
trivializable. The potential dependence of the invariants on a choice of trivialization
has not been explored.

\item In the case of $SU(2)$ $\CN=2^*$ investigated in the present paper
 $\rho_a(\zeta_{\alpha\beta\gamma}^{\rm gauge})=1$ since the center of $SU(2)$
 acts trivially in the adjoint representation. Thus  the twisting must be canceled by
a different mechanism. In our case $G^f$ contains a $U(1)$ factor, but the $U(1)\times SU(2)_R$
action on $\CR$ descends to a $\left( U(1) \times SU(2)_R \right)/\IZ_2$-action. But now
we can choose a homomorphism
\be\label{eq:keyhom}
{\rm Spin}^c(4) \to \left( U(1) \times SU(2)_R \right)/\IZ_2
\ee
given by
\be
(u_1, u_2) \rightarrow [ \left( \sqrt{\det u_1}, \frac{u_2}{\sqrt{\det u_2}} ) \right) ] ~ .
\ee
This allows us to define $\CR$ as an honest
representation of ${\rm Spin}^c(4)$.
When we make a choice of ${\rm Spin}^c(4)$-structure and use the topological
twisting there is an honest associated bundle to the $SO(3)_{\rm gauge} \times {\rm Spin}^c(4)$
principal bundle over $X$ with fiber $\CR$.
\end{enumerate}

It is clear that there are many generalizations of the two ideas mentioned above.
For example, if $G^f = U(1) \times { \tilde G^f } $ we could choose a principal ${ \tilde G^f } /Z({ \tilde G^f } )$
bundle with 't Hooft flux for the external gauge fields and decompose $\CR$
as a representation of $G\times { \tilde G^f } $:
\be
\CR \cong \oplus_{\alpha,a} D_{\alpha,a}\otimes \rho_\alpha \otimes \rho_a^f
\ee
where $\rho_a^f$ are the irreps of ${ \tilde G^f } $. We twisting of $\rho_a^f $
could cancel that of $\rho_\alpha$ while we use the $U(1)$ factor
and the homomorphism \eqref{eq:keyhom} to cancel the twisting from
$w_2(X)$. One could generalize further.   On top of this
one could consider other topological twistings based on
other homomorphisms $SU(2) \to G^f \times SU(2)_+$, as in the multiple
twists of $\CN=4$ SYM. We will not
try to give a comprehensive treatment of all such twistings, although it
would be interesting to do so.
We will content ourselves with some brief remarks on the class S theories
of $A_1$ type, all of which are Lagrangian. For simplicity we restrict
attention to regular punctures.

The class S theories of $A_1$-type on a curve $\CC$ with genus $g$
with puncture data $D$ having support at $n$ punctures $p_\alpha$
can be written as Lagrangian theories in different ways in different
weak coupling regimes. These regimes are determined by a choice of pants
decomposition of $\CC$ \cite{Gaiotto:2009we}. The (four-dimensional) gauge group is
 $G=SU(2)^{3g-3+n}$. The matter fields   are a collection of
half-hypermultiplets $\Phi_{c_1,c_2,c_3}$ labeled by unordered triples of
cutting curves of the pants decomposition. (We refer to the curves
surrounding punctures as ``cutting curves'' as well.) It is useful
to distinguish internal cutting curves $c_i$ associated with gauge groups
from external curves $c_\alpha$ associated with punctures.

The principal $SO(3)$ bundles $P_{c_i}$ associated with internal cuts have
twistings $\tau_i = w_2(P_{c_i})$. For each trinion with three internal
cutting curves $\Phi_{c_i, c_j, c_k}$ we require
\be
\tau_i + \tau_j + \tau_k + w_2(X) = 0  \in  H^2(X,\IZ_2)
\ee
For each trinion with a single external puncture $\Phi_{c_{\alpha, c_i, c_j}}$ there is a global $U(1)$ symmetry (with nonzero mass) so
we choose a ${\rm Spin}^c$-structure $\fs_\alpha$ and use the homomorphism
\eqref{eq:keyhom}. Then we require $\tau_i = \tau_j$ to cancel
the twisting. For trinions with two external punctures there are a variety of
options for canceling the twisting. Putting these rules together one can
see that there is a large set of choices for topologically twisting a class S
theory of type $A_1$.  How these different twistings interact with the S-duality group is an interesting question
beyond the scope of the present paper.

As far as we know, the topological twisting of general class S theories
has not been written down. The above considerations should make it plain that in general we do expect extra data needs to be specified to define the topological twisting of a general $d=4, \CN=2$ theory. Filling this gap in the literature presents an interesting challenge since
in general these theories are not Lagrangian.

\subsection{The class S Coulomb branch integral}

A central idea we are proposing here is that the partition vector of the M5 tensor multiplet on $M_6 =   \Sigma\times X $,
where we perform the Donaldson-Witten twist on $X$ (generalized, as per above) and the class $S$ twist on $\Sigma $,  will
define the Coulomb branch measure on $\CB$. (This raises a natural question: Why should the partition vector
of the tensor multiplet naturally define a \underline{measure} on $\CB$ ? )  Thus we expect
a generalized expression of the form
\be\label{eq:Class-S-CB-integral}
\Phi_\mu = \int_{\CB} [du d \bar u]\, \CH \cdot \Psi_\mu ~ .
\ee
Here $[du d\bar u]$ is the natural translation-invariant measure on $\CB$, while
  $\mu$ labels a basis for a Stone-von Neumann representation
of a Heisenberg extension of $H^3(M_6; \Lambda_{wt}(\fg)/\Lambda_{rt}(\fg))$.
We expect $\CH$ to be metric-independent and holomorphic, although duality-frame dependent. We also expect $\Psi_\mu$
to be metric-dependent, non-holomorphic, and duality-frame dependent. The measure will be singular at the discriminant locus in
$\CB$, where physical BPS states become massless. Let $\CB^*$ denote the complement in $\CB$ of this discriminant locus.
A  crucial and highly nontrivial criterion on \eqref{eq:Class-S-CB-integral}
is that the measure should be single-valued, without having to pass to some cover of $\CB^*$.
\footnote{In this paper we have used the criterion of single-valuedness as a crucial check on our formula for the Coulomb branch measure. But, given the measure, it could be turned around as a  potentially useful criterion on the physical validity of a proposed Coulomb branch in general $d=4,\CN=2$ theories. In this way the criterion of single-valuedness might be a useful supplement to the criteria used in the program of \cite{Argyres:2015ffa, Argyres:2020nrr}. See \cite{Martone:2020nsy} for some preliminary comments along these lines. }
We will now review a few of the
more obvious expected properties of $\CH$ and $\Psi_\mu$ from the six-, four-, and two-dimensional viewpoints on $M_6$, $X$, and
$\Sigma$, respectively. 

\subsection{Six-dimensional perspective}

 The ``partition vector''  of the tensor-multiplet theory on $X \times \Sigma$ has been discussed in
 \cite{Belov:2006jd,Henningson:2012xk,Hopkins:2002rd,Monnier:2010ww,Monnier:2011mv,Monnier:2011rk,Moore:2004jv,Witten:1996hc,Witten:1999vg, Tachikawa:2013hya}
 and elsewhere. It has the form:
\be
Z^{TM}_\mu = \cD \sum_{x \in \Lambda}  \Omega_{\mu}(x) e^{-\frac{\pi \kappa}{2}  \int H_x * H_x }
\ee 
 where $\Lambda \subset H^3(M_6; \IZ)$ is a maximal Lagrangian sublattice,
 \footnote{We are assuming here that the relevant generalized cohomology
 theory for the quantization of the self-dual tensormultiplet is the Eilenberg-Maclane spectrum.}
$\Omega_\mu$ is a quadratic refinement of the intersection form (QRIF),
defining the holomorphic line bundle over the space of flat external $C$-fields. Also,   
$H_x$ is a harmonic representative
of the class $x$ and $\cD$ is the ``determinant of quadratic fluctuations'' - a suitable section of a
product of determinant line bundles. The contribution from the self-dual tensor field is
given by Cheeger's half-torsion \cite{Belov:2006jd,Monnier:2010ww} and the scalar and fermion fields give standard determinant
line bundles and sections of those line bundles. Finally,
$\kappa$ is a constant defining the ``level'' of the self-dual system, analogous to the
level of the chiral scalar field in two dimensions. (Roughly speaking, the theta function
will have level $\kappa/2$.)

The most subtle component of the partition function is the QRIF.
The quadratic refinement for the smallest level (sometimes called ``level 1/2'')
depends on a choice of Wu structure of $M_6$, i.e. a trivialization of $w_4$ on
$W_8$ where $\p W_8 = S^1 \times M_6$. Descriptions can be found in
\cite{Belov:2006jd,Hopkins:2002rd,Witten:1996hc,Witten:1999vg}. Here we are considering the
case where $M_6 = \Sigma \times X$ with $X$ not necessarily spin. To define the full tensormultiplet
we do not need a spin structure on $M_6$, but we do require a spin structure on
$TM_6 \oplus N$ where $N$ is the R-symmetry bundle with structure group $SO(5)$.  It suffices to make a
choice of ${\rm Spin^c}$ structure on $TM_6$ and $N$ so that $w_2(TM_6 \oplus N)=0$.
 Fortunately a choice of ${\rm Spin^c}$
structure on $M_6$ induces a choice of Wu structure, which we can then use to
define the self-dual field.
\footnote{In \cite{Monnier:2018cfa,Monnier:2018nfs}
the authors went to some length to define non-anomalous six-dimensional
supergravity theories \underline{without} making a choice of Wu structure. It is possible similar considerations might be useful in the present context.}
We will now attempt to give a generalization of the discussion
 in section 5.2 of   \cite{Witten:1999vg} to the  present case.

In section 5.2 of   \cite{Witten:1999vg}  one considers the integral class $x_4 = \theta \cup x$ on
$S^1 \times M_6$.\footnote{Here the cup product is taken on integral cocycles. For a definition see, for examples, \cite{Bredon, Massey,Vick}.} One then evaluates the level $1/2$ Wu-Chern-Simons theory on an 8-manfiold
$W_8$ with $\p W_8 = S^1 \times M_6$. This is done by extending $x_4$ to $W_8$ and taking
\be
\Omega(x) = \exp  \left[ \I \pi   \int_{W_8} x_4 \cup (x_4 + \lambda) \right]
\ee
for a suitable lift $\lambda$ of the $4^{\rm th}$ Wu class. One needs to choose a trivialization of
$\lambda$ on $\p W_8$ and this leads to a dependence on Wu-structure (much like 3d level 1/2
Chern-Simons theory depends on spin structure). A ${\rm Spin}^c$ structure on $W_8$
is enough to define a choice of $\lambda$ so that $2\lambda = p_1 - \alpha^2$ where $\alpha$ is
the characteristic class of the ${\rm Spin}^c$ structure. (See footnote 11 of \cite{Witten:1999vg}
for a proof.)

In the present case it is natural to fill in $\Sigma$ with a handlebody $\CH_3$
so that $\p \CH_{3} = \Sigma$. (Such a choice of handlebody represents a weak-coupling
limit of the theory.) The kernel of the inclusion
\be
\iota_*: H_1(\Sigma;\IZ) \to H_1(\CH_3;\IZ)
\ee
defines a maximal Lagrangian sublattice of  $H_1(\Sigma;\IZ)$.
One then chooses a  complementary maximal Lagrangian sublattice in $H_1(\Sigma;\IZ)$.  Then the
sublattice of $H^1(\Sigma;\IZ)$ dual to the kernel will be denoted $L_B$.
These are the degree one cohomology classes which have nontrivial periods on $\Sigma$ and
extend to $\CH_3$. Similarly $L_A$ will be the complementary maximal Lagrangian
lattice of classes which do not extend.

We can now choose $W_8 = S^1 \times \CH_{3} \times X$ and our theta function will be a sum over
classes in
\be
\Lambda =L_B \otimes H^2(X;\IZ)
\ee
If $x \in \Lambda$ is  of the from $\sum_i b_i \cup k_i $ where $b_i \in L_B$
and $k_i\in H^2(X;\IZ)$ we can extend $b_i$ to degree one classes $\tilde b_i$ in $\CH_3$.  Then $x_4 \cup x_4=0$
because $\theta\cup \theta=0$ and Witten's definition reduces to
\be
\Omega(x) = \exp  \left[ \I \pi \int_{\CH_{3} \times X}  (\sum_i \tilde b_i \cup k_i) \cup \lambda \right]
\ee
Now, a choice of ${\rm Spin}^c$ structure on $\CH_3 \times M_6$ allows us to define a class $\tilde \rho$ on
$\CH_3$ with the property that
\be
\begin{split}
%\int_{H_3} \tilde \rho \cup \alpha^I & = 0 ~ \mod 2  \\
\int_{\CH_3} \tilde \rho \cup \beta_I & = 1  ~ \mod 2,
\end{split}
\ee
where $\{\alpha^I, \beta_I \}$ is a symplectic basis for $L_A, L_B$.
To prove this we note that a ${\rm Spin^c}$ structure on the disk allows us to
define a characteristic class  $\tilde \alpha $ with $\int_D \tilde \alpha = 1$.
Then we extend to the solid torus $D\times S^1$ finally we choose such a class on
each torus and take a connected sum.

Now, following \cite{Witten:1999vg} we can write
$\lambda(W) = \lambda(X) - \tilde \rho \cup c_\uv $.
Next, recall that in the class $S$ theory, $L_B$ will be,
roughly speaking, the co-root lattice of the gauge group while $L_A$ will be the weight lattice.
(See \cite{Longhi:2016rjt} for more precise statements.)
Thus $\tilde \rho$ functions as a kind of ``Weyl vector,'' $\rho\in L_A $
 and  the above quadratic refinement becomes
\be\label{eq:w2-phase}
\exp[ \I \pi  x \cdot (\rho \otimes w_2(X)) ],
\ee
where we use the dual pairing between $L_A$ and $L_B$ and the 
usual intersection pairing on $H^2(X;\IZ_2)$. 
This is the 6d derivation
of the phase discussed in \cite{LoNeSha,Marino:1998bm,Witten:1995gf}.

It is important to note that the quadratic refinement $\Omega$ used in Witten's prescription, and in the generalizations
discussed in \cite{Belov:2006jd} are all $\IZ_2$-valued, whereas the natural quadratic refinement coming from decoupling
the M5-brane is $U(1)$ valued and can vary continuously with metric \cite{Moore:2004jv}. This will be important below.

In equation \eqref{ZSWSdual} we gave the transformation properties of the partition function under S-duality.
It would be interesting to reproduce this from the 6-dimensional perspective using gravitational anomalies.
This is perhaps one of the more accessible of the many open problems suggested in this section.

\subsection{Four-dimensional perspective}

A careful reduction of an action for the interacting Abelian self-dual tensor multiplet from $M_6$ to $X$
(properly accounting for self-duality using, for example, the PST action or the action described in
\cite{Belov:2006jd}), with class S twist, indeed yields the Seiberg-Witten LEET on $X$. (Without the
topological couplings $A,B,C,...$.)
Roughly speaking, the flux-lattice of the $U(1)^r$ gauge fields is $\Lambda = L_B \otimes H^2(X;\IZ)$
and the reduction of the fieldstrength $H_x$ is of the form $H_x = \beta_I \wedge F^I$ where $\beta_I$ is
a basis of harmonic forms representing a basis in $L_B$. Then
\be
\int_{M_6} H_x * H_x  \sim   \int_X  \mathrm{Im}(\tau_{IJ}) F^I * F^J \qquad .
\ee
(For one derivation see section 3.1.5 in \cite{Gaiotto:2009hg}. The result for the remainder of the
action follows from supersymmetry. )

This raises two important issues which need to be understood more clearly:

\begin{enumerate}

\item In this paper it has been very important that $\Psi_\bfmu^J$ is not precisely a theta function. Rather,
it has a nontrivial kernel $\CK$, as in equation \eqref{defPsi3} above. It arises from   fermion zeromodes and interaction
terms. This kernel becomes somewhat more intricate in the higher rank case. See, for example,
\cite{Marino:1998bm} equation (4.17). The higher-rank generalization of the error functions we use in defining
$\Psi_\bfmu^J$ needs to be understood better. It is probably best to start by finding the higher rank
analog of $\widehat{\Theta}^{J,J'}$ of equation \eqref{ThetaComplete}. Explicit results for the holomorphic and non-holomorphic parts of the $SU(N)$ VW partition function \cite{Manschot:2014cca, Manschot:2017xcr, Alexandrov:2019rth, Dabholkar:2020fde, Alexandrov:2020bwg} could be a useful benchmark result for the analysis of the $SU(N)$, $\CN=2^*$ theory. 

\item As we have just discussed, when defining the 5-brane partition function a crucial ingredient is the choice of QRIF.  As we have shown above, using Witten's
prescription for $\Omega(x)$ for the level 1/2 case,  this descends to Witten's famous term
$(-1)^{B(c_1, w_2(X))}$ where $c_1$ is the first Chern class of the $U(1)$ gauge field in the LEET on the Coulomb branch.
On the other hand, in this paper no such factor appears in the lattice sum $\Psi_\bfmu^J$. Rather there is a complex weighting
factor
\be
\exp\!\left[-2\pi i (B(\bfk_+,c_{\uv})\, v(\tau,\tau_{\uv}) + B(\bfk_-, c_{\uv})\, \bar v(\tau,\tau_{\uv}))\right]
\ee
which, in the $N_f=0$ limit becomes Witten's phase. (Recall the discussion from section 4.7 above.)
This strongly suggests that the appropriate QRIF
relevant to the general class S problem is in fact $\IC^*$-valued, and it is natural to speculate that it
is related to the metric-dependent, $\IC^*$-valued QRIF discussed in \cite{Moore:2004jv}.
\footnote{In the 2018 String-Math lecture by one of us, in Sendai, 
equation \eqref{eq:w2-phase} was cited as evidence that twisted class $S$ theories will not lead to new four-manifold invariants. In view of the present remark, that claim should be retracted.}

\end{enumerate}

Another important lesson from the 4-dimensional perspective concerns the factor $A^\chi B^\sigma$ in the
measure. The arguments of \cite{Moore:1997pc,Shapere:2008zf,Witten:1995gf}   based on R-symmetry
should generalize to all class S examples. Thus the coupling $A^\chi$, which is frame-dependent, is given by
\be
A = \alpha \left( \det \frac{\p u^I }{\p a^J} \right)^{1/2} \quad . 
\ee
To write $B$ we should find a holomorphic function $\Delta_{\rm phys}$ on the Hitchin base $\CB$
with first order zeroes at the discriminant locus and take
\be
B = \beta \Delta_{\rm phys}^{1/8} \quad . 
\ee
Now, in class S theory, where there is a nontrivial conformal manifold of superconformal couplings:
It is the  Teichm\"uller space of pairs $(C,D)$ \cite{Gaiotto:2009we,Witten:1997sc}. Just the way we found
the couplings $A,B,C$ to be interesting bi-modular forms, we expect $A,B$ to be interesting bi-automorphic
objects  in the general case. (For Teichm\"uller space and for the moduli space of Abelian varieties, respectively.)  Moreover, it
is natural to ask if $\alpha,\beta$ must be interesting automorphic forms on this Teichm\"uller space.
The results of this paper strongly suggest that the answer is affirmative
since, in our example, the proper couplings to $\chi$ and $\sigma$ are given by the functions
$\tilde A$ and $\tilde B$ of equation \eqref{ABCextra}. In this example then, $\alpha,\beta$ are
nontrivial powers of $\eta(\tau_\uv)$. 
Note that there are weak-coupling limits of class S theories
where $C_{g,n}$ degenerates to a connected sum of several punctured tori connected by long thin tubes.
This shows that $\alpha,\beta$ must be nontrivial in general.

\subsection{Two-dimensional perspective}

Finally, we can also gain insight by reducing the 6d Abelian interacting TM theory on $X$
with Donaldson-Witten twist to obtain a $(0,2)$ QFT on $\Sigma$, and we can identify the
partition function, and therefore the Coulomb branch measure with the partition function
of this $d=2$, $(0,2)$ QFT on $\Sigma$ (with the class $S$ twist). We should note that a special case of this computation goes
back to the case of the M5 brane wrapped on an ample divisor in a Calabi-Yau manifold
\cite{Maldacena:1997de,Minasian:1999qn}  in which case it has $(0,4)$ supersymmetry.

The field content of the $d=2$  $(0,2)$ theory on $\Sigma$ is 
a model with the following fields \cite{Dabholkar:2020fde,Gadde:2013sca,NidaievPhD}:

From the noncompact normal bundle scalars we get

\begin{enumerate}

\item A noncompact scalar field $dY \in \Omega^1(\Sigma) \otimes H^{2,+}(X;\IR)$.
In deriving the $u$-plane measure these will play a role analogous to the
$\CN=2$ auxiliary fields $D$.

\item A noncompact scalar field $Y_z dz \in \Omega^{1,0}(\Sigma)$.
When deriving the $u$-plane measure the zeromodes of these fields will
play the role of the special K\"ahler coordinates $a^I$.

\end{enumerate}

From the self-dual 2-form gauge potential we get

\begin{enumerate}

\item Self-dual, compact, chiral bosons with fieldstrength
\be
F^+ \in \Omega^{1,0}(\Sigma) \otimes H^{2,+}(X;\IR)
\ee
satisfying $\bar\partial F^+=0$.

\item Anti-self-dual, compact, chiral bosons with fieldstrength
\be
F^- \in \Omega^{0,1}(\Sigma) \otimes H^{2,-}(X;\IR)
\ee
satisfying $\partial F^-=0$.

\end{enumerate}

The sum over the zeromodes of these fields becomes the
sum over the fluxes in the ``theta function''  $\Psi_\mu$.
Viewed this way $\Psi_\mu$ is, roughly speaking, a Siegel-Narain  theta function
for the zero-modes of the 2d model.

Finally, we have fermionic fields

\begin{enumerate}

\item $\eta, \eta_{\bar z}d\bar z $. These play the role of the fermion $\eta$ in the 4d
twisted theory.

\item $\chi, \chi_{\bar z} d \bar z$. These are valued in $H^{2,+}(X)$ and play the role
of the fermion $\chi$ in the 4d twisted theory.

\end{enumerate}

The action one obtains from reducing - for example - the PST action is that of a nontrivial
interacting theory.
\footnote{The $(0,2)$ field multiplets and the interacting theory up to 2-fermion terms
and some of the 4-fermion terms was worked out in unpublished work by G.M. and I. Nidaiev in 2012. A few
details are mentioned in \cite{NidaievPhD}.}
 Some of the interactions are very important and account, for example
for the crucial difference between $\Psi_\mu$ and an actual Siegel-Narain theta function.

Of particular interest from this point of view are the $A,B,C$ couplings that appear in the Coulomb branch measure.
We expect that the defects $D\subset \CC$ in the 6d (0,2) theory will turn into operator insertions in the 2d (0,2)
theory on $\Sigma$. It is possible the suitable operators can be deduced
simply from the singularity structure of the
Higgs fields near the defects. Certainly \eqref{CphijC} is very suggestive of a correlation function in a 2d CFT.
This point of view is also probably the only practical approach to deriving explicit formulae for the automorphic functions $\alpha$ and $\beta$.

Probably the first practical exercise for carrying out this 2d approach is to account for the modular weight
 $-4\ell - \chi/2$  of the $u$-plane integral in the $SU(2)$ $\CN=2^*$ case.

\section{Further open problems}
\label{sec:Discussion}
 
We have given a quite extensive discussion on the partition function of the $\CN=2^*$ theory on smooth compact four manifolds. Nevertheless, a number of aspects remain to be better understood. We list a few here:
\begin{enumerate}
\item A concrete open problem is a proof of the non-trivial identities (\ref{Cttv2}) and (\ref{tauvtauUV}). We also expect similar identities for theories with higher rank gauge groups.
\item For manifolds with $b_2^+=1$, we evaluated the $u$-plane integral $\Phi^J_\bfmu$ only for specific  Spin$^c$ structures. It would be interesting to have a solution for arbitrary $\bfk_m$. To this end, one needs to determine a suitable anti-derivative $\widehat G(\tau,z;\bfk_m)$ with poles at appropriate positions. We have seen that
for $\bfk_m$ an almost complex structure, the $G(\tau,z;\bfk_m)$ are
related to the refined VW partition function, enumerating Betti
numbers (or better $\chi_y$ genera) of instanton moduli spaces.  This raises the questions, whether there is a geometric meaning of $G(\tau,z;\bfk_m)$
for general $\bfk_m$.
\item Throughout this paper we have restricted attention to the case where $X$ is simply-connected. It should be possible to extend the discussion to the non-simply connected case. (See \cite{Marino:1998rg, Marino:1998eg} for the case of the original Donaldson-Witten theory.) It is possible that such an extension can teach us lessons about three-dimensional manifolds, and 3d-3d correspondences \cite{Dimofte:2011ju, Eckhard:2018raj}. 
\item We have demonstrated that when we choose the Spin$^c$ structure canonically associated to an ACS, the $\CN=2^*$ partition functions satisfy the same $S$-duality relations as VW partition functions. Moreover, for the canonical ${\rm Spin}^c$ structure both partition functions formally compute the generating function of the Euler characters of instanton moduli spaces. 
This strongly suggests that they are identical for non-K\"ahler manifolds as well as K\"ahler manifolds. Our discussion of the relation of the VW and adjoint-SW equations shows that they are not the same in general, but rather are related by a compact deformation. It would be good to have a rigorous mathematical proof that the moduli spaces of these two equations are nevertheless topologically equivalent.  

%We  have  demonstrated  that  the $\CN=2^∗$ partition  functions  for  a  canonically determined Spin$^c$ structure satisfy the same $S$-duality relations as VW partition functions.  Moreover, when we choose the Spin$^c$ structure canonically associated to  an  ACS  both  partition  functions  formally  compute  the  generating  function of the Euler characters of instanton moduli spaces.  This strongly suggests that they  are  identical  for  non-K\"̈ahler  manifolds.   Our  discussion  of  the  relation  of the VW and adjoint-SW equations shows that they are not the same in general, but rather are related by a compact deformation.  It would be good to have a rigorous mathematical proof that the moduli spaces of these two equations are nevertheless topologically equivalent.

\item The $\CN=2^*$ $u-$plane integral reproduces and generalizes the generating function of Euler numbers of instanton moduli spaces. In the mathematical literature, generating functions for  $\chi_y$-genera are available \cite{Gottsche:2019vbi, Yoshioka1994, Gottsche:1999, Thomas:2018lvm}. This involves functions of McMahon type \cite{Gottsche:2019vbi}, which are related to quantum modular forms rather than modular forms.
\footnote{The relation to the quantum modular forms arises as follows. 
In Appendix E of \cite{Dabholkar:2005dt} it was noted that the 
logarithmic derivative of the McMahon function gives a ``half-Eisenstein function'' $E_3$. But $E_3$ is one of the standard examples of the new classes of quantum modular forms described in \cite{Zagier2010} .}

It would be interesting to the understand the refinement parameter $y$ from the perspective of the $\CN=2^*$ $u$-plane integral.
\item Various string-theoretic descriptions are available for the $\CN=2^*$ theory \cite{Witten:1997sc, Donagi:1995cf}. It would be interesting to connect our work to such D-brane and M-brane systems, as well as to integrable systems. 

\item As discussed extensively in Section \ref{SecClassS}, an interesting avenue is to develop the $u$-plane integral for other $\CN=2$ supersymmetric theories, such as theories with higher rank gauge groups and more generally theories of class $S$. Trying to do this raises many open 
questions, some of which were pointed out in Section \ref{SecClassS}. 

\item Finally, it has been emphasized over the years by Edward Witten that a promising approach to finding new four-manifold invariants using supersymmetric gauge theory is to categorify the VW invariants. In the realm of algebraic surfaces an important step forward (using a mathematical viewpoint) was taken in \cite{Thomas:2018lvm}. Our discussion of the relation of the VW with the adjoint-Seiberg-Witten equations shows that one can deform the VW equations in such a way as to introduce an $SO(2)$ symmetry. We hope this deformation, and the resulting $SO(2)$ symmetry will be  useful in dealing with the noncompact branches where the $C$ and $B$ fields are nonzero.
\footnote{As shown in \cite{Vafa:1994tf} the field $C$ is generically zero. However $B$ need not be zero.}
The results of \cite{Thomas:2018lvm} suggest that it should be interesting 
to investigate the equivariant $L^2$-kernel of certain Dirac-type operators on the moduli space of the adjoint Seiberg-Witten equations. 

\end{enumerate} 

\section*{Acknowledgements}
We thank Johannes Aspman, Emanuel Diaconescu, Simon Donaldson, Elias Furrer, Dan Freed, Davide Gaiotto, Martijn Kool, Georgios Korpas, Peter Kronheimer, Marcos Mari\~no,  Hiraku Nakajima, Nikita Nekrasov, Samson Shatashvili, Yuji Tachikawa, Richard Thomas, Edward Witten and Xinyu Zhang for discussions and useful comments on the draft. JM is supported
by the Laureate Award 15175 “Modularity in Quantum Field Theory and
Gravity” of the Irish Research Council. GM is supported by the
US Department of Energy under grant DE-SC0010008. This research was
supported in part by the National Science Foundation under Grant
No. NSF PHY-1748958 through the KITP program ``Modularity in Quantum Systems''.

\appendix
 
\section{Modular forms}
We review in this appendix aspects of modular forms used in the main text. After discussing classical modular forms, we discuss the Appell-Lerch sum and bi-modular forms. A few references for modular forms are \cite{Eichler:1985, Diamond, Bruinier08, Bringmann:2309148}.

\label{AModForms}
\subsection*{Eisenstein series}
We let $\tau\in \mathbb{H}$ and define $q=e^{2\pi i \tau}$. The Eisenstein series $E_k:\mathbb{H}\to \mathbb{C}$ for even $k\geq 2$ are defined as the $q$-series 
\be
\label{Ek}
E_{k}(\tau)=1-\frac{2k}{B_k}\sum_{n=1}^\infty \sigma_{k-1}(n)\,q^n,
\ee
with $\sigma_k(n)=\sum_{d|n} d^k$ the divisor sum. For $k\geq 4$, $E_{k}$ is a modular form of
$\operatorname{SL}(2,\mathbb{Z})$ of weight $k$. In other words, it transforms under $\operatorname{SL}(2,\mathbb{Z})$ as
\be
E_k\!\left( \frac{a\tau+b}{c\tau+d}\right)=(c\tau+d)^kE_k(\tau).
\ee
The first terms of the expansions of $E_4$ and $E_6$ are
\begin{eqnarray}
E_4(\tau) &=& 1+240\, q+\dots,\\
E_6(\tau) &=& 1-504\, q + \dots.
\end{eqnarray}

On the other hand $E_2$ is a quasi-modular form, which means that the $\operatorname{SL}(2,\mathbb{Z})$ transformation of $E_2$ includes a shift in addition to the weight,
\be 
\label{E2trafo}
E_2\!\left(\frac{a\tau+b}{c\tau+d}\right) =(c\tau+d)^2E_2(\tau)-\frac{6i}{\pi}c(c\tau+d).
\ee
The non-holomorphic modular form
\be
\label{whE2}
\widehat E_2(\tau,\bar \tau)=E_2(\tau)-\frac{3}{\pi y},
\ee
does transform as a modular form of weight 2.

\subsection*{Dedekind eta function}
The Dedekind eta function $\eta:\mathbb{H}\to\mathbb{C}$ is defined as
\be
\label{etafunction}
\eta(\tau)=q^{\frac{1}{24}}\prod_{n=1}^\infty (1-q^n).
\ee
It is a modular form of weight $\frac{1}{2}$ under SL$(2,\BZ)$ with a
non-trivial multiplier system\footnote{For an unambiguous value of the square root, we define the phase of $z\in \mathbb{C}^*$ by $\log(z):=\log|z|+i\,\mathrm{arg}(z)$ with $-\pi<\arg(z)\leq \pi$.}
\be
\label{vareps}
\varepsilon(\gamma)=\frac{\eta((a\tau+b)/(c\tau+d))}{(c\tau+d)^{1/2}\,\eta(\tau)},
\ee
The phase $\varepsilon(\gamma)$ can be explicitly evaluated \cite{Apostol}.
$\eta$ transforms under the generators of
SL$(2,\BZ)$ as
\be
\begin{split}
&\eta(-1/\tau)=\sqrt{-i\tau}\,\eta(\tau),\\
&\eta(\tau+1)=e^{\frac{\pi i}{12}}\, \eta(\tau).
\end{split}
\ee

\subsection*{Jacobi theta functions}
The four Jacobi theta functions $\vartheta_j:\BH\times \BC\to \BC$ are defined by,
\be
\label{Jacobitheta}
\begin{split}
&\vartheta_1(\tau,z)=i \sum_{n\in
  \mathbb{Z}+\frac12}(-1)^{n-\frac12}q^{\frac{n^2}{2}}e^{2\pi i
  nz}, \\
&\vartheta_2(\tau,z)= \sum_{n\in
  \mathbb{Z}+\frac12}q^{\frac{n^2}{2}}e^{2\pi i
  nz},\\
&\vartheta_3(\tau,z)= \sum_{n\in
  \mathbb{Z}}q^{\frac{n^2}{2}}e^{2\pi i
  n z}, \\
&\vartheta_4(\tau,z)= \sum_{n\in 
  \mathbb{Z}} (-1)^nq^{\frac{n^2}{2}}e^{2\pi i
  n z}. 
  \end{split}
\ee
$\vartheta_1$ is odd as function of $v$, while the $\vartheta_j$ are even
for $j=2,3,4$. We have $\partial_v\vartheta_1(\tau,v)\vert_{v=0}=-2\pi
\eta(\tau)^3$.  We use $\vartheta_j(\tau,0)=\vartheta_j(\tau)$. These functions
satisfy various identities. For their sum, we have
\be
\label{t3=t2pt4}
\begin{split}
&\vartheta_3(\tau)^4=\vartheta_2(\tau)^4 + \vartheta_4(\tau)^4,\\
\end{split}
\ee
and 
\be
\label{sumstheta}
\begin{split}
\vartheta_3(\tau/2,z/2)=\vartheta_3(2\tau,z)+\vartheta_2(2\tau,z),\\
\vartheta_4(\tau/2,z/2)=\vartheta_3(2\tau,z)-\vartheta_2(2\tau,z).
\end{split}
\ee
For their product, we have
\be
\label{tsproduct}
\begin{split}
&\vartheta_2(\tau)\,\vartheta_3(\tau)\,\vartheta_4(\tau)=2\,\eta(\tau)^3,\\
&\vartheta_1(\tau,2z)\,\vartheta_2(\tau)\,\vartheta_3(\tau)\,\vartheta_4(\tau)=2\,\vartheta_1(\tau,z)\,\vartheta_2(\tau,z)\,\vartheta_3(\tau,z)\,\vartheta_4(\tau,z).
\end{split}
\ee

The function $\vartheta_1$ transforms for an element $\gamma=\left( \begin{array}{cc}
    a & b \\ c & d \end{array} \right) \in SL(2,\mathbb{Z})$ as
\be
\vartheta_1\!\left(\frac{a\tau+b}{c\tau+d},\frac{z}{c\tau+d}\right)=
\varepsilon(\gamma)^3\, (c\tau+d)^{1/2}\, e^{\pi i cz^2/(c\tau+d)}\,
\vartheta_1(\tau,z), 
\ee
where $\varepsilon(\gamma)$ is the multiplier system of the
$\eta$-function (\ref{vareps}).

For the two generators $S$ and $T$ of $SL(2,\mathbb{Z})$, we have
\be
\begin{split}
&S:\qquad \vartheta_1\!\left(-1/\tau,z/\tau \right)=-i\sqrt{-i\tau}\, e^{\pi
  i z^2/\tau}\,\vartheta_1(\tau,z),\\
&T:\qquad \vartheta_1\!\left(\tau+1,z\right)=e^{2\pi i/8}\, \vartheta_1(\tau,z).
\end{split}
\ee

These have the following transformations for modular inversion
\be
\label{JacobiSdual}
\begin{split}
&\vartheta_2(-1/\tau)=\sqrt{-i\tau}\,\vartheta_4(\tau),\\
&\vartheta_3(-1/\tau)=\sqrt{-i\tau}\,\vartheta_3(\tau),\\
&\vartheta_4(-1/\tau)=\sqrt{-i\tau}\,\vartheta_2(\tau),
\end{split}
\ee
and for the shift
\be
\label{JacobiTdual}
\begin{split}
&\vartheta_2(\tau+1)=e^{2\pi i /8}\,\vartheta_2(\tau),\\
&\vartheta_3(\tau+1)=\vartheta_4(\tau),\\
&\vartheta_4(\tau+1)=\vartheta_3(\tau).
\end{split}
\ee
The generators of the congruence subgroup $\Gamma(2)$ are $\tau\to
\tau+2$ and $\tau\to \frac{\tau}{-2\tau+1}$. Using the above two, we
find for the second transformation
\be
\begin{split}
&\vartheta_2\!\left(\frac{\tau}{-2\tau+1}\right)= \sqrt{-2\tau+1}\, \vartheta_2(\tau),\\
&\vartheta_3\!\left(\frac{\tau}{-2\tau+1}\right)= \sqrt{-2\tau+1}\, \vartheta_3(\tau),\\
&\vartheta_4\!\left(\frac{\tau}{-2\tau+1}\right)= e^{2\pi i /4} \sqrt{-2\tau+1}\, \vartheta_4(\tau). \\
\end{split} 
\ee
Using these transformations, we can derive the behavior at the cusps of $\mathbb{H}^2/\Gamma(2)$. 
\begin{itemize}
\item For $\tau\to i\infty$,
\be
\begin{split}
&\vartheta_2(\tau)\to 0, \\
&\vartheta_3(\tau)\to1, \\
&\vartheta_4(\tau)\to 1.
\end{split}
\ee 
\item For $\tau\to 0$,
\be
\begin{split}
&\vartheta_2(\tau)\to \frac{1}{\sqrt{-i\tau}}, \\
&\vartheta_3(\tau)\to \frac{1}{\sqrt{-i\tau}}, \\
&\vartheta_4(\tau)\to 0.
\end{split}
\ee 
\item For $\tau\to 1$,
\be
\begin{split}
&\vartheta_2(\tau)\to \frac{e^{2\pi i/8}}{\sqrt{-i(\tau-1)}}, \\
&\vartheta_3(\tau)\to 0, \\
&\vartheta_4(\tau)\to \frac{1}{\sqrt{-i(\tau-1)}}.
\end{split}
\ee 
\end{itemize}
We also have the following quasi-periodicity 
\be
\begin{split}
\vartheta_1(\tau,z+n\tau)&=(-1)^n q^{-n^2/2}e^{-2\pi i n
  z}\vartheta_1(\tau,z),\\
\vartheta_2(\tau,z+n\tau)&= q^{-n^2/2}e^{-2\pi i n
  z}\vartheta_2(\tau,z).
\end{split}
\ee
For the other two theta functions, $\vartheta_3$ transforms as
$\vartheta_2$, and $\vartheta_4$ as $\vartheta_1$.

We list moreover the expression for the expression for
$\vartheta^{-1}$, as a sum over its poles
\be
\label{InvThetaId}
\frac{i\eta(\tau)^3}{\vartheta_1(\tau,z)}=-e^{\pi i z}\sum_{n\in \mathbb{Z}}
\frac{(-1)^n q^{n(n+1)/2}}{1-e^{2\pi i z}q^n}.
\ee
The Eisenstein series can be expressed in terms of Jacobi theta functions
\begin{eqnarray}
E_4(\tau) &=&  \frac{\vartheta_2(\tau)^8+\vartheta_3(\tau)^8+\vartheta_4(\tau)^8}{2}, \\
E_6(\tau) &=&  \frac{(\vartheta_2(\tau)^4+\vartheta_4(\tau)^4)(\vartheta_3(\tau)^4+\vartheta_4(\tau)^4)(\vartheta_4(\tau)^4-\vartheta_2(\tau)^4)}{2}.
\end{eqnarray}
The half-periods $e_j$ are defined in terms of the Jacobi theta
functions by
\be
\label{halfperiods}
\begin{split}
&e_1(\tau)=\wp(\tfrac{1}{2},\tau)/\pi^2=\frac{1}{3} (\vartheta_3(\tau)^4+\vartheta_4(\tau)^4)=\frac{2}{3}+16q+\dots,\\
&e_2(\tau)=\wp(\tfrac{\tau}{2},\tau)/\pi^2=-\frac{1}{3}(\vartheta_2(\tau)^4+\vartheta_3(\tau)^4)=-\frac{1}{3}-8q^{\frac{1}{2}}+\dots,\\
&e_3(\tau)=\wp(\tfrac{1+\tau}{2},\tau)/\pi^2=\frac{1}{3}(\vartheta_2(\tau)^4-\vartheta_4(\tau)^4)=-\frac{1}{3}+8q^{\frac{1}{2}}+\dots,
\end{split}
\ee
where $\wp(z,\tau)$ is the Weierstrass $\wp$-function.

The half-periods $e_j$ transform as modular forms of weight 2 under the congruence
subgroup $\Gamma(2)$. They are exchanged under the generators $S$ and
$T$ of $SL(2,\mathbb{Z})$. For $S$,
\be
\begin{split}
e_1(-1/\tau)&= \tau^2 e_2(\tau),\\
e_2(-1/\tau)&= \tau^2 e_1(\tau),\\
e_3(-1/\tau)&= \tau^2 e_3(\tau).
\end{split}
\ee
and for $T$,
\be
\begin{split}
e_1(\tau+1)&= e_1(\tau),\\
e_2(\tau+1)&= e_3(\tau),\\
e_3(\tau+1)&= e_2(\tau).
\end{split}
\ee
The $e_i$ satisfy
\be
\label{eiEj}
\begin{split}
&\sum_{i=1}^3 e_i = 0, \\
&\sum_{i=1}^3 e_i^2 = \frac{2}{3} E_4, \\
&\sum_{i=1}^3 e_i^3 = \frac{2}{9} E_6,\\
&e_1 e_2 e_3=\frac{2}{27}E_6.
\end{split}
\ee 
and they give a three dimensional representation of $\SL$ of weight two.

Let $D=\frac{1}{2\pi i} \frac{d}{d\tau}$. We list the following derivatives
\be
\label{Dtheta4}
\begin{split}
D\vartheta_2(\tau)^4&=\frac{1}{6}\left(
  E_2(\tau)\vartheta_2(\tau)^4+\vartheta_3(\tau)^8-\vartheta_4(\tau)^8\right)\\
&=\frac{\vartheta_2(\tau)^4}{6} \left(E_2(\tau)+\vartheta_3(\tau)^4+\vartheta_4(\tau)^4 \right),\\
D\vartheta_3(\tau)^4&=\frac{1}{6}\left(
  E_2(\tau)\vartheta_3(\tau)^4-\vartheta_4(\tau)^8+\vartheta_2(\tau)^8\right)\\
&= \frac{\vartheta_3(\tau)^4}{6} \left(E_2(\tau)+\vartheta_2(\tau)^4-\vartheta_4(\tau)^4 \right), \\
D\vartheta_4(\tau)^4&=\frac{1}{6}\left( E_2(\tau)
  \vartheta_4(\tau)^4-\vartheta_3(\tau)^8+\vartheta_2(\tau)^8\right)\\
&= \frac{\vartheta_4(\tau)^4}{6} \left(E_2(\tau)-\vartheta_2(\tau)^4-\vartheta_3(\tau)^4 \right), \\
\end{split}
\ee
and 
\be
\label{DE2}
DE_2(\tau)=\frac{1}{12}\left( E_2(\tau)^2-E_4(\tau)\right).
\ee
 
\subsection*{The Appell-Lerch sum}
\label{subsecZwmu}
We recall the definition and properties of the Appell-Lerch sum
following Zwegers \cite{ZwegersThesis}. We
define  
\be
\label{Mtuv}
M(\tau,u,v):=\frac{e^{\pi i u}}{\vartheta_1(\tau,v)}\sum_{n\in
  \mathbb{Z}} \frac{(-1)^n q^{n(n+1)/2}e^{2\pi i nv}}{1-e^{2\pi i u}q^n}.
\ee
With $a=\mathrm{Im}(u)/y$, we introduce also 
\be
\label{muR}
R(\tau,\bar \tau, u,\bar u):=\sum_{n\in \mathbb{Z}+\frac{1}{2}} \left( \sgn(n) -
  E\!\left( (n+a)\sqrt{2y}
     \right) \right) (-1)^{n-\frac{1}{2}}\,e^{-2\pi i un}q^{-n^2/2},
\ee
where $E(t)$ is the rescaled error function
\be
\label{Eerror}
E(t)=2 \int_0^t e^{-\pi u^2}du=\mathrm{Erf}(\sqrt{\pi}t).
\ee
This function is anti-symmetric, $E(-t)=-E(t)$.

The non-holomorphic completion of $M(\tau,u,v)$ is then given by
\be
\label{mucomplete}
\widehat M(\tau,\bar \tau, u,\bar u,v,\bar v) =M(\tau,u,v)+\frac{i}{2} R(\tau,\bar \tau,u-v,\bar u-\bar v). 
\ee
For an element $\gamma\in SL(2,\mathbb{Z})$, it transforms as
\be
\label{whMtrafo}
\begin{split}
&\widehat M\!\left(\frac{a\tau+b}{c\tau+d},\frac{a\bar \tau+b}{c\bar \tau+d},\frac{u}{c\tau+d},\frac{\bar u}{c\bar \tau+d},\frac{v}{c\tau+d},\frac{\bar v}{c\bar \tau+d}\right)\\
&\qquad =\varepsilon(\gamma)^{-3}
(c\tau+d)^{\frac{1}{2}} e^{-\pi i c(u-v)^2/(c\tau+d)}\,\widehat M(\tau,\bar \tau,u,\bar u, v,\bar v),
\end{split}
\ee
where $\varepsilon(\gamma)$ is the multiplier of the $\eta$-function (\ref{vareps}).

The anti-holomorphic derivative of $\widehat M$ is given by
\be
\begin{split}
&\partial_{\bar \tau} \widehat M(\tau, \bar \tau, u,\bar u, v,\bar v)=-i \left(\partial_{\bar \tau}
  \sqrt{2y}\right) e^{-2\pi (a-b)^2} \\
&\qquad \times \sum_{n\in \mathbb{Z}+\frac{1}{2}} (n+a-b)
(-1)^{n-\frac{1}{2}} \bar q^{n^2/2}e^{-2\pi i (\bar u-\bar v) n}.
\end{split}
\ee

A number of other useful properties are \cite{ZwegersThesis}:
\begin{enumerate} 
\item $M$ is anti-periodic under shifts by 1:
\be  
M(\tau,u+1,v)=M(\tau,u,v+1)=-M(\tau,u,v).
\ee
\item Inversion of the elliptic arguments leaves $M$ invariant:
\be 
\label{M-u-v}  
M(\tau,-u,-v)=M(\tau,u,v).
\ee
\item Simultaneous shifts of $u$ and $v$ by $z$
lead to a simple transformation of $M(\tau,u,v)$. For $u,v,u+z,v+z\neq
\mathbb{Z}\tau+\mathbb{Z}$, one has 
\be
\label{mushifts} 
M(\tau,u+z,v+z)-M(\tau,u,v)=\frac{i\,\eta^3(\tau)\,\vartheta_1(\tau,u+v+z)\,\vartheta_1(\tau,z)}{\vartheta_1(\tau,u)\,\vartheta_1(\tau,v)\,\vartheta_1(\tau,u+z)\,\vartheta_1(\tau,v+z)}.
\ee 
\item Quasi-periodicity property in $u$ and $v$:
\be
\label{Mqutr}
\begin{split}
&\widehat M(\tau,\bar \tau,u+k\tau+l,\bar u+k\bar \tau+l, v+m\tau+n,\bar v+m\bar \tau+n)\\
&\qquad = (-1)^{k+l+m+n} q^{(k-m)^2/2}e^{2\pi i (k-m)(u-v)}\widehat M(\tau,\bar\tau,u,\bar u,v,\bar v).
\end{split}
\ee   
\end{enumerate} 

\subsection*{Bi-modular forms}
The physical quantities of the $\CN=2^*$ theory are examples of
so-called bi-modular forms. See in particular \eqref{expdaduDelta*}. We review the definition due to Stienstra and Zagier of these
functions following \cite{2020arXiv200302572W}. See
also \cite{yang2007}. Let $f:\mathbb{H}\times \mathbb{H}\to
\mathbb{C}$ be a function in two variables, $\tau_1\in \mathbb{H}$ and
$\tau_2\in \mathbb{H}$. Let $\Gamma$ be a congruence subgroup, and
$\chi_1,\chi_2$ characters on $\Gamma$. The function $f$
is a bi-modular form of weight $(w_1,w_2)$ on
$(\Gamma,SL(2,\mathbb{Z}))$, if 
\begin{enumerate}
\item for $\gamma_1=\left( \begin{array}{cc}  a_1 & b_1 \\ c_1 &
      d_1 \end{array}  \right)$ and $\gamma_2=\left( \begin{array}{cc}  a_2 & b_2 \\ c_2 &
      d_2 \end{array}  \right) \in \Gamma$
\be f\!\left( \frac{a_1\tau_1+b_1}{c_1\tau_1+d_1},
    \frac{a_2\tau_2+b_2}{c_2\tau_2+d_2}
  \right)=\chi_1(\gamma_1)\,\chi_2(\gamma_2)\,(c_1\tau_1+d_1)^{w_1}(c_2\tau_2+d_2)^{w_2}\,
  f(\tau_1,\tau_2). \ee
\item for $\gamma=\left( \begin{array}{cc}  a & b \\ c &
      d \end{array}  \right) \in SL(2,\mathbb{Z})$
\be f\!\left( \frac{a\tau_1+b}{c\tau_1+d},
    \frac{a\tau_2+b}{c\tau_2+d}
  \right)=(c\tau_1+d)^{w_1}(c\tau_2+d)^{w_2}\,
  f(\tau_1,\tau_2). \ee
\end{enumerate} 

For the functions we encounter for the $\CN=2^*$ theory, $\Gamma=\Gamma(2)$.
Some of these functions have a pole for $\tau=\tau_{\rm uv}$ modulo
images of $\Gamma(2)$. Their proper domain is therefore
$\mathbb{H}^2 - \{(\tau, \tau_{\rm uv})\vert \tau=\gamma(\tau_{\rm uv}),\,\, \gamma\in \Gamma(2)\}$.

\section{Derivation of $\Delta_{\rm phys}$ and $da/du$}
\label{daduDelta}
This appendix gives a derivation of the closed
expressions for $u$, the discriminant $\Delta_{\rm phys}$, and $da/du$
for the $\CN=2^*$ theory. We will follow the normalization of the
variables as in \cite{Seiberg:1994aj} and \cite{Labastida:1998sk}, which differs from the one used
commonly in instanton counting \cite{Manschot:2019pog}. For large $u$, $a$ behaves as $a=\sqrt{u}+\dots$. 

To determine the IR gauge coupling $\tau$,  we first expand the
product on the rhs of (\ref{N2s2}). The SW curve $\Sigma$ then reads,
 \be \label{EllCurve}
\Sigma:\quad y^2=x^3-\tfrac{1}{6} E_4(\tau_{\rm uv})\, m^2\,  x^2- f\,x-g,
 \ee
 where $E_4$ is the weight 4 Eisenstein series, and $f$ and $g$ are functions of $u$, $\tau_{\rm uv}$ and $m$ obtained by expanding out the product in equation (\ref{N2s2})
 \be
 \begin{split}
 &f(\tau_{\rm uv},u,m)= -\sum_{i<j} (e_i(\tau_{\rm uv})\, u +\tfrac{1}{4}e_i(\tau_{\rm uv})^2\,m^2) (e_j(\tau_{\rm uv}) u +\tfrac{1}{4}e_j^2(\tau_{\rm uv})\,m^2), \\
 &g(\tau_{\rm uv},u,m)=\prod_{i=1}^3 (e_i(\tau_{\rm uv})\, u +\tfrac{1}{4}e_i(\tau_{\rm uv})^2\,m^2).
 \end{split}
 \ee
We then bring the curve (\ref{EllCurve}) to the curve $\tilde \Sigma$ in Weierstrass form
\be \label{Weierstrass}
\tilde \Sigma:\quad \tilde{y}^2 = 4\,\tilde{x}^3 - \tilde g_2\,\tilde{x} - \tilde g_3.
\ee
Whereas $x$ and $y$ have dimension $\Lambda^2$ and $\Lambda^3$, $\tilde{x}$ and $\tilde y$ are dimensionless. The coefficients are proportional to the Eisenstein
series $E_4$ and $E_6$: $\tilde g_2 = \frac{4\pi^4}{3}E_4(\tau)$ and $\tilde
g_3=\frac{8\pi^6}{27}E_6(\tau)$. This is achieved by the transformation
$x=4Z^{-2}\, \tilde x + \tfrac{1}{18}m^2 E_4(\tau_{\rm uv})$ and $y=4Z^{-3}\,
\tilde y$, where $Z$ is a dimensionful parameter which we will solve
for below. The coefficients $\tilde g_2$ and $\tilde g_3$ are given
in terms of $Z$ and the original parameters as
\be
\begin{split}
\tilde g_2(\tau_{\rm uv},u,m)&=\frac{Z^4}{4}\left( \frac{1}{108} E_4(\tau_{\rm uv})^2 m^4+f(u,\tau_{\rm uv},m) \right), \\
\tilde g_3(\tau_{\rm uv},u,m)&=  \frac{Z^6}{16}\left( g(u,\tau_{\rm uv},m)+ \frac{m^2}{18} E_4(\tau_{\rm uv}) f(u,\tau_{\rm uv},m) +\frac{2}{18^3}m^6 E_4(\tau_{\rm uv})^3  \right).
\end{split} 
\ee
Using these expressions, we can determine the (mathematical) discriminant of the
curve $(2\pi)^{12}\,\Delta(\tau) \equiv \tilde g_2^3-27 \tilde g_3^2$, which factorizes as
\be
(2\pi)^{12}\,\Delta(\tau)=Z^{12}\Delta(\tau_{\rm uv})\,(u-\tfrac{m^2}{4}e_1(\tau_{\rm uv}))^2\, (u-\tfrac{m^2}{4}e_2(\tau_{\rm uv}))^2\, (u-\tfrac{m^2}{4}e_3(\tau_{\rm uv}))^2.
\ee 
The discriminant vanishes at the three cusps of
$\mathbb{H}/\Gamma(2)$, where $u$ approaches
$\tfrac{m^2}{4}e_j(\tau_{\rm uv})$, $j=1,2,3$. To determine which value of $u$
corresponds to which cusp, note that in the scaling limit the
singularities for $\tau\to 0,1$ should correspond to $u\to
\tfrac{m^2}{4}e_{j}(\tau_{\rm uv})$ with $j$ either $2$ or 3, since these become the strong coupling
singularities at $u_0=\pm\Lambda^2$ of the pure $SU(2)$ theory by
(\ref{scaling_limit}). Thus for $\tau\to i\infty$, we necessarily have
$u\to \tfrac{m^2}{4}e_{1}(\tau_{\rm uv})$. Modular transformations fix
the other cusps, giving us
\be
\label{cuspstauu}
\begin{split}
&\lim_{\tau\to i\infty} u(\tau,\tau_{\rm uv})=\frac{m^2}{4}e_1(\tau_{\rm uv}),\\
&\lim_{\tau\to 0} u(\tau,\tau_{\rm uv})=\frac{m^2}{4}e_2(\tau_{\rm uv}),\\
&\lim_{\tau\to 1} u(\tau,\tau_{\rm uv})=\frac{m^2}{4}e_3(\tau_{\rm uv}).
\end{split}
\ee

Using these constraints, we can determine $u$ and $Z$ explicitly. The
equation for $\tilde g_2$ is a quadratic equation in $u$, which is
solved by
\be 
\label{abcforz}
u=\frac{-\frac{1}{12}E_6(\tau_{\rm uv})m^2\pm \sqrt{-12\Delta(\tau_{\rm uv})m^4+12E_4(\tau_{\rm uv})\tilde g_2(\tau)/Z^4}}{E_4(\tau_{\rm uv})}.
\ee
We will find later that the $-$ sign of $\pm$ matches with the convention used for $u$ in the main text. Taking the scaling limit for $u_0=u+\frac{1}{8} e_1(\tau_{\rm uv})m^2$
indeed reproduces $u_0$ of the pure Seiberg-Witten theory. Moreover, the zero of $E_4(\tau_\uv)$ for $\tau_\uv=i$ is a removable singularity.

Equation (\ref{abcforz}) together with the expressions for $u$ at the cusps
(\ref{cuspstauu}), give us expressions for $Z$ at the cusps:
\begin{itemize}
\item For $\tau\to \tau_{\rm uv}$, $u\to \infty$. Then 
\be \label{tautotau0} \lim_{\tau\to \tau_{\rm uv}} Z=0.\ee
\item For $\tau\to i\infty$, $\tilde g_2\to 4\pi^4/3$, and $u\to e_1(\tau_{\rm uv})\,m^2/4 $. Then 
\be
\label{tautoinfL}
\lim_{\tau\to i\infty} Z=\pm\frac{4\pi}{m}
\frac{1}{\vartheta_3(\tau_{\rm uv})^2\vartheta_4(\tau_{\rm uv})^2},
\ee
\item For $\tau\to 0$, we have $\tilde g_2\to 4\pi^4/3\,\tau^{-4}$ and $u\to e_2(\tau_{\rm uv})\,m^2/4$. Then
\be
\label{tauto0}
\lim_{\tau\to 0} Z\sim\frac{\pm}{\tau}\frac{4\pi}{m} \frac{1}{\vartheta_2(\tau_{\rm uv})^2 \vartheta_3(\tau_{\rm uv})^2},
\ee
\item For $\tau\to 1$, we have $\tilde g_2\to 4\pi^4/3\,(\tau-1)^{-4}$ and $u\to e_3(\tau_{\rm uv})\,m^2/4$. Then
\be \label{tauto1}
\lim_{\tau\to 1} Z\sim \frac{\pm}{\tau-1}\frac{4\pi}{m} \frac{1}{\vartheta_2(\tau_{\rm uv})^2 \vartheta_4(\tau_{\rm uv})^2}.
\ee
\end{itemize}
  
Next we make an Ansatz for $Z$ with unknown $K_i$'s, 
\be 
Z^4=\left(\frac{4\pi}{m}\right)^4
\left(\frac{K_1}{\vartheta_3(\tau_{\rm uv})^8\vartheta_4(\tau_{\rm uv})^8}
  +\frac{K_2}{\vartheta_2(\tau_{\rm uv})^8\vartheta_3(\tau_{\rm uv})^8}+\frac{K_3}{\vartheta_2(\tau_{\rm uv})^8\vartheta_4(\tau_{\rm uv})^8}\right),
\ee
where we assume that in each limit $\tau\to 0,1,i\infty$, one of the $K_i$ dominates.
To solve for the $K_i$'s, we make a further Ansatz: $K_i=\alpha_i \vartheta_2(\tau)^8+\beta_i \vartheta_3(\tau)^8+\gamma_i \vartheta_4(\tau)^8$. From the limits (\ref{tautotau0})-(\ref{tauto1}), we deduce the following relations for the $\alpha_i$, $\beta_i$ and $\gamma_i$'s:
\begin{itemize}
\item The scaling limit gives that these constants must satisfy $\alpha_2+\alpha_3= 1$, $\beta_2=-\beta_3$ and $\gamma_2=-\gamma_3$. 
\item The $\tau\to i\infty$ limit shows that $\beta_1+\gamma_1=1$, and $\beta_2+\gamma_2=0$. 
\item The $\tau\to 0$ limit shows that $\alpha_1+\beta_1=0$, $\alpha_2+\beta_2=1$ and $\alpha_3+\beta_3=0$.
\item The $\tau\to 1$ limit shows that $\alpha_1+\gamma_1=0$,
  $\alpha_2+\gamma_2=0$, $\alpha_3+\gamma_3=1$. 
\end{itemize}
These conditions are solved by
\be
\begin{split}
&\alpha_1=-1/2, \quad \beta_1=1/2,\quad \gamma_1=1/2,\\
&\alpha_2=1/2, \quad \beta_2=1/2, \quad \gamma_2=-1/2,\\
&\alpha_3=1/2,\quad \beta_3=-1/2,\quad \gamma_3=1/2.
\end{split}
\ee
Using the identity (\ref{t3=t2pt4}), we arrive for the $K_i$'s at
\be
K_1=\vartheta_3(\tau)^4 \vartheta_4(\tau)^4,\quad K_2=\vartheta_2(\tau)^4 \vartheta_3(\tau)^4, \quad K_3=-\vartheta_2(\tau)^4 \vartheta_4(\tau)^4.
\ee
Multiplying each term by $\vartheta_i(\tau_{\rm uv})^8$, we arrive at
$$
Z^4=\frac{(\pi/m)^4}{\Delta(\tau_{\rm uv})} \left( \vartheta_3(\tau)^4 \vartheta_4(\tau)^4\vartheta_2(\tau_{\rm uv})^8+ \vartheta_2(\tau)^4 \vartheta_3(\tau)^4\vartheta_4(\tau_{\rm uv})^8-\vartheta_2(\tau)^4 \vartheta_4(\tau)^4\vartheta_3(\tau_{\rm uv})^8\right).
$$
We can simplify this expression by taking the square root
\be 
\label{Z2}
Z^2=\frac{\pi^2}{m^2}\frac{1}{\eta(\tau_{\rm uv})^{12}} \left( \vartheta_4(\tau)^4\, \vartheta_3(\tau_{\rm uv})^4-\vartheta_3(\tau)^4\, \vartheta_4(\tau_{\rm uv})^4\right).
\ee
Using (\ref{Dtheta4}), we find that in the limit $\tau\to \tau_{\rm uv}$,
$Z^2$ vanishes as
\be
\label{Ztau0}
\lim_{\tau\to \tau_{\rm uv}} Z^2=-\frac{\pi^2}{m^2} (16\pi i (\tau-\tau_{\rm uv})+O((\tau-\tau_{\rm uv})^3).
\ee
To avoid ambiguities in taking a square root, we will assume that
$\mathrm{Im}(\tau-\tau_{\rm uv})>0$.

Substitution of (\ref{Z2}) in (\ref{abcforz}), gives $u$ as a
bi-modular form with arguments $\tau$ and $\tau_\uv$,
\be
\label{uBiMod}
\begin{split}
&u(\tau,\tau_{\rm uv})=\\
&\frac{-\frac{m^2}{12}E_6(\tau_{\rm
    uv})-2m^2\eta(\tau_{\rm uv})^{12}\,
  \frac{\vartheta_3(\tau)^4(\vartheta_3(\tau_{\rm
      uv})^4+\vartheta_2(\tau_{\rm
      uv})^4)+\vartheta_4(\tau)^4(\vartheta_4(\tau_{\rm
      uv})^4-\vartheta_2(\tau_{\rm uv})^4)}{\vartheta_3(\tau)^4
    \vartheta_4(\tau_{\rm uv})^4- \vartheta_4(\tau)^4
    \vartheta_3(\tau_{\rm uv})^4}}{E_4(\tau_{\rm uv})}.
\end{split}
\ee 
When $\SL$ acts on $\tau_{\rm uv}$ and $\tau$ simultaneously, $u$ is a
bi-modular form of weight two for $\tau_\uv$ and weight 0 for
$\tau$. On the other hand, $u$ transforms as a modular form for
$\Gamma(2)$, if the modular transformation acts on $\tau_{\rm uv}$ or $\tau$ separately. 
 
By comparing expansions, we find that expression \eqref{uBiMod} agrees with expressions for $u$ in the literature \cite{Huang:2011qx, Bonelli:2019boe}. The form of $u$ in \cite[Eq. (5.10)]{Huang:2011qx}
is\footnote{This expression differs by a sign from
  \cite{Huang:2011qx} to match our convention.}
\be 
\label{u2star}
\begin{split}
&u(\tau,\tau_{\rm uv})=-\frac{m^2}{4}\\
&\times \frac{e_1(\tau_{\rm
    uv})^2(e_2(\tau)-e_3(\tau))+e_2(\tau_{\rm
    uv})^2(e_3(\tau)-e_1(\tau))+e_3(\tau_{\rm
    uv})^2(e_1(\tau)-e_2(\tau))}{e_1(\tau_{\rm
    uv})(e_2(\tau)-e_3(\tau))+e_2(\tau_{\rm
    uv})(e_3(\tau)-e_1(\tau))+e_3(\tau_{\rm
    uv})(e_1(\tau)-e_2(\tau))},
\end{split}
\ee  
Reference \cite[Eq. (4.30)]{Bonelli:2019boe} gives yet
another expression for $u$ as a bi-modular form, if we identify $\tilde u$ in
\cite[Eq. (4.30)]{Bonelli:2019boe} with $4\pi^2\,u-m^2 \pi^2\,e_1(\tau_\uv)$ in
our notation. The equivalence of the expressions can be verified by
making $(q,q_{\uv})$-expansions. These expressions confirm moreover
 the leading behavior $u\sim a^2$ using the expression for $\tau$ (\ref{tauvxiexp}).  
   
In the scaling limit (\ref{scaling_limit}), we reproduce 
\be  
u+\frac{m^2}{8}\,e_1(\tau_{\rm uv}) \to u_0.
\ee  
The large $\tau_{\rm uv}$ expansion is 
\be
u=-\frac{m^2}{12}-2\,m^2\left(2\frac{\vartheta_3(\tau)^4}{\vartheta_2(\tau)^4}-1 \right)\,q_{\rm uv}^{1/2}+O(q_{\rm uv})
\ee
The expressions for $u$ allow to write the physical discriminant
$\Delta_{\rm phys}$ explicitly as a bi-modular form,
\be
\label{Deltaphys}
\begin{split}
\Delta_{\rm phys}&=\prod_{i=1}^3
(u-\tfrac{m^2}{4}e_i(\tau_{\rm uv}))=\left(\frac{2\pi}{Z}\right)^6\,\frac{\eta(\tau)^{12}}{\eta(\tau_{\rm uv})^{12}}\\
&= (2m)^6\,\frac{\eta(\tau_{\rm uv})^{24} \,\eta(\tau)^{12}}{\left(\vartheta_4(\tau)^4 \vartheta_3(\tau_{\rm uv})^4-\vartheta_3(\tau)^4 \vartheta_4(\tau_{\rm uv})^4\right)^{3}}.
\end{split}
\ee

Finally, we can show that the explicit expression for $Z$ (\ref{Z2}) provides
$da/du$. To this end, let $a$ be given by a contour integral along a one-cycle with
coordinate $z\in [0,1]$. Then,
\be
\label{daduZ}
\begin{split}
\frac{da}{du}&= \frac{1}{4\pi}\oint \frac{dx}{y}\\
&= \frac{1}{4\pi}\int_0^1 \frac{dx/dz}{y}\,dz\\
&=\frac{Z}{4\pi} \int_0^1 \frac{d\tilde x(z)}{\tilde y(z)}\\
&= \frac{Z}{4\pi},
\end{split}
\ee 
where the $\tilde x$ and $\tilde y$ on the third line are the variables of the
Weierstrass curve \eqref{Weierstrass}. Reference \cite{Labastida:1998sk} expressed $da/du$ in terms of $u$,
$m$, $\tau$ and $\tau_{\rm uv}$ as
\be  
\frac{da}{du}=\frac{1}{2\sqrt{u-\frac{1}{4} m^2
    e_3(\tau_{\rm uv})}}\frac{\vartheta_3(\tau)^2}{\vartheta_3(\tau_{\rm uv})^2}.
\ee  
Using the explicit expression for $da/du$ (\ref{daduZ}), this provides another equivalent expression for $u$ as
a bi-modular form.

We determine also $\frac{da}{d\tau}$, which is useful for the analysis
of the measure of the $u$-plane integral. We have
\be
\label{dadtau}
\begin{split}
\frac{1}{2\pi i}\frac{da}{d\tau}&=\frac{1}{2\pi i}\frac{da}{du} \frac{du}{d\tau}\\
& = \frac{8\,\Delta_{\rm phys}}{m^2} \,\left(\frac{da}{du}\right)^3\\
& =  \frac{8\,m\,\eta(\tau_{\rm uv})^6\,\eta(\tau)^{12}}{(\vartheta_4(\tau)^4\vartheta_3(\tau_{\rm uv})^4-\vartheta_3(\tau)^4\vartheta_4(\tau_{\rm uv})^4)^{3/2}},
\end{split} 
\ee 
where we substituted (\ref{Matone2*}). The equations demonstrate that $a$
has (approximate) weight 1 in $\tau$ and weight 0 in $\tau_{\rm
  uv}$. It furthermore confirms that the expansion of $2a/m$ in $\tau - \tau_\uv$ starts as
\be
\label{aAsymp}
\frac{2a}{m} = \frac{1}{\sqrt{-\pi i(\tau-\tau_{\rm uv})}} + O\!\left((\tau-\tau_{\rm uv})^{1/2}\right),
\ee
which is consistent with the leading term of (\ref{atautau0}).

\section{Almost complex four-manifolds}
\label{ACS4folds}
This appendix reviews a few aspects of almost complex four-manifolds. We refer to \cite{Scorpan, draghici2010, Gauduchon} for further reading. A four-manifold $X$ is almost complex if $X$ admits an almost complex structure, i.e. a fiber-preserving automorphism $\CJ:T_X\to T_X$ of the tangent bundle $T_X$ such that $\CJ^2=-1$. A metric $g$ is compatible with $\CJ$ if $g(\CJ u,\CJ v)=g(u,v)$, and such metrics always exist. An almost Hermitian manifold is the triple $(X,g,\CJ)$.
We let $\omega$ be the fundamental (1,1)-form compatible with the metric $\omega(u,v)=g(\CJ u,v)$. $\omega$ is self-dual but not necessarily closed for an ACS manifold. It defines the Lee form $\theta$, $d\omega=\theta\wedge \omega$ \cite{Gauduchon}. 
%Since $d^2\omega=0=d\theta\wedge \omega$, we find that $d\theta=0$ or $d\theta\perp\omega$. 

For an ACS four-manifold $X$, let $\pi_{p,q}$ be the canonical projection from $\Omega^{p+q}(X)$ to $\Omega^{p,q}$. We can write $\pi_{0,1}$ explicitly as $\pi_{0,1}=\frac{1}{2}(1+i\CJ)$ and $\pi_{1,0}=\frac{1}{2}(1-i\CJ)$. We define furthermore the Nijenhuis tensor $N_\CJ$ by
\be 
\label{Nijenhuis}
N_\CJ(u,v)=[\CJ u,\CJ v]-\CJ[u,\CJ v]-\CJ [\CJ u, v]-[u,v],
\ee 
where $u, v\in TX$ are two vector fields. This tensor gives the anti-holomorphic component of the commutator of two holomorphic vector fields plus its complex conjugate,
\be 
N_\CJ(u,v)=-4\left(\pi_{0,1}[\pi_{1,0}u,\pi_{1,0}v]+\pi_{1,0}[\pi_{0,1}u,\pi_{0,1}v]\right),
\ee 
or equivalently,
\be 
\pi_{0,1}[\pi_{1,0}u,\pi_{1,0}v]=-\frac{1}{4}\pi_{0,1}N(u,v).
\ee 
Thus $N_\CJ$ is the sum of a $(2,-1)$ and a $(-1,2)$ tensor, and $N_\CJ$ vanishes if $\CJ$ is integrable, i.e. if $X$ is a complex manifold.

The exterior derivative $d:\Omega^r(X)\to \Omega^{r+1}(X)$, can be written as the sum
\be 
d=\sum_{r+s=p+q+1} \pi_{p,q+1}\circ d.
\ee
In fact, for fixed $r$ and $s$, there are only four terms on the rhs
\be
\label{dACS}
d: \Omega^{r,s}\to \Omega^{r+2,s-1} \oplus \Omega^{r+1,s}\oplus \Omega^{r,s+1}\oplus \Omega^{r-1,s+2}.
\ee 
We define Dolbeault operators,
\be 
\bar \partial: \Omega^{p,q}(X)\to \Omega^{p,q+1}(X),\qquad \bar \partial=\pi_{p,q+1}\circ d,
\ee
which together with $\partial$ give the projections to the second and third  terms on the rhs of \eqref{dACS}. The projection to the first and last term is given in terms of the Nijenhuis tensor $N_\CJ$ (\ref{Nijenhuis}). To see this for a $(0,1)$-form $\alpha\in \Omega^{0,1}(X)$, we calculate for $u,v\in TX$
\be 
\pi_{2,0}\,d\alpha(u,v)=-\alpha(\pi_{0,1}[\pi_{1,0}u,\pi_{1,0}v])=\frac{1}{4}\alpha(N_\CJ(u,v)),
\ee
where we used $d\alpha(u,v)=u\alpha(v)-v\alpha(u)-\alpha([u,v])$. With respect to an orthonormal frame $e^{a}$, we have in components,
\be 
d\alpha=\partial\alpha+\bar \partial \alpha + (N_\CJ)^a_{bc}\,\alpha_a\,e^b\wedge e^c.
\ee 
If $N_\CJ$ is non-vanishing, $\bar\partial^2\neq 0$, and $d\neq \partial +\bar \partial$ acting on generic $(r,s)$-forms. For an ACS manifold, $d=\partial +\bar \partial$ only when acting $\Omega^0(X)$. 

We also introduce the adjoint Dolbeault operators, $\bar \partial^\dagger=-*\partial\,*: \Omega^{p,q}(X)\to \Omega^{p,q-1}(X)$ and $\partial^\dagger=-*\bar \partial\,*: \Omega^{p,q}(X)\to \Omega^{p-1,q}(X)$ with $*$ the Hodge $*:\Omega^{p,q}\to \Omega^{2-q,2-p}$. 
%A connection $\nabla$ is Hermitian if it preserves the metric and the ACS $\CJ$, $\nabla g=\nabla \CJ=0$. The Levi-Civita connection preserves $\CJ$ if $X$ is K\"ahler, but not more generally.

%\newpage
\section{Symbol list}
\label{AppSymbolList}

\vskip1in
\begin{longtable}{|p{2cm}|p{12cm}|}
\hline 
  $a,a_D$ &  Special K\"ahler coordinates in the canonical weak coupling frame at $u \to \infty$   \\
\hline 
$\bfb$  & Quantity used in the definition of
$\Psi^J_\mu(\tau,\bar\tau, \bfz, \bar \bfz)$ and $\widehat
\Theta^{J,J'}$, $\bfb={\rm Im}(\bfz)/{\rm Im}(\tau)$. Equation \eqref{Defb}. \\
\hline
$B$ &  Intersection product or bilinear form on $H^2(X)$. \\
\hline
    $\CB$ &  The Coulomb branch of the $N=2^*$ theory for $G=SU(2)$. In this theory it can be identified with $\BC - \{u_1,u_2,u_3\}$.    \\
\hline 
$c_{\rm ir}$ & Characteristic class of the ${\rm Spin}^c$ structure on $X$ in one of the low energy effective theories valid near $u_j$.   \\
\hline 
$c_{\uv}$ & Characteristic class of the ${\rm Spin}^c$ structure $\mathfrak{s}_\uv$ on $X$ needed for topological twisting.   \\
\hline 
$C$ & A bi-modular form needed to construct the measure in the
presence of a background ${\rm Spin}^c$ connection for $\mathfrak{s}_{\uv}$. See
equations (\ref{Cexpxi}) and \eqref{CphijC}; Also, a field in the Vafa-Witten equations.  \\
\hline
$\chi$ &  Euler character of $X$.  \\
\hline
$\chi_{\rm h}$ & Holomorphic Euler characteristic of $X$, equal to
$\frac{1}{4}(\chi+\sigma)$. Eq. \eqref{holchi}. \\
\hline
$\CF$ &  Prepotential in the canonical weak coupling frame at $u\to
\infty$. See Eqs \eqref{FPertplusInst} and \eqref{Fmassexp}. \\
\hline 
$\tilde \CF$ &  Modified prepotential taking into account $U(1)$ coupling to mass. Needed for modular properties and proper $m\to 0,\infty$ limits.
See Eqs \eqref{tildeFeps} and \eqref{tildeF}.   \\
\hline  
$f_\mu(\tau,\bar\tau, \rho, \bar \rho)$ &  Unary theta series. See Equation \eqref{fmutbart}. \\
\hline
$\Phi_\bfmu^J$ & The Coulomb branch (a.k.a. $u$-plane)
integral. Equations \eqref{uintegral} and \eqref{UplanetvPsi}. \\
\hline
$\BH$ & Upper half of the complex plane.  \\
\hline
$J$ &  Period point $J\in H^2(X, \BR)$ when $b_2^+=1$. Note $*J=J$ and $J^2=1$ and $J$ is in the forward lightcone. (This requires a choice of orientation on $H^2(X,\BR)$). \\
\hline
$\CJ$ & Almost complex structure.\\
\hline 
$K_0$ &  Choice of ${\rm Spin}^c$ class that arises in comparing the $m\to \infty$ limit of the $\CN=2^*$-theory with Donaldson theory. See equation (\ref{PsiSW}). \\
\hline
$\bfk$ &  element of $\bfmu+ L$ \\
\hline
$\bfk_+,\bfk_- $ &   Self-dual, anti-self-dual projections of $\bfk$. \\
\hline 
$\bfk_m$ &   $2\bfk_m = c_\uv$ \\
\hline
$\kappa$  & An unknown numerical constant in the normalization of the
Coulomb branch integral. By comparing with mathematical results we
will find $\kappa=1$. \\
\hline
$\ell$ & Complex dimension of moduli space of Abelian monopoles \\
 & Equal to $(c_\uv^2-2\chi-3\sigma)/8$ \\
\hline 
$L$ &  $L= H^2(X,\BZ)$ \\
\hline 
$\lambda$ & Short hand for $2\chi+3\sigma$. \\
\hline
$\Lambda$ & Scale in the $\CN=2^*$ theory. See Equation
\eqref{scaling_limit} and \eqref{Fpert}.   \\
\hline 
$\Lambda_0$ & Scale in the pure $N_f=0$ SYM theory. See Equation
\eqref{N=2curve}.  \\
\hline 
$m$ &  Mass parameter in the $\CN=2^*$ theory.   \\
\hline 
$M$ &  Monopole field in the Seiberg-Witten equations. It is a section of $W^+$. It arises from the scalar fields in the adjoint-valued hypermultiplet.   \\
\hline
$\bfmu$ &   $2\bfmu$ is an integral lift of the 't Hooft flux of the gauge bundle. (i.e. $2\bfmu$ is an integral lift of $w_2(P)$ for a principal $SO(3)$ bundle $P\to X$. \\
\hline
$N_\CJ$ & Nijenhuis tensor\\
\hline 
$\nu$ &  holomorphic part of the Coulomb branch measure. See Equation \eqref{tnu}.
\\
\hline 
$p$ &  The fugacity for the point class. See Equation \eqref{epu}. \\
\hline
$\Psi_\bfmu^J$ &  The ``theta'' function entering the Coulomb branch
measure. See Equations \eqref{defPsi3} and \eqref{defPsi4}. \\ 
\hline
$q_\uv$ &  The UV coupling constant, $q_\uv = e^{2\pi i \tau_\uv}$. \\
\hline
$R^X_\bfmu$ & Multiplicative factor for the limit from $\CN=2^*$ to
$N_f=0$. \\
\hline
$s$ & Shorthand for $-\pi i(\tau-\tau_\uv)$. See Equation \eqref{atautau0}.  \\
\hline
$S$ & A generic homology class. Used for surface observables defined
in Equations \eqref{UVSobs} and \eqref{tildeIMx}. \\
\hline 
$\sigma$ &  The signature of $X$ \\
\hline 
$SW(c_{\rm ir})$ &  Seiberg-Witten invariant of the IR ${\rm Spin}^c$
structure with class $c_{\rm ir}$. \\
\hline
$t$ & Shorthand for the ratio $m/\Lambda$.\\
\hline 
$\tau$ & A generic modular parameter $\tau \in \BH$. Also used for the modular parameter of the Seiberg-Witten elliptic 
curve, in which case it is a function of $u,m,\tau_\uv$. \\
\hline
 $\tau_{\uv}$ & The UV coupling constant, $\tau_{\uv} =
 \frac{\theta_\uv}{2\pi} + i\frac{ 4 \pi}{e_{\uv}^2}$.\\
\hline 
$\hat \Theta^{J,J'}$ &   Indefinite theta series, useful for writing
the difference of measures as a total derivative. See Equation \eqref{ThetaComplete}.\\
\hline 
$u$ &  General coordinate on the Coulomb branch $u\in \CB$. But it can
also be regarded as a function of $m, \tau, \tau_\uv$. See Equation \eqref{u2star}. \\
\hline
$u_1, u_2, u_3$ &   Points in the discriminant locus of the Seiberg-Witten family of elliptic curves. \\
\hline 
$v, v(\tau, \tau_\uv)$ &   $ v= \frac{\partial^2 \CF}{\p a \p m} =
(a_D-\tau a)/m$. Coupling of low energy dynamical $U(1)$ fieldstrength
to background ${\rm Spin}^c$ connection fieldstrength. See Equations
\eqref{2*couplings} and \eqref{vaDat}. \\ 
\hline 
$W^\pm$ & Rank two chiral spin bundles associated to a ${\rm Spin}^c$ structure on $X$. \\
\hline 
$X$ &  A compact, oriented, connected, simply-connected smooth four-manifold with $b_2^+$ odd and positive. \\
\hline 
$\xi$ &   $\xi = \frac{\p^2 \CF}{\p^2 m}$ \eqref{2*couplings}. \\
\hline 
$Z^J_{SW,j,\mu}$ & Contribution of the LEET near $u_j$ to the
partition function $Z^J_{\mu}$. See equations \eqref{SW1Final},
\eqref{u2DPSfinal} and \eqref{u3DPSfinal}. \\
\hline 
\end{longtable}   
\vskip1in

%\bibliography{N2star} 

\begin{thebibliography}{100}

\bibitem{Witten:1988ze}
E.~Witten, \emph{{Topological Quantum Field Theory}},
  \href{http://dx.doi.org/10.1007/BF01223371}{\emph{Commun. Math. Phys.} {\bf
  117} (1988) 353}.

\bibitem{Seiberg:1994rs}
N.~Seiberg and E.~Witten, \emph{{Electric - magnetic duality, monopole
  condensation, and confinement in N=2 supersymmetric Yang-Mills theory}},
  \href{http://dx.doi.org/10.1016/0550-3213(94)90124-4,
  10.1016/0550-3213(94)00449-8}{\emph{Nucl. Phys.} {\bf B426} (1994) 19--52},
  [\href{https://arxiv.org/abs/hep-th/9407087}{{\tt hep-th/9407087}}].

\bibitem{Seiberg:1994aj}
N.~Seiberg and E.~Witten, \emph{{Monopoles, duality and chiral symmetry
  breaking in N=2 supersymmetric QCD}},
  \href{http://dx.doi.org/10.1016/0550-3213(94)90214-3}{\emph{Nucl. Phys.} {\bf
  B431} (1994) 484--550}, [\href{https://arxiv.org/abs/hep-th/9408099}{{\tt
  hep-th/9408099}}].

\bibitem{Witten:1994cg}
E.~Witten, \emph{{Monopoles and four-manifolds}},
  \href{http://dx.doi.org/10.4310/MRL.1994.v1.n6.a13}{\emph{Math. Res. Lett.}
  {\bf 1} (1994) 769--796}, [\href{https://arxiv.org/abs/hep-th/9411102}{{\tt
  hep-th/9411102}}].

\bibitem{Moore:1997pc}
G.~W. Moore and E.~Witten, \emph{{Integration over the u plane in Donaldson
  theory}}, {\emph{Adv. Theor. Math. Phys.} {\bf 1} (1997) 298--387},
  [\href{https://arxiv.org/abs/hep-th/9709193}{{\tt hep-th/9709193}}].

\bibitem{Scorpan}
A.~Scorpan, \emph{The wild world of 4-manifolds}.
\newblock Wiley Classics Library. American Mathematical Society, 2005.

\bibitem{Labastida:1997rg}
J.~Labastida and M.~Marino, \emph{{Twisted baryon number in N=2 supersymmetric
  QCD}}, \href{http://dx.doi.org/10.1016/S0370-2693(97)00376-6}{\emph{Phys.
  Lett. B} {\bf 400} (1997) 323--330},
  [\href{https://arxiv.org/abs/hep-th/9702054}{{\tt hep-th/9702054}}].

\bibitem{LoNeSha}
A.~Losev, N.~Nekrasov and S.~L. Shatashvili, \emph{{Issues in topological gauge
  theory}}, \href{http://dx.doi.org/10.1016/S0550-3213(98)00628-2}{\emph{Nucl.
  Phys.} {\bf B534} (1998) 549--611},
  [\href{https://arxiv.org/abs/hep-th/9711108}{{\tt hep-th/9711108}}].

\bibitem{Atiyah:1990tm}
M.~Atiyah and L.~Jeffrey, \emph{{Topological Lagrangians and cohomology}},
  \href{http://dx.doi.org/10.1016/0393-0440(90)90023-V}{\emph{J.\ Geom.\ Phys.}
  {\bf 7} (1990) 119--136}.

\bibitem{Cordes:1994fc}
S.~Cordes, G.~W. Moore and S.~Ramgoolam, \emph{{Lectures on 2-d Yang-Mills
  theory, equivariant cohomology and topological field theories}},
  \href{http://dx.doi.org/10.1016/0920-5632(95)00434-B}{\emph{Nucl. Phys. Proc.
  Suppl.} {\bf 41} (1995) 184--244},
  [\href{https://arxiv.org/abs/hep-th/9411210}{{\tt hep-th/9411210}}].

\bibitem{Laba05}
J.~Labastida and M.~Marino, \emph{{Topological quantum field theory and four
  manifolds}}.
\newblock Springer Netherlands, 2005.

\bibitem{MooreNotes2017}
G.~W. Moore, \emph{{Lectures On The Physical Approach To Donaldson And
  Seiberg-Witten Invariants}}, {\emph{Item 78 at
  http://www.physics.rutgers.edu/~gmoore/} (2017) }.

\bibitem{Tanaka:2017jom}
Y.~Tanaka and R.~P. Thomas, \emph{{Vafa-Witten invariants for projective
  surfaces I: stable case}},  \href{https://arxiv.org/abs/1702.08487}{{\tt
  1702.08487}}.

\bibitem{Gholampour:2017bxh}
A.~Gholampour, A.~Sheshmani and S.-T. Yau, \emph{{Localized Donaldson-Thomas
  theory of surfaces}}, {\emph{Am. J. Math.} {\bf 142} (2020) 2},
  [\href{https://arxiv.org/abs/1701.08902}{{\tt 1701.08902}}].

\bibitem{Vafa:1994tf}
C.~Vafa and E.~Witten, \emph{{A Strong coupling test of S duality}},
  \href{http://dx.doi.org/10.1016/0550-3213(94)90097-3}{\emph{Nucl. Phys.} {\bf
  B431} (1994) 3--77}, [\href{https://arxiv.org/abs/hep-th/9408074}{{\tt
  hep-th/9408074}}].

\bibitem{Dijkgraaf:1997ce}
R.~Dijkgraaf, J.-S. Park and B.~J. Schroers, \emph{{N=4 supersymmetric
  Yang-Mills theory on a Kahler surface}},
  \href{https://arxiv.org/abs/hep-th/9801066}{{\tt hep-th/9801066}}.

\bibitem{Labastida:1998sk}
J.~M.~F. Labastida and C.~Lozano, \emph{{Duality in twisted N=4 supersymmetric
  gauge theories in four-dimensions}},
  \href{http://dx.doi.org/10.1016/S0550-3213(98)00653-1}{\emph{Nucl. Phys.}
  {\bf B537} (1999) 203--242},
  [\href{https://arxiv.org/abs/hep-th/9806032}{{\tt hep-th/9806032}}].

\bibitem{Gottsche:2017vxs}
L.~G{\"o}ttsche and M.~Kool, \emph{{Virtual refinements of the Vafa-Witten
  formula}}, \href{http://dx.doi.org/10.1007/s00220-020-03748-7}{\emph{Commun.
  Math. Phys.} {\bf 376} (2020) 1--49},
  [\href{https://arxiv.org/abs/1703.07196}{{\tt 1703.07196}}].

\bibitem{Gottsche:2019vbi}
L.~G{\"o}ttsche, M.~Kool and R.~A. Williams, \emph{{Verlinde formulae on
  complex surfaces: K-theoretic invariants}},
  \href{http://dx.doi.org/10.1017/fms.2020.50}{\emph{Forum of Mathematics,
  Sigma} {\bf 9} (2021) e5}.

\bibitem{Gottsche:2020ale}
L.~Göttsche and M.~Kool, \emph{{Sheaves on surfaces and virtual invariants}},
  \href{https://arxiv.org/abs/2007.12730}{{\tt 2007.12730}}.

\bibitem{Witten:1995gf}
E.~Witten, \emph{{On S duality in Abelian gauge theory}},
  \href{http://dx.doi.org/10.1007/BF01671570}{\emph{Selecta Math.} {\bf 1}
  (1995) 383}, [\href{https://arxiv.org/abs/hep-th/9505186}{{\tt
  hep-th/9505186}}].

\bibitem{Shapere:2008zf}
A.~D. Shapere and Y.~Tachikawa, \emph{{Central charges of N=2 superconformal
  field theories in four dimensions}},
  \href{http://dx.doi.org/10.1088/1126-6708/2008/09/109}{\emph{JHEP} {\bf 09}
  (2008) 109}, [\href{https://arxiv.org/abs/0804.1957}{{\tt 0804.1957}}].

\bibitem{Nelson:1993nf}
A.~E. Nelson and N.~Seiberg, \emph{{R symmetry breaking versus supersymmetry
  breaking}}, \href{http://dx.doi.org/10.1016/0550-3213(94)90577-0}{\emph{Nucl.
  Phys. B} {\bf 416} (1994) 46--62},
  [\href{https://arxiv.org/abs/hep-ph/9309299}{{\tt hep-ph/9309299}}].

\bibitem{Korpas:2019cwg}
G.~Korpas, J.~Manschot, G.~W. Moore and I.~Nidaiev, \emph{{Mocking the
  $u$-plane integral}},  \href{https://arxiv.org/abs/1910.13410}{{\tt
  1910.13410}}.

\bibitem{Manschot:2007ha}
J.~Manschot and G.~W. Moore, \emph{{A Modern Farey Tail}},
  \href{http://dx.doi.org/10.4310/CNTP.2010.v4.n1.a3}{\emph{Commun. Num. Theor.
  Phys.} {\bf 4} (2010) 103--159}, [\href{https://arxiv.org/abs/0712.0573}{{\tt
  0712.0573}}].

\bibitem{Troost:2010ud}
J.~Troost, \emph{{The non-compact elliptic genus: mock or modular}},
  \href{http://dx.doi.org/10.1007/JHEP06(2010)104}{\emph{JHEP} {\bf 06} (2010)
  104}, [\href{https://arxiv.org/abs/1004.3649}{{\tt 1004.3649}}].

\bibitem{Eguchi:2010cb}
T.~Eguchi and Y.~Sugawara, \emph{{Non-holomorphic Modular Forms and
  SL(2,R)/U(1) Superconformal Field Theory}},
  \href{http://dx.doi.org/10.1007/JHEP03(2011)107}{\emph{JHEP} {\bf 03} (2011)
  107}, [\href{https://arxiv.org/abs/1012.5721}{{\tt 1012.5721}}].

\bibitem{Murthy:2013mya}
S.~Murthy, \emph{{A holomorphic anomaly in the elliptic genus}},
  \href{http://dx.doi.org/10.1007/JHEP06(2014)165}{\emph{JHEP} {\bf 06} (2014)
  165}, [\href{https://arxiv.org/abs/1311.0918}{{\tt 1311.0918}}].

\bibitem{Gaiotto:2019gef}
D.~Gaiotto and T.~Johnson-Freyd, \emph{{Mock modularity and a secondary
  elliptic genus}},  \href{https://arxiv.org/abs/1904.05788}{{\tt 1904.05788}}.

\bibitem{Dabholkar:2020fde}
A.~Dabholkar, P.~Putrov and E.~Witten, \emph{{Duality and Mock Modularity}},
  \href{http://dx.doi.org/10.21468/SciPostPhys.9.5.072}{\emph{SciPost Phys.}
  {\bf 9} (2020) 072}, [\href{https://arxiv.org/abs/2004.14387}{{\tt
  2004.14387}}].

\bibitem{Gottsche:1996}
L.~G\"ottsche, \emph{Modular forms and donaldson invariants for 4-manifolds
  with $b_2^+= 1$}, {\emph{Journal of the American Mathematical Society} {\bf
  9} (1996) 827--843}, [\href{https://arxiv.org/abs/alg-geom/9506018}{{\tt
  alg-geom/9506018}}].

\bibitem{Minahan:1998vr}
J.~Minahan, D.~Nemeschansky, C.~Vafa and N.~Warner, \emph{{E strings and N=4
  topological Yang-Mills theories}},
  \href{http://dx.doi.org/10.1016/S0550-3213(98)00426-X}{\emph{Nucl. Phys. B}
  {\bf 527} (1998) 581--623}, [\href{https://arxiv.org/abs/hep-th/9802168}{{\tt
  hep-th/9802168}}].

\bibitem{Alim:2010cf}
M.~Alim, B.~Haghighat, M.~Hecht, A.~Klemm, M.~Rauch and T.~Wotschke,
  \emph{{Wall-crossing holomorphic anomaly and mock modularity of multiple
  M5-branes}}, \href{http://dx.doi.org/10.1007/s00220-015-2436-3}{\emph{Commun.
  Math. Phys.} {\bf 339} (2015) 773--814},
  [\href{https://arxiv.org/abs/1012.1608}{{\tt 1012.1608}}].

\bibitem{Manschot:2011dj}
J.~Manschot, \emph{{BPS invariants of $\CN=4$ gauge theory on Hirzebruch
  surfaces}}, \href{http://dx.doi.org/10.4310/CNTP.2012.v6.n2.a4}{\emph{Commun.
  Num. Theor. Phys.} {\bf 6} (2012) 497--516},
  [\href{https://arxiv.org/abs/1103.0012}{{\tt 1103.0012}}].

\bibitem{Yoshioka1994}
K.~Yoshioka, \emph{The betti numbers of the moduli space of stable sheaves of
  rank2 on p2.}, {\emph{Journal fur die reine und angewandte Mathematik} {\bf
  453} (1994) 193--220}.

\bibitem{Klyachko1991}
A.~Klyachko, \emph{Moduli of vector bundles and numbers of classes},
  {\emph{Funct. Anal. and Appl.} {\bf 25} (1991) }.

\bibitem{Zagier:1975}
D.~Zagier, \emph{{Nombres de classes et formes modulaires de poids 3/2}},
  {\emph{C. R. Acad. Sc. Paris} {\bf 281} (1975) 883--886}.

\bibitem{Alexandrov:2016tnf}
S.~Alexandrov, S.~Banerjee, J.~Manschot and B.~Pioline, \emph{{Multiple
  D3-instantons and mock modular forms I}},
  \href{http://dx.doi.org/10.1007/s00220-016-2799-0}{\emph{Commun. Math. Phys.}
  {\bf 353} (2017) 379--411}, [\href{https://arxiv.org/abs/1605.05945}{{\tt
  1605.05945}}].

\bibitem{Alexandrov:2019rth}
S.~Alexandrov, J.~Manschot and B.~Pioline, \emph{{S-duality and refined BPS
  indices}}, \href{http://dx.doi.org/10.1007/s00220-020-03854-6}{\emph{Commun.
  Math. Phys.} {\bf 380} (2020) 755--810},
  [\href{https://arxiv.org/abs/1910.03098}{{\tt 1910.03098}}].

\bibitem{Bonelli:2020xps}
G.~Bonelli, F.~Fucito, J.~F. Morales, M.~Ronzani, E.~Sysoeva and A.~Tanzini,
  \emph{{Gauge theories on compact toric manifolds}},
  \href{https://arxiv.org/abs/2007.15468}{{\tt 2007.15468}}.

\bibitem{Dorey:1996ez}
N.~Dorey, V.~V. Khoze and M.~P. Mattis, \emph{{On mass deformed N=4
  supersymmetric Yang-Mills theory}},
  \href{http://dx.doi.org/10.1016/S0370-2693(97)00102-0}{\emph{Phys. Lett.}
  {\bf B396} (1997) 141--149},
  [\href{https://arxiv.org/abs/hep-th/9612231}{{\tt hep-th/9612231}}].

\bibitem{2020arXiv200302572W}
L.~{Wang} and Y.~{Yang}, \emph{{Ramanujan-type $1/\pi$-series from bimodular
  forms}}, {\emph{arXiv e-prints} (Mar., 2020) arXiv:2003.02572},
  [\href{https://arxiv.org/abs/2003.02572}{{\tt 2003.02572}}].

\bibitem{yang2007}
Y.~Yang and N.~Yui, \emph{Differential equations satisfied by modular forms and
  $k3$ surfaces},
  \href{http://dx.doi.org/10.1215/ijm/1258138437}{\emph{Illinois J. Math.} {\bf
  51} (04, 2007) 667--696}.

\bibitem{Huang:2011qx}
M.-x. Huang, A.-K. Kashani-Poor and A.~Klemm, \emph{{The $\Omega$ deformed
  B-model for rigid $\mathcal{N}=2$ theories}},
  \href{http://dx.doi.org/10.1007/s00023-012-0192-x}{\emph{Annales Henri
  Poincare} {\bf 14} (2013) 425--497},
  [\href{https://arxiv.org/abs/1109.5728}{{\tt 1109.5728}}].

\bibitem{Bonelli:2019boe}
G.~Bonelli, F.~Del~Monte, P.~Gavrylenko and A.~Tanzini, \emph{{${\mathcal {N}}$
  = $2^*$ Gauge Theory, Free Fermions on the Torus and Painlev\'e VI}},
  \href{http://dx.doi.org/10.1007/s00220-020-03743-y}{\emph{Commun. Math.
  Phys.} {\bf 377} (2020) 1381--1419},
  [\href{https://arxiv.org/abs/1901.10497}{{\tt 1901.10497}}].

\bibitem{Minahan:1997if}
J.~A. Minahan, D.~Nemeschansky and N.~P. Warner, \emph{{Instanton expansions
  for mass deformed N=4 superYang-Mills theories}},
  \href{http://dx.doi.org/10.1016/S0550-3213(98)00314-9}{\emph{Nucl. Phys.}
  {\bf B528} (1998) 109--132},
  [\href{https://arxiv.org/abs/hep-th/9710146}{{\tt hep-th/9710146}}].

\bibitem{DHoker:1997hut}
E.~D'Hoker and D.~H. Phong, \emph{{Calogero-Moser systems in SU(N)
  Seiberg-Witten theory}},
  \href{http://dx.doi.org/10.1016/S0550-3213(97)00763-3}{\emph{Nucl. Phys.}
  {\bf B513} (1998) 405--444},
  [\href{https://arxiv.org/abs/hep-th/9709053}{{\tt hep-th/9709053}}].

\bibitem{DHoker:1999hmo}
E.~D'Hoker and D.~H. Phong, \emph{{Seiberg-Witten theory and Calogero-Moser
  systems}}, \href{http://dx.doi.org/10.1143/PTPS.135.75}{\emph{Prog. Theor.
  Phys. Suppl.} {\bf 135} (1999) 75--93},
  [\href{https://arxiv.org/abs/hep-th/9906027}{{\tt hep-th/9906027}}].

\bibitem{Manschot:2019pog}
J.~Manschot, G.~W. Moore and X.~Zhang, \emph{{Effective gravitational couplings
  of four-dimensional $ \mathcal{N} $ = 2 supersymmetric gauge theories}},
  \href{http://dx.doi.org/10.1007/JHEP06(2020)150}{\emph{JHEP} {\bf 06} (2020)
  150}, [\href{https://arxiv.org/abs/1912.04091}{{\tt 1912.04091}}].

\bibitem{Nakajima:2003uh}
H.~Nakajima and K.~Yoshioka, \emph{{Lectures on instanton counting}},  in
  \emph{{CRM Workshop on Algebraic Structures and Moduli Spaces Montreal,
  Canada, July 14-20, 2003}}, 2003.
\newblock \href{https://arxiv.org/abs/math/0311058}{{\tt math/0311058}}.

\bibitem{Ne}
N.~A. Nekrasov, \emph{{Seiberg-Witten prepotential from instanton counting}},
  \href{http://dx.doi.org/10.4310/ATMP.2003.v7.n5.a4}{\emph{Adv. Theor. Math.
  Phys.} {\bf 7} (2003) 831--864},
  [\href{https://arxiv.org/abs/hep-th/0206161}{{\tt hep-th/0206161}}].

\bibitem{Moore:1997dj}
G.~W. Moore, N.~Nekrasov and S.~Shatashvili, \emph{{Integrating over Higgs
  branches}}, \href{http://dx.doi.org/10.1007/PL00005525}{\emph{Commun. Math.
  Phys.} {\bf 209} (2000) 97--121},
  [\href{https://arxiv.org/abs/hep-th/9712241}{{\tt hep-th/9712241}}].

\bibitem{Alday:2009aq}
L.~F. Alday, D.~Gaiotto and Y.~Tachikawa, \emph{{Liouville Correlation
  Functions from Four-dimensional Gauge Theories}},
  \href{http://dx.doi.org/10.1007/s11005-010-0369-5}{\emph{Lett. Math. Phys.}
  {\bf 91} (2010) 167--197}, [\href{https://arxiv.org/abs/0906.3219}{{\tt
  0906.3219}}].

\bibitem{Nekrasov:2015wsu}
N.~Nekrasov, \emph{{BPS/CFT correspondence: non-perturbative Dyson-Schwinger
  equations and qq-characters}},
  \href{http://dx.doi.org/10.1007/JHEP03(2016)181}{\emph{JHEP} {\bf 03} (2016)
  181}, [\href{https://arxiv.org/abs/1512.05388}{{\tt 1512.05388}}].

\bibitem{Harvey:1996ir}
J.~A. Harvey and G.~W. Moore, \emph{{Five-brane instantons and R**2 couplings
  in N=4 string theory}},
  \href{http://dx.doi.org/10.1103/PhysRevD.57.2323}{\emph{Phys. Rev.} {\bf D57}
  (1998) 2323--2328}, [\href{https://arxiv.org/abs/hep-th/9610237}{{\tt
  hep-th/9610237}}].

\bibitem{Bershtein:2015xfa}
M.~Bershtein, G.~Bonelli, M.~Ronzani and A.~Tanzini, \emph{{Exact results for $
  \mathcal{N} $ = 2 supersymmetric gauge theories on compact toric manifolds
  and equivariant Donaldson invariants}},
  \href{http://dx.doi.org/10.1007/JHEP07(2016)023}{\emph{JHEP} {\bf 07} (2016)
  023}, [\href{https://arxiv.org/abs/1509.00267}{{\tt 1509.00267}}].

\bibitem{Eichler:1985}
M.~M. Eichler and D.~Zagier, \emph{The theory of Jacobi forms}.
\newblock Progress in mathematics v. 55. Birkhauser, Boston, 1st ed. 1985.~ed.,
  1985.

\bibitem{Gaiotto:2010be}
D.~Gaiotto, G.~W. Moore and A.~Neitzke, \emph{{Framed BPS States}},
  \href{http://dx.doi.org/10.4310/ATMP.2013.v17.n2.a1}{\emph{Adv. Theor. Math.
  Phys.} {\bf 17} (2013) 241--397},
  [\href{https://arxiv.org/abs/1006.0146}{{\tt 1006.0146}}].

\bibitem{Aharony:2013hda}
O.~Aharony, N.~Seiberg and Y.~Tachikawa, \emph{{Reading between the lines of
  four-dimensional gauge theories}},
  \href{http://dx.doi.org/10.1007/JHEP08(2013)115}{\emph{JHEP} {\bf 08} (2013)
  115}, [\href{https://arxiv.org/abs/1305.0318}{{\tt 1305.0318}}].

\bibitem{Seiberg:2013}
N.~Seiberg, ``Reading between the lines of four-dimensional gauge theories.''
  Talk at Strings 2013, Korea, 2013.

\bibitem{Ang:2019txy}
J.~Ang, K.~Roumpedakis and S.~Seifnashri, \emph{{Line Operators of Gauge
  Theories on Non-Spin Manifolds}},
  \href{http://dx.doi.org/10.1007/JHEP04(2020)087}{\emph{JHEP} {\bf 04} (2020)
  087}, [\href{https://arxiv.org/abs/1911.00589}{{\tt 1911.00589}}].

\bibitem{Donaldson90}
S.~K. Donaldson and P.~B. Kronheimer, \emph{The geometry of four-manifolds /
  S.K. Donaldson and P.B. Kronheimer}.
\newblock Clarendon Press ; Oxford University Press Oxford : New York, 1990.

\bibitem{draghici2010}
T.~{Draghici}, T.~{Li} and W.~{Zhang}, \emph{Symplectic forms and cohomology
  decomposition of almost complex four-manifolds},
  \href{http://dx.doi.org/10.1093/imrn/rnp113}{\emph{International Mathematics
  Research Notices} {\bf 2010} (2010) 1--17}.

\bibitem{Morgan}
J.~Morgan, \emph{The Seiberg-Witten equations and applications to the topology
  of smooth four- manifolds}, vol.~44 of \emph{Math. Notes}.
\newblock Princeton University Press, 1996.

\bibitem{TeichnerSpinc}
P.~Teichner and E.~Vogt, ``All four-manifolds have spin$^c$ structures.''
  https://math.berkeley.edu/~teichner/Papers/spin.pdf, 1994.

\bibitem{Hyun:1995hz}
S.~Hyun, J.~Park and J.-S. Park, \emph{{N=2 supersymmetric QCD and four
  manifolds: 1. The Donaldson and Seiberg-Witten invariants}},
  \href{https://arxiv.org/abs/hep-th/9508162}{{\tt hep-th/9508162}}.

\bibitem{Labastida:1996tz}
J.~M.~F. Labastida and M.~Marino, \emph{{Twisted N=2 supersymmetry with central
  charge and equivariant cohomology}},
  \href{http://dx.doi.org/10.1007/s002200050081}{\emph{Commun. Math. Phys.}
  {\bf 185} (1997) 37--71}, [\href{https://arxiv.org/abs/hep-th/9603169}{{\tt
  hep-th/9603169}}].

\bibitem{Pidstrigach:1995dg}
V.~{Pidstrigach} and A.~{Tyurin}, \emph{{Localisation of the Donaldson's
  invariants along Seiberg-Witten classes}}, {\emph{eprint arXiv:dg-ga/9507004}
  (July, 1995) dg--ga/9507004},
  [\href{https://arxiv.org/abs/dg-ga/9507004}{{\tt dg-ga/9507004}}].

\bibitem{Bradlow:1996alg}
S.~{Bradlow} and O.~{Garcia-Prada}, \emph{{Non-abelian monopoles and
  vortices}}, {\emph{arXiv e-prints} (Feb., 1996) alg--geom/9602010},
  [\href{https://arxiv.org/abs/alg-geom/9602010}{{\tt alg-geom/9602010}}].

\bibitem{Teleman:1996}
A.~{Teleman}, \emph{{Non-abelian Seiberg-Witten theory and projectively stable
  pairs}}, {\emph{arXiv e-prints} (Sept., 1996) alg--geom/9609020},
  [\href{https://arxiv.org/abs/alg-geom/9609020}{{\tt alg-geom/9609020}}].

\bibitem{Okonek:1996}
C.~{Okonek} and A.~{Teleman}, \emph{{Recent Developments in Seiberg-Witten
  Theory and Complex Geometry}}, {\emph{arXiv e-prints} (Dec., 1996)
  alg--geom/9612015}, [\href{https://arxiv.org/abs/alg-geom/9612015}{{\tt
  alg-geom/9612015}}].

\bibitem{Feehan:1997gj}
P.~M. Feehan and T.~G. Leness, \emph{{$\rm PU(2)$ monopoles. I. Regularity,
  Uhlenbeck compactness, and transversality}}, {\emph{J. Diff. Geom.} {\bf 49}
  (1998) 265--410}, [\href{https://arxiv.org/abs/dg-ga/9710032}{{\tt
  dg-ga/9710032}}].

\bibitem{Gauduchon}
P.~Gauduchon, \emph{{Hermitian connections and Dirac operators}}, {\emph{Boll.
  Un. Mat. Ital. B} {\bf 7} (1997) 257--288}.

\bibitem{MaresThesis}
B.~A. Mares, \emph{Some Analytic Aspects of Vafa-Witten Twisted N = 4
  Supersymmetric Yang-Mills Theory}.
\newblock PhD thesis, Brown University, 2008.

\bibitem{Tanaka:2014}
Y.~{Tanaka}, \emph{{A perturbation and generic smoothness of the Vafa-Witten
  moduli spaces on closed symplectic four-manifolds}}, {\emph{arXiv e-prints}
  (Oct., 2014) arXiv:1410.1691}, [\href{https://arxiv.org/abs/1410.1691}{{\tt
  1410.1691}}].

\bibitem{GHOLAMPOUR2020107046}
A.~Gholampour, A.~Sheshmani and S.-T. Yau, \emph{Nested hilbert schemes on
  surfaces: Virtual fundamental class},
  \href{http://dx.doi.org/https://doi.org/10.1016/j.aim.2020.107046}{\emph{Advances
  in Mathematics} {\bf 365} (2020) 107046}.

\bibitem{Gholampour:2017dcp}
A.~Gholampour and R.~P. Thomas, \emph{{Degeneracy loci, virtual cycles and
  nested Hilbert schemes I}},  \href{https://arxiv.org/abs/1709.06105}{{\tt
  1709.06105}}.

\bibitem{Laarakker_2020}
T.~Laarakker, \emph{Monopole contributions to refined vafa–witten
  invariants}, \href{http://dx.doi.org/10.2140/gt.2020.24.2781}{\emph{Geometry
  \& Topology} {\bf 24} (Dec, 2020) 2781–2828}.

\bibitem{Gottsche:2018meg}
L.~Göttsche and M.~Kool, \emph{{Refined $\mathrm{SU}(3)$ Vafa-Witten
  invariants and modularity}}, {\emph{Pure Appl. Math. Quart.} {\bf 14} (2018)
  467--513}, [\href{https://arxiv.org/abs/1808.03245}{{\tt 1808.03245}}].

\bibitem{Atiyah:1978}
M.~F. Atiyah, N.~J. Hitchin and I.~M. Singer, \emph{Self-duality in
  four-dimensional riemannian geometry}, {\emph{Proceedings of the Royal
  Society of London. Series A, Mathematical and Physical Sciences} {\bf 362}
  (1978) 425--461}.

\bibitem{Labastida:1995zj}
J.~M.~F. Labastida and M.~Marino, \emph{{NonAbelian monopoles on four
  manifolds}},
  \href{http://dx.doi.org/10.1016/0550-3213(95)00300-H}{\emph{Nucl. Phys.} {\bf
  B448} (1995) 373--398}, [\href{https://arxiv.org/abs/hep-th/9504010}{{\tt
  hep-th/9504010}}].

\bibitem{Feehan:2001jc}
P.~M.~N. Feehan and T.~G. Leness, \emph{{SO(3) monopoles, level one
  Seiberg-Witten moduli spaces, and Witten's conjecture in low degrees}},
  \href{http://dx.doi.org/10.1016/S0166-8641(01)00233-4}{\emph{Topology Appl.}
  {\bf 124} (2002) 221--326}, [\href{https://arxiv.org/abs/math/0106238}{{\tt
  math/0106238}}].

\bibitem{Mathai:1986tc}
V.~Mathai and D.~G. Quillen, \emph{{Superconnections, Thom classes and
  equivariant differential forms}},
  \href{http://dx.doi.org/10.1016/0040-9383(86)90007-8}{\emph{Topology} {\bf
  25} (1986) 85--110}.

\bibitem{Deligne:1999}
P.~Deligne and J.~W. Morgan, \emph{Quantum Fields and Strings: A Course for
  Mathematicians}, vol.~1, ch.~Notes on Supersymmetry (following Joseph
  Bernstein), pp.~41--96.
\newblock AMS, 1999.

\bibitem{Leites_1980}
D.~A. Leites, \emph{{INTRODUCTION} {TO} {THE} {THEORY} {OF} {SUPERMANIFOLDS}},
  \href{http://dx.doi.org/10.1070/rm1980v035n01abeh001545}{\emph{Russian
  Mathematical Surveys} {\bf 35} (feb, 1980) 1--64}.

\bibitem{Witten:2012bg}
E.~Witten, \emph{{Notes On Supermanifolds and Integration}},
  \href{http://dx.doi.org/10.4310/PAMQ.2019.v15.n1.a1}{\emph{Pure Appl. Math.
  Quart.} {\bf 15} (2019) 3--56}, [\href{https://arxiv.org/abs/1209.2199}{{\tt
  1209.2199}}].

\bibitem{Marino:1998bm}
M.~Marino and G.~W. Moore, \emph{{The Donaldson-Witten function for gauge
  groups of rank larger than one}},
  \href{http://dx.doi.org/10.1007/s002200050494}{\emph{Commun. Math. Phys.}
  {\bf 199} (1998) 25--69}, [\href{https://arxiv.org/abs/hep-th/9802185}{{\tt
  hep-th/9802185}}].

\bibitem{Moore:2017cmm}
G.~W. Moore and I.~Nidaiev, \emph{{The Partition Function Of Argyres-Douglas
  Theory On A Four-Manifold}},  \href{https://arxiv.org/abs/1711.09257}{{\tt
  1711.09257}}.

\bibitem{Matone:1995rx}
M.~Matone, \emph{{Instantons and recursion relations in N=2 SUSY gauge
  theory}}, \href{http://dx.doi.org/10.1016/0370-2693(95)00920-G}{\emph{Phys.
  Lett.} {\bf B357} (1995) 342--348},
  [\href{https://arxiv.org/abs/hep-th/9506102}{{\tt hep-th/9506102}}].

\bibitem{Cordova:2018acb}
C.~C\'ordova and T.~T. Dumitrescu, \emph{{Candidate Phases for SU(2) Adjoint
  QCD$_4$ with Two Flavors from $\mathcal{N}=2$ Supersymmetric Yang-Mills
  Theory}},  \href{https://arxiv.org/abs/1806.09592}{{\tt 1806.09592}}.

\bibitem{Wang:2018qoy}
J.~Wang, X.-G. Wen and E.~Witten, \emph{{A New SU(2) Anomaly}},
  \href{http://dx.doi.org/10.1063/1.5082852}{\emph{J. Math. Phys.} {\bf 60}
  (2019) 052301}, [\href{https://arxiv.org/abs/1810.00844}{{\tt 1810.00844}}].

\bibitem{Marino:1997gj}
M.~Marino and G.~W. Moore, \emph{{Integrating over the Coulomb branch in N=2
  gauge theory}},
  \href{http://dx.doi.org/10.1016/S0920-5632(98)00168-6}{\emph{Nucl. Phys. B
  Proc. Suppl.} {\bf 68} (1998) 336--347},
  [\href{https://arxiv.org/abs/hep-th/9712062}{{\tt hep-th/9712062}}].

\bibitem{Dixon:1990pc}
L.~J. Dixon, V.~Kaplunovsky and J.~Louis, \emph{{Moduli dependence of string
  loop corrections to gauge coupling constants}},
  \href{http://dx.doi.org/10.1016/0550-3213(91)90490-O}{\emph{Nucl. Phys.} {\bf
  B355} (1991) 649--688}.

\bibitem{Harvey:1995fq}
J.~A. Harvey and G.~W. Moore, \emph{{Algebras, BPS states, and strings}},
  \href{http://dx.doi.org/10.1016/0550-3213(95)00605-2}{\emph{Nucl. Phys.} {\bf
  B463} (1996) 315--368}, [\href{https://arxiv.org/abs/hep-th/9510182}{{\tt
  hep-th/9510182}}].

\bibitem{Borcherds:1996uda}
R.~E. Borcherds, \emph{Automorphic forms with singularities on
  {G}rassmannians}, {\emph{Invent.\ Math.\ {\textbf {132}} (1998) 491
  doi:10.1007/s002220050232} (1998) },
  [\href{https://arxiv.org/abs/alg-geom/9609022}{{\tt alg-geom/9609022}}].

\bibitem{Korpas:2017qdo}
G.~Korpas and J.~Manschot, \emph{{Donaldson-Witten theory and indefinite theta
  functions}}, \href{http://dx.doi.org/10.1007/JHEP11(2017)083}{\emph{JHEP}
  {\bf 11} (2017) 083}, [\href{https://arxiv.org/abs/1707.06235}{{\tt
  1707.06235}}].

\bibitem{Korpas:2019ava}
G.~Korpas, J.~Manschot, G.~Moore and I.~Nidaiev, \emph{{Renormalization and
  BRST Symmetry in Donaldson--Witten Theory}},
  \href{http://dx.doi.org/10.1007/s00023-019-00835-x}{\emph{Annales Henri
  Poincare} {\bf 20} (2019) 3229--3264},
  [\href{https://arxiv.org/abs/1901.03540}{{\tt 1901.03540}}].

\bibitem{Gukov:2020btk}
S.~Gukov, P.-S. Hsin and D.~Pei, \emph{{Generalized Global Symmetries of $T[M]$
  Theories. I}},  \href{https://arxiv.org/abs/2010.15890}{{\tt 2010.15890}}.

\bibitem{atiyah_patodi_singer_1975}
M.~F. Atiyah, V.~K. Patodi and I.~M. Singer, \emph{Spectral asymmetry and
  riemannian geometry. i},
  \href{http://dx.doi.org/10.1017/S0305004100049410}{\emph{Mathematical
  Proceedings of the Cambridge Philosophical Society} {\bf 77} (1975) 43--69}.

\bibitem{Cecotti:1992qh}
S.~Cecotti, P.~Fendley, K.~A. Intriligator and C.~Vafa, \emph{{A New
  supersymmetric index}},
  \href{http://dx.doi.org/10.1016/0550-3213(92)90572-S}{\emph{Nucl. Phys. B}
  {\bf 386} (1992) 405--452}, [\href{https://arxiv.org/abs/hep-th/9204102}{{\tt
  hep-th/9204102}}].

\bibitem{Bershadsky:1993ta}
M.~Bershadsky, S.~Cecotti, H.~Ooguri and C.~Vafa, \emph{{Holomorphic anomalies
  in topological field theories}},
  \href{http://dx.doi.org/10.1016/0550-3213(93)90548-4}{\emph{AMS/IP Stud. Adv.
  Math.} {\bf 1} (1996) 655--682},
  [\href{https://arxiv.org/abs/hep-th/9302103}{{\tt hep-th/9302103}}].

\bibitem{Dabholkar:2019nnc}
A.~Dabholkar, D.~Jain and A.~Rudra, \emph{{APS $\eta$-invariant, path
  integrals, and mock modularity}},
  \href{http://dx.doi.org/10.1007/JHEP11(2019)080}{\emph{JHEP} {\bf 11} (2019)
  080}, [\href{https://arxiv.org/abs/1905.05207}{{\tt 1905.05207}}].

\bibitem{Alexandrov:2014wca}
S.~Alexandrov, G.~W. Moore, A.~Neitzke and B.~Pioline, \emph{{$\mathbb R^3$
  Index for Four-Dimensional $N=2$ Field Theories}},
  \href{http://dx.doi.org/10.1103/PhysRevLett.114.121601}{\emph{Phys. Rev.
  Lett.} {\bf 114} (2015) 121601}, [\href{https://arxiv.org/abs/1406.2360}{{\tt
  1406.2360}}].

\bibitem{Pioline:2015wza}
B.~Pioline, \emph{{Wall-crossing made smooth}},
  \href{http://dx.doi.org/10.1007/JHEP04(2015)092}{\emph{JHEP} {\bf 04} (2015)
  092}, [\href{https://arxiv.org/abs/1501.01643}{{\tt 1501.01643}}].

\bibitem{Yoshioka_1999}
K.~Yoshioka, \emph{Euler characteristics of su (2) instanton moduli spaces on
  rational elliptic surfaces},
  \href{http://dx.doi.org/10.1007/s002200050687}{\emph{Communications in
  Mathematical Physics} {\bf 205} (Sep, 1999) 501–517}.

\bibitem{Klemm:2012sx}
A.~Klemm, J.~Manschot and T.~Wotschke, \emph{{Quantum geometry of elliptic
  Calabi-Yau manifolds}}, {\emph{Comm. Number Theor. Phys.} {\bf 6} (2012)
  849--917}, [\href{https://arxiv.org/abs/1205.1795}{{\tt 1205.1795}}].

\bibitem{Haghighat:2008gw}
B.~Haghighat, A.~Klemm and M.~Rauch, \emph{{Integrability of the holomorphic
  anomaly equations}},
  \href{http://dx.doi.org/10.1088/1126-6708/2008/10/097}{\emph{JHEP} {\bf 10}
  (2008) 097}, [\href{https://arxiv.org/abs/0809.1674}{{\tt 0809.1674}}].

\bibitem{Alexandrov:2020bwg}
S.~Alexandrov, \emph{{Vafa-Witten invariants from modular anomaly}},
  \href{https://arxiv.org/abs/2005.03680}{{\tt 2005.03680}}.

\bibitem{Yoshioka:1996}
K.~Yoshioka, \emph{{The chamber structure of polarizations and the moduli of
  stable sheaves on a ruled surface}}, {\emph{Int. J. of Math.} {\bf 7} (1996)
  411--431}, [\href{https://arxiv.org/abs/9409008}{{\tt 9409008}}].

\bibitem{Li_1999}
W.-p. Li and Z.~Qin, \emph{On blowup formulae for the s -duality conjecture of
  vafa and witten},
  \href{http://dx.doi.org/10.1007/s002220050316}{\emph{Inventiones
  Mathematicae} {\bf 136} (Apr, 1999) 451--482}.

\bibitem{Gottsche:1999}
L.~Göttsche, \emph{{Theta Functions and Hodge Numbers of Moduli Spaces of
  Sheaves on Rational Surfaces}},
  \href{http://dx.doi.org/10.1007/s002200050699}{\emph{Communications in
  Mathematical Physics} {\bf 206} (Sep, 1999) 105–136}.

\bibitem{ZwegersThesis}
S.~P. Zwegers, \emph{Mock Theta Functions}.
\newblock PhD thesis, Utrecht University, 2008.

\bibitem{MR2605321}
D.~Zagier, \emph{Ramanujan's mock theta functions and their applications (after
  {Z}wegers and {O}no-{B}ringmann)}, {\emph{Ast\'erisque} (2009) Exp. No. 986,
  vii--viii, 143--164 (2010)}.

\bibitem{2012arXiv1207.5600B}
K.~{Bringmann}, M.~{Raum} and O.~{Richter}, \emph{{Harmonic Maass-Jacobi forms
  with singularities and a theta-like decomposition}}, {\emph{arXiv e-prints}
  (July, 2012) arXiv:1207.5600}, [\href{https://arxiv.org/abs/1207.5600}{{\tt
  1207.5600}}].

\bibitem{Dabholkar:2012nd}
A.~Dabholkar, S.~Murthy and D.~Zagier, \emph{{Quantum Black Holes, Wall
  Crossing, and Mock Modular Forms}},
  \href{https://arxiv.org/abs/1208.4074}{{\tt 1208.4074}}.

\bibitem{Bringmann:2010sd}
K.~Bringmann and J.~Manschot, \emph{{From sheaves on $P^2$ to a generalization
  of the Rademacher expansion}},
  \href{http://dx.doi.org/10.1353/ajm.2013.0031}{\emph{Am. J. Math.} {\bf 135}
  (2013) 1039--1065}, [\href{https://arxiv.org/abs/1006.0915}{{\tt
  1006.0915}}].

\bibitem{Ramanujan:1988}
S.~Ramanujan, \emph{The lost notebook and other unpublished papers}.
\newblock Narosa Pub. House New Delhi, 1988.

\bibitem{ellingsrud1995wall}
G.~Ellingsrud and L.~G{\"o}ttsche, \emph{Wall-crossing formulas, bott residue
  formula and the donaldson invariants of rational surfaces}, {\emph{Quart. J.
  Math. Oxford Ser.} {\bf 49} (1998) 307--329},
  [\href{https://arxiv.org/abs/alg-geom/9506019}{{\tt alg-geom/9506019}}].

\bibitem{Haghighat:2012bm}
B.~Haghighat, J.~Manschot and S.~Vandoren, \emph{{A 5d/2d/4d correspondence}},
  \href{http://dx.doi.org/10.1007/JHEP03(2013)157}{\emph{JHEP} {\bf 03} (2013)
  157}, [\href{https://arxiv.org/abs/1211.0513}{{\tt 1211.0513}}].

\bibitem{Yoshioka:1995}
K.~Yoshioka, \emph{{The Betti numbers of the moduli space of stable sheaves of
  rank 2 on a ruled surface}}, {\emph{Mathematische Annalen} {\bf 302} (1995)
  519--540}.

\bibitem{Manschot:2016gsx}
J.~Manschot and S.~Mozgovoy, \emph{Intersection cohomology of moduli spaces of
  sheaves on surfaces},
  \href{http://dx.doi.org/10.1007/s00029-018-0431-1}{\emph{Selecta Mathematica}
  {\bf 24} (2018) 3889--3926}.

\bibitem{Gottsche1990}
L.~G\"ottsche, \emph{{The Betti numbers of the Hilbert scheme of points on a
  smooth projective surface.}}, {\emph{Mathematische Annalen} {\bf 286} (1990)
  193--208}.

\bibitem{Manschot:2010nc}
J.~Manschot, \emph{{The Betti numbers of the moduli space of stable sheaves of
  rank 3 on $P^2$}},
  \href{http://dx.doi.org/10.1007/s11005-011-0490-0}{\emph{Lett. Math. Phys.}
  {\bf 98} (2011) 65--78}, [\href{https://arxiv.org/abs/1009.1775}{{\tt
  1009.1775}}].

\bibitem{Manschot:2014cca}
J.~Manschot, \emph{{Sheaves on $\mathbb{P}^2$ and generalized Appell
  functions}}, \href{http://dx.doi.org/10.4310/ATMP.2017.v21.n3.a3}{\emph{Adv.
  Theor. Math. Phys.} {\bf 21} (2017) 655--681},
  [\href{https://arxiv.org/abs/1407.7785}{{\tt 1407.7785}}].

\bibitem{Mozgovoy:2013zqx}
S.~Mozgovoy, \emph{{Invariants of moduli spaces of stable sheaves on ruled
  surfaces}},  \href{https://arxiv.org/abs/1302.4134}{{\tt 1302.4134}}.

\bibitem{Beaujard:2020sgs}
G.~Beaujard, J.~Manschot and B.~Pioline, \emph{{Vafa-Witten invariants from
  exceptional collections}},  \href{https://arxiv.org/abs/2004.14466}{{\tt
  2004.14466}}.

\bibitem{Manschot:2017xcr}
J.~Manschot, \emph{{Vafa\textendash{}Witten Theory and Iterated Integrals of
  Modular Forms}},
  \href{http://dx.doi.org/10.1007/s00220-019-03389-5}{\emph{Commun. Math.
  Phys.} {\bf 371} (2019) 787--831},
  [\href{https://arxiv.org/abs/1709.10098}{{\tt 1709.10098}}].

\bibitem{Li:1995}
T.~J. Li and A.~Liu, \emph{{General wall-crossing formula}}, {\emph{Math. Res.
  Lett.} {\bf 2} (1995) 797--810}.

\bibitem{park_2004}
J.~Park, \emph{Non-complex symplectic 4-manifolds with
  $\lowercase{b}_{2}^+=1$},
  \href{http://dx.doi.org/10.1112/S0024609303002893}{\emph{Bulletin of the
  London Mathematical Society} {\bf 36} (2004) 231--240}.

\bibitem{Gottsche:2010ig}
L.~Gottsche, H.~Nakajima and K.~Yoshioka, \emph{{Donaldson = Seiberg-Witten
  from Mochizuki's formula and instanton counting}}, {\emph{Publ. Res. Inst.
  Math. Sci. Kyoto} {\bf 47} (2011) 307--359},
  [\href{https://arxiv.org/abs/1001.5024}{{\tt 1001.5024}}].

\bibitem{Mochizuki}
T.~Mochizuki, \emph{Donaldson Type Invariants for Algebraic Surfaces}.
\newblock Lecture Notes in Math. 1972. Springer-Verlag Berlin Heidelberg, 2009.

\bibitem{Milnor}
J.~W. Milnor, \emph{On simply-connected four-manifolds},  Symposium
  Internacional de Topologia Alg., (Mexico), pp.~122--128, 1958.

\bibitem{Marino:1998uy}
M.~Marino, G.~W. Moore and G.~Peradze, \emph{{Four manifold geography and
  superconformal symmetry}},  \href{https://arxiv.org/abs/math/9812042}{{\tt
  math/9812042}}.

\bibitem{Marino:1998tb}
M.~Marino, G.~W. Moore and G.~Peradze, \emph{{Superconformal invariance and the
  geography of four manifolds}},
  \href{http://dx.doi.org/10.1007/s002200050694}{\emph{Commun. Math. Phys.}
  {\bf 205} (1999) 691--735}, [\href{https://arxiv.org/abs/hep-th/9812055}{{\tt
  hep-th/9812055}}].

\bibitem{Feehan:1997rt}
P.~M. Feehan and T.~G. Leness, \emph{{PU(2) monopoles and relations between
  four manifold invariants}},
  \href{http://dx.doi.org/10.1016/S0166-8641(97)00201-0}{\emph{Topology and its
  Applications} {\bf 88} (1998) 111--145},
  [\href{https://arxiv.org/abs/dg-ga/9709022}{{\tt dg-ga/9709022}}].

\bibitem{Dedushenko:2017tdw}
M.~Dedushenko, S.~Gukov and P.~Putrov, \emph{{Vertex algebras and 4-manifold
  invariants}},  in \emph{{Nigel Hitchin's 70th Birthday Conference}}, vol.~1,
  pp.~249--318, 5, 2017.
\newblock \href{https://arxiv.org/abs/1705.01645}{{\tt 1705.01645}}.
\newblock \href{http://dx.doi.org/10.1093/oso/9780198802013.003.0011}{DOI}.

\bibitem{Gadde:2013sca}
A.~Gadde, S.~Gukov and P.~Putrov, \emph{{Fivebranes and 4-manifolds}},
  \href{http://dx.doi.org/10.1007/978-3-319-43648-7_7}{\emph{Prog. Math.} {\bf
  319} (2016) 155--245}, [\href{https://arxiv.org/abs/1306.4320}{{\tt
  1306.4320}}].

\bibitem{Gukov:2017zao}
S.~Gukov, \emph{{Trisecting non-Lagrangian theories}},
  \href{http://dx.doi.org/10.1007/JHEP11(2017)178}{\emph{JHEP} {\bf 11} (2017)
  178}, [\href{https://arxiv.org/abs/1707.01515}{{\tt 1707.01515}}].

\bibitem{Gukov:2018iiq}
S.~Gukov, D.~Pei, P.~Putrov and C.~Vafa, \emph{{4-manifolds and topological
  modular forms}},  \href{https://arxiv.org/abs/1811.07884}{{\tt 1811.07884}}.

\bibitem{kronheimer2005}
P.~Kronheimer, \emph{Four-manifold invariants from higher-rank bundles},
  \href{http://dx.doi.org/10.4310/jdg/1143572014}{\emph{J. Differential Geom.}
  {\bf 70} (05, 2005) 59--112}.

\bibitem{Gaiotto:2009we}
D.~Gaiotto, \emph{{N=2 dualities}},
  \href{http://dx.doi.org/10.1007/JHEP08(2012)034}{\emph{JHEP} {\bf 08} (2012)
  034}, [\href{https://arxiv.org/abs/0904.2715}{{\tt 0904.2715}}].

\bibitem{Gaiotto:2009hg}
D.~Gaiotto, G.~W. Moore and A.~Neitzke, \emph{{Wall-crossing, Hitchin Systems,
  and the WKB Approximation}},  \href{https://arxiv.org/abs/0907.3987}{{\tt
  0907.3987}}.

\bibitem{Klemm:1996bj}
A.~Klemm, W.~Lerche, P.~Mayr, C.~Vafa and N.~P. Warner, \emph{{Selfdual strings
  and N=2 supersymmetric field theory}},
  \href{http://dx.doi.org/10.1016/0550-3213(96)00353-7}{\emph{Nucl. Phys. B}
  {\bf 477} (1996) 746--766}, [\href{https://arxiv.org/abs/hep-th/9604034}{{\tt
  hep-th/9604034}}].

\bibitem{Witten:1997sc}
E.~Witten, \emph{{Solutions of four-dimensional field theories via M theory}},
  \href{http://dx.doi.org/10.1016/S0550-3213(97)00416-1}{\emph{Nucl. Phys. B}
  {\bf 500} (1997) 3--42}, [\href{https://arxiv.org/abs/hep-th/9703166}{{\tt
  hep-th/9703166}}].

\bibitem{Moore-FelixKlein}
G.~W. Moore, ``Lecture notes for felix klein lectures.''
  http://www.physics.rutgers.edu/~gmoore/FelixKleinLectureNotes.pdf, 2012.

\bibitem{Tachikawa:2013kta}
Y.~Tachikawa, \emph{{N=2 supersymmetric dynamics for pedestrians}}, vol.~890 of
  \emph{Lect. Notes Phys.}
\newblock Springer International Publishing, 2014,
  \href{http://dx.doi.org/10.1007/978-3-319-08822-8}{10.1007/978-3-319-08822-8}.

\bibitem{Seiberg:1996vs}
N.~Seiberg and E.~Witten, \emph{{Comments on string dynamics in
  six-dimensions}},
  \href{http://dx.doi.org/10.1016/0550-3213(96)00189-7}{\emph{Nucl. Phys. B}
  {\bf 471} (1996) 121--134}, [\href{https://arxiv.org/abs/hep-th/9603003}{{\tt
  hep-th/9603003}}].

\bibitem{Seiberg:1996qx}
N.~Seiberg, \emph{{Nontrivial fixed points of the renormalization group in
  six-dimensions}},
  \href{http://dx.doi.org/10.1016/S0370-2693(96)01424-4}{\emph{Phys. Lett. B}
  {\bf 390} (1997) 169--171}, [\href{https://arxiv.org/abs/hep-th/9609161}{{\tt
  hep-th/9609161}}].

\bibitem{Witten:1995zh}
E.~Witten, \emph{{Some comments on string dynamics}},  in \emph{{STRINGS 95:
  Future Perspectives in String Theory}}, pp.~501--523, 7, 1995.
\newblock \href{https://arxiv.org/abs/hep-th/9507121}{{\tt hep-th/9507121}}.

\bibitem{Balasubramanian:2018pbp}
A.~Balasubramanian and J.~Distler, \emph{{Masses, Sheets and Rigid SCFTs}},
  \href{https://arxiv.org/abs/1810.10652}{{\tt 1810.10652}}.

\bibitem{Chacaltana:2012zy}
O.~Chacaltana, J.~Distler and Y.~Tachikawa, \emph{{Nilpotent orbits and
  codimension-two defects of 6d N=(2,0) theories}},
  \href{http://dx.doi.org/10.1142/S0217751X1340006X}{\emph{Int. J. Mod. Phys.
  A} {\bf 28} (2013) 1340006}, [\href{https://arxiv.org/abs/1203.2930}{{\tt
  1203.2930}}].

\bibitem{Gaiotto:2011xs}
D.~Gaiotto, G.~W. Moore and Y.~Tachikawa, \emph{{On 6d $\mathcal N=$(2,0)
  theory compactified on a Riemann surface with finite area}},
  \href{http://dx.doi.org/10.1093/ptep/pts047}{\emph{PTEP} {\bf 2013} (2013)
  013B03}, [\href{https://arxiv.org/abs/1110.2657}{{\tt 1110.2657}}].

\bibitem{Argyres:2015ffa}
P.~Argyres, M.~Lotito, Y.~L\"u and M.~Martone, \emph{{Geometric constraints on
  the space of $ \mathcal{N} $ = 2 SCFTs. Part I: physical constraints on
  relevant deformations}},
  \href{http://dx.doi.org/10.1007/JHEP02(2018)001}{\emph{JHEP} {\bf 02} (2018)
  001}, [\href{https://arxiv.org/abs/1505.04814}{{\tt 1505.04814}}].

\bibitem{Argyres:2020nrr}
P.~Argyres and M.~Martone, \emph{{Construction and classification of Coulomb
  branch geometries}},  \href{https://arxiv.org/abs/2003.04954}{{\tt
  2003.04954}}.

\bibitem{Martone:2020nsy}
M.~Martone, \emph{{Towards the classification of rank-r$ \mathcal{N} $ = 2
  SCFTs. Part I. Twisted partition function and central charge formulae}},
  \href{http://dx.doi.org/10.1007/JHEP12(2020)021}{\emph{JHEP} {\bf 12} (2020)
  021}, [\href{https://arxiv.org/abs/2006.16255}{{\tt 2006.16255}}].

\bibitem{Belov:2006jd}
D.~Belov and G.~W. Moore, \emph{{Holographic Action for the Self-Dual Field}},
  \href{https://arxiv.org/abs/hep-th/0605038}{{\tt hep-th/0605038}}.

\bibitem{Henningson:2012xk}
M.~Henningson, \emph{{What is the partition bundle?}},
  \href{http://dx.doi.org/10.1134/S1063779612050164}{\emph{Phys. Part. Nucl.}
  {\bf 43} (2012) 621--624}.

\bibitem{Hopkins:2002rd}
M.~Hopkins and I.~Singer, \emph{{Quadratic functions in geometry, topology, and
  M theory}}, {\emph{J. Diff. Geom.} {\bf 70} (2005) 329--452},
  [\href{https://arxiv.org/abs/math/0211216}{{\tt math/0211216}}].

\bibitem{Monnier:2010ww}
S.~Monnier, \emph{{Geometric quantization and the metric dependence of the
  self-dual field theory}},
  \href{http://dx.doi.org/10.1007/s00220-012-1525-9}{\emph{Commun. Math. Phys.}
  {\bf 314} (2012) 305--328}, [\href{https://arxiv.org/abs/1011.5890}{{\tt
  1011.5890}}].

\bibitem{Monnier:2011mv}
S.~Monnier, \emph{{The Anomaly line bundle of the self-dual field theory}},
  \href{http://dx.doi.org/10.1007/s00220-013-1844-5}{\emph{Commun. Math. Phys.}
  {\bf 325} (2014) 41--72}, [\href{https://arxiv.org/abs/1109.2904}{{\tt
  1109.2904}}].

\bibitem{Monnier:2011rk}
S.~Monnier, \emph{{The global gravitational anomaly of the self-dual field
  theory}}, \href{http://dx.doi.org/10.1007/s00220-013-1845-4}{\emph{Commun.
  Math. Phys.} {\bf 325} (2014) 73--104},
  [\href{https://arxiv.org/abs/1110.4639}{{\tt 1110.4639}}].

\bibitem{Moore:2004jv}
G.~W. Moore, \emph{{Anomalies, Gauss laws, and Page charges in M-theory}},
  \href{http://dx.doi.org/10.1016/j.crhy.2004.12.005}{\emph{Comptes Rendus
  Physique} {\bf 6} (2005) 251--259},
  [\href{https://arxiv.org/abs/hep-th/0409158}{{\tt hep-th/0409158}}].

\bibitem{Witten:1996hc}
E.~Witten, \emph{{Five-brane effective action in M theory}},
  \href{http://dx.doi.org/10.1016/S0393-0440(97)80160-X}{\emph{J. Geom. Phys.}
  {\bf 22} (1997) 103--133}, [\href{https://arxiv.org/abs/hep-th/9610234}{{\tt
  hep-th/9610234}}].

\bibitem{Witten:1999vg}
E.~Witten, \emph{{Duality relations among topological effects in string
  theory}}, \href{http://dx.doi.org/10.1088/1126-6708/2000/05/031}{\emph{JHEP}
  {\bf 05} (2000) 031}, [\href{https://arxiv.org/abs/hep-th/9912086}{{\tt
  hep-th/9912086}}].

\bibitem{Tachikawa:2013hya}
Y.~Tachikawa, \emph{{On the 6d origin of discrete additional data of 4d gauge
  theories}}, \href{http://dx.doi.org/10.1007/JHEP05(2014)020}{\emph{JHEP} {\bf
  05} (2014) 020}, [\href{https://arxiv.org/abs/1309.0697}{{\tt 1309.0697}}].

\bibitem{Monnier:2018cfa}
S.~Monnier and G.~W. Moore, \emph{{A Brief Summary Of Global Anomaly
  Cancellation In Six-Dimensional Supergravity}},
  \href{https://arxiv.org/abs/1808.01335}{{\tt 1808.01335}}.

\bibitem{Monnier:2018nfs}
S.~Monnier and G.~W. Moore, \emph{{Remarks on the Green\textendash{}Schwarz
  Terms of Six-Dimensional Supergravity Theories}},
  \href{http://dx.doi.org/10.1007/s00220-019-03341-7}{\emph{Commun. Math.
  Phys.} {\bf 372} (2019) 963--1025},
  [\href{https://arxiv.org/abs/1808.01334}{{\tt 1808.01334}}].

\bibitem{Bredon}
G.~E. Bredon, \emph{Topology and geometry}.
\newblock Graduate texts in mathematics; 139. Springer-Verlag, New York, 1993.

\bibitem{Massey}
W.~S. Massey, \emph{Singular homology theory}.
\newblock Springer-Verlag New York, 1980.

\bibitem{Vick}
J.~W. Vick, \emph{Homology theory; an introduction to algebraic topology}.
\newblock Academic Press New York, 1973.

\bibitem{Longhi:2016rjt}
P.~Longhi and C.~Y. Park, \emph{{ADE Spectral Networks}},
  \href{http://dx.doi.org/10.1007/JHEP08(2016)087}{\emph{JHEP} {\bf 08} (2016)
  087}, [\href{https://arxiv.org/abs/1601.02633}{{\tt 1601.02633}}].

\bibitem{Maldacena:1997de}
J.~M. Maldacena, A.~Strominger and E.~Witten, \emph{{Black hole entropy in M
  theory}}, \href{http://dx.doi.org/10.1088/1126-6708/1997/12/002}{\emph{JHEP}
  {\bf 12} (1997) 002}, [\href{https://arxiv.org/abs/hep-th/9711053}{{\tt
  hep-th/9711053}}].

\bibitem{Minasian:1999qn}
R.~Minasian, G.~W. Moore and D.~Tsimpis, \emph{{Calabi-Yau black holes and
  (0,4) sigma models}}, {\emph{Commun. Math. Phys.} {\bf 209} (2000) 325--352},
  [\href{https://arxiv.org/abs/hep-th/9904217}{{\tt hep-th/9904217}}].

\bibitem{NidaievPhD}
I.~Nidaiev, \emph{Cohomological Field Theories and Four-Manifold Invariants}.
\newblock PhD thesis, Rutgers University, 2019.

\bibitem{Marino:1998rg}
M.~Marino and G.~W. Moore, \emph{{Donaldson invariants for nonsimply connected
  manifolds}}, \href{http://dx.doi.org/10.1007/s002200050611}{\emph{Commun.
  Math. Phys.} {\bf 203} (1999) 249},
  [\href{https://arxiv.org/abs/hep-th/9804104}{{\tt hep-th/9804104}}].

\bibitem{Marino:1998eg}
M.~Marino and G.~W. Moore, \emph{{Three manifold topology and the
  Donaldson-Witten partition function}},
  \href{http://dx.doi.org/10.1016/S0550-3213(99)00105-4}{\emph{Nucl. Phys. B}
  {\bf 547} (1999) 569--598}, [\href{https://arxiv.org/abs/hep-th/9811214}{{\tt
  hep-th/9811214}}].

\bibitem{Dimofte:2011ju}
T.~Dimofte, D.~Gaiotto and S.~Gukov, \emph{{Gauge Theories Labelled by
  Three-Manifolds}},
  \href{http://dx.doi.org/10.1007/s00220-013-1863-2}{\emph{Commun. Math. Phys.}
  {\bf 325} (2014) 367--419}, [\href{https://arxiv.org/abs/1108.4389}{{\tt
  1108.4389}}].

\bibitem{Eckhard:2018raj}
J.~Eckhard, S.~Sch\"afer-Nameki and J.-M. Wong, \emph{{An $\mathcal{N}=1$ 3d-3d
  Correspondence}},
  \href{http://dx.doi.org/10.1007/JHEP07(2018)052}{\emph{JHEP} {\bf 07} (2018)
  052}, [\href{https://arxiv.org/abs/1804.02368}{{\tt 1804.02368}}].

\bibitem{Thomas:2018lvm}
R.~P. Thomas, \emph{{Equivariant $K$-Theory and Refined Vafa\textendash{}Witten
  Invariants}},
  \href{http://dx.doi.org/10.1007/s00220-020-03821-1}{\emph{Commun. Math.
  Phys.} {\bf 378} (2020) 1451--1500},
  [\href{https://arxiv.org/abs/1810.00078}{{\tt 1810.00078}}].

\bibitem{Dabholkar:2005dt}
A.~Dabholkar, F.~Denef, G.~W. Moore and B.~Pioline, \emph{{Precision counting
  of small black holes}},
  \href{http://dx.doi.org/10.1088/1126-6708/2005/10/096}{\emph{JHEP} {\bf 10}
  (2005) 096}, [\href{https://arxiv.org/abs/hep-th/0507014}{{\tt
  hep-th/0507014}}].

\bibitem{Zagier2010}
D.~Zagier, \emph{Quantum modular forms},  in \emph{Quanta of maths}, vol.~11 of
  \emph{Clay Math. Proc.}, pp.~659 -- 675.
\newblock Amer. Math. Soc., 2010.

\bibitem{Donagi:1995cf}
R.~Donagi and E.~Witten, \emph{{Supersymmetric Yang-Mills theory and integrable
  systems}}, \href{http://dx.doi.org/10.1016/0550-3213(95)00609-5}{\emph{Nucl.
  Phys. B} {\bf 460} (1996) 299--334},
  [\href{https://arxiv.org/abs/hep-th/9510101}{{\tt hep-th/9510101}}].

\bibitem{Diamond}
F.~Diamond and J.~Shurman, \emph{A First Course in Modular Forms}, vol.~228 of
  \emph{Graduate Texts in Mathematics}.
\newblock Springer-Verlag New York, 1~ed., 2005.

\bibitem{Bruinier08}
G.~H. J.H.~Bruinier, G. van der~Geer and D.~Zagier, \emph{The 1-2-3 of Modular
  Forms}.
\newblock Springer-Verlag Berlin Heidelberg, 2008,
  \href{http://dx.doi.org/10.1007/978-3-540-74119-0}{10.1007/978-3-540-74119-0}.

\bibitem{Bringmann:2309148}
K.~Bringmann, A.~Folsom, K.~Ono and L.~Rolen, \emph{{Harmonic Maass forms and
  mock modular forms}}.
\newblock Colloquium publications. American Mathematical Society, Providence,
  RI, 2017.

\bibitem{Apostol}
T.~M. Apostol, \emph{Modular functions and Dirichlet series in number theory /
  Tom M. Apostol.}
\newblock Graduate texts in mathematics ; 41. Springer-Verlag, New York, 1976.

\end{thebibliography}
%\bibliographystyle{JHEP}  
%\bibliographystyle{unsrt}
 
\providecommand{\href}[2]{#2}\begingroup\raggedright\endgroup

\end{document}